\documentclass[openany,12pt,a4paper]{book}
\usepackage{graphicx} 

\usepackage[normalem]{ulem}
\usepackage{cancel}

\makeatletter
\newcommand{\enableopenany}{%
  \@openrightfalse%
}
\newcommand{\disableopenany}{%
  \@openrighttrue%
}
\makeatother

\newcommand\summaryname{\normalsize Abstract}
\newenvironment{Abstract}%
    {\normalsize\begin{center}%
    \bfseries{\summaryname} \end{center}}

\newenvironment{changemargin}[2]{%
\begin{list}{}{%
\setlength{\topsep}{0pt}%
\setlength{\leftmargin}{#1}%
\setlength{\rightmargin}{#2}%
\setlength{\listparindent}{\parindent}%
\setlength{\itemindent}{\parindent}%
\setlength{\parsep}{\parskip}%
}%
\item[]}{\end{list}}

\usepackage{color}
\usepackage{xcolor}
\usepackage{setspace}
\usepackage[utf8]{inputenc}
\usepackage[top=2.5cm, bottom=2.5cm, inner=3.cm, outer=2.cm]{geometry}
\usepackage[colorlinks=true]{hyperref}
\definecolor{dark-gray}{gray}{0.13}
\hypersetup{urlcolor=dark-gray,linkcolor=dark-gray,citecolor=dark-gray}
\usepackage[numbers, square, comma, sort&compress]{natbib}
\usepackage{fancyhdr}
\usepackage[Rejne]{fncychap}
\usepackage{chemformula}
\usepackage{enumerate}
\usepackage{comment}
\usepackage{multirow}

\usepackage{mathtools}
\usepackage{mathrsfs} 
\usepackage{amsfonts}
\usepackage{amssymb}
\usepackage{amsmath}
\usepackage{amsthm}
\usepackage{relsize} 
\usepackage{bbm}
\usepackage{nccmath}
\usepackage{tikz}
\usetikzlibrary{arrows.meta, positioning}

\usepackage[framemethod=TikZ]{mdframed}
\mdfsetup{skipabove=15pt, innertopmargin=10pt, skipbelow=5pt}

\usepackage{lipsum} 

\newcounter{equ}[section]
\newenvironment{equ}[1][]{%
\stepcounter{equ}%
\ifstrempty{#1}%
{\mdfsetup{%
frametitle={%
\tikz[baseline=(current bounding box.east),outer sep=0pt]
\node[anchor=east,rectangle,fill=blue!20]
{\strut };}}
}%
{\mdfsetup{%
frametitle={%
\tikz[baseline=(current bounding box.east),outer sep=0pt]
\node[anchor=east,rectangle,fill=blue!20]
{\strut ~#1};}}%
}%
\mdfsetup{innertopmargin=10pt,linecolor=blue!20,%
middlelinewidth=2pt,topline=true,
frametitleaboveskip=\dimexpr-\ht\strutbox\relax,}
\begin{mdframed}[]\relax%
}{\end{mdframed}}

\usepackage{makeidx}
\makeindex

\newcommand{\bZ}{{\mathbb Z}}
\newcommand{\bR}{{\mathbb R}}
\newcommand{\bC}{{\mathbb C}}

\newcommand{\cM}{{\mathcal M}}
\newcommand{\cN}{{\mathcal N}}

\newcommand{\ov}{\overline}
\newcommand{\I}{\mathrm{i}}
\newcommand{\p}{\partial}
\newcommand{\dif}{\,\mathrm{d}}
\newcommand{\kom}{\, ,\;} 
\def\vol{\operatorname{Vol}}

\usepackage{orcidlink}

\usepackage{fontawesome5}
\usepackage{hyperref}

\makeatletter
\newcommand{\github}[1]{%
   \href{#1}{{\normalsize \color{black}\faGithub}}%
}
\makeatother

\counterwithout{footnote}{chapter}

\usepackage{diagbox}

\title{\textbf{TASI Lectures on de Sitter Vacua}}
\author{Liam McAllister and Andreas Schachner}
\date{\today} 

\begin{document}

\pagenumbering{roman}

\newgeometry{top=2.5cm, bottom=2.5cm, inner=2.5cm, outer=2.5cm}

\begin{titlepage}

\thispagestyle{empty} 

\begin{center}

\vspace*{4cm}

{\LARGE \textbf{TASI Lectures on de Sitter Vacua}}

\vspace*{2cm}

{\large {Liam McAllister \orcidlink{0000-0001-5830-5577} and Andreas Schachner \orcidlink{0000-0002-7287-1476}}}

\vspace{0.5 cm}
{
\textsl{Department of Physics, Cornell University, Ithaca, NY 14853, USA}
}

\vspace*{3.cm}

\end{center}


\begin{Abstract}
\begin{changemargin}{0.75cm}{0.75cm}

These lectures provide a self-contained introduction to flux compactifications of type IIB string theory on Calabi-Yau orientifolds. The first lecture begins with geometric foundations, then presents vacuum solutions in Calabi-Yau compactifications, as well as the geometry and physics of the moduli problem. The second lecture develops the classical theory of type IIB flux compactifications, both in ten dimensions and in the four-dimensional effective theory. The third lecture turns to the quantum theory of flux compactifications, including  perturbative and non-perturbative corrections.  With this foundation, in the fourth lecture we give a detailed treatment of the candidate de Sitter vacua recently constructed in \cite{McAllister:2024lnt}. These notes are intended to be accessible to graduate students working in adjacent fields, and so extensive background material is included throughout.

\vfill
\today
\end{changemargin}
\end{Abstract}

\end{titlepage}

\newpage

\thispagestyle{empty} 

\vphantom{Tasi}

\newpage

\restoregeometry

{
\hypertarget{toc}{}
\onehalfspacing
\setcounter{tocdepth}{2}
\tableofcontents
}

\pagenumbering{arabic}

\chapter{Introduction}

The goal of these lectures is to present and explain a class of compactifications of string theory that yield four-dimensional cosmologies.

The cosmologies arise as solutions to the equations of motion of certain four-dimensional supergravity theories: specifically, of the low-energy effective theories obtained from the long-wavelength limit of flux compactifications of type IIB string theory on orientifolds of Calabi-Yau threefold hypersurfaces in toric varieties.  At the classical level, these configurations have massless moduli, but the inclusion of quantum effects lifts the moduli space, making it possible in principle to find isolated solutions, or \emph{vacua}. However, establishing the existence of a vacuum in any particular compactification requires detailed and explicit knowledge of the effective theory.

A series of recent works \cite{Demirtas:2019sip,Demirtas:2020ffz,Demirtas:2021nlu,Demirtas:2021ote} constructed a special class of compactifications and computed the effective theory therein with enough accuracy to give strong evidence for the existence of $\mathcal{N}=1$ supersymmetric AdS$_4$ vacua \cite{Demirtas:2021nlu}. Building on this starting point, in collaboration with J.~Moritz and R.~Nally, we have found compactifications for which the \emph{leading-order} effective theory admits de Sitter vacua \cite{McAllister:2024lnt}.  

In the long run, to better understand cosmology in quantum gravity, one should work to obtain cosmological solutions of string theory beyond leading order.  This task --- whether in type IIB orientifolds, or in another corner of the duality web --- will require fundamental advances in our ability to compute in string theory.

In these lectures, we will begin by explaining the state of string compactifications, highlighting the capabilities in hand in 2025, as well as the most important limitations.  We will then focus on the case of type IIB flux compactifications on Calabi-Yau orientifolds, and lay the groundwork for computing the effective theory in this setting. Next we will describe the KKLT mechanism \cite{Kachru:2003aw} for moduli stabilization in AdS$_4$ and dS$_4$, and the compactification technology that has made it possible to construct vacua along these lines. Finally, we will survey a host of open problems in this area, and strategies for approaching them.

\vfill

\newpage

\section{Cosmology and string theory}

\begin{figure}[t!]
    \centering
    \includegraphics[width=0.7\linewidth]{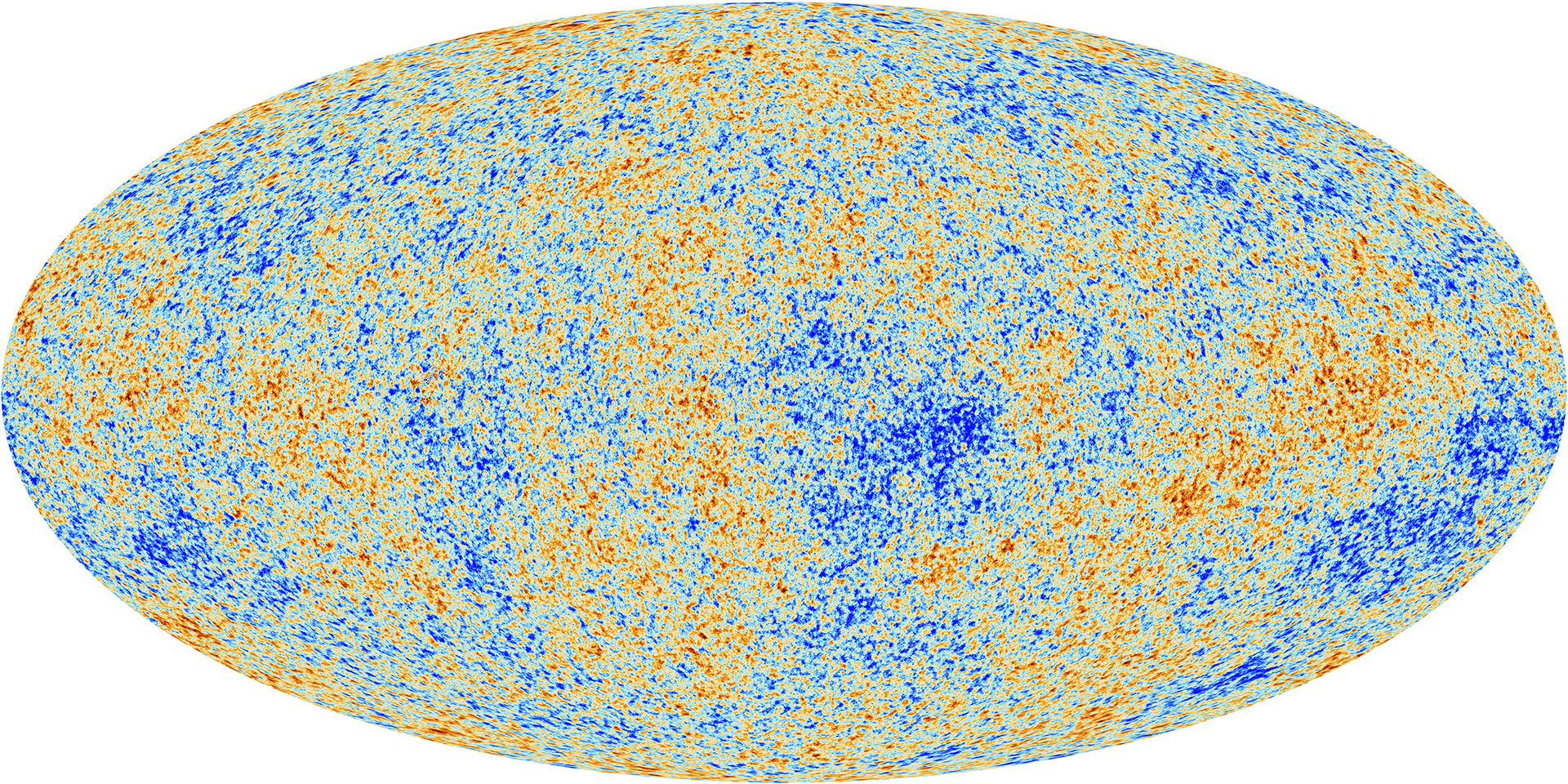}
    \caption{The CMB temperature map measured by Planck \cite{Planck:2013pxb}.
     Figure courtesy of ESA and the Planck Collaboration.}
    \label{fig:CMB}
\end{figure}

Sixty years after the discovery of the cosmic microwave background, cosmology provides a spectacularly accurate  description of the observable universe: the $\Lambda$CDM model, with initial conditions provided by quantum fluctuations during inflation.  But progress in measurement and modeling have not 
been paired with corresponding advances in answering fundamental questions about the nature of inflation, dark energy, and dark matter.  

The puzzle of inflation in particular is entangled with that of understanding quantum-gravitational phenomena.\footnote{For a comprehensive discussion of this issue, see \cite{Baumann:2014nda}.} The temperature anisotropies in the CMB, which were predicted by inflationary theory more than a decade before they were measured, result from quantum fluctuations of the scalar curvature mode of the metric during inflation.  A suitable effective theory of a quantum scalar field coupled to general relativity suffices to model this phenomenon, but the structure of such an effective theory depends on the nature of quantum gravity: operators suppressed by the Planck scale can change whether inflation occurs, and what it imprints in the CMB. This issue, known as the eta problem, can be viewed as a hierarchy problem: why is the inflaton mass small enough for prolonged inflation, despite contributions from unknown Planck-scale physics?

A far deeper question, inextricable from quantum gravity, concerns yet-unmeasured tensor perturbations in the CMB, which are the imprint of quantum gravity waves.  An inflaton field displacement large enough to leave a detectable tensor signal requires a symmetry of the Planck-scale theory.  Mechanisms for such symmetries in string theory have been proposed, and criticized: for displacements close to the Planck scale, and energy densities not far below, it is extremely challenging to carry out a controlled computation using early 21st-century theory. Exhibiting a inflationary solution of string theory that imprints a bright tensor signal would require an unprecedented tour de force of computational string theory, at the Planck scale, and might be impossible without fundamental advances in our understanding of quantum gravity. Nevertheless, this connection is a remarkable opportunity to use observable phenomena to probe the physics of the Planck scale, and even if it eludes our grasp now, it may be taken up by a future generation.

Understanding the impact of Planck-scale physics on inflation is by no means the only motivation for studying cosmology in string theory. One can hope that the resolution of the cosmological constant problem can be reached through string theory.  To date the most significant step in this direction is the proposal of anthropic selection in a landscape of flux vacua \cite{Bousso:2000xa}. To assess this proposal, and to understand if dynamical selection in this setting could provide an alternative to anthropic reasoning, one likely needs to first construct parts of the landscape in full detail, a task that has just begun.  In these lectures we will encounter a new mechanism for generating exponentially small (and negative) vacuum energy in supersymmetric solutions, but at present this mechanism is of no help in solving the actual cosmological constant problem in our non-supersymmetric universe.

String theory also serves as a source of new mechanisms, models, and observational signatures in cosmology. Ideas that were inspired by string theory, or have found their sharpest realization therein, include extra dimensions, moduli, cosmic superstrings, brane worlds, and the axiverse.  Inflationary mechanisms such as Dirac-Born-Infeld inflation, D-brane inflation, and axion monodromy inflation emerged from string theory, and furnished novel signatures.

Another use of string theory in cosmology is as a filter for potential inconsistencies: if a class of effective field theories has not been found in string theory, there is very often a reason.  The reason can sometimes be a fundamental incompatibility with quantum gravity, but it can also reflect a failure by string theorists to look in the necessary regime, or a previously-unexamined limitation of the effective theory that can nonetheless be understood in purely low-energy terms.  In each of these cases, something is learned by applying string theory as a test. An aim of the Swampland program is to demarcate the space of consistent theories, which if achieved will provide important lessons for cosmology.  However, we caution that string theory is a work in progress: we lack a fundamental, non-perturbative formulation of the theory that is valid off-shell and in arbitrary spacetimes.  Correspondingly, much of what we can say about the space of solutions of string theory is provisional.  For this reason, most conclusions about structures that ``cannot'' be found in string theory should be read as statements about what has not been found \emph{to date} in string theory.

Perhaps the ultimate application of string theory to cosmology would be the resolution of fundamental questions in quantum cosmology: for example, computing the wavefunction of the Universe, defining measures in eternal inflation, explaining the microscopic origin of the de Sitter entropy, and understanding cosmological singularities. Indeed, the principal motivation for constructing cosmological solutions of string theory is to lay the foundations for addressing such questions: one constructs vacua not in order to tack them to a wall, but to use them as stepping stones to understand quantum-gravitational effects in our universe. Although the compactifications described in these lectures have not yet supported progress on any of the grand questions of cosmology, they do provide a mathematically sound framework for building cosmological solutions of string theory.  In time, this setting may lead to illumination.

\section{Hierarchies}\label{sec:hier}

These lectures, and the associated research works, revolve around achieving controlled computations of the vacuum structure in four-dimensional effective quantum gravity theories obtained from string compactifications.  A central theme is the use of mechanisms to generate hierarchies of scales, or equivalently to generate small dimensionless numbers, starting from integer inputs. Many of the structures that arise in the study of compactifications, and the choices made by workers in the field, can be understood as driven by the (non)existence of hierarchy mechanisms.  Let us explain this issue now in very broad terms, and delve into the specific mechanisms in the later lectures.

Human understanding of the natural world is possible only through approximations, but in a finite landscape the small parameters that measure the corrections in approximation schemes are necessarily finite.  We should therefore examine the degree of theoretical control one can hope to achieve in a finite landscape.

The definition of an isolated vacuum, as contrasted with a moduli space of nonzero dimension, is that there are no continuously-variable parameters, or massless scalar fields, in the solution.  In particular, the string coupling $g_s$ takes a specific value in each vacuum, and --- unless the landscape of vacua is itself infinite, a possibility that we will set aside for now --- there is a smallest nonzero value $g_s^{\mathrm{min}}>0$ attained by the string coupling.  Formally, then, the loop expansion parameter in string theory, and in the corresponding effective field theories, cannot be arbitrarily small.  Likewise, all other approximation schemes necessarily have finite control parameters.

For this reason, one should be wary of objections that a particular construction of a string vacuum is ``not parametrically controlled'': none can ever be parametrically controlled, by a strict definition of the term. Having parametrically controlled --- or even exact --- results in quantum field theory can give one a feeling of calm security, and in highly supersymmetric settings, this experience can be habit-forming.  But in our view this luxury was not essential in twentieth-century physics: the Standard Model gets by with asymptotic series, and whole fields are build on ad-hoc approximations that do not even aspire to become systematic series.

Even so, we certainly wish to achieve sufficient theoretical control, at least in the old-fashioned sense of being able to estimate the next correction in some small parameter and determine whether we must compute to yet higher order.

There are two complementary approaches to the problem of theoretical control. The first is to carry out fundamental computations to higher and higher accuracy, over ever-larger regions of parameter space: for example, to compute string amplitudes at higher genera. This is a worthy approach, but in string theory it has proved to be slow and difficult. In these lectures we will not report any progress in string amplitudes, but on the other hand we will make heavy use of advances in the practical computation of worldsheet instanton corrections (at string tree level) via mirror symmetry.

The other approach is to identify and exploit strategies that lead to solutions in regions where theoretical knowledge is already sufficient, or can be made so. Because the degree of control of an effective theory can be represented by the smallness of one or more dimensionless numbers, strategies for finding controlled solutions are equivalent to strategies for finding small numbers, or hierarchies.  Such strategies are often called \emph{mechanisms}. 

The constructions described in these lectures rest on two such mechanisms. Mathematically, the small numbers used here all arise from exponentials, and the exponentials are engineered by a choice of quantized fluxes.  However, there are two physical mechanisms at work.  

The first mechanism is a structure that allows the flux superpotential to be exponentially small: for quantized fluxes obeying a certain Diophantine equation, all the perturbative terms in the flux superpotential exactly vanish along a complex line pointing toward the large complex structure limit in moduli space.  The terms that remain can be understood as arising from worldsheet instantons of type IIA on the mirror threefold, and are exponentially small.  Thus, by solving a Diophantine problem, one can find vacua with exponentially small flux superpotential \cite{Demirtas:2019sip}. Such vacua are called \emph{perturbatively flat vacua}, or PFVs.

The second mechanism is warping, along the lines of the Randall-Sundrum model \cite{Randall:1999ee}. A suitable choice of quantized fluxes leads to an exponential warp factor in a region of the compactification, and correspondingly suppresses the supersymmetry-breaking effects of an anti-D3-brane in this region.  Specifically, we will make use of the Klebanov-Strassler warped deformed conifold solution \cite{Klebanov:2000hb} of type IIB supergravity: this solution has a smooth infrared region, and in an appropriate parameter regime it can support metastable supersymmetry breaking at an energy scale that is exponentially smaller than the fundamental (unwarped) string scale.  This is the essential structure that will allow for controlled supersymmetry breaking and a long-lived vacuum, as opposed to rapid instabilities.  

Many other mechanisms for breaking supersymmetry at an exponentially small scale have been proposed, but to date it has proved difficult to realize these other mechanisms in stabilized compactifications of string theory, and hence to controllably break supersymmetry in a four-dimensional effective quantum gravity theory, as opposed to in a supersymmetric gauge theory or in an ad-hoc supergravity theory.  Moreover, no adequate analogue of the Klebanov-Strassler solution is available in other corners of the duality web. Put together, these facts partially explain the comparatively advanced state of moduli stabilization in type IIB string theory as compared to dual descriptions.

The tasks before us can now be stated more precisely. We seek to first construct compactifications of type IIB string theory in which choices of quantized flux lead to an exponentially small flux superpotential, and also yield a Klebanov-Strassler region in the internal space.  For each such compactification we will compute the effective theory, making a series of approximations detailed below, and we will select examples in which the approximate effective theory has a de Sitter vacuum.  Finally, we will evaluate the potential corrections to our approximations, and use this information both to assess the status of our constructions, and to direct future work aimed at improvements.

\section{Plan of these lectures}

In Lecture I, we will introduce vacuum solutions from Calabi-Yau compactifications.  We will begin with mathematical foundations, work out the moduli of the Ricci-flat metric on a Calabi-Yau threefold, and explain the cosmological moduli problem (Chapter \ref{chap:vac}).

\medskip

\noindent In Lecture II, we will present non-vacuum solutions: flux compactifications of type IIB string theory on orientifolds of Calabi-Yau threefolds (Chapter  \ref{chap:gkp}). After examining these in ten dimensions, we will obtain the four-dimensional effective theory of flux compactifications, at the classical level (Chapter  \ref{chap:classicalEFT}).

\medskip

\noindent In Lecture III, we will lay out the quantum theory of flux compactifications  (Chapter \ref{chap:quantumEFT}), leading to a mechanism for moduli stabilization (Chapter \ref{chap:modulistabilization}).

\medskip

\noindent In Lecture IV, we will present candidate de Sitter vacua of the quantum theory of flux compactifications (Chapter \ref{chap:deSitter}), and discuss the outlook for future work (Chapter \ref{chap:future}).

\medskip

\noindent Broadly speaking, the content of Lecture I was obtained in the 1980s, while the results of Lecture II and much of Lecture III were foundational for the understanding of the string landscape that was achieved in the early 2000s.  By late in Lecture III, though, we will be discussing new results, setting the stage for a synthesis in Lecture IV, where --- at the frontier of our ability to compute quantum effects in string compactifications --- we will encounter uncertainty and open questions.

\chapter{Vacuum Solutions}\label{chap:vac}

\section{Preliminaries}

String theory furnishes our only working model of quantum gravity. As we will see, certain solutions of string theory yield weakly-coupled ultraviolet completions of general relativity. To understand quantum-gravitational effects in cosmology, we will therefore study cosmology in string theory.

Much of what we know about string theory is 
\begin{itemize}
    \item obtained from perturbative on-shell string amplitudes,
    \item formulated in flat or AdS spacetimes, with spacetime dimensions $D\neq 4$, and
    \item based on \emph{extended supersymmetry}, i.e.,~$\ge 8$ supercharges, corresponding to $\mathcal{N}\geq 2$ supersymmetry in $D=4$.
\end{itemize}
To describe the observed universe, we must have $D=4$ and $\mathcal{N}=0$ at low energies, even if there may be more dimensions and more supersymmetry at high energies.

Fortunately, one can productively study $\mathcal{N}=0$ theories by building on $\mathcal{N}=1$ and $\mathcal{N}=2$ theories.  Similarly, one can learn about $D=4$ by studying certain cases with $D \neq 4$, particularly $D=10$.

Much of the work in string cosmology occurs in the following

\begin{equ}[Primary setting] 

Solutions of critical superstring theories with $\mathcal{N}=1$ or $\mathcal{N}=0$ supersymmetry, in maximally-symmetric $D=4$ spacetimes, i.e.,~$\mathcal{M}^{3,1}$, AdS$_4$, or dS$_4$.
\end{equ}

Such solutions involve an internal CFT with $c=\tilde{c}=9$, since
\begin{equation}
    D_{\mathrm{crit}}\left(1+\frac{1}{2}\right)-26+11=0\,,
\end{equation} so that $D_{\mathrm{crit}}=10$.
Four large dimensions of spacetime contribute $4 \times \left(1+\frac{1}{2}\right)$, so that the remaining internal contribution must be $6 \times \left(1+\frac{1}{2}\right) = 9$.

The internal CFT could perfectly well be an abstract CFT with no obvious relation to a $\sigma$-model.  However, we know much more about geometric solutions, i.e.,~solutions in which $c=9$ arises from strings propagating on a compact 6-manifold $X$.  One reason for this difference in our state of knowledge is that we evolved in three large spatial dimensions, and so have specialized abilities for navigating in such spaces, whereas we have not yet evolved any CFT-specific circuitry.  Perhaps our AI successors will have a different preference.

For simplicity we are considering solutions in the critical dimension $D_{\mathrm{crit}}=10$.  But at the cost of Lorentz invariance of the field configuration, one can consider the linear dilaton theories $\Phi(X) = V_M X^M$, with $M=0,\ldots, D-1$, and with $V_M$ a constant vector. The $\sigma$-model $\beta$-function (for other fields vanishing) then reads (in the conventions of \cite{Blumenhagen:2013fgp})
\begin{equation}
    \beta(\Phi) = \frac{D-10}{4}+\alpha' \partial_M \Phi\, \partial^M \Phi\,,
\end{equation} 
and vanishes if
\begin{equation}
    D=10- 4\alpha' V_M V^M\,.
\end{equation}
For $V_M$ spacelike one has subcritical strings with $D<D_{\mathrm{crit}}$, while for $V_M$ timelike one has supercritical strings with $D>D_{\mathrm{crit}}$.  But in the remainder of these lectures, \emph{string theory} will mean \emph{superstring theory in the critical dimension}.

The low-energy effective theories resulting from superstring theories are supergravity theories, which contain the gravitational field, as well as a collection of other bosonic and fermionic fields. We will be particularly concerned with type IIB supergravity in $D=10$.

The classical equations of motion of this theory include the equations of motion for the gravitational field, i.e.,~the Einstein equations, which in $D=10$ take the form
\begin{equation}\label{eq:EEinit}
    R_{MN}=T_{MN}-\dfrac{1}{8}g_{MN}T^{K}\,_{K}\,,
\end{equation} 
in units where the ten-dimensional gravitational coupling is $\kappa_{10} =1$. The sources in the stress-energy tensor $T_{MN}$ are given by the configurations of the other bosonic fields.

\begin{equ}[Vacuum solutions and string vacua]\index{Vacuum solutions}\index{String vacua}

Vacuum solutions are solutions of the Einstein equations with
\begin{equation}
    T_{MN}=0\,,
\end{equation}
corresponding to  pure geometry. \emph{String vacua} are solutions of string theory, and may be vacuum ($T_{MN}=0$) or non-vacuum ($T_{MN}\neq 0$) solutions of the Einstein equations.
    
\end{equ}

\noindent The flux compactifications that are the main subject of these lectures are string vacua, but are not vacuum solutions: the stress-energy of $p$-form fluxes plays a crucial role.

We seek solutions of this theory by starting from the ansatz
\begin{equation}
    \dif s^{2}=g_{MN}^{(10)}\dif x^{M}\dif x^{N}=g_{\mu\nu}^{(4)}\dif x^{\mu}\dif x^{\nu}+g_{mn}^{(6)}\dif y^{m}\dif y^{n}
\end{equation}
where $M,N=0,\ldots,9$, $\mu,\nu=0,\ldots,3$, and $m,n=4,\ldots,9$. Here, $g_{\mu\nu}^{(4)}$ is the metric on four-dimensional spacetime given by one of our choices: $\cM^{3,1}$, $\mathrm{AdS}_{4}$, and $\mathrm{dS}_{4}$. The metric $g_{mn}^{(6)}$ describes a compact Riemannian manifold. 

Later it will be important to consider a \emph{warped} ansatz,
\begin{equation}
    \dif s^{2}=g_{MN}^{(10)}\dif x^{M}\dif x^{N}=f(y)g_{\mu\nu}^{(4)}\dif x^{\mu}\dif x^{\nu}+g_{mn}^{(6)}\dif y^{m}\dif y^{n}\,,
\end{equation}
where the warp factor $f(y)$ multiplying the four-dimensional metric depends on the internal coordinates $y^{m}$. When $f(y)$ is non-trivial, physical length and energy scales measured in the external four dimensions vary across the compact space: for instance, the local four-dimensional Planck scale or particle masses can be redshifted depending on the location in the internal manifold. In this way, warping allows local physics in certain regions (such as the warped throats described in \S\ref{sec:KS}) to be parametrically different from that in the bulk. The flux compactifications studied in Chapter~\ref{chap:gkp} will be warped, but for now we will simplify the discussion by setting $f(y)=1$.

The vacuum Einstein equations for the product ansatz are
\begin{equation}
    R_{MN}=0\,,
\end{equation}
so that
\begin{equation}
    R_{\mu\nu}=0\kom R_{mn}=0\, .
\end{equation}
The first relation implies that $\mathrm{AdS}_{4}$ and $\mathrm{dS}_{4}$ are not vacuum solutions, so that we can only have $\cM^{3,1}$. The right hand side means that our compact manifold has to be \emph{Ricci-flat}\index{Ricci-flatness}. Thus, we need to better understand compact, Ricci-flat 6-manifolds. In general, $R_{mn}=0$ is a complicated PDE, and no analytic solutions have been derived to date, except in trivial cases such as the six-torus.\footnote{However, see \cite{Kachru:2020tat} for analytic results for compact Ricci-flat 4-manifolds, i.e.,~K3 surfaces.  The non-compact problem is far more tractable, and analytic Ricci-flat metrics on non-compact $k$-folds were obtained in \cite{Greene:1989ya} for $k=2$ and in \cite{MR1068113} for $k=3$.}

\section{Mathematical foundations}
 
To construct and understand compact Ricci-flat  $6$-manifolds, we will need to review some geometry.  We will begin by setting notation for differential forms on real manifolds, then introduce complex manifolds and Dolbeault cohomology classes, and finally discuss K\"ahler manifolds.

\subsection{Differential calculus on real manifolds}

Let $M$ be a real manifold of dimension $D$ and let $\varphi_{\alpha} :\, U_{\alpha} \to \mathbb{R}^D$ be coordinate charts on patches $U_{\alpha} \subset M$. Provided that all transition functions $\varphi_{\alpha}\circ\varphi_{\beta}^{-1}$ are differentiable, we call $M$ a {\emph{differentiable manifold}}. We can then formulate the calculus on manifolds most naturally in terms of \emph{differential forms}.\index{Differential forms} Writing $T_p M$ for the tangent space to $M$ at a point $p \in M$, we recall that an \emph{$r$-form} $\omega$ assigns to each point $p$ a map
\begin{equation}\label{apA:eqnA.6}
    \omega_p :\,\underbrace{T_p M \times \cdots \times T_p M}_{r} \, \to\, \mathbb{R}
\end{equation}
that is linear and totally antisymmetric in each variable. The tangent bundle $TM=\cup_{p\in M}\, T_p M$ is obtained by taking the union of all tangent spaces $T_p M$. We denote the vector space of all rank $(0,r)$ antisymmetric tensor fields at a point $p$ by $\Omega^r_p(M)$, and the vector space of $r$-forms on $M$ as $\Omega^r(M)$. Locally, an element $\omega \in \Omega^r_p(M)$ can be defined as
\begin{equation}
    \omega = \frac{1}{r!} \hskip 1pt \omega_{m_1 \ldots m_r} \, \mathrm{d} x^{m_1} \wedge \cdots \wedge \mathrm{d} x^{m_r}\ ,
\label{apA:eqnA.8}
\end{equation}
in terms of totally antisymmetric components $\omega_{m_1 \ldots m_r}$.

We define a map $\mathrm{d}:\,\Omega^r(M) \to \Omega^{r+1}(M)$, called the \emph{exterior derivative}\index{exterior derivative}, such that for an $r$-form $\omega\in \Omega^r(M)$ 
\begin{equation}
    \mathrm{d} \omega = \frac{1}{r!} \frac{\partial \omega_{m_1 \ldots m_r}}{\partial x^{n}} \hskip 2pt \mathrm{d} x^{n} \wedge \mathrm{d} x^{m_1} \wedge \cdots \wedge \mathrm{d} x^{m_r} \ . \label{apA:eqnA.12}
\end{equation}
We call an $r$-form $\omega\in \Omega^r(M)$ \emph{closed} if $\mathrm{d} \omega = 0$, and \emph{exact} if it can be written as $\omega = \mathrm{d} \alpha$ for some (globally defined) \hbox{$(r-1)$-form} $\alpha$. We denote the set of closed $r$-forms by $Z^{r}(M)$ and of exact $r$-forms by $B^r(M)$.\index{de Rham cohomology} Their quotient defines the $r$th \emph{de Rham cohomology group} 
\begin{equation}
    H^r(M,\mathbb{R}) \equiv Z^r(M)/B^r(M)\ .\label{apA:eqnA.16}
\end{equation}
The number of non-trivial cohomology classes of $r$-forms on $M$ is given by the $r$th \emph{Betti number},\index{Betti number}
\begin{equation}\label{apA:eqnA.27}
    b^r \equiv \dim(H^r(M,\mathbb{R}))\ ,
\end{equation}
which, due to Poincar\'{e} duality, satisfies $b^r = b^{D-r}$. The \emph{Euler characteristic} of $M$ is defined in terms of the Betti numbers as the alternating sum\index{Euler characteristic}
\begin{equation}\label{apA:eqnA.29}
    \chi(M) \equiv \sum_{r=0}^{D} (-1)^r \, b^r\ .
\end{equation}

Next, let us review the basic concepts of \emph{Hodge theory}.

\begin{equ}[Hodge star]\index{Hodge theory}

The \emph{Hodge star} defines an isomorphism $\star :\,\Omega^r(M) \to \Omega^{D-r}(M)$ acting on an $r$-form $\omega\in \Omega^r(M)$ as\index{Hodge star}
\begin{equation}
    (\star\hskip 2pt \omega_r)_{i_1 \ldots i_{D-r}} = \frac{\sqrt{|g|}}{r!(D-r)!} \, \epsilon_{i_1 \ldots i_{D-r}}{}^{j_1 \ldots j_r} \omega_{j_1 \ldots j_r}\ .
    \label{apA:eqnA.32}
\end{equation}
where $g \equiv {\rm{det}}(g_{ij})$, and $\epsilon_{i_1\ldots i_D}$ denotes the totally antisymmetric Levi-Civita symbol.\footnotemark

\end{equ}
\footnotetext{In our conventions, $\epsilon_{i_1\cdots i_{D}}=+1$ if $i_1\cdots i_{D}$ is an even permutation of $012\cdots D$.}

\noindent The Hodge star provides us with an explicit (but metric-dependent) isomorphism between the cohomology groups $H^r(M , \mathbb{R})$ and $H^{D-r}(M , \mathbb{R})$.
The action of $\star$ on the basis elements of $\Omega^r(M)$ is given by
\begin{align}\label{apA:eqnA.30}
    &\star \left( \,\mathrm{d} x^{i_1} \wedge \ldots \wedge \,\mathrm{d} x^{i_r} \right) = \frac{\sqrt{|g|}}{(D-r)!}\, \epsilon^{i_1 \ldots i_r}{}_{j_{r+1} \ldots j_D} \,\mathrm{d} x^{j_{r+1}} \wedge \ldots \wedge \,\mathrm{d} x^{j_D}\ .
\end{align}
It follows that the invariant volume element is related to $\star \hskip 1pt1$, i.e.,
\begin{equation}
    \star \hskip 1pt1 = \sqrt{|g|}\, \,\mathrm{d} x^{j_{1}} \wedge \cdots \wedge \,\mathrm{d} x^{j_D}\ .
\label{apA:eqnA.31}
\end{equation}
Moreover, the Hodge star has the property 
\begin{equation}
    \star \star \omega_r = (-1)^{r(D-r)+s}\, \omega_r\ , \label{starstar}
\end{equation}
where $s=0,1$ for $M$ Riemannian or Lorentzian. The Hodge star is particularly useful since it can be used to construct a $D$-form from two $r$-forms $\omega,\xi \in \Omega^r(M)$
\begin{equation}\label{apA:eqnA.34}
    \omega \wedge \star\hskip 2pt \xi = \frac{1}{r!} \omega_{i_1 \ldots i_r} \xi^{i_1 \ldots i_r} \sqrt{|g|}\, \,\mathrm{d} x^{1} \wedge \cdots \wedge \,\mathrm{d} x^D\ .
\end{equation}
This leads to the natural definition of the \emph{inner product} of two $r$-forms $\omega,\xi \in \Omega^r(M)$
\begin{equation}\label{apA:eqnA.35}
    \langle \omega, \xi \rangle \equiv \int \omega \wedge \star\hskip 2pt \xi = \int \,\mathrm{d}^D x \sqrt{|g|}\, \omega_{i_1 \ldots i_r} \xi^{i_1 \ldots i_r}\ .
\end{equation}

\begin{equ}[Adjoint exterior derivative and Laplacian]\index{adjoint exterior derivative}

The \emph{adjoint exterior derivative} corresponds to the map $\mathrm{d}^{\dagger}:\,\Omega^r(M) \to \Omega^{r-1}(M)$
defined as  
\begin{equation}\label{apA:eqnA.36}
    \,\mathrm{d}^{\dagger} = (-1)^\mathscr{S} \star \,\mathrm{d}\, \star\ , 
\end{equation}
where $\mathscr{S}=r(D-r + 1)+s$, with $s=0,1$ for $M$ Riemannian or Lorentzian.  For $D$ even and $M$ Riemannian, we have simply $\mathrm{d}^{\dagger} = \star \,\mathrm{d}\, \star$.

The map $\Delta :\,\Omega^r(M) \to \Omega^{r}(M)$ given by
\begin{equation}\label{apA:eqnA.38}
    \Delta = \mathrm{d} \,\mathrm{d}^{\dagger} + \mathrm{d}^{\dagger} \,\mathrm{d} \ ,
\end{equation}
defines the \emph{Laplacian}\index{Laplacian} on $M$.

\end{equ}

\noindent For $\eta \in \Omega^{r-1}(M)$ and $\omega \in \Omega^r(M)$, we find
\begin{equation}\label{apA:eqnA.37}
    \langle \,\mathrm{d}\eta , \omega \rangle = \langle \eta , \,\mathrm{d}^{\dagger} \omega \rangle \ ,
\end{equation}
so that $\,\mathrm{d}^{\dagger}$ is the adjoint of $\,\mathrm{d}$. An $r$-form $\omega$ is called \emph{co-closed} if $\,\mathrm{d}^{\dagger} \omega = 0$, and \emph{co-exact} if $\omega = \,\mathrm{d}^{\dagger} \beta$ for a globally defined $(r+1)$-form $\beta$. 

\begin{equ}[Harmonic forms]\index{harmonic form}

An $r$-form $h$ is \emph{harmonic} if it satisfies
\begin{equation}
    \Delta h = 0\, .
\end{equation}
We denote the vector space of harmonic $r$-forms on $M$ by ${\cal H}^r(M)$.

\end{equ}

The \emph{Hodge decomposition}\index{Hodge decomposition} states that, for every $r$-form $\omega\in\Omega^r(M)$ on a compact Riemannian manifold, there exists a unique global decomposition of the form
\begin{equation}
    \omega = h + \,\mathrm{d} \alpha + \,\mathrm{d}^{\dagger} \beta\ , \label{Hdecomposition}
\end{equation}
where $h \in {\cal H}^r(M)$ is harmonic. For elements $\xi \in H^r(M , \mathbb{R})$ of de Rham cohomology groups, \emph{Hodge's theorem} asserts that the Hodge decomposition \eqref{Hdecomposition} becomes
\begin{equation}
    \xi = h+ \,\mathrm{d} \alpha\ , \label{apA:eqnA.42}
\end{equation}
where  $h \in {\cal H}^r(M)$ is a \emph{unique} harmonic form. Said differently, each de Rham cohomology class has a unique harmonic representative, thereby establishing an isomorphism\index{Hodge's theorem}
\begin{equation}\label{Hodgetheorem}
    H^r(M , \mathbb{R}) \cong {\cal H}^r(M)\ . 
\end{equation}

\subsection{Complex manifolds and Dolbeault cohomology}
\label{apA:secA.4}

Up to this point, our discussion has involved real manifolds. Next, we discuss complex manifolds: these are even-dimensional real manifolds that locally look like $\mathbb{C}^k$. Just as complex analysis on $\mathbb{C}$ leads to highly constraining structures like holomorphy, transitioning from real manifolds to complex manifolds brings about powerful refinements. The role of complex manifolds in string theory is twofold. For one, moduli spaces in many supersymmetric theories correspond to complex manifolds. Second, as we will see below, the analysis of Ricci-flat compactification manifolds in the context of vacuum solutions necessitates a certain toolkit from complex geometry. Indeed, Ricci-flat manifolds that are compact and complex are generally easier to construct and study in practice.\footnote{Further details on complex manifolds can be found e.g.~in the lecture notes by Candelas~\cite{Candelas:1987is}.}

\begin{equ}[Complex manifolds]

A {\emph{complex manifold}}\index{complex manifold} $M$ is a real manifold of dimension $D=2k$, $k\in \mathbb{Z}_+$, for which the coordinate charts are maps $\varphi_{\alpha} : U_{\alpha} \to \mathbb{C}^k$, with holomorphic transition functions.

\end{equ}

\noindent A useful notion on complex manifolds is that of \emph{complex structures}. An \emph{almost complex structure} ${\cal J}$ on a real manifold $M$ of dimension $n=2k$ is a $(1,1)$ tensor field satisfying ${\cal J}^2 = -1$. It is useful to think of ${\cal J}$ as a map ${\cal J} : TM \to TM$, with $TM$ the tangent bundle of $M$. In local coordinates $z^{i_1},\ldots, z^{i_k}$ and their complex conjugates, the components of ${\cal J}$ take the form
\begin{equation}
    {\cal J}^{i}_{~j} = \I \delta^{i}_{~j}\ \ ,\quad {\cal J}^{\bar{\imath}}_{~\bar{\jmath}} = - \I \delta^{\bar{\imath}}_{~\bar{\jmath}} \ . \label{canonicalACS}
\end{equation}
We call ${\cal J}$ a \emph{complex structure} if there exists a suitable choice of local holomorphic coordinates in each patch $U_\alpha$ of $M$ such that the components of ${\cal J}$ take the form \eqref{canonicalACS}. This leads us to the following alternative definition of complex manifolds.

\begin{equ}[Complex manifolds and complex structures]\index{complex structure}

A complex manifold $M$ is a real manifold of dimension $D=2k$ equipped with a \emph{complex structure} ${\cal J}$ corresponding to a tensor field of type $(1,1)$ satisfying \eqref{canonicalACS} in each patch $U_\alpha$ of $M$.\index{complex structure}

\end{equ}

For any real manifold $M$ of dimension $D$ (even when $D$ is not even), we can introduce a \emph{complexified tangent space} $T_p M^{\mathbb{C}}$ at a point $p$ defined as\index{complexified tangent space}
\begin{equation}
    T_p M^{\mathbb{C}} = \{X+\I Y|X,Y \in T_p M\} \ .\label{apA:eqnA.51}
\end{equation} 
Similarly, the complexifications of general tensors can be defined. If $M$ is in addition a complex manifold, the underlying complex structure ${\cal J}$ can be used to split $T_p M^{\mathbb{C}}$ into eigenspaces of ${\cal J}$. Locally, a basis for $T_p M^{\mathbb{C}}$ is furnished by the vectors $\partial/\partial z^{i}$ and $\partial/\partial \bar{z}^{\bar{\imath}}$ in terms of local holomorphic coordinates $z^{i} \equiv x^{i} + \I y^{i}$. Let the action of ${\cal J}$ on $T_p M^{\mathbb{C}}$ be given by 
\begin{equation}
    {\cal J}(X+\I Y) = {\cal J}(X) + \I {\cal J}(Y)\quad \forall X,Y \in T_p M\, .
\end{equation}
Then, at any point $p\in M$, this implies
\begin{equation}
    {\cal J}(\partial/\partial z^{i}) = \I \partial/\partial z^{i}\ \ , \quad {\cal J}(\partial/\partial \bar{z}^{\bar{\imath}}) = - \I \partial/\partial \bar{z}^{\bar{\imath}}\ . \label{apA:eqnA.52}
\end{equation}
This induces a natural decomposition of $T_p M^{\mathbb{C}}$ into the positive and negative eigenspaces spanned by $\{\partial/\partial z^{i}\}$ and $\{\partial/\partial \bar{z}^{\bar{\imath}}\}$ respectively. Correspondingly, the elements of these eigenspaces are called holomorphic and anti-holomorphic vectors, respectively.
 
On a complex manifold, differential forms can be defined analogously by considering linear combinations of real $r$-forms with complex coefficients.

\begin{equ}[Complex differential forms]
 
Given two real $r$-forms $\alpha_r$ and $\beta_r$ on a manifold $M$,  
the complex-valued combination
\begin{equation}
    \gamma_r \coloneqq \alpha_r + \I\, \beta_r\,,
    \label{apA:eqnA.53}
\end{equation}
is called a \emph{complex $r$-form}. Its complex conjugate is $\bar{\gamma}_r = \alpha_r - \I\, \beta_r$.  
Complex $r$-forms can be decomposed according to their holomorphic and anti-holomorphic index structure. 
A \emph{$(p,q)$-form} is a complex $(p+q)$-form with $p$ holomorphic and $q$ anti-holomorphic indices.  
\end{equ}
The space of all complex $r$-forms on $M$ is denoted by $\Omega^r_{\mathbb{C}}(M)$, and that of $(p,q)$-forms is denoted by $\Omega^{p,q}(M)$.

In local coordinates, a basis for $(p,q)$-forms is
\begin{equation}\label{apA:eqmA.54}
    \,\mathrm{d} z^{i_1} \wedge \ldots \wedge \,\mathrm{d} z^{i_p} \wedge \,\mathrm{d} \bar z^{\bar{\jmath}_1} \wedge \ldots \wedge \,\mathrm{d} \bar z^{\bar{\jmath}_q} \ ,
\end{equation}
and any element $\gamma_{p,q}$ of $\Omega^{p,q}(M)$ can be written locally as
\begin{equation}\label{apA:eqnA.55}
    \gamma_{p,q} = \frac{1}{p! q!} \gamma_{i_1 \ldots i_p\, \bar{\jmath}_1 \ldots \bar{\jmath}_q}  \,\mathrm{d} z^{i_1} \wedge \ldots \wedge \,\mathrm{d} z^{i_p} \wedge \,\mathrm{d} \bar z^{\bar{\jmath}_1} \wedge \ldots \wedge \,\mathrm{d} \bar z^{\bar{\jmath}_q} \ ,
\end{equation} 
Similarly, any complex $r$-form $\gamma_r\in \Omega^r_{\mathbb{C}}(M)$ can be decomposed uniquely into a sum of $(p,q)$-forms
\begin{equation}\label{apA:eqnA.56}
    \gamma_r = \sum_{p+q=r} \gamma_{p,q}\ ,
\end{equation} 
so that
\begin{equation}\label{Omegadecomposition}
    \Omega^r_{\mathbb{C}}(M) = \bigoplus_{p+q=r} \Omega^{p,q}(M)\ . 
\end{equation}

The next step is to extend the definition of the exterior derivative $\mathrm{d}:\, \Omega^{r}(M)\to \Omega^{r+1}(M)$ as defined in Eq.~\eqref{apA:eqnA.12} such that it respects the complex structure.
\newpage

\begin{equ}[Dolbeault operators]\index{Dolbeault operators}
 
The \emph{Dolbeault operators} are differential operators that act on complex differential forms by increasing either the holomorphic or anti-holomorphic degree. More precisely, they are defined as linear maps $\partial :\,\Omega^{p,q} \to \Omega^{p+1,q}$ and $\bar \partial :\,\Omega^{p,q} \to \Omega^{p,q+1}$ that act on
a $(p,q)$-form $\gamma_{p,q}$ as
\begin{align}
    \partial \gamma_{p,q} &= \frac{1}{p! q!} \left( \frac{\partial}{\partial z^{k}} \gamma_{i_1 \ldots i_p\, \bar{\jmath}_1 \ldots \bar{\jmath}_q}\right) \,\mathrm{d} z^{k} \wedge \,\mathrm{d} z^{i_1} \wedge \ldots \wedge \,\mathrm{d} z^{i_p} \wedge \,\mathrm{d} \bar z^{\bar{\jmath}_1} \wedge \ldots \wedge \,\mathrm{d} \bar z^{\bar{\jmath}_q} \ , \label{apA:eqnA.58}\\
    \bar \partial \gamma_{p,q} &= \frac{1}{p! q!} \left(\frac{\partial }{\partial \bar z^{\bar{k}}} \gamma_{i_1 \ldots i_p\, \bar{\jmath}_1 \ldots \bar{\jmath}_q} \right) \,\mathrm{d} \bar z^{\bar{k}} \wedge \,\mathrm{d} z^{i_1} \wedge \ldots \wedge \,\mathrm{d} z^{i_p} \wedge \,\mathrm{d} \bar z^{\bar{\jmath}_1} \wedge \ldots \wedge \,\mathrm{d} \bar z^{\bar{\jmath}_q} \ .\label{apA:eqnA.59}
\end{align}

\end{equ}

\noindent With this definition, $\mathrm{d} = \partial + \bar \partial$ and 
\begin{align}\label{apA:eqnA.61}
    \partial^2 = \bar \partial^2 = \partial \bar \partial + \bar \partial \partial = 0\ .
\end{align}
We call a $(p,0)$-form $\gamma_{p,0}$ \emph{holomorphic} if and only if it satisfies
\begin{equation}\label{apA:eqnA.62}
    \bar \partial \gamma_{p,0} = 0\ .
\end{equation}
In this sense, holomorphic functions correspond to holomorphic $0$-forms.

Just as the exterior derivative $\mathrm{d}$ defines de Rham cohomology classes, we can use the Dolbeault operators to define Dolbeault cohomology classes.

\begin{equ}[Dolbeault cohomology groups]\index{Dolbeault cohomology}
 
Denote by $Z^{p,q}_{\bar{\partial}}(M)$ the space of $(p,q)$-forms on $M$ that are $\bar{\partial}$-closed,   
and by $B^{p,q}_{\bar{\partial}}(M)$ the subspace consisting of those forms that are $\bar{\partial}$-exact. The \emph{Dolbeault cohomology group} $H^{p,q}_{\bar \partial}(M , \mathbb{C})$ is defined as the quotient
\begin{equation}\label{apA:eqnA.63}
    H^{p,q}_{\bar \partial}(M , \mathbb{C}) \equiv Z^{p,q}_{\bar \partial}(M)/B^{p,q}_{\bar \partial}(M)\ .
\end{equation}

\end{equ}

\noindent The generalization of Hodge theory to complex manifolds proceeds in parallel to the real case. By letting the Hodge star \eqref{apA:eqnA.32} act on the complexified tangent space, one again finds an inner product on $r$-forms, and defines adjoints $\partial^{{\dagger}}$ and $\bar{\partial}^{{\dagger}}$ of $\partial$ and $\bar{\partial}$. In contrast to the real case, there are two types of Laplacians,
\begin{equation}\label{apA:eqnA.64}
    \Delta_{\partial} = \partial\partial^{{\dagger}}+\partial^{{\dagger}}\partial \ , \quad
    \Delta_{\bar{\partial}} = \bar{\partial}\bar{\partial}^{{\dagger}}+\bar{\partial}^{{\dagger}}\bar{\partial} \ .
\end{equation}
We say $(p,q)$-forms are $\bar{\partial}$-harmonic if they are annihilated by $\Delta_{\bar{\partial}}$. We denote the set of all $\bar{\partial}$-harmonic $(p,q)$-forms as ${\cal{H}}^{p,q}_{\bar \partial}(M)$. Due to Hodge's theorem, we find an isomorphism $H^{p,q}_{\bar \partial}(M , \mathbb{C}) \cong {\cal{H}}^{p,q}_{\bar \partial}(M)$. The complex dimensions of the Dolbeault cohomology groups are referred to as \emph{Hodge numbers},\index{Hodge numbers}
\begin{equation}\label{apA:eqnA.65}
    h^{p,q} = \dim\bigl(H^{p,q}_{\bar \partial}(M , \mathbb{C})\bigr) \ .
\end{equation}
The Hodge star can be used to show that $h^{p,q} = h^{k-p,k-q}$ on a manifold $M$ of complex dimension $k$.

\subsection{Hermitian and K\"ahler manifolds}\label{apA:secA.5}

Thus far, we have remained agnostic about the type of metric defined on complex manifolds. It turns out that there is a special class of metrics satisfying a compatibility condition with respect to the complex structure ${\cal J}$ on $M$.\footnote{A metric $g$ on a complex manifold $M$ need not necessarily be compatible with the complex structure ${\cal J}$. In this case, the metric and the complex structure are not ``aligned'' in the way needed to define the rich geometry associated with Hermitian (and especially K\"ahler) manifolds.}

\begin{equ}[Hermitian metrics and manifolds]\index{Hermitian manifold}

A \emph{Hermitian metric} on a complex manifold $M$ is a Riemannian metric $g\, :\, TM \times TM \to \mathbb{R}$ that is compatible with the complex structure ${\cal J}$ in the sense that
\begin{equation}\label{apA:eqnA.66}
    g({\cal J}X, {\cal J} X) = g(X,Y)\ .
\end{equation}
A complex manifold $M$ equipped with a Hermitian metric is called a \emph{Hermitian manifold}.

\end{equ}

\noindent In local coordinates, the condition that the metric is Hermitian becomes $g_{i j} = g_{\bar{\imath}\bar{\jmath}} = 0$, and hence the metric can be expressed as
\begin{equation}\label{apA:eqnA.67}
    g = g_{i \bar{\jmath}}\hskip 2pt \,\mathrm{d} z^{i}  \,\mathrm{d} \bar z^{\bar{\jmath}} + g_{\bar{\imath} j} \hskip 2pt \,\mathrm{d} \bar z^{\bar{\imath}}  \,\mathrm{d} z^j\ .
\end{equation}
The metric components further satisfy $g_{i\bar{\jmath}} = \overline{g_{j\bar{\imath}}}$ as required by reality. 

On a Hermitian manifold, we can define a 2-form $J$, called the \emph{K\"ahler form}, as \index{K\"ahler form}
\begin{equation}\label{apA:eqnA.68}
    J = \I g_{i \bar{\jmath}}\, \,\mathrm{d} z^{i} \wedge \,\mathrm{d} \bar z^{\bar{\jmath}}\ .
\end{equation}
By imposing closedness of $J$, we find an important subclass of complex manifolds.

\begin{equ}[K\"ahler manifolds]\index{K\"ahler manifold}

A \emph{K\"ahler manifold} is a complex manifold equipped with a Hermitian metric whose associated K\"ahler form $J$ is closed,
\begin{equation}\label{apA:eqnA.69}
    \mathrm{d} J = 0\,.
\end{equation}
This condition ensures that the metric, complex structure, and symplectic structure are mutually compatible. Equivalently, a complex manifold $M$ of complex dimension $k$ is K\"ahler if its holonomy group is contained in $\mathrm{U}(k)$, that is, ${\mathrm{Hol}}(M) \subseteq \mathrm{U}(k)$.\footnotemark

\end{equ}
\footnotetext{See \S\ref{ss:holonomy} for background on holonomy.}

Because the K\"ahler form $J$ is closed, the associated Hermitian metric $g_{i\bar{\jmath}}$ obeys an additional integrability condition. In particular, one finds $\partial_k g_{i\bar{\jmath}}=\partial_i g_{k\bar{\jmath}}$ which follows directly from the definition of the K\"ahler form in \eqref{apA:eqnA.68}. As a result, the metric can be written locally, i.e., within a single coordinate patch $U_\alpha$, in terms of a real scalar function known as the \emph{K\"ahler potential} $k$\index{K\"ahler potential}
\begin{equation}
    J = \I \partial \bar \partial k \quad \mathrm{or} \quad g_{i\bar \jmath} = \partial_i \partial_{\bar \jmath} k \ ,\label{equ:kmetric}
\end{equation}
and we refer to $g_{i\bar \jmath}$ as a \emph{K\"ahler metric}.\index{K\"ahler metric} Due to the definition \eqref{equ:kmetric}, the K\"ahler metric $g_{i\bar \jmath}$ is invariant under \emph{K\"ahler transformations}\index{K\"ahler transformation}
\begin{equation}
    k(z^i,\bar z^{\bar \imath}) \mapsto k(z^i,\bar z^{\bar \imath}) + f(z^i) + \bar f(\bar z^{\bar \imath})\,,  \label{ktransform}
\end{equation}
for arbitrary holomorphic functions $f(z^i)$. Further, we note that the K\"ahler potential on a non-trivial compact K\"ahler manifold is not a globally~defined function, but is instead patched together via suitable K\"ahler transformations.

On a K\"ahler manifold, the Laplacians constructed from the exterior derivative $\mathrm{d}$ and from the Dolbeault operators $\partial$ and $\bar{\partial}$ are not independent. Rather, they are related by the identities (see, for example, \cite{schwartz1955lectures})
\begin{equation}
    \Delta = 2\Delta_{\partial} = 2\Delta_{\bar{\partial}} \ . \label{KLaplacian}
\end{equation}
As a consequence, the Hodge numbers of a K\"ahler manifold satisfy the conjugation symmetry\index{Hodge numbers}
\begin{equation}
    h^{p,q} = h^{q,p} \ , \label{Hodgeconjugation}
\end{equation}
which reflects the equivalence of $\partial$- and $\bar{\partial}$-harmonic forms.
Moreover, using \eqref{KLaplacian} together with the decomposition of differential forms into $(p,q)$-types, cf.~\eqref{Omegadecomposition}, the Betti numbers $b^r \coloneqq \dim(H^{r}(M , \mathbb{R}))$ can be expressed in terms of the Hodge numbers $h^{p,q}$ as
\begin{equation}
    b^r = \sum_{p+q=r} h^{p,q}\ .
\label{apA:eqnA.82}
\end{equation}
An immediate implication is that all odd Betti numbers are even on a K\"ahler manifold,\index{Betti number}
\begin{equation}\label{apA:eqnA.83}
    b^{2r-1} =\ \ 2 \times \hspace{-0.2cm} \sum_{\substack{ p+q=2r-1,\\[1pt] \hspace*{-6pt}p<q}} \hspace{-0.2cm} h^{p,q} \ .
\end{equation}  
Since the K\"ahler form $J$ is closed but not exact, the even Betti numbers are always non-zero, $b^{2k}\geq 1$.  
Finally, combining \eqref{apA:eqnA.29} with \eqref{apA:eqnA.82}, the Euler characteristic of a K\"ahler manifold can be written in terms of Hodge numbers as
\begin{equation}\label{apA:eqnA.84}
    \chi(M) = \sum_{p,q} (-1)^{p+q} h^{p,q}\ .
\end{equation}

The special structure of a K\"ahler metric also simplifies its connection and curvature. The only non-vanishing Christoffel symbols are\index{Christoffel symbols}
\begin{equation}\label{apA:eqnA.73}
    \Gamma^{i}_{kl} = g^{i\bar{\jmath}} \partial_{k} g_{l\bar{\jmath}}
\end{equation} 
together with their complex conjugates $\Gamma^{\bar{\imath}}_{\bar{k}\bar{l}}$.  
Correspondingly, up to index symmetries and complex conjugation, the only nonzero components of the Riemann tensor are\index{Riemann tensor}
\begin{equation}\label{apA:eqnA.74}
    R^{i}_{~kl\bar{\jmath}} = - \partial_{\bar{\jmath}} \Gamma^{i}_{kl}\ .
\end{equation}
The \emph{Ricci form} is defined as\index{Ricci form}
\begin{equation}
    {\cal{R}} = \I R_{i\bar{\jmath}} \hskip 1pt \,\mathrm{d} z^{i} \wedge \,\mathrm{d} \bar{z}^{\bar{\jmath}} \ ,
\label{apA:eqnA.75}
\end{equation} 
with $R_{i\bar{\jmath}} = R^{k}_{~k i\bar{\jmath}}$.
On a K\"ahler manifold,  the identity
\begin{equation}\label{apA:eqnA.76}
    \Gamma^{i}_{kl} = \partial_{k}\, \mathrm{ln\, det}\, (g_{m\bar{n}})\ ,
\end{equation} allows us
to express the Ricci tensor and Ricci form as
\begin{equation}
    R_{i\bar{\jmath}} = - \partial_{\bar{\jmath}} \Gamma^{k}_{ik} = - \partial_{\bar{\jmath}} \partial_{i} \,\mathrm{ln} \,\mathfrak{g}\; , \quad {\cal{R}} = \I \partial\bar\partial \,\mathrm{ln } \,\mathfrak{g} \ , \label{Ricciform}
\end{equation}
with $\mathfrak{g} = \det(g_{m\bar{n}})$. Using the relation $\mathrm{d}(\bar{\partial}-\partial)=2\,\partial\bar{\partial}$, it follows immediately that the Ricci form is closed, $\mathrm{d}\mathcal{R}=0$.

Although $\mathcal{R}$ is closed, it is not generally exact, since $\mathfrak{g}$ does not transform as a scalar under coordinate changes. However, under a smooth variation of the metric $\delta g$, one has $\delta \mathfrak{g} = \mathfrak{g}\, g^{\mu\nu}\delta g_{\mu\nu}$, which leads to
\begin{equation}
    \delta \mathcal{R} =  \I \partial\bar\partial \, \bigl(g^{m\bar{n}}\delta g_{m\bar{n}}\bigr)=-\frac{\I}{2}\mathrm{d}\Bigl\{ \bigl(\partial-\bar{\partial}\bigr)g^{m\bar{n}}\delta g_{m\bar{n}}\Bigr\}\, .
\end{equation}
Since $g^{m\bar{n}}\delta g_{m\bar{n}}$ is a scalar, the variation $\delta\mathcal{R}$ is exact. 

We conclude that $\mathcal{R}$ defines a de Rham cohomology class, and this class is invariant under smooth changes of the metric:

\begin{equ}[First Chern class]

On a K\"ahler manifold, 
the 
Ricci form defines a de Rham cohomology class, the \emph{first Chern class}:\index{Chern class}
\begin{equation}
    c_1 \equiv \frac{1}{2\pi}[{\cal{R}}] \in H^2(M ,\mathbb{R})\ . \label{apA:eqnA.79}
\end{equation}

\end{equ}

For a K\"ahler manifold $M$ to admit a Ricci-flat metric, $c_1(M)=0$ is a \emph{necessary} condition. To see this, we suppose that $g$ and  $g'=g + \delta g$  are two metrics on $M$. We have shown above that
\begin{equation}
    [\mathcal{R'}] = [\mathcal{R}] + [\delta \mathcal{R}]\,,
\end{equation} 
for any $g'$, and also that $[\delta\mathcal{R}]=[d\omega]$ for some $\omega$.

If we now suppose that $g'$ is a Ricci-flat metric, then $[\mathcal{R'}]=0$, and so $[\mathcal{R}]=-[d\omega]$, and it follows that $[\mathcal{R}]=0$, and so $c_1(M)=0$, as claimed. The \emph{sufficiency} of $c_1(M)=0$ is part of the content of Yau's theorem, as we will explain further below.

\subsection{Berger classification}\label{ss:holonomy}
\index{Berger classification}

Suppose $X_{n}$ is a Riemannian manifold, $\mathrm{dim}_{\bR}(X_{n})=n$, with metric $g$. Let us fix a point $p\in X_{n}$ and consider a closed loop $\gamma$ starting at $p$. We consider the parallel transport of a tangent vector $v\in T_{p}X_{n}$ around the closed loop $\gamma$. The set of transformations resulting from all such loops is the \emph{holonomy group}\index{Holonomy group} $\mathrm{Hol}_{p}(\Gamma)$, where $\Gamma$ is the connection used on the tangent bundle.

Now, $\mathrm{Hol}_{p}(\Gamma)$ and $\mathrm{Hol}_{p^{\prime}}(\Gamma)$ are related by conjugation in $\mathrm{GL}(n,\bR)$, so we do not care about the base point, since we consider connected manifolds anyway. If $\Gamma$ is the Levi-Civita connection $\nabla$, then we write
\begin{equation}
    \mathrm{Hol}(X_{n})\equiv \mathrm{Hol}(\nabla)\,,
\end{equation}
which is called the \emph{Riemannian holonomy group}\index{Riemannian holonomy group}.

\begin{equ}[Berger classification]

Suppose $X_{n}$ is orientable and simply-connected. Then $X_{n}$ is either a product, a symmetric space $[G/H]$, or one of the following holds:
\begin{enumerate}
\item $\mathrm{Hol}(X_{n})=\mathrm{SO}(n)$
\item $n=2m$, $\mathrm{Hol}(X_{n})=\mathrm{U}(m)\subset\mathrm{SO}(2m)$
\item $n=2m$, $\mathrm{Hol}(X_{n})=\mathrm{SU}(m)\subset\mathrm{U}(m)\subset\mathrm{SO}(2m)$ 
\item $n=4m$, $\mathrm{Hol}(X_{n})=\mathrm{Sp}(m) \subset\mathrm{SO}(4m)$
\item $n=4m$, $\mathrm{Hol}(X_{n})=\mathrm{Sp}(m)\mathrm{Sp}(1)\subset\mathrm{SO}(4m)$, but $\mathrm{Sp}(m)\mathrm{Sp}(1)\nsubseteq \mathrm{U}(2m)$
\item $n=7$, $\mathrm{Hol}(X_{n})=\mathrm{G}_{2}\subset\mathrm{SO}(7)$
\item $n=8$, $\mathrm{Hol}(X_{n})=\mathrm{Spin}(7)\subset\mathrm{SO}(8)$
\end{enumerate}
\end{equ}

There are various definitions of symplectic groups.  What we have denoted by $\mathrm{Sp}(m)$ is known as the compact symplectic group, or the quaternionic unitary group, and is sometimes written $\mathrm{USp}(2m)$. A key point is that $\mathrm{Sp}(1)$ is equivalent to $\mathrm{SU}(2)$, so in case (4) with $m=1$, we have $\mathrm{Hol}(X_{n})=\mathrm{SU}(2)$: this is the case of a K3 surface, which is hyper-K\"ahler and is a Calabi-Yau twofold.

\begin{table}[t!]    
    \centering
    \resizebox{0.99\linewidth}{!}{
    \begin{tabular}{|c|c|c|c|c|c|c}
    \hline 
    & & &  &  &  \\[-1em]
    Dimension $n$ & Holonomy & Type  & K\"ahler & Calabi-Yau & Ricci-flat   \\[0.15em]
    \hline 
    \hline 
     & & &  &  &    \\[-1em]
    any & $\mathrm{SO}(n)$ & generic   & $\times$ & $\times$ & $\times$   \\[0.15em]
    \hline 
    & & &  &  &    \\[-1em]
    $n=2m$ & $\mathrm{U}(n/2)$ & special   & $\checkmark$ & $\times$ & $\times$   \\[0.15em]
    \hline 
    & & &  &  &    \\[-1em]
    $n=2m$ & $\mathrm{SU}(n/2)$ & special    & $\checkmark$ & $\checkmark$ & $\checkmark$   \\[0.15em] 
    \hline 
    & & &  &  &    \\[-1em]
    $n=4m$ & $\mathrm{Sp}(n/4)$ & special; hyper-K\"ahler    & $\checkmark$ & $\checkmark$ & $\checkmark$  \\[0.15em] 
    \hline 
    & & &  &  &    \\[-1em]
    $n=4m$ & $\mathrm{Sp}(n/4)\,\mathrm{Sp}(1)$   & special; quaternionic-K\"ahler & $\times$ & $\times$ & $\times$   \\[0.15em] 
    \hline 
    & & &  &  &   \\[-1em]
    $n=7$ & $\mathrm{G}_{2}$ & exceptional    & $\times$ & $\times$ & $\checkmark$  \\[0.15em] 
    \hline 
    & & &  &  &    \\[-1em]
    $n=8$ & $\mathrm{Spin}(7)$ & exceptional    & $\times$ & $\times$ & $\checkmark$   \\[0.15em] 
    \hline 
    \end{tabular} 
    }
    \caption{Properties of $X_n$ with different holonomy groups as classified by Berger.}
    \label{tab:BergerClassification}
\end{table}

We want $n=6$ and Ricci-flatness, so our main interest is in case 3.  Now, $\mathrm{Hol}(X)=\mathrm{SU}(3)$ implies $R_{mn}=0$, but the reverse is not necessarily true.\\ 

\begin{equ}[Special holonomy and Ricci-flatness]\index{Ricci-flatness}\index{Holonomy}\index{Special holonomy}

\noindent Special holonomy implies Ricci-flatness. All \emph{known} Ricci-flat $X_{n}$ have  special holonomy.

\end{equ}

Very conveniently, in the case $\mathrm{Hol}(X)=\mathrm{SU}(3)$, i.e.,~for Calabi-Yau threefolds, the manifolds $X$ are complex and K\"ahler, and can be analyzed using complex, K\"ahler, and algebraic geometry, a massive advantage.

\section{Calabi-Yau manifolds}\label{apA:secA.5.2}

With the above preliminaries, we can now give a precise definition of a
Calabi-Yau manifold. \index{Calabi-Yau manifold}

\begin{equ}[Calabi-Yau manifolds]
 
A \emph{Calabi-Yau $n$-fold} is a simply connected, compact K\"ahler manifold $M$ of complex dimension $n$ that satisfies the following equivalent conditions:
\begin{enumerate}[1.]
    \item[1.] $M$ admits a K\"ahler metric with holonomy contained in $\mathrm{SU}(n)$.
    \item[2.] $M$ admits a nowhere-vanishing $(n,0)$-form $\Omega$.
    \item[3.] $M$ admits a Ricci-flat K\"ahler metric.\index{Ricci-flatness}
    \item[4.] The first Chern class $c_1(M)$ vanishes.
\end{enumerate}

\end{equ}

\noindent Let us point out that different variants of these definitions can be found in the literature. 
The requirement of compactness is sometimes relaxed, while other conditions are tightened: for example, requiring ${\mathrm{Hol}}(M) = \mathrm{SU}(n)$ instead of allowing ${\mathrm{Hol}}(M)$ to be a subgroup of $\mathrm{SU}(n)$. Similarly, some authors do not require that $M$ is simply connected, but in this case the above conditions are not all equivalent. Moreover, certain physically understood singularities are allowed, as we discuss in Chapter~\ref{chap:modulistabilization}.

On a Calabi-Yau $n$-fold, there is a unique $(n,0)$-form, and so $h^{0,n}=h^{n,0}=1$, \index{Hodge numbers} while $h^{0,m}=h^{m,0}=0$ for $m<n$. Then, because $h^{p,q} = h^{n-p,n-q}$ and $h^{p,q} = h^{q,p}$, we conclude that the non-vanishing Hodge numbers a Calabi-Yau threefold are $h^{0,0}=h^{3,3}=h^{3,0}=h^{0,3}=1$; $h^{1,1}$; and $h^{2,1}=h^{1,2}$. The two numbers $h^{1,1}$ and $h^{2,1}$ thus suffice to specify the dimensions of the cohomology groups, and we can write the Euler characteristic as
\begin{equation}\label{apA:eqnA.86}
    \chi(\mathrm{CY}_3) = 2(h^{1,1} - h^{2,1})\ .
\end{equation}

In the remainder, we implicitly assume $n\le 4$. Compactifications of string theory on Calabi-Yau $n$-folds $M$ automatically satisfy the vacuum Einstein equations $R_{MN}=0$ due to the existence of a Ricci-flat metric on $M$. Actually, Calabi-Yau manifolds make up the overwhelming majority of known examples of compact manifolds satisfying the vacuum Einstein equations.\footnote{Joyce's constructions of seven-manifolds of holonomy $G_2$, and of eight-manifolds of holonomy $\mathrm{Spin}(7)$, are the only existing constructions of compact manifolds admitting Ricci-flat metrics without being Calabi-Yau $n$-folds \cite{moroianu2006lectureskahlergeometry}.}\index{Reduced holonomy} In all existing examples of Calabi-Yau manifolds, the explicit Ricci-flat metrics are \emph{unknown}. Instead, their existence is inferred indirectly by e.g. demonstrating that the first Chern class is trivial as a cohomology class. The existence of a Ricci-flat metric then follows from Yau's theorem \cite{YauThm}. In explicit compactifications of string theory on Calabi-Yau manifolds, we are therefore mostly limited to working with their topological data only. However, recent years have seen tremendous progress in constructing Calabi-Yau metrics numerically using either Donaldson's algorithm~\cite{Donaldson:2005hvr,Douglas:2006hz,Douglas:2006rr,Braun:2007sn,Braun:2008jp,Ashmore:2019wzb} or energy functional minimization~\cite{Headrick:2009jz,Anderson:2020hux,Douglas:2020hpv,Jejjala:2020wcc,Larfors:2022nep,Berglund:2022gvm,Gerdes:2022nzr} (see also e.g. \cite{Ahmed:2023cnw, Constantin:2024yxh, Butbaia:2024xgj, Fraser-Taliente:2024etl} for applications), mostly with the help of neural networks. Such novel approaches in computational string theory pave the way towards computing various crucial aspects of the low-energy theory.
 
Calabi-Yau $k$-folds $M$ play an outstanding role in string compactifications because they preserve some supersymmetry. The underlying reasoning is that the holonomy groups $\mathrm{SU}(k) \subset \mathrm{SO}(2k)$ leave invariant one or more spinors. As a concrete example, let us consider the case $k=3$. Through the basic Lie algebra relationship $\mathfrak{so}(6)\cong \mathfrak{su}(4)$, we deduce that spinors $\eta$ and $\bar{\eta}$ on $M$ transform under the representations $\mathbf{4}$ and $\mathbf{\bar{4}}$ of $\mathrm{SU}(4)$, respectively. We decompose $\eta$ into irreducible representations of the holonomy group $\mathrm{SU}(3)$ \index{Calabi-Yau compactification}
\begin{equation}\label{apA:eqnA.85}
    \textbf{4}=\textbf{3}+\textbf{1} \, .
\end{equation}
This decomposition contains the singlet $\textbf{1}$, corresponding to the unique covariantly constant spinor (of positive chirality) on $M$.\index{invariant spinor} This implies that Calabi-Yau threefold compactifications of a ten-dimensional string theory with ${\cal N}=1$ supersymmetry lead to four-dimensional theories that preserve ${\cal N}=1$ supersymmetry, due to the existence of a single (four-real-component) spinor.
 
One is often told that taking $X$ to be a Calabi-Yau threefold is motivated for two reasons:
\begin{enumerate}
    \item $X$ a CY$_3 \Rightarrow R_{mn}=0$\,.
    \item $X$ a CY$_3 \Rightarrow$ SUSY\,.
\end{enumerate}
Both implications are true, as we have explained. However, it is important to recognize that Calabi-Yau threefolds preserve some supersymmetry at the \emph{Kaluza-Klein scale}, which is not necessarily related to low-energy supersymmetry, as in the supersymmetric solution for the electroweak hierarchy problem.  This point is often misunderstood, so let us be very explicit. We are not studying supersymmetric compactifications of the superstring to explain why the Higgs mass is much smaller than the Planck mass, though that would be extremely important if it could be achieved. Instead, we have selected the superstring to have maximal theoretical control on the string worldsheet.  We have selected Calabi-Yau compactifications because these are the only known examples of vacuum solutions of the critical superstring, and our plan is to start with vacuum solutions and add controllably small sources of stress-energy. The fact that Calabi-Yau compactifications necessarily preserve supersymmetry at the Kaluza-Klein scale is a bonus, as it allows us further theoretical control.

\section{Low energy spectrum of type II superstrings}\label{sec:TypeIIspectrum} 
\index{Type II superstring}

With the mathematical preliminaries in place, we now turn to a brief review of type II superstring theory and its low-energy limits. Our main focus throughout these lectures will be on compactifications of type II supergravity, and so it is important to recall how this effective description emerges from the underlying string theory. At energies well below the string scale, the massive excitations of the string decouple, and the dynamics of the massless sector are captured by ten-dimensional type IIA or type IIB supergravity. These supergravity theories therefore provide the starting point for studying compactifications to four dimensions, while higher-order $\alpha'$ and string loop corrections encode the effects of integrating out the heavy modes. Understanding this correspondence between string theory and its supergravity limit will be essential for the discussions to follow.
 
In the Ramond-Neveu-Schwarz (RNS) formulation \cite{Ramond:1971gb,Neveu:1971fz}\index{RNS formalism} of the superstring, one introduces Majorana-Weyl fermions $\psi_{M}$ 
and $\tilde{\psi}_{M}$, and the worldsheet gravitino, in such a way that one has $\cN=1$ supergravity in the two-dimensional worldsheet theory.

The action for superstrings in flat spacetime is 
\begin{equation}\label{eq:superstringaction} 
    S=\dfrac{1}{2\pi\alpha^{\prime}}\int_{\Sigma}\,\dif^{2}z\,\Bigl( \p X^{M}\,\bar{\p} X_{M} + \tfrac{\alpha'}{2}\,\psi^M\bar{\p}\psi_M + \tfrac{\alpha'}{2}\,\tilde{\psi}^M\p \tilde{\psi}_M \Bigr)\,.
\end{equation}
Here $\Sigma$ denotes the string worldsheet, $X^M(z,\bar{z})$ are the embedding coordinates, and we have fixed a convenient gauge and used complex coordinates $z, \bar{z}$ on the worldsheet. The fields $\psi^M$ and $\tilde{\psi}^M$ are left-moving (i.e., holomorphic) and right-moving (i.e., anti-holomorphic) Majorana-Weyl fermions, respectively.

We need to specify the boundary conditions that the fields obey upon traversing the string in the spatial direction, denoted by a coordinate $\sigma \in [0,\ell)$. We will take $X^{M}(\tau,0)=X^{M}(\tau,\ell)$, but we distinguish
\begin{equation}
    \psi^{M}(\tau,\sigma+\ell)=\begin{cases}
    \psi^{M}(\tau,\sigma) & \text{Ramond (R)} \, ,\\[0.4em]
    -\psi^{M}(\tau,\sigma) & \text{Neveu-Schwarz (NS)}\,, 
    \end{cases}
\end{equation} for the left-movers, and similarly
\begin{equation}
    \tilde{\psi}^{M}(\tau,\sigma+\ell)=\begin{cases}
    \tilde{\psi}^{M}(\tau,\sigma) & \text{Ramond (R)} \, ,\\[0.4em]
    -\tilde{\psi}^{M}(\tau,\sigma) & \text{Neveu-Schwarz (NS)}\,, 
    \end{cases}
\end{equation}
for the right-movers.

Since we can choose the boundary conditions independently for left- and right-moving modes, we find four sectors for the closed superstring: (R,R), (R,NS), (NS,R) and (NS,NS). The notation records the boundary condition of the left-moving sector, then that of the right-moving sector, so (NS,R) has Neveu-Schwarz boundary conditions for the left-moving fermions, and Ramond boundary conditions for the right-moving fermions.

For $\psi^{M}$, the solutions to the two-dimensional Dirac equation give rise to mode expansions in terms of raising and lowering  
operators. 
Upon acting on a vacuum state, the raising operators generate an infinite tower of states corresponding to suitable representations of $\mathrm{SO}(8)$.
The lowest-lying states of string excitations are
\begin{equation}\label{eq:LowestStringExcitations} 
\text{NS}=\begin{cases}
\text{NS}_{-}:\; \mathbf{1}&m^{2}<0\, ,\\
\text{NS}_{+}:\; \mathbf{8}_{v}&m^{2}=0\, ,
\end{cases}\qquad \text{R}=\begin{cases}
\text{R}_{-}:\; \mathbf{8}_{c}&m^{2}=0\, ,\\
\text{R}_{+}:\; \mathbf{8}_{s}&m^{2}=0\, ,
\end{cases}
\end{equation}
where $\mathbf{8}_{c}$, $\mathbf{8}_{s}$ are the two spinor representations of $\mathrm{Spin}(8)$ and $\mathbf{8}_{v}$ is the vector representation of $\mathrm{SO}(8)$. The $\pm$ denotes the transformation under the left-moving fermion number operator $(-1)^{F_{L}}$. 
The $\mathrm{NS}_{-}$ sector contains a tachyon that is removed by the \emph{GSO projection}\index{GSO projection} \cite{Gliozzi:1976qd}.
We will state only the final result of the projection; see e.g.~\cite{Blumenhagen:2013fgp} for details.

The outcome of the GSO projection is that 
\begin{enumerate}[i)]
    \item The $\mathrm{NS}_{-}$ sectors on the left and right, denoted
($\mathrm{NS}_{-}$, $\bullet$) and
($\bullet$, $\mathrm{NS}_{-}$), are removed. 
\item The sectors
($\mathrm{NS}_{+}$, $\bullet$) and
($\bullet$, $\mathrm{NS}_{+}$) remain.
\item The Ramond sectors are either 
($\mathrm{R}_{+}$, $\bullet$) and
($\bullet$, $\mathrm{R}_{+}$), 
or
($\mathrm{R}_{+}$, $\bullet$) and
($\bullet$, $\mathrm{R}_{-}$).

\end{enumerate} 

Thus, only one choice remains: whether the left-moving and right-moving Ramond sectors have the same chirality.

\begin{table}[t!]
\centering
\begin{tabular}{|c||c|c|c|}
\hline 
sector &  statistics & $\mathrm{SO}(8)$ & irreps \\ 
\hline 
\hline 
$(\text{NS}_{+},\text{NS}_{+})$ & boson & $ \mathbf{8}_{v}\otimes \mathbf{8}_{v}$ & $\mathbf{1}+\mathbf{28}+\mathbf{35}_{v}$ \\ 
\hline 
$(\text{R}_{+},\text{R}_{+})$ & boson & $ \mathbf{8}_{s}\otimes \mathbf{8}_{s}$ & $\textbf{1}+\textbf{28}+\mathbf{35}_{c}$  \\ 
\hline 
$(\text{R}_{+},\text{R}_{-})$ & boson & $ \mathbf{8}_{s}\otimes \mathbf{8}_{c}$ & $\mathbf{8}_{v}+\mathbf{56}_{t}$\\ 
\hline 
$(\text{R}_{-},\text{R}_{-})$ & boson & $\mathbf{8}_{c}\otimes \mathbf{8}_{c}$ & $\textbf{1}+\textbf{28}+\mathbf{35}_{\tilde{c}}$ \\ 
\hline 
$(\text{NS}_{+},\text{R}_{+})$ & fermion & $ \mathbf{8}_{v}\otimes \mathbf{8}_{s}$ & $\mathbf{8}_{c}+\mathbf{56}_{s}$ \\ 
\hline 
$(\text{NS}_{+},\text{R}_{-})$ & fermion & $ \mathbf{8}_{v}\otimes \mathbf{8}_{c}$ & $\mathbf{8}_{s}+\mathbf{56}_{c}$ \\ 
\hline 
\end{tabular} 
\vspace*{0.2cm}
\caption{Massless closed string states from boundary conditions of the string, see e.g. \cite{Blumenhagen:2013fgp} for equivalent tables including open strings and massive states. The statistics column refers to the \emph{spacetime} statistics of the resulting field.}\label{tab:WeigandSectorString} 
\end{table}

Following \eqref{eq:LowestStringExcitations}, the spacetime spectrum of massless states is determined by certain product representations of $\mathbf{8}_{v}, \mathbf{8}_{c}, \mathbf{8}_{s}$ listed in Table~\ref{tab:WeigandSectorString}. The boldface notation again represents the dimension of the individual representations. Here, $\mathbf{35}_{c}$ ($\mathbf{35}_{\tilde{c}}$) is the self-dual (anti-self-dual) anti-symmetric $4$-tensor representation $C_{4}$ of $\mathrm{SO}(8)$. Similarly, $\textbf{56}_{t}$ is a $3$-form $C_{3}$, while $\mathbf{56}_{s}$ and $\mathbf{56}_{c}$ are helicity $3/2$ gravitini of opposite chirality. Lastly, $\mathbf{35}_{v}$ is the symmetric $2$-tensor representation corresponding to the massless ten-dimensional graviton, and $\mathbf{28}$ is a $2$-form $C_{2}$.

We can now state the field content of the  
type II closed superstring theories.  The two theories are:
\begin{itemize}
    \item \emph{type IIB}:
        \begin{align}
            (\text{NS}_{+},\text{NS}_{+}) &:\quad \phi,\, B_{2},\, G_{MN}\, ,\\[0.3em]
            (\text{R}_{+},\text{R}_{+}) &:\quad C_0,\, C_2,\, C_4 \, ,\\[0.3em]
            (\text{NS}_{+},\text{R}_{+}) &:\quad \lambda,\,\Psi^M\, ,\\[0.3em]
            (\text{R}_{+},\text{NS}_{+}) &:\quad \lambda,\,\Psi^M\,.
        \end{align}
        After using Table~\ref{tab:WeigandSectorString}, the bosonic spectrum is given by a symmetric 2-tensor $G_{MN}$,
        two 
        spin-0 fields $\phi$ and $C_0$,
        a self-dual $4$-form $C_{4}$, and two $2$-forms $B_2$ and $C_{2}$ coming from the NS-NS sector and R-R sector, respectively. A characteristic feature of the type IIB theory is its chiral spectrum, which stems from the fact that left- and right-movers have the same chirality. Indeed, we find two  gravitini\index{Gravitini} $\Psi^M$ of the same chirality, and two Majorana-Weyl spinors $\lambda$, the dilatini\index{Dilatino}, of the same chirality.  Even though it is chiral, type IIB supergravity in ten dimensions is anomaly-free \cite{AlvarezGaume:1983ig}.
    \item \emph{type IIA}:
        \begin{align}
            (\text{NS}_{+},\text{NS}_{+}) &:\quad \phi,\, B_{2},\, G_{MN}\, ,\\[0.3em]
            (\text{R}_{+},\text{R}_{-}) &:\quad C_1,\, C_3\, ,\\[0.3em]
            (\text{NS}_{+},\text{R}_{-}) &:\quad \tilde{\lambda}, \, \widetilde{\Psi}^M\, ,\\[0.3em]
            (\text{R}_{+},\text{NS}_{+}) &:\quad \lambda,\,\Psi^M\,.
        \end{align}
        The bosonic spectrum consists of a symmetric 2-tensor $G_{MN}$, a scalar $\phi$, a $3$-form $C_{3}$, a $1$-form $C_{1}$, and the NS-NS $2$-form $B_2$. Consulting Table~\ref{tab:WeigandSectorString}, we observe that the spectrum is non-chiral, with two gravitini $\Psi^M,\widetilde{\Psi}^M$  of opposite chirality, and two dilatini $\lambda,\tilde{\lambda}$ of opposite chirality.
\end{itemize}

The corresponding supergravity actions for the massless spectrum of these two theories are schematically given by\index{Type II supergravity}
\begin{align}\label{eq:TypeIIAct} 
    S_{\text{IIA/B}} &= \frac{1}{2\kappa_{10}^2}\int\, \biggl \{\mathrm{e}^{-2\phi} \biggl (\mathcal{R}^{(10)}+4(\p\phi)^{2}-\frac{1}{2} |H_{3}|^2\biggl )\nonumber\\[0.4em]
    &\hphantom{= \frac{1}{2\kappa_{10}^2}\int\, \biggl \{\mathrm{e}^{-2\phi}} +\sum_{p \text{ odd/even}} \, a_p\, |F_{p+1}|^2+\ldots \biggl \}\,\sqrt{-g^{(10)}}\dif^{10}x\,,
\end{align}
where $\kappa_{10}^2 = \ell_s^8/4\pi$
is the gravitational coupling constant in ten dimensions, 
$\ell_s$ is the string length, and $\mathcal{R}^{(10)}$ is the ten-dimensional Ricci scalar.
We have introduced the standard notation for the R-R-sector field strength tensors $F_{p+1}=\dif C_{p}$, with kinetic terms normalized by some constants $a_{p}$, and for the NS-NS-sector field strength tensor $H_3=\dif B_2$, as well as the normalization
\begin{equation}\label{eq:defPFB}
    |F_{p+1}|^2 \coloneqq   F_{N_{1}\ldots N_{p+1}} F^{N_{1}\ldots N_{p+1}}\, .
\end{equation}
We omit the fermionic terms that make the full action supersymmetric.

The ellipsis in the action \eqref{eq:TypeIIAct} indicates additional contributions that can broadly be divided into two classes. First, there are \emph{Chern-Simons couplings}, which are topological interaction terms that involve wedge products of differential forms, typically combining potentials such as $C_p$ or $B_2$ with field strengths such as $F_p$ or $H_3$. One finds couplings of the form $\int \, C_4 \wedge H_3 \wedge F_3$ in type IIB and $\int \, B_2\wedge F_4\wedge F_4$ in type IIA.  Importantly, the Chern-Simons couplings enforce consistency conditions such as tadpole cancellation and anomaly cancellation, and also encode the charges carried by extended objects in the theory. As we will see in Chapter~\ref{chap:gkp}, these couplings dictate in flux compactifications how background fluxes source lower-dimensional effective charges, and thus strongly influence moduli stabilization and the structure of vacua.

Second, there are \emph{higher-derivative corrections} to the two-derivative supergravity action. These originate from integrating out the massive modes of string theory and appear as $\alpha'$ corrections in the low-energy effective theory. A prototypical example arises from the scattering of four closed strings: at low energies, this process reduces to a four-graviton interaction, which in the effective action is captured by an operator of the schematic form $R^4$, involving four copies of the Riemann tensor. Such terms modify the dynamics of gravity and moduli fields at higher orders, and their effects will be further elaborated in \S\ref{sec:BBHL}.

Starting from the type II action \eqref{eq:TypeIIAct}, one can construct compactifications to lower dimensions by putting the theory on a compact manifold.
Below, we will try to construct four-dimensional universes by compactifying the type IIB theory on Ricci-flat $6$-manifolds.

\vfill

\newpage

\section{Massless fields in Calabi-Yau reductions}

We now turn to examining the massless fields that arise in four dimensions from 
compactification of supergravity actions of the form \eqref{eq:TypeIIAct} on a Calabi-Yau threefold $X$.  Such fields fall into two classes: 
the geometric moduli of the Ricci-flat metric on $X$, and the fields descending from $p$-form potentials of the ten-dimensional theory reduced on cycles of $X$.  The moduli spring from the gravitational sector of supergravity, while the $p$-forms are thought of as (bosonic) matter fields.  As we will see, supersymmetry combines the moduli and $p$-forms into multiplets.

We will begin with the $p$-forms, because the mathematics is somewhat simpler than for moduli.

\subsection{Harmonic $p$-forms}\label{apA:secA.3.5}

To understand how $p$-form fields in ten dimensions appear in the four-dimensional effective theory, we consider the Maxwell-type action for a $r$-form potential $C_r$,
\begin{equation}\label{Hodgeaction}
    S_{r+1} = \frac{c}{(r+1)!} \int \, F_{M_1 \ldots M_{r+1}} F^{M_1 \ldots M_{r+1}}\, \sqrt{-g^{(10)}}\, \mathrm{d}^{10} x = c\, \langle F_{r+1}, F_{r+1} \rangle \ , 
\end{equation} 
where $F_{r+1} = \mathrm{d} C_{r}$ is the $(r+1)$-form field strength resulting from an $r$-form potential $C_{r}$, and $c$ is a constant.
The equation of motion is 
\begin{equation}
    \mathrm{d}^{\dagger} \,\mathrm{d} C_{r} = 0 \,, \label{ddagB}
\end{equation}
and upon imposing the gauge condition $\,\mathrm{d}^{\dagger} C_{r} = 0$,
(\ref{ddagB}) takes the form
\begin{equation}
    \Delta C_{r} = 0\ . \label{DeltaC}
\end{equation}
For a product spacetime ${\cal M}_{10} = {\cal M}_{4} \times M$, which we take to have coordinates $x^\mu$ and $y^m$, respectively,
we can write the Laplacian $\Delta$ as
\begin{equation}
\Delta = \Delta_4 + \Delta_M \ ,
\label{apA:eqnA.47}
\end{equation}
where $\Delta_M$ is the Laplacian on $M$, and $\Delta_4 \equiv \Box$ is the d'Alembertian.
Taking the $r$-form ansatz
\begin{equation}
C_r = A_p(x) \wedge B_{r-p}(y) \,,
\label{apA:eqnA.48}
\end{equation}  
with $B_{r-p}$ an eigenfunction of $\Delta_M$, 
\begin{equation}
\Delta_M B_{r-p} = \lambda B_{r-p}\,,
\end{equation}
the equation of motion (\ref{DeltaC}) takes the form
\begin{equation}
(\Box + \lambda) A_p = 0 \ .
\end{equation}
Thus, in a theory with an $r$-form potential, massless $p$-forms $A_p$ in four dimensions result from harmonic $(r-p)$-forms.

Let us expand on this key result. 
We make use of the relations between massless fields and harmonic forms, and between harmonic forms and cohomology classes, as follows.
To find massless fields given an $r$-form $C_r$, we take the ansatz
\eqref{apA:eqnA.48}, but now with $B_{r-p}$ a \emph{harmonic} $(r-p)$ form $\omega_{r-p}$,
\begin{equation}
C_r = A_p(x) \wedge \omega_{r-p}(y) \,,
\end{equation}  
and we take $\Sigma_{r-p}$ to be an $(r-p)$-cycle in the class dual to $\omega_{r-p}$.
Then
\begin{equation}\label{eq:intoncyc}
   \int_{\Sigma_{r-p}} C_{r} = \int_{\Sigma_{r-p}} A_p \wedge \omega_{r-p} = A_p\,.
\end{equation}
In view of \eqref{eq:intoncyc}, we say that the massless $p$-form $A_p$ results from integrating $C_r$ over (or `reducing $C_r$ on') an $(r-p)$-cycle, or equivalently from
expanding $C_r$ in terms of a harmonic $(r-p)$-form.
 
We conclude that
\begin{equ}[Harmonic forms and massless fields]
    An $r$-form potential in ten dimensions furnishes one massless $p$-form in four dimensions for each of the $b^{r-p}$ harmonic $(r-p)$-forms on $M$.\index{harmonic form}
\end{equ}
\noindent
Thus, the topological data of $M$, encoded in its Betti numbers, fixes the number of massless fields in the four-dimensional effective theory that result from
the ten-dimensional $r$-form potential $C_r$.
For example, a $2$-form potential $B_2$ expanded in harmonic $1$-forms of $M$ yields $b^1$ massless vectors in four dimensions, while expansion in harmonic $2$-forms yields $b^2$ axion-like scalars.

To find all massless fields resulting from $C_r$, we expand in terms of all harmonic $k$-forms with $k \le r$,
\begin{equation}\label{eq:Crdecomp}
   C_r(x,y)=\sum_{p=0}^{r}\sum_{I=1}^{b^{\,r-p}} A_p^{\,I}(x)\wedge \omega_{r-p,I}(y)\;+\;\text{non-harmonic modes}\, ,
\end{equation}
where $\{\omega_{r-p,I}\}$ is a basis of harmonic $(r-p)$-forms on $M$. The omitted modes are eigenmodes of $\Delta_M$ with $\lambda>0$ and give rise to massive fields in four dimensions, which decouple at low energies.  

To compute the kinetic terms for   the four-dimensional $p$-form fields $A_p^{\,I}$, we reduce the kinetic term $\langle F_{r+1},F_{r+1}\rangle$ appearing in the ten-dimensional action \eqref{Hodgeaction}.
The field-space metric in the effective theory is then determined by overlap integrals over the internal manifold $M$. 
For example, for pseudo-scalar 
fields $a^I \coloneqq A^{I}_0$ arising from $C_r$ with $p=0$, one finds
\begin{equation}\label{eq:zeroform}
    S_{\rm kin}^{(4)} \supset -\frac{1}{2} \int \; G_{IJ}\,\partial_\mu a^I\partial^\mu a^J\, \sqrt{-g^{(4)}}\,\mathrm{d}^4x
    \; ,\quad 
    G_{IJ}\coloneqq \int_M \omega_{r,I}\wedge \star\,\omega_{r,J}\, .
\end{equation}
The field-space metric $G_{IJ}$ is crucial for understanding the couplings, dynamics, and potential terms for these fields. The result \eqref{eq:zeroform} is readily generalized to $p$-form fields with $p>0$: the resulting four-dimensional fields $A_p^I$ inherit kinetic terms and possible topological couplings from the structure of $M$. Later, in \S\ref{sec:kahlercoord}, we will present an explicit computation of the metric $G_{IJ}$ in the setting of type IIB Calabi-Yau compactifications with $C_4$ axions. There we also clarify how this metric is related to the Weil-Petersson metric on K\"ahler moduli space.

\begin{table}[t!]
\centering
\begin{tabular}{|c|c|c|c|c|}
\hline 
& & & & \\[-1em]
& 10D & scalars & $1$-forms & $2$-forms  \\[0.2em]
\hline 
\hline
\multirow{6}{*}{\vspace{0.3cm}type IIB} 
& & & & \\[-1em]
& $C_4$ &  $C_{i \bar{\jmath} k \bar{l}}$ & $C_{\mu i j \bar{k}}$, $C_{\mu i j k}$, $C_{\mu \bar{\imath}\bar{\jmath} k}$, $C_{\mu \bar{\imath}\bar{\jmath} \bar{k}}$ & $C_{\mu\nu i\bar{\jmath}}$   \\[0.2em]
\cline{2-5}
& & & & \\[-1em]
& $C_2$ &    $C_{i\bar{\jmath}}$ & -- & $C_{\mu\nu}$  \\[0.2em] 
\cline{2-5}
& & & & \\[-1em]
& $B_2$ &    $B_{i\bar{\jmath}}$ & -- & $B_{\mu\nu}$ \\[0.2em] 
\cline{2-5}
& & & & \\[-1em]
& $C_0$ &  $C_0$ & -- & --\\[0.2em]
\hline 
\hline
\multirow{4}{*}{\vspace{-0.2cm}type IIA} & & & & \\[-1em]
& $C_3$ & $C_{i j \bar{k}}$, $C_{\bar{\imath} \bar{\jmath}k}$, $C_{i j k}$, $C_{\bar{\imath} \bar{\jmath} \bar{k}}$ & $C_{\mu i \bar{\jmath}}$ & --  \\[0.2em] 
\cline{2-5}
& & & & \\[-1em]
& $B_2$ &  $B_{i\bar{\jmath}}$ & -- & $B_{\mu\nu}$   \\[0.2em]
\cline{2-5}
& & & & \\[-1em]
&$C_1$ &  -- & $C_{\mu}$ & --  \\[0.2em] 
\hline 
\end{tabular} 
\caption{Independent modes of $p$-form potentials obtained by expanding them in harmonic forms on the Calabi–Yau threefold $X$.}\label{tab:IIABPforms} 
\end{table}

Let us now specialize to the case of Calabi–Yau threefold compactifications of type II superstring theory.
As discussed in \S\ref{sec:TypeIIspectrum}, the massless spectra of type IIA and type IIB contain different $p$-form potentials in the R–R sector. Expanding the ten-dimensional fields in terms of harmonic forms on the Calabi–Yau threefold $X$ yields the four-dimensional fields summarized in Table~\ref{tab:IIABPforms}; see, for example, \cite{Grimm:2004uq,Grimm:2004ua} for further details. A more precise counting of these components will be provided later in \S\ref{sec:N2multiplets}, after the construction of the relevant harmonic forms on $X$ in the following section.

\subsection{Moduli of Calabi-Yau manifolds}\label{sec:moduliofCY}

Ricci-flat metrics on Calabi–Yau threefolds do not exist in isolation --- they come in continuous families. In other words, there is a \emph{moduli space} of solutions to the equation
\begin{equation}
    R_{mn} = 0\, ,
\end{equation}
parametrizing the different Ricci-flat metrics compatible with a given Calabi-Yau threefold topology.

To see this concretely in the simplest example, suppose we are given a topological manifold $X$ admitting a metric $g$ with $\mathrm{Hol}_{g}(X)=\mathrm{SU}(3)$, i.e.,~such that 
\begin{equation}
    R_{mn}(g)=0\,.    
\end{equation}
One simple way to generate a family of Ricci-flat metrics is through an overall rescaling of the metric. For any positive real number $\lambda \in \mathbb{R}_+$, define $g^{\prime} \coloneqq \lambda g$. Then the Ricci tensor remains zero:
\begin{equation}
    R_{mn}(g^{\prime})=0\,.    
\end{equation}
This rescaling corresponds to a global dilation of the manifold and is often referred to as the `breathing mode.’ It controls the overall size of the Calabi-Yau and plays an important role in the low-energy effective theory arising from string compactifications. 

However, the breathing mode is just one example. In general, Calabi–Yau manifolds admit many more moduli, corresponding to continuous deformations of the metric that preserve Ricci-flatness and $\mathrm{SU}(3)$ holonomy. These additional moduli capture variations in the shape, rather than the overall size, of the manifold and lead to a rich moduli space structure that we will explore throughout these lectures.

In terms of coordinates $z^1, z^2, z^3 \equiv \{z^i\}$ and $\bar{z}^{\bar{1}}, \bar{z}^{\bar{2}}, \bar{z}^{\bar{3}} \equiv \{\bar{z}^{\bar{\imath}}\}$, suppose that $g_{i\bar{\jmath}}$ is a Hermitian Ricci-flat metric on $X$, i.e.
\begin{equation}
    g_{ij}=g_{\bar{\imath}\bar{\jmath}}=0\,,
\end{equation} and
\begin{equation}
    R_{i\bar{\jmath}}(g)=0\,.
\end{equation} 
Let us consider an infinitesimal change $g \to g + \delta g$, such that
\begin{equation}
    (g+\delta g)_{ij}=(g+\delta g)_{\bar{\imath}\bar{\jmath}}=0\,,
\end{equation} and
\begin{equation} \label{eq:stillricci}
    R_{i\bar{\jmath}}(g+\delta g)=0\,.
\end{equation} 
We fix a convenient gauge,
\begin{equation}
    \nabla^m \delta g_{mn} - \frac{1}{2} \nabla_n \delta g^{m}\,_{m}=0\,,
\end{equation}
in which the condition \eqref{eq:stillricci} linearizes to give the \emph{Lichnerowicz equation}\index{Lichnerowicz equation},
\begin{equation}\label{eq:Lichnerowicz_equation}
    \mathscr{L}\delta g \coloneqq \nabla^m \nabla_m \delta g_{pq} + 2 R_{p}^{~r}\,_{q}^{~s}\,\delta g_{rs} = 0\,.
\end{equation}
We write the general metric fluctuations $\delta g$ as
\begin{equation}
    \delta g = \delta g_{i\bar{\jmath}} \dif z^i \dif\bar{z}^{\bar{\jmath}} + \delta g_{ij} \dif z^i \dif z^{j} + \text{c.c.} 
\end{equation} 
Because the metric $g$ is Hermitian, the only non-vanishing components of the Riemann tensor, up to symmetries, are of the form $R_{k\bar{\imath}l\bar{\jmath}}$. It follows that in \eqref{eq:Lichnerowicz_equation} the terms involving $i\bar{\jmath}$ decouple from the terms involving $ij$. Thus, there are two independent classes of deformations to the Ricci-flat metric satisfying the
\begin{equ}[Lichnerowicz equations]
    \vspace*{-0.5cm}
    \begin{subequations}\label{eq:Lichnerowicz_equation_complex}
    \begin{equation}
        \mathscr{L}\delta g_{i\bar{\jmath}}\coloneqq g^{k\bar{\ell}}\nabla_{k} \nabla_{\bar{\ell}} \delta g_{i\hspace{1pt}\bar{\jmath}} +  R_{i}^{~k}\,_{\bar{\jmath}}^{~\bar{\ell}}\,\delta g_{k\hspace{1pt}\bar{\ell}} = 0\; ,\label{eq:Lichnerowicz_equation_kahler}
    \end{equation}
    \begin{equation}
        \mathscr{L}\delta g_{ij}\coloneqq g^{k\bar{\ell}}\nabla_{k} \nabla_{\bar{\ell}} \delta g_{i\hspace{1pt}j} +  R_{i}^{~k}\,_{j}^{~\ell}\,\delta g_{k\hspace{1pt}{\ell}} = 0\,.\label{eq:Lichnerowicz_equation_csm}
    \end{equation}
    \end{subequations}
\end{equ}

Let us first consider the deformation $\delta g_{i\bar{\jmath}}$ obeying $\mathscr{L}\delta g_{i\bar{\jmath}}=0$.
From such a deformation we can construct a harmonic $(1,1)$ form,
\begin{equation}\label{eq:kahlerdeformations}
    \delta J \coloneqq \delta g_{i\bar{\jmath}}\,\dif z^i \wedge \dif\bar{z}^{\bar{\jmath}} \,.
\end{equation}
By Hodge's theorem, harmonic $(p,q)$-forms are in one-to-one correspondence with Dolbeault cohomology groups $H^{p,q}_{\bar{\partial}}$, which are counted by the \emph{Hodge numbers} $h^{p,q}\coloneqq\mathrm{dim}\,H^{p,q}_{\bar{\partial}}$, see \S\ref{apA:secA.4}.
So the number of deformations given by \eqref{eq:kahlerdeformations} is $h^{1,1}$.  We call these deformations \emph{K\"ahler moduli}. One can show the following useful corollary.

\begin{equ}[K\"ahler deformations]

$\delta J\coloneqq  \delta g_{i\bar{\jmath}}\,\dif z^i \wedge \dif\bar{z}^{\bar{\jmath}}$ is harmonic if and only if $\mathscr{L}\delta g_{i\bar{\jmath}}=0$.

\end{equ}

\begin{proof}

    Using the definition of $\delta J$ in \eqref{eq:kahlerdeformations} and the action of the Laplacian \eqref{apA:eqnA.38} on $(1,1)$-forms,
    we can write
    \begin{align}
        \Delta_d\, \delta g_{i\bar{\jmath}} = -g^{k\bar{\ell}}\nabla_{k} \nabla_{\bar{\ell}}\,\delta g_{i\bar{\jmath}}-R_{i}\,^{ k}\,_{\bar{\jmath}}^{~\bar{\ell}}\,\delta g_{k\bar{\ell}}\,,
    \end{align}
    which vanishes due to \eqref{eq:Lichnerowicz_equation_kahler}.
\end{proof}

One likewise finds $h^{2,1}$ deformations $\delta g_{ij}$ obeying $\mathscr{L}\delta g_{ij}=0$. These are infinitesimal deformations of the complex structure such that $g+\delta g$ becomes Hermitian. 
The corresponding harmonic forms are
\begin{equation}\label{eq:delcs}
    \delta_{\mathrm{cs}} = \Omega_{ijk} \delta g_{\bar{h}\bar{\ell}}g^{k\bar{h}}\, \dif z^{i}\wedge \dif z^{j}\wedge \dif\bar{z}^{\bar{\ell}}\,.
\end{equation}
These deformations are called \emph{complex structure moduli} and have the following property.

\begin{equ}[Harmonic $(2,1)$-forms]

$\delta_{\text{cs}}\coloneqq \Omega_{ijk} \delta g_{\bar{h}\bar{\ell}}g^{k\bar{h}}\, \dif z^{i}\wedge \dif z^{j}\wedge \dif\bar{z}^{\bar{\ell}}$ is harmonic if and only if $\mathscr{L}\delta g_{ij}=0$.

\end{equ}

\begin{proof}

Let us define (see e.g.~\cite{Candelas:1987is,CANDELAS1988458})
    \begin{equation}\label{eq:def1formscs}
        \delta g^i = \delta g^i_{\bar{\jmath}}\, \mathrm{d}\bar{z}^{\bar{\jmath}}\kom \delta g^i_{\bar{\jmath}} = g^{i\bar\ell}\delta g_{\bar{\ell}\bar{\jmath}} 
    \end{equation}
    which is a $TM$-valued $1$-form. Using the definition \eqref{apA:eqnA.64} for the Laplacian $\Delta_{\bar{\partial}}$, we have
    \begin{align}\label{eq:harmdg}
        \Delta_{\bar{\partial}} \delta g^i_{\bar{\jmath}} &= \left (\bar{\partial}\bar{\partial}^{{\dagger}}+\bar{\partial}^{{\dagger}}\bar{\partial}\right )\delta g^i_{\bar{\jmath}}\nonumber\\
        &=-g^{k\bar{\ell}}\nabla_{k} \nabla_{\bar{\ell}}\, \delta g^{i}_{\bar{\jmath}}+R^{\bar{k}}\,_{\bar{\jmath}}^{~i}\,_{\ell}\delta g^{\ell}_{\bar{k}}\nonumber\\
        &=-g^{i\bar{m}}\biggl [g^{k\bar{\ell}}\nabla_{k} \nabla_{\bar{\ell}}\, \delta g_{\bar{m}\bar{\jmath}}+R_{\bar{\jmath}}^{~\bar{k}}\,_{\bar{m}}^{~\bar{h}}\delta g_{\bar{h}\bar{k}}\biggl ]\,,
    \end{align}
    where the sum of terms in the brackets vanishes due to \eqref{eq:Lichnerowicz_equation_csm}, and thus $\delta g^i\in H^{1}(M,TM)$.
    Since $\Omega$ is holomorphic, there is an isomorphism between $H^{1}(M,TM)$ and $H^{2,1}_{\bar{\partial}}(M)$ by defining the $(2,1)$-forms
    \begin{equation}\label{eq:IDH1H21}
        \Omega_{ijk}\, \dif z^{i}\wedge \dif z^{j}\wedge \delta g^k =\Omega_{ijk} \delta g_{\bar{h}\bar{\ell}}g^{k\bar{h}}\, \dif z^{i}\wedge \dif z^{j}\wedge \dif\bar{z}^{\bar{\ell}}\,.
    \end{equation}
    These $(2,1)$-forms are clearly harmonic due to \eqref{eq:harmdg}.
\end{proof}

\begin{equ}[Complex structure deformations]
$\delta_{\text{cs}}$ is a change of complex structure $\mathcal{J}$ on $M$.
Equivalently, deformations of the complex structure $\mathcal{J}$ on $M$ correspond to $\bar{\partial}$-closed elements of $H^{1}(M,TM)$.

\end{equ}
To see that $\delta_{\text{cs}}$ changes the complex structure, we provide a heuristic, and then a proof.  The heuristic is
that having a form $\delta g_{\bar{\imath}}^k\,\mathrm{d}\bar{z}^{\bar{\imath}}$
allows one to write
\begin{equation}
        \mathrm{d} {z^\prime}^i = \mathrm{d} z^i -\delta g_{\bar{\imath}}^k\,\mathrm{d}\bar{z}^{\bar{\imath}}\, ,
    \end{equation}
which ``changes the $z^i$ coordinates''.
Here is the more detailed argument.
\begin{proof}

    Recall from above that, locally in each patch $U_{\alpha}\subset M$ with coordinates $z^{1},z^2,z^3$, the complex structure ${\cal J}$ can be written in the form \eqref{canonicalACS}.

    For any point $p\in U_{\alpha}\subset M$, the complex structure ${\cal J}$ defines a decomposition $T_p M^{\mathbb{C}}  = T_p M^{(1,0)}\oplus T_p M^{(0,1)}$ into the positive and negative eigenspaces spanned by $\lbrace \partial/\partial z^i\rbrace$ and $\lbrace \partial/\partial \bar{z}^{\bar{\imath}}\rbrace$, respectively.
    
    Following \cite{todorov1989weil}, we then consider a continuous family ${\cal J}(t)$ of almost complex structures on $M$ such that ${\cal J}(0)={\cal J}$. For every $t$, there is a decomposition of the tangent and co-tangent bundle as
    \begin{equation}
        TM^{\bC}=T_{t}^{1,0}\oplus T_{t}^{0,1}\kom T^{*}M^{\bC}=\Omega_{t}^{1,0}\oplus \Omega_{t}^{0,1}\, .
    \end{equation}
    The deformation ${\cal J}(t)$ of ${\cal J}$ is encoded, for small $t$, in a map
    \begin{equation}
        \phi(t):\, T^{0,1}\rightarrow T^{1,0}\kom \phi(t)=- \mathrm{pr}_{T^{1,0}}\circ \mathrm{pr}_{T_{t}^{0,1}}^{-1}
    \end{equation}
    in terms of projection maps
    \begin{equation}
        \mathrm{pr}_{T^{1,0}}:\, T_{\bC}M\rightarrow T^{1,0}\kom \mathrm{pr}_{T_{t}^{0,1}}:\, T^{0,1}_{t}\rightarrow T^{0,1}\, .
    \end{equation}
    Let us now consider the expansion of $\phi(t)$ in $t$
    \begin{equation}\label{eq:ExpansionPhi} 
        \phi(t)=\phi_{1} t+\phi_{2}t^{2}+\ldots
    \end{equation}
    where $\phi_{i}\in \Gamma(\Omega^{0,1}\otimes T^{1,0})$, $i>1$, are smooth sections of the bundle $\Omega^{0,1}\otimes T^{1,0}$.
    
    When promoting this family of ACSs ${\cal J}(t)$ to complex structures, they have to be integrable in the sense that (see e.g. Prop.~2.6.27 in \cite{DT1})
    \begin{equation}\label{eq:IntegrableCS} 
        [T_{t}^{0,1},T_{t}^{0,1}]\subset T_{t}^{0,1}\, ,
    \end{equation}
    in terms of the Lie bracket of vector fields which must  preserve $T_{t}^{0,1}$. In fact, \eqref{eq:IntegrableCS} is equivalent to the \emph{Maurer-Cartan equation} for $\phi(t)$
    \begin{equation}\label{eq:MCEPhi} 
        \bar{\p}\phi(t)+[\phi(t),\phi(t)]=0\, .
    \end{equation}
    In terms of the expansion \eqref{eq:ExpansionPhi} one finds the recursive system of equations
    \begin{align}\label{eq:RecursiveMCEPhi} 
        0&=\bar{\p}\phi_{k}+\sum_{i=1}^{k-1}\, [\phi_{i},\phi_{k-i}]\kom k=1,2,\ldots\, .
    \end{align}
    We now focus on the first equation in \eqref{eq:RecursiveMCEPhi} corresponding to
    \begin{equation}
        \bar{\p}\phi_{1}=0\, .
    \end{equation}
    That is, the Maurer-Cartan equation \eqref{eq:RecursiveMCEPhi} proves that the first-order deformation $\phi_{1}$ of the complex structure ${\cal J}$ is described by a $\bar{\p}$-closed $(0,1)$-form $\phi_{1}$ with values in the holomorphic tangent bundle $TM$.
    
    More generally, the \emph{Kodaira-Spencer class} of deformations ${\cal J}(t)$ of the complex structure ${\cal J}$ is given by the induced cohomology class
    \begin{equation}
        [\phi_{1}]\in H^{1}(M,TM)\, .
    \end{equation}
    A basis of $H^{1}(M,TM)$ is given by $\lbrace \delta g^i\rbrace$ defined in \eqref{eq:def1formscs} for which the standard basis on eigenspaces of $T_p M^{\mathbb{C}}$ can be locally expressed as
    \begin{equation}
        \dfrac{\partial}{\partial \bar{z}^{\bar{\imath}}}+\delta g_{\bar{\imath}}^k\dfrac{\partial}{\partial z^{k}}\,,
    \end{equation}
    with associated basis 1-forms
    \begin{equation}
        \mathrm{d} z^i -\delta g_{\bar{\imath}}^k\,\mathrm{d}\bar{z}^{\bar{\imath}}\, .
    \end{equation}
    Thus, after contracting with the holomorphic $3$-form, we see that indeed $\delta_{\text{cs}}$ corresponds to a change in complex structure.
    
\end{proof}

\subsection{Field content of $\mathcal{N}=2$ multiplets}\label{sec:N2multiplets}

\begin{table}[t!]
\centering
\begin{tabular}{|c|c|c|c|c|c|c|}
\hline 
&&&&&&\\[-1em]
SUSY&\backslashbox{multiplet}{spin} &   $0$ & $1/2$ & $1$&$3/2$& $2$ \\ 
\hline 
\hline
\multirow{3}{*}{\vspace{-0.5cm}$\cN=1$} 
&&&&&&\\[-1em]
& gravity   & 0& 0& 0 & 1& 1  \\[0.1em]
\cline{2-7}
&&&&&&\\[-1em]
&vector   &   0 & 1 &1 & 0 & 0 \\[0.1em]
\cline{2-7}
&&&&&&\\[-1em]
&chiral   & 1 & 1 & 0 & 0& 0\\[0.1em]
\hline 
\hline
\multirow{3}{*}{\vspace{-0.5cm}
$\cN=2$} &&&&&&\\[-1em]
&gravity  &  0 & 0 &1 & 2& 1  \\[0.1em]
\cline{2-7}
&&&&&&\\[-1em]
&vector  &     1 &2 & 1  & 0 &0\\[0.1em] 
\cline{2-7}
&&&&&&\\[-1em]
&hyper &    2 & 1+1 & 0& 0 &0 \\[0.1em] 
\hline 
\hline
\multirow{2}{*}{\vspace{-0.3cm}$\cN=4$}&&&&&&\\[-1em]
&gravity  &   2 & 4+4 &6+6 & 4+4 & 1+1 \\[0.1em] 
\cline{2-7}
&&&&&&\\[-1em]
&vector &  6 &4+4 & 1+1 & 0& 0 \\[0.1em] 
\hline 
\hline
&&&&&&\\[-1em]
$\cN=8$ &gravity  &   70 & 56+56 & 28+28 & 8+8 & 1+1 \\[0.1em] 
\hline 
\end{tabular} 
\caption{Massless supermultiplets in four dimensions for $\cN\leq 8$.  For each  multiplet type and spin, the number of states is listed.
}\label{tab:MasslessMultFourDim} 
\end{table}

Let us now bring together the results of \S\ref{apA:secA.3.5} and \S\ref{sec:moduliofCY} by assembling the fields into appropriate supermultiplets. For convenience, we briefly review some terminology here and refer the reader to standard textbooks such as \cite{Weinberg:2000cr} for a more comprehensive discussion of supersymmetric quantum field theories. We will focus on massless multiplets in a theory with $\mathcal{N}$ supercharges, classifying them according to their maximal spin $s_{\text{max}}$:
\begin{itemize}
    \item \emph{gravity multiplets}, $s_{\text{max}}=2$: these contain the graviton, $\cN$ gravitini of 
    spin $3/2$, and additional fields determined by the supersymmetry algebra. The presence of the graviton implies that the theory contains gravitational interactions, resulting in \emph{supergravity} (SUGRA),\index{Supergravity}\index{SUGRA} where supersymmetry is promoted to a local gauge symmetry.
    \item \emph{vector multiplets}, $s_{\text{max}}=1$: these are supersymmetric generalizations of gauge bosons, always containing a spin-$1$ boson paired with fermions and, depending on $\cN$, also scalars.
    \index{Vector multiplet}
    \item \emph{chiral multiplets} and \emph{hypermultiplets}, $s_{\text{max}}=1/2$: these are often  referred to as matter multiplets since they are made up of only scalars and fermions. Moreover, they generally exist only at small enough $\cN$ and low spacetime dimensions. \index{Chiral multiplet}
\end{itemize}
We summarize the relevant multiplets of four-dimensional supergravity theories in Table~\ref{tab:MasslessMultFourDim}. The case with $\mathcal{N}=2$ supersymmetry arises from compactifications on Calabi-Yau threefolds, while $\mathcal{N}=1$ supersymmetry is obtained from orientifolds thereof, which we introduce in \S\ref{sec:orientifolds}. For completeness, we have also included the multiplets in supersymmetric theories with $\mathcal{N}=4$ and $\mathcal{N}=8$.\footnote{Compactifications of type II superstring theory on e.g.~SU$(2)$ structure manifolds yield $\mathcal{N}=4$ theories (see e.g. \cite{Reid-Edwards:2008fql}), and compactifications on six-tori $T^6$ give rise to $\mathcal{N}=8$ theories (see e.g. \cite{Blumenhagen:2013fgp}).}

To reiterate, focusing here on compactifications of type II superstring theory on Calabi-Yau threefolds, the resulting four-dimensional theory is an $\mathcal{N}=2$ supergravity. We described the dimensional reduction of $p$-form fields in \S\ref{apA:secA.3.5} and the additional moduli from deformations of the Calabi-Yau metric in \S\ref{sec:moduliofCY}. We can now assemble the K\"ahler moduli, the complex structure moduli, and the scalars originating from $p$-forms listed in Table~\ref{tab:IIABPforms} into the following multiplets, summarized in Table~\ref{tab:fieldcontentTypeII}.

\begin{table}[t!]
    \centering
    \begin{tabular}{c|c|c|c}
         &  &  & \\[-1em]
         & & IIA & IIB \\[0.3em]
         \hline
         \hline
          & &  & \\[-1em]
         \multirow{ 2}{*}{\vspace{-0.25cm}vector multiplets} & number & $h^{1,1}(X)$ & $h^{2,1}(X)$ \\[0.3em]\cline{2-4}
          & &  & \\[-0.8em]
          & fields & $\delta g_{i\bar{\jmath}}$, $B_{i\bar{\jmath}}$, $C_{\mu i \bar{\jmath}}$ & $\delta g_{ij}$, $\delta g_{\bar{\imath}\bar{\jmath}}$, $C_{\mu i j \bar{k}}$ \\[0.3em]
         \hline
         & &  & \\[-1em]
         \multirow{ 2}{*}{\vspace{-0.25cm}hypermultiplets} & number & $h^{2,1}(X)$ & $h^{1,1}(X)$ \\[0.3em]\cline{2-4}
          & &  & \\[-0.8em]
          & fields & $\delta g_{ij}$, $\delta g_{\bar{\imath}\bar{\jmath}}$, $C_{i j \bar{k}}$, $C_{\bar{\imath} \bar{\jmath}k}$ & $\delta g_{i\bar{\jmath}}$, $B_{i\bar{\jmath}}$, $C_{i\bar{\jmath}}$, $C_{\mu\nu i\bar{\jmath}}$ \\[0.3em]
          \hline
          & &  & \\[-1em]
          universal hypermultiplet & fields & $\phi$, $B_{\mu\nu}$, $C_{i j k}$, $C_{\bar{\imath} \bar{\jmath} \bar{k}}$ & $\phi$, $C_0$, $B_{\mu\nu}$, $C_{\mu\nu}$\\[0.3em]
    \end{tabular}
    \caption{Field content of $\mathcal{N}=2$ multiplets in type II compactifications on Calabi-Yau threefolds.}
    \label{tab:fieldcontentTypeII}
\end{table}

In a type IIB compactification\footnote{In type IIB, the self-duality condition $F_5 = \star_{10} F_5$ eliminates half of the degrees of freedom of $C_4$, which we choose to be $C_{i \bar{\jmath} k \bar{l}}$, $C_{\mu \bar{\imath}\bar{\jmath} k}$, and $C_{\mu \bar{\imath}\bar{\jmath} \bar{k}}$.} on a Calabi-Yau threefold $X$, the complex structure moduli $\delta g_{ij}$ and $\delta g_{\bar{\imath}\bar{\jmath}}$ and $C_{\mu i j \bar{k}}$ form the bosonic content of a vector multiplet.  There are $h^{2,1}(X)$ such multiplets.

The K\"ahler moduli $\delta g_{i\bar{\jmath}}$ combine with $B_{i\bar{\jmath}}$, $C_{i\bar{\jmath}}$, and $C_{\mu\nu i\bar{\jmath}}$ to form the bosonic content of a hypermultiplet.   There are $h^{1,1}(X)$ such multiplets.  The bosons of the universal hypermultiplet consist of\footnote{For an antisymmetric tensor in four dimensions with two spacetime indices, such as $B_{\mu\nu}$,
we dualize to an axion $\vartheta$ using $\dif \vartheta = \star_{4}\dif B_2$.}
$B_{\mu\nu}$, $C_{\mu\nu}$, $C_0$, and the dilaton $\phi$, related to the string coupling by $g_s = \mathrm{e}^{\phi}$. In addition, the gravity multiplet in type IIB consists of the metric $g_{\mu\nu}$ and the $1$-form $C_{\mu i j k}$.

In a type IIA compactification, one has instead $h^{1,1}(X)$ vector multiplets from 
the K\"ahler moduli $\delta g_{i\bar{\jmath}}$, complexified by $B_{i\bar{\jmath}}$, and $C_{\mu i \bar{\jmath}}$.
There are $h^{2,1}(X)$ hypermultiplets consisting of $\delta g_{ij}$, $\delta g_{\bar{\imath}\bar{\jmath}}$, $C_{i j \bar{k}}$, and $C_{\bar{\imath} \bar{\jmath}k}$.
The universal hypermultiplet consists of $\phi$, $B_{\mu\nu}$, $C_{i j k}$, and 
$C_{\bar{\imath} \bar{\jmath} \bar{k}}$. The type IIA gravity multiplet consists of the metric $g_{\mu\nu}$ and the $1$-form $C_{\mu}$.

The multiplet assignments of the moduli have important implications for quantum corrections, as we now explain.
The K\"ahler moduli, which measure the geometric volumes of even-dimensional cycles in the Calabi–Yau threefold, 
correspondingly control the size of perturbative $\alpha'$ corrections, as well as worldsheet instanton and D-brane instanton effects. The dilaton, which governs the string coupling $g_s$, similarly serves as the counting parameter for string loop corrections and D$(-1)$-instanton contributions. Together, the K\"ahler moduli and the dilaton control the magnitude of all relevant quantum corrections and appear explicitly in the exponentials of non-perturbative effects. In other words, any perturbative or non-perturbative contribution to the four-dimensional effective  theory can be expressed as an expansion in terms of these fields, making them the natural expansion parameters of the quantum-corrected moduli space.

With this in mind, we observe
that in type IIA, the K\"ahler moduli are in vector multiplets and  the dilaton is in a hypermultiplet, whereas in type IIB both the K\"ahler moduli and the dilaton are in hypermultiplets.  
As a result, the vector multiplet moduli space in type IIA receives corrections in the $\alpha'$ expansion but not the $g_s$ expansion, while the hypermultiplet moduli space in type IIA receives corrections in the $g_s$ expansion but not the $\alpha'$ expansion, see e.g. \cite{Antoniadis:1997eg,Antoniadis:2003sw}.  But in type IIB, the hypermultiplet moduli space receives \emph{both} types of corrections, while the vector multiplet moduli space receives \emph{neither}. In other words, the vector multiplet moduli space in type IIB string theory compactified on a Calabi-Yau threefold is classically exact.

\subsection{Kinetic terms for moduli}\label{sec:kintermex}

To understand the kinetic terms for the moduli, we consider a simple example: the dilatation or `breathing mode' $g \to \lambda g$ mentioned in \S\ref{sec:moduliofCY}. We parametrize the ten-dimensional background as
\begin{equation}
    \mathrm{d}s^2 = g^{(4)}_{\mu\nu}\dif x^\mu \dif x^\nu +\mathrm{e}^{2u(x)}g^{(6)}_{mn}\dif y^m \dif y^n\,,
\end{equation} 
where $g^{(6)}_{mn}$ is a unit-volume metric and the modulus $u(x)$ determines the size of the compact space.
For this background, the ten-dimensional Ricci scalar $\mathcal{R}^{(10)}$ can be written as
\begin{equation}
    \mathcal{R}^{(10)} = \mathcal{R}^{(4)}- 42(\nabla u)^2-12\nabla^{\mu}\nabla_{\mu} u \, ,
\end{equation}
where $(\nabla u)^2\coloneqq  g^{\mu\nu}_{(4)}\nabla_{\mu} u\, \nabla_{\nu} u$, and
we used that the Ricci scalar for the Calabi-Yau metric vanishes.
Then, the ten-dimensional Einstein Hilbert term
\begin{equation}
    S_{\text{EH}} = \frac{1}{2\kappa_{10}^2}\int\,\mathcal{R}^{(10)}\, \sqrt{-g^{(10)}}\,\dif^{10}x
\end{equation} 
reduced on this background becomes
\begin{equation}
    S_{\text{EH}}=\frac{\mathcal{V}}{2\kappa_{10}^2}\int \dif^4x \,\sqrt{-g^{(4)}}\, \mathrm{e}^{6u} \Big(\mathcal{R}^{(4)}- 42(\nabla u)^2-12\nabla^{\mu}\nabla_{\mu} u \Bigr)\,,
\end{equation}
with
\begin{equation}
    \mathcal{V} \coloneqq \int_X\, \sqrt{g^{(6)}} \dif^6 y\, .
\end{equation}
Integrating the last term by parts, we obtain
\begin{equation}\label{eq:EH4DBW}
    S_{\text{EH}}=\frac{\mathcal{V}}{2\kappa_{10}^2}\int \dif^4x \, \sqrt{-g^{(4)}}\, \mathrm{e}^{6u} \Big(\mathcal{R}^{(4)}+30(\nabla u)^2\Bigr)\, .
\end{equation}
To canonically normalize the Einstein-Hilbert term in four dimensions, we have to perform a Weyl rescaling of the metric by writing
\begin{equation}
    \tilde{g}_{\mu\nu}^{(4)} = \mathrm{e}^{6u}\, g_{\mu\nu}^{(4)}\, .
\end{equation} 
In general, given two $D$-dimensional metrics related by a Weyl rescaling, 
\begin{equation}
    \tilde{g}_{MN} = \mathrm{e}^{-2\omega} g_{MN}\,,
\end{equation}
we have the relation 
\begin{equation}\label{eq:weylformula}
    \mathrm{e}^{2\omega} \mathcal{R}^{(D)} = \tilde{\mathcal{R}}^{(D)} + 2(D-1)\nabla^2 \omega - (D-2)(D-1)(\nabla^M\omega)(\nabla_M \omega)\,.
\end{equation} 
Using \eqref{eq:weylformula} with $\omega=-3u$ and $D=4$, we find 
\begin{equation}
    \sqrt{-g^{(4)}}\, \mathrm{e}^{6u} \mathcal{R}^{(4)} = \sqrt{-\tilde{g}^{(4)}}\biggl (\tilde{\mathcal{R}}^{(4)} -18\nabla^2 u - 54(\nabla u)^2\biggl )\,.
\end{equation}
and thus \eqref{eq:EH4DBW} becomes\footnote{Note that the term $\nabla^2 u$ does not contribute, since it vanishes upon integration by parts.}
\begin{equation}\label{eq:EH4DBW1}
    S_{\text{EH}}=\frac{\mathcal{V}}{2\kappa_{10}^2}\int \dif^4x \,\sqrt{-\tilde{g}^{(4)}}\,\Big(\tilde{\mathcal{R}}^{(4)}-24(\nabla u)^2\Bigr)\, .
\end{equation}
Defining the four-dimensional reduced Planck mass as
\begin{equation}\label{eq:Mplred}
    M_{\mathrm{pl}}^2\coloneqq  \frac{\mathcal{V}}{\kappa_{10}^2}\, ,
\end{equation}
we arrive at a canonical kinetic term for the field
\begin{equation}
    \varphi\coloneqq u \cdot \sqrt{24} M_{\mathrm{pl}}\, .
\end{equation}

The kinetic term that we have just obtained,
\begin{equation}\label{eq:12u}
    \mathcal{L} \supset  -12\, g^{\mu\nu}_{(4)}\nabla_\mu u \,\nabla_{\nu} u = -12\, \partial_\mu u \,\partial^{\mu} u\,,
\end{equation}
follows from a \emph{K\"ahler potential}, as we will explain more fully in Chapter \ref{chap:classicalEFT}. Taking $h^{1,1}=1$ for simplicity, so that $H_4(X,\mathbb{Z})$ has a single generator, $\Sigma_4$,
we define
\begin{equation}
\theta\coloneqq  \int_{\Sigma_4} C_4\,,
\end{equation} 
which is a specific instance of the dimensional reduction of $p$-forms explained above in \S\ref{apA:secA.3.5}. Further defining the complexified K\"ahler modulus
\begin{equation}\label{eq:cxka}
    T\coloneqq  \mathrm{e}^{4u} + \I \theta\,,
\end{equation} 
and writing
\begin{equation}\label{eq:K3lg}
    \mathcal{K} = -3 \,\mathrm{log}\,\bigl(T+\overline{T}\bigr)\,,
\end{equation}
we have
\begin{equation}
    \mathcal{K}_{T\overline{T}} \coloneqq \partial_{T} \partial_{\overline{T}} \mathcal{K}(T,\overline{T}) = \frac{3}{(T+\overline{T})^2}\,,
\end{equation}
and thus
\begin{align}
    \mathcal{K}_{T\overline{T}}\,\partial_\mu T \partial^{\mu} \overline{T} &=
     \frac{3}{(T+\overline{T})^2} \bigl (16\mathrm{e}^{8u} \partial_\mu u\,\partial^{\mu} u +\partial_\mu \theta\,\partial^{\mu} \theta\bigl )\nonumber\\
     &= 12 \partial_\mu u\,\partial^{\mu} u +\dfrac{3}{4}\mathrm{e}^{-8u}\partial_\mu \theta\,\partial^{\mu} \theta\,.
\end{align}
In other words, the complexified K\"ahler coordinate $T$ in \eqref{eq:cxka} combined with the K\"ahler potential \eqref{eq:K3lg} yields the kinetic term \eqref{eq:12u} that we derived directly.
The kinetic terms for more general cases are treated in Chapter \ref{chap:classicalEFT}: see in particular Eq.~\eqref{eq:4DSugra_classical}.

\vfill

\newpage

\section{The moduli problem}
\index{Moduli problem}

We have just seen that compactifications on Calabi-Yau threefolds have moduli --- often hundreds.  This is a problem in cosmology, but to explain why, we will first have to proliferate definitions.

So far we have used the following notion of a modulus:
\begin{equ}[Modulus: definition 1]
\noindent A \emph{modulus} in a Calabi-Yau compactification is an infinitesimal deformation of the metric preserving Ricci-flatness.
\end{equ}
However, we also have a broader meaning of the term:
\begin{equ}[Modulus: definition 2]
\noindent A \emph{modulus} is a spin-0 field with exactly vanishing potential.\\
\noindent  A \emph{pseudomodulus} is a spin-0 field whose potential is small in some sense: for example, vanishing in perturbation theory, or vanishing before supersymmetry breaking.
\end{equ}
The geometric moduli discussed so far --- the K\"ahler moduli and complex structure moduli --- fulfill both of these definitions: they are deformations preserving Ricci-flatness, they are spin-0,
and their potentials vanish exactly before supersymmetry breaking.

The massless spin-0 fields obtained from reducing $p$-forms on $p$-cycles are unrelated to deformations of the Ricci-flat metric, though some are paired with geometric moduli to form \emph{complexified} moduli, but they also have vanishing potential in the $\mathcal{N}=2$ theory, and so they fulfill the second definition but not the first.

A useful third definition is: 
\begin{equ}[Modulus: definition 3]
\noindent A \emph{modulus} is a scalar field with gravitational-strength interactions that do not vanish at zero momentum.\\ 
\end{equ}
The couplings envisaged in this definition take the 
form
\begin{equation}\label{eq:modint}
    \Delta \mathcal{L} \supset c\,\frac{\varphi}{M_{\mathrm{pl}}}\,\mathcal{O}\,,
\end{equation} with $\mathcal{O}$ some operator constructed of other fields, and with the dimensionless Wilson coefficient $c$ being of order unity.
The interaction \eqref{eq:modint} is not invariant under $\varphi \to \varphi +\text{const.}$, and the corresponding Feynman vertex does not involve the momentum of the $\varphi$ particle. 

The utility of definition 3 is that ultralight fields with couplings of the form \eqref{eq:modint} can mediate long-range forces with strength comparable to that of gravity.
Such forces can violate the Equivalence Principle through dependence on elemental composition.

In contrast, shift-symmetric pseudoscalar fields, called \emph{axions},
generally do not mediate significant long-range forces.  For an axion whose couplings preserve CP, the leading interaction induced at tree-level is a dipole-dipole coupling.  The resulting force between unpolarized bodies vanishes, because the dipole couplings average out, and even for a pair of polarized objects the force falls of at least as $1/r^3$ (see e.g.~\cite{Ferrer:1998ue}).  One can consider axion forces by violating CP \cite{OHare:2020wah}, or working beyond tree level, for example in axion backgrounds \cite{Grossman:2025cov}.  However, the limits from such considerations are not nearly as strong as those on definition-3 moduli. 

The geometric moduli --- the real K\"ahler moduli and the complex structure moduli --- arise from the ten-dimensional gravitational action,
\begin{equation}\label{eq:s10grav}
    S_{10} \supset \frac{1}{2\kappa_{10}^2}\int\,\mathcal{R}^{(10)} \sqrt{-g^{(10)}}\dif^{10}x\,,
\end{equation} 
and so their couplings as four-dimensional fields are of gravitational strength.   On the other hand, the $0$-forms obtained from reducing R-R $p$-forms on $p$-cycles generally enjoy shift symmetries, though these symmetries may be broken in certain flux backgrounds.  In the class of compactifications we will discuss, the definition-3 moduli arise from geometric moduli.

Working now with definition 3, we will describe a collection of challenges to realistic cosmology in theories with such moduli: the
\emph{cosmological moduli problem(s)} \cite{deCarlos:1993wie,Banks:1993en}.

\begin{table}[t!]
    \centering
    \begin{tabular}{c|c|c}
          &  & \\[-1em]
          name & mass range & constraints \\[0.3em]
         \hline
         \hline
          &  & \\[-1em]
          ultralight & $m\ll 10^{-3}$ eV $\Leftrightarrow$ $\lambda \gg 1$mm & fifth forces \\[0.3em]
         \hline
         &  & \\[-1em]
          intermediate & $10^{-3}\, \text{eV}\lesssim m\lesssim 30\,$TeV & BBN/DM \\[0.3em]
         \hline
         &  & \\[-1em]
          heavy & $30\,\text{TeV}\lesssim m$ & very early Universe (e.g. inflation) \\[0.3em]
         \hline
    \end{tabular}
    \caption{Different mass ranges for a modulus $\varphi$ of mass $m$.}
    \label{tab:massrange}
\end{table}

\subsection{Tests of fifth forces}
\index{Fifth forces}

Consider a modulus $\varphi$ with couplings of the form
\begin{equation}\label{eq:violco}
\delta\mathcal{L} = c_i \frac{\varphi}{M_{\mathrm{pl}}}\,\mathcal{O}_{\mathrm{SM}}^{i}  \,,
\end{equation}
where the $\mathcal{O}_{\mathrm{SM}}^{i}$ are gauge-invariant operators made of Standard Model fields, and the $c_i$ are some Wilson coefficients.  By the hypothesis of gravitational-strength couplings, the $c_i$ are generally $\mathcal{O}(1)$.

The couplings \eqref{eq:violco} will in general violate the Equivalence Principle: for example, nuclear binding energies, gravitational binding energies, and fermion masses will couple differently to $\varphi$, barring some special structure.

Tests of couplings of the form \eqref{eq:violco} include the E\"ot-Wash torsion balance, which measures relative forces on beryllium and titanium (Be/Ti) \cite{Adelberger:2003zx}. 
Lunar Laser Ranging determines the Earth-Moon distance to millimeter precision \cite{Hofmann:2018myc}, from which one can show that the gravitational and inertial masses of the Earth and Moon obey
\begin{equation}
\left(\frac{m_g}{m_i}\right)_{\text{Earth}} - \left(\frac{m_g}{m_i}\right)_{\text{Moon}} = -3.0 \pm 5.0 \times 10^{-14}\,.\end{equation}
This is a grand extension of Galileo's apocryphal experiment from the Leaning Tower, which was replicated by the Apollo 15 demonstration of dropping an aluminum hammer and a falcon feather on the Moon.

To express the constraints from the torsion balance and LLR experiments, one can replace the simple expression \eqref{eq:violco} with the sophisticated parameterization of Damour and Donoghue \cite{Damour:2010rp}, who find constraints at the $10^{-9}$ level.  Constraints from the asteroid Bennu were obtained in \cite{Tsai:2023zza},
and an analysis of the potential for tests via nuclear clocks appears in \cite{Delaunay:2025lgk}.

One can also constrain light fields that evolve over cosmic time.
The Oklo natural reactor in Gabon is a natural uranium deposit that underwent fission starting around two billion years ago.  
Natural uranium is now $0.72\%$ 
\ch{^{235}U}
 globally, but $0.60\%$ 
\ch{^{235}U}
in the Oklo mine in Gabon.  Two billion years ago, uranium was $3\%$ 
\ch{^{235}U} globally, about the level to which fuel is enriched.  Oxygen in the atmosphere, resulting from early photosynthetic life, made uranium more soluble in water.  The water slowed neutrons, allowing for criticality over a period of around 30 minutes, until the water boiled off.  After a cooling period of about 2.5 hours, reinfiltration of water restarted the reaction.  The cycle continued for more than $100{,}000$ years, until the uranium was appreciably depleted. 

The isotopic abundance of samarium in the Oklo mine implies that the neutron capture cross section of \ch{^{149}SM} has been extremely stable, leading to the limit  
\cite{Shlyakhter:1982yr, Damour:1996zw}
\begin{equation}
    -6.7 \times 10^{-17}\, \mathrm{yr}^{-1} < \frac{\dot \alpha}{\alpha} <5.0\times10^{-17}\, \mathrm{yr}^{-1}
\end{equation}
Thus, ultralight moduli that couple to the Standard Model and evolve over gigayears are very strongly constrained.

An even more stringent constraint on contemporary time-variation has been obtained with atomic clocks \cite{Lange_2021}:
\begin{equation}
    \frac{\dot{\alpha}}{{\alpha}} = 1.0(1.1) \times 10^{-18}\,\mathrm{yr}^{-1}\,.  
\end{equation}
Moreover, quasar absorption lines provide limits 
at the part per million level on the variation of $\alpha$ between Earth and the absorption systems at $z \sim 1$ \cite{quasardipole,Kotu2016}, which for a model of constant rate of variation in time give 
\begin{equation}
    \frac{|\dot{\alpha}|}{{\alpha}} \lesssim  10^{-15}\,\mathrm{yr}^{-1}\,.  
\end{equation}

\subsection{Heavier moduli: BBN and dark matter} 

Moduli that are too heavy to mediate long-range forces can still be problematic.  The nature of the problem depends on the mass/lifetime of the modulus.
Very slow or very fast decays are often less dangerous than decays around the time of Big Bang Nucleosynthesis, as we will explain.

Moduli that have cosmologically-long lifetimes can be dark matter, and could potentially contribute more than the observed abundance.  
Moduli that are very heavy, and so decay well before BBN, are fairly safe: at least, there are plausible cosmologies that accommodate such moduli, though there can often be an impact on inflationary model-building. 

Moduli that decay during or shortly after BBN can upset the observed light element abundances (D, T, \ch{^{3}He}, \ch{^{4}He}, \ch{^{7}Li})  by photodissociation or hadrodissociation \cite{Kawasaki:2004qu}.

To understand the masses for which the BBN constraint applies,
we consider couplings of the form
\begin{equation}
    \delta \mathcal{L} = \mu^4 \Biggl(\frac{D_{ij}}{M_{\mathrm{pl}}^2} \Phi^i \Phi^j + \frac{C_{ijk}}{M_{\mathrm{pl}}^3} \Phi^i \Phi^j \Phi^k+\ldots\Biggr)\,,
\end{equation}
where $\Phi^i$ are a collection of scalars, out of which we single out two:
$\Phi^i = \{\varphi, \chi, \ldots \}$.  The dimensionless couplings $D_{ij}$ and $C_{ijk}$ are taken to be of order unity.

We have $m_{\varphi} \sim \mu^2/M_{\mathrm{pl}}$, and there is a cubic interaction $\lambda \varphi \chi^2$, with $\lambda \sim \mu^4/M_{\mathrm{pl}}^3 \sim m_{\varphi}^2/M_{\mathrm{pl}}$.
For the interaction $\lambda \varphi \chi^2$ and for $m_{\varphi} \gg m_{\chi}$, the rate for $\varphi \to \chi\chi$ is $\Gamma \sim \frac{\lambda^2}{m_{\varphi}} $.
We conclude that the decay rate for a scalar $\varphi$ of mass $m$, with gravitational-strength couplings, is  
\begin{equation}
    \Gamma \sim \frac{m^3}{M_{\mathrm{pl}}^2}\,.
\end{equation} 
For $m \approx 30\,\mathrm{TeV}$, we have $\Gamma \approx 4.7 \times 10^{-24}\,\mathrm{GeV}$, corresponding to a lifetime of order $0.1 s$.  

We conclude that moduli with $m \gg 30\,\mathrm{TeV}$ decay early enough to avoid BBN constraints.  Such heavy moduli can potentially impact phenomena in the very early universe, e.g.~by coupling to the inflaton, but the space of models is vast and there are few sharp constraints at present.  
We arrive at:
\begin{equ}[Constraints on moduli]
Sufficiently heavy moduli, $m \gg 30\,\mathrm{TeV}$, are generally  consistent with experimental limits.  Somewhat lighter moduli are constrained by BBN, and ultralight moduli are constrained by fifth force tests.
\end{equ}

\section{Recap}

Geometric vacuum solutions of critical superstring theory are Ricci-flat 6-manifolds, and the known examples, Calabi-Yau threefolds, have many moduli.
Unless the moduli acquire large masses, or very special couplings, such fields are excluded by cosmology --- especially BBN --- and by fifth-force bounds.

The possible ways forward include solutions that have one or more of the following characteristics:
\begin{enumerate}
    \item non-geometric, 
    \item non-vacuum, 
    \item non-critical,
    \item non-superstring\,.
\end{enumerate}
For reasons of theoretical convenience, we will stick to geometric solutions of critical superstring theory, but give up on the vacuum condition, i.e.~we choose option (2).\footnote{Moduli stabilization in supercritical string theories was studied in \cite{Maloney:2002rr}, and non-geometric backgrounds are reviewed in \cite{Plauschinn:2018wbo}.  For recent work on vacua of non-supersymmetric string theories, see e.g.~\cite{Blaszczyk:2014qoa,Blaszczyk:2015zta,Raucci:2024fnp,Abel:2015oxa}.}  We will see that in suitable non-vacuum solutions, called \emph{flux compactifications}, the moduli acquire large masses.

\chapter{Flux Compactifications} \label{chap:gkp}
\index{Flux Compactifications}

To address the problem of unfixed moduli, we now turn to studying non-vacuum solutions. Specifically, we will consider compactifications on Calabi-Yau threefolds, but including three sources of stress-energy: D-branes, orientifold planes, and quanta of the NS-NS and R-R $3$-form fluxes. As shown in the seminal work of Giddings, Kachru, and Polchinski (GKP), the presence of $3$-form flux generates a non-trivial potential for the complex structure moduli and the axio-dilaton, while preserving the underlying Calabi-Yau geometry at leading order in $\alpha'$ and $g_s$. The fluxes not only stabilize a large subset of the moduli at tree level, but also backreact on the ten-dimensional geometry, inducing a warped metric and modifying the four-dimensional  effective supergravity theory. The presence of orientifold planes turns out to be necessary for a consistent compactification with fluxes (see \S\ref{sec:nogoMN}), and D-branes are very often present as well. Even though fluxes are not the only sources, they play a critical role, and the corresponding solutions are called \emph{flux compactifications}.

\medskip

Type IIB flux compactifications provide a controlled starting point for constructing semi-realistic vacua. In this chapter we examine flux compactifications in ten-dimensional supergravity, including localized sources introduced in \S\ref{sec:locs}. In Chapter \ref{chap:classicalEFT} we will construct the four-dimensional effective theory at the classical level, and in Chapter \ref{chap:quantumEFT} we will incorporate quantum corrections, which turn out to be essential for the vacuum structure.

\vfill

\newpage

\section{Fluxes as generalized charges}

We begin by recalling from \eqref{eq:TypeIIAct} that the field strength $F_{p+1} = \mathrm{d}C_p$ of a $p$-form gauge potential $C_p$ enters the supergravity action via a kinetic (Maxwell) term of the form
\begin{equation}\label{eq:Fpp1Act}
    S = c\, \int \, F_{p+1}\wedge \star F_{p+1} = \frac{c}{(p+1)!} \int \, F_{M_1 \ldots M_{p+1}} F^{M_1 \ldots M_{p+1}}\, \sqrt{-g^{(d)}}\, \mathrm{d}^{d} x\, ,
\end{equation}
where $\star$ denotes the Hodge star on the background manifold $M$, and $c$ is a normalization constant. 

Let $\Sigma_{p+1}\subset M$ be a $(p+1)$-dimensional cycle in $M$.
Due to the Dirac quantization condition, we have
\begin{equation}\label{eq:DiracQuant}
    \int_{\Sigma_{p+1}}\, F_{p+1}\in\mathbb{Z}
\end{equation}
in appropriate units: the flux is quantized.
Now if $\{\Sigma_i\}, i=1,\ldots, b^{p+1}(M)$ is a basis of $H_{p+1}(M,\mathbb{Z})$, with $b^{p+1}$ the corresponding Betti number, we can define a vector of integers, with components
\begin{equation}\label{eq:DiracQuantvec}
    f_i\coloneqq  \int_{\Sigma_i} F_{p+1} \in \mathbb{Z}\,,
\end{equation}
Thus, for each $(p+1)$-form field strength $F_{p+1}$ in the theory, the configuration of quantized fluxes is specified by $b^{p+1}(M)$ integers, the \emph{flux quanta}.

The field strength $F_{p+1}$ satisfies a Bianchi identity of the form   
\begin{equation}
    \dif F_{p+1}=0\, ,   
\end{equation}
which is generally modified in the presence of additional couplings in the full action, such as contributions from localized sources. Specifically, in the presence of sources, the total action can be written as
\begin{equation}
    S_{\text{tot}} = S+S_{\text{loc}}\; ,\quad  S_{\text{loc}}=\int\, C_{d-p-2}\wedge Q_{p+2}
\end{equation}
for some $(p+2)$-form $Q_{p+2}$. 
Computing the equations of motion for the Hodge dual field  
${F}_{d-p-1} = \star F_{p+1}$ then yields the modified Bianchi identity
\begin{equation}
    \dif F_{p+1}=\dfrac{1}{2c}\, Q_{p+2}\, .
\end{equation}
Integrating this equation over the compact manifold $M$ and applying Stokes’ theorem yields
\begin{equation}\label{eq:ccvh}
    \int_M \, \dif F_{p+1} = 0 \quad \Rightarrow \quad \int_M \, Q_{p+2} = 0\, .
\end{equation}
This relation expresses Gauss's
law, 
extended to higher-degree differential forms. 
The form $Q_{p+2}$ encodes the distribution of certain localized sources or fluxes, and the integral constraint \eqref{eq:ccvh} implies that their net contribution must vanish globally. 
Such global consistency conditions restrict the possible combinations of fluxes, sources, and topology that can coexist in a given solution of string theory. 
Correspondingly, these constraints play a central role in determining the admissible compactifications and the overall structure of the theory’s solution space.

Our goal now is to study compactifications in which the  quantized fluxes are non-zero, and so  
contribute a stress-energy tensor $T^{\mathrm{flux}}_{MN}$, thereby sourcing a non-trivial ten-dimensional geometry through Einstein's equations \eqref{eq:EEinit}. The resulting compactifications are no longer vacuum solutions in the strict sense, but they provide a controlled and calculable way to generate scalar potentials for moduli and break part of the supersymmetry.

\section{Local sources: D-branes and orientifolds}\label{sec:locs}

We begin by reviewing the key properties of the localized sources that will appear in flux compactifications.

\subsection{D-branes}\label{sec:Dbranes}
\index{D-branes}

String theory is not just a theory of strings: it also includes a variety of extended, solitonic objects. The most prominent among these are D$p$-branes: these are $(p+1)$-dimensional dynamical boundary surfaces for open strings, which carry charge under the gauge symmetry of the $(p+1)$-form Ramond-Ramond field \cite{Polchinski:1995mt,Polchinski:1996na}. D-branes provide a crucial link between the perturbative and non-perturbative sectors of the theory, and play a central role in the constructions described in these lectures.

\subsubsection{String boundary conditions}

D-branes can be defined as surfaces on which open strings can end. 
Recall the boundary conditions that arise from varying the worldsheet action \eqref{eq:superstringaction}
with respect to $X_{M}$, i.e., $X_{M} \rightarrow X_{M} + \delta X_{M}$, with
\begin{equation}\label{eq:CondVarX} 
    \delta X_{M} = 0 \quad \text{for } \tau = \pm \infty \, .
\end{equation}
Working in flat gauge with worldsheet metric $h_{ab} = \eta_{ab}$, one finds
\begin{align}
    2\pi\alpha^{\prime}\,\delta S &= \int \mathrm{d}^{2}\xi\, (\partial_{\tau}^{2} - \partial_{\sigma}^{2})X_{M}\, \delta X^{M} 
    - \int_{0}^{\ell} d\sigma\, \dot{X}_{M}\delta X^{M} \big|_{\tau=-\infty}^{\tau=\infty} \nonumber \\
    &\quad + \int_{-\infty}^{\infty} d\tau\, X'_{M}\delta X^{M} \big|_{\sigma=0}^{\sigma=\ell} \, .
\end{align}
The first term yields the equations of motion, and the second vanishes due to \eqref{eq:CondVarX}. The third term must vanish through boundary conditions on the string endpoints.

For closed strings, the periodicity condition $X^{M}(\sigma=0) = X^{M}(\sigma=\ell)$ ensures cancellation of boundary terms. For open strings, the variation vanishes if one imposes either:\index{String boundary conditions}
\begin{itemize}
\item \textit{Neumann boundary conditions}:\index{Neumann boundary conditions}
\begin{equation}\label{eq:NBC} 
    \partial_{\sigma}X^{M}(\tau, \sigma \in \{0, \ell\}) = 0 \, .
\end{equation}
This condition allows the endpoints to move freely since $\delta X^{M}$ is unrestricted, but ensures no momentum flows off the string: 
\begin{equation}
    P^{\sigma}_{M} = \frac{1}{2\pi\alpha'} \, X'_{M}(\tau, \sigma \in \{0, \ell\}) = 0 \, ,
\end{equation}
where $P^{a}_{M} = \partial \mathcal{L}_{P} / \partial (\partial_{a}X^{M}) = \frac{1}{2\pi\alpha'}\, \partial^{a}X_{M}$.
\item \textit{Dirichlet boundary conditions}:\index{Dirichlet boundary conditions}
\begin{equation}\label{eq:DBC} 
    \delta X^{M}(\tau, \sigma \in \{0, \ell\}) = 0 \quad \Longrightarrow \quad \partial_{\tau}X^{M}(\tau, \sigma \in \{0, \ell\}) = 0 \, .
\end{equation}
Here the string endpoints are fixed in spacetime, and the associated canonical momentum 
\begin{equation}
    P^{\tau}_{M} = \frac{1}{2\pi\alpha'}\, \dot{X}_{M}(\tau, \sigma \in \{0, \ell\}) = 0 
\end{equation}
vanishes, breaking spacetime Poincar\'e invariance.
\end{itemize}
The name \emph{D-brane} is a contraction of \textit{Dirichlet-brane}: an open string ending on a D-brane has its endpoints fixed to the D-brane, and satisfies Dirichlet boundary conditions in the directions transverse to the brane. Along the spatial directions parallel to a D$p$-brane ($p>0$), the strings satisfy Neumann boundary conditions, allowing their endpoints to move freely along the brane’s worldvolume.

\subsubsection{D-branes in string theory}

A D$p$-brane is a dynamical $p$-dimensional membrane carrying R-R charge and tension, and it appears as a solitonic state preserving a fraction of the spacetime supersymmetry. For a D$p$-brane with  worldvolume  $\Sigma_{p+1}$, the electric coupling to the $(p+1)$-form potential $C_{p+1}$ occurs through the \textit{Chern-Simons} (CS) action
\begin{equation}
    S_{\rm CS}  = \mu_p \int_{\Sigma_{p+1}} C_{p+1}\ , \label{equ:CS}
\end{equation}
where $\mu_p$ is the brane charge.  
In writing the coupling \eqref{equ:CS}, we have used the standard shorthand notation that suppresses the pullback of the bulk field $C_{p+1}$ onto the worldvolume. 
Explicitly, if $\iota: \Sigma \rightarrow \mathbb{R}^{1,d-1}$ denotes the embedding of the D$p$-brane into $d$-dimensional Minkowski space, then the pullback $\iota^{*}(C_{p+1})$ gives
\begin{equation}
    \int_{\Sigma_{p+1}} C_{p+1} \equiv \int_{\Sigma_{p+1}} \iota^{*}(C_{p+1}) = \dfrac{1}{(p+1)!}\int_{\Sigma_{p+1}}\, C_{M_{1}\ldots M_{p+1}}\, \dfrac{\p x^{M_{1}}}{\p \xi^{0}}\cdots\, \dfrac{\p x^{M_{p+1}}}{\p \xi^{p}}\dif^{p+1}\xi
\end{equation}
where $\xi^{a}$ ($a = 0, \ldots, p$) are local coordinates on the worldvolume.

The coupling \eqref{equ:CS} is the higher-dimensional analog of the interaction between a charged point particle and a gauge potential,
\begin{equation}
    e\int_{\gamma} A_1 = e\int \dif x^\mu A_\mu=e\int\, \dif \tau \dfrac{\dif x^{\mu}}{\dif \tau}\, A_\mu\,,
\end{equation}
integrated along the worldline $\gamma$ of the particle.  

Later, we will consider D-branes embedded in product spaces of the form $\mathbb{R}^{1,3} \times X_{\mathcal{O}}$, where $X_{\mathcal{O}}$ denotes a Calabi-Yau orientifold, introduced in detail in \S\ref{sec:orientifolds}. In such settings, spacetime-filling Dp-branes with $p>3$ are not merely flat hypersurfaces extending along Minkowski space.  
Instead, they are intrinsically curved objects  
that wrap topologically-non-trivial cycles in $X_{\mathcal{O}}$.

Type IIA string theory contains stable D$p$-branes with even $p$, i.e.~$p=0,2,4,6,8$, while type IIB string theory contains stable D$p$-branes with odd $p$, i.e.~$p=1,3,5,7,9$.
In type I string theory, D$p$-branes with $p = 1$, $5$, and $9$ are stable. 
Here, ``stable'' means that the D-branes preserve part of the underlying supersymmetry and cannot decay continuously into other configurations without violating charge conservation or supersymmetry. A D$p$-brane and a D$(6-p)$-brane are electrically charged under the R-R gauge potentials $C_{p+1}$ and $C_{7-p}$, respectively. The field strengths $F_{p+2}$ and $F_{8-p}$ are related by Hodge duality in ten dimensions,
\begin{equation}
	\star F_{8-p} = F_{p+2} \, .
\end{equation}
Thus, D$(6-p)$-branes carry magnetic charge with respect to the same potential $C_{p+1}$ that couples electrically to D$p$-branes.\footnote{In string theories that include the NS-NS $2$-form $B_2$, fundamental strings couple electrically to $B_2$ and are likewise stable, as they carry a conserved NS-NS charge. An additional solitonic object, the NS5-brane, appears as the magnetic dual of the fundamental string and carries magnetic charge under $B_2$.}

\subsubsection{D-brane actions}

The open strings whose endpoints lie on a D-brane give rise to quantum fields confined to the brane’s worldvolume. Among these fields are scalars describing transverse fluctuations of the D-brane, a gauge field $A_a$ with field strength $F_{ab}$, and their supersymmetric partners. 

We now seek to understand D$p$-branes propagating in general solutions of the type II and type I supergravity theories that are the low-energy limits of the corresponding string theories. Such a solution can involve non-trivial profiles for each of the massless bosonic closed-string fields.

The low-energy dynamics of such a D-brane is captured by the effective action for the massless worldvolume fields. The D-brane worldvolume action can be obtained by computing open string scattering amplitudes, while mixed open-closed string amplitudes determine how the D-brane couples to background closed string fields such as the metric, the NS-NS $2$-form $B_{2}$, and the dilaton $\phi$.  The result takes the form given in \eqref{DBIplusCS} below, but to explain this we first recall a few simpler actions.

We begin with the Dirac action. Consider a $p$-dimensional membrane of tension $T_p$, propagating in a spacetime with metric $G_{MN}$. The induced metric on the worldvolume $\Sigma_{p+1}$ is obtained by pulling back the spacetime metric $G_{MN}$ under the embedding $X^M(\xi)$:
\begin{equation}
    G_{ab} \equiv \frac{\partial X^M}{\partial \xi^a} \frac{\partial X^N}{\partial \xi^b} G_{MN}\,.
\end{equation}
The Dirac action, defined as
\begin{equation}
    S_{\rm D} = -  T_p \int_{\Sigma_{p+1}} \dif^{p+1} \xi \,  \sqrt{- \det(G_{ab} )}\,,
\end{equation}
is the natural generalization of the Polyakov action~\eqref{eq:superstringaction} to higher-dimensional extended objects.

To introduce the dynamics of gauge fields living on the brane, it is useful to recall the \textit{Born-Infeld} theory, a nonlinear extension of Maxwell theory.
In $p{+}1$ flat spacetime dimensions, the Born-Infeld action for an Abelian gauge field $A_a$ with field strength $F_{ab} = \partial_a A_b - \partial_b A_a$ takes the form
\begin{align}
    S_{\rm BI} &\, =\, - Q_p \int_{\Sigma_{p+1}}  \dif^{p+1} \xi \,  \sqrt{- \det(\eta_{ab} + 2\pi \alpha' F_{ab})} \, ,
\end{align}
where $\eta_{ab}$ is the flat worldvolume metric.
The constant $Q_p$ has the same dimensions as the tension $T_p$ of a $p$-brane.
By expanding the square root in powers of $\alpha'$,
we obtain to the first non-trivial order
\begin{equation}
    S_{\rm BI} = - Q_p \int_{\Sigma_{p+1}} \dif^{p+1} \xi \,  \left( 1 + \frac{(2\pi \alpha')^2}{4} F_{ab} F^{ab} + \cdots \right) \ ,
\end{equation}
where the second term reproduces the familiar kinetic term for a $\mathrm{U}(1)$ gauge field on the worldvolume. Thus, the Born-Infeld action encapsulates the full nonlinear dynamics of the gauge field, but reduces to Maxwell theory at leading order, i.e.~when the field strength is small in units of $\alpha'$.

The effective action describing a D$p$-brane in a general closed string background combines the Dirac and Born-Infeld terms into a single nonlinear structure known as the \textit{Dirac-Born-Infeld}\index{DBI action} (DBI) action:
\begin{equation}
    S_{\mathrm{DBI}} = - g_{s} T_p \int_{\Sigma_{p+1}} \dif^{p+1}\xi \, \mathrm{e}^{-\Phi} \sqrt{-\det(G_{ab} + 2\pi \alpha'\mathcal{F}_{ab})} \, , 
\label{equ:DBIXX}
\end{equation}
where $G_{ab}$ is the induced worldvolume metric, and 
\begin{equation}\label{eq:WVF}
    2\pi \alpha'{\cal F}_{ab} \equiv  2\pi \alpha' F_{ab}-B_{ab}\ ,
\end{equation}
is the gauge-invariant field strength. 
Here, $B_{ab}$ denotes the pullback of the Neveu--Schwarz $2$-form $B_{MN}$ onto the D-brane, and $F_{ab}$ is the field strength of the worldvolume $\mathrm{U}(1)$ gauge field. 
The combination $\mathcal{F}_{ab}$ ensures invariance under both the Kalb-Ramond gauge transformations and the $\mathrm{U}(1)$ gauge symmetry on the brane.

The Chern-Simons action that couples target-space background fields to D-brane worldvolume fields is given by
\begin{equation}\label{eq:NCS}
S_{\rm CS} =  \mu_p \int_{\Sigma_{p+1}} \, \sum_{n}\, C_n \wedge \mathrm{e}^{2\pi \alpha'{\cal F}} \, .
\end{equation}
The sum in \eqref{eq:NCS} is over all the R-R $n$-form potentials $C_p$ of the theory, and the integral in~\eqref{eq:NCS} picks out $(p+1)$-form contributions.

Assembling the above pieces, the full bosonic action for a D$p$-brane 
takes the Dirac-Born-Infeld plus Chern-Simons form
\begin{equation}
S_{{\rm D}p} = S_{\rm DBI} + S_{\rm CS}\,  \label{DBIplusCS}
\end{equation}
where the D$p$-brane tension is 
\begin{equation}
T_p \equiv \frac{1}{(2\pi)^p \, g_{\mathrm{s}} \, (\alpha')^{(p+1)/2}} \,,
\label{deftp}
\end{equation}
and the R-R charge $\mu_p$ is related to the tension by
\begin{equation} \label{eq:bpsq}
\mu_p = g_{\mathrm{s}} T_p \, .
\end{equation}
The relation \eqref{eq:bpsq} follows from the fact that stable D$p$-branes are BPS objects:\index{BPS states} they
preserve half the supersymmetries of the parent theory, and
extremize the \emph{Bogomolny'i-Prasad-Sommerfeld (BPS) bound}\index{BPS bound} $M\geq Q$ on the mass $M$ and charge $Q$ of a state \cite{Bogomolny:1975de,Prasad:1975kr}.

The action \eqref{DBIplusCS} governs a single D$p$-brane, which carries an Abelian gauge field.
On a stack of $N$ coincident D$p$-branes, the gauge theory is non-Abelian, and the action acquires additional terms involving commutators of the scalar fields describing transverse fluctuations.  
These interactions lead to a rich structure of non-Abelian potentials and are captured in the Myers action~\cite{Myers:1999ps}, which is crucial for the considerations of \S\ref{sec:KPV}.

\subsubsection{D-branes as sources and $p$-brane solutions in supergravity}\label{sec:DbraneSources} 
 
D-branes are not merely probes of the supergravity solutions in which they reside: they are themselves sources for massless closed string fields.
In particular, a D-brane contributes localized sources of stress-energy and R-R charge, thereby generating curvature and R-R fluxes in proportion to its tension and charge.  
This phenomenon, known as \emph{backreaction}, captures the mutual interaction between D-branes and the surrounding supergravity fields.
 
The supergravity solution generated by a stack of D$p$-branes describes a spatially extended, extremal gravitational object. Such configurations are   called \emph{extremal $p$-branes} \cite{Horowitz:1991cd}\index{Extremal $p$-brane}; comprehensive reviews and further discussion can be found in \cite{Duff:1994an,Stelle:1998xg,Blumenhagen:2013fgp}. To construct $p$-brane solutions in higher-dimensional supergravity theories, one must solve the equations of motion that govern the background configurations of the relevant $p$-form gauge fields and the spacetime metric. We therefore begin by considering a $d$-dimensional field theory containing a $(p+1)$-form $C_{p+1}$ and a scalar $\phi$, called the dilaton, coupled to gravity. The general form of the low-energy effective action for this theory can be expressed as
\begin{equation}\label{eq:actionfieldsforbranes} 
S= \int\dif^{d}x\sqrt{-g}\left [R-\dfrac{1}{2}(\p_{M}\phi)^{2}-\dfrac{1}{2(p+2)!}\mathrm{e}^{a_{p}\phi}F_{M_{1}\ldots M_{p+2}} F^{M_{1}\ldots M_{p+2}}\right ]\,,
\end{equation}
where $F_{p+2}=\dif C_{p+1}$ and $a_{p}=(3-p)/2$ for R-R potentials.\footnote{We note that $a_{p}=-1$ for the NS-NS $2$-form.}
The equations of motion for this action are given by
\begin{align}
    \p_{M}\p^{M}\phi &= \dfrac{a_{p}}{2(p+2)!}\, \mathrm{e}^{a_{p}\phi}\,  F_{p+2}^{2}\nonumber\, ,\\
    \nabla_{M}\left (\mathrm{e}^{a_{p}\phi}F^{MM_{1}\ldots M_{p+1}}\right )&=0\label{eq:EOMbrane}\, ,\\
    R_{MN}-\dfrac{1}{2}\p_{M}\phi\p_{N}\phi&=\dfrac{\mathrm{e}^{a_{p}\phi}}{2(p+1)!}\left (F_{MM_{1}\ldots M_{p+1}} F_{N}\,^{M_{1}\ldots M_{p+1}} - \dfrac{(p+1) g_{MN}}{(d-2)(p+2)} F_{p+2}^{2}\right )\nonumber\, ,
\end{align}
where $F_{p+2}^{2}=F_{M_{1}\ldots M_{p+2}} F^{M_{1}\ldots M_{p+2}}$.

We want to show that the above system of equations \eqref{eq:EOMbrane} admits solitonic brane solutions with a non-trivial background value for $C_{p+1}$. Let us set $n=p+1$ and $\tilde{n}=\tilde{p}+1=d-p-3$, corresponding to the dimensions of the worldvolume of a $p$-brane and its dual $(d-p-4)$-brane. We make the following ansatz for the $d$-dimensional metric
\begin{equation}\label{eq:ansatzMetricPBrane} 
	\dif s^{2}=\mathrm{e}^{2A(r)}\eta_{\mu\nu}\dif x^{\mu}\dif x^{\nu}+\mathrm{e}^{2B(r)}\delta_{mn}\dif y^{m}\dif y^{n}\kom \mathrm{e}^{\phi}=\mathrm{e}^{\phi(r)}\kom r^{2}=y^{m}y^{n}\delta_{mn}\, .
\end{equation}
For the $(p+2)$-form field strength, it turns out to be convenient to choose
\begin{equation}
	F_{p+2}^{e}=\alpha_{e}\star\left (\mathrm{e}^{-a_{p}\phi}\epsilon_{S^{d-p-2}}\right )\kom F_{p+2}^{m}=\alpha_{m}\epsilon_{S^{p+2}}
\end{equation}
where we introduced two constants $\alpha_{e}$, $\alpha_{m}$ and defined the volume form of an $n$-sphere as
\begin{equation}
	\epsilon_{S^{n}} = \dfrac{1}{n!}\dfrac{1}{r^{n+1}}\epsilon_{m_{0}\ldots m_{n}}y^{m_{0}}\dif y^{m_{1}}\wedge\ldots\wedge\dif y^{m_{n}}\, .
\end{equation}
Plugging in this ansatz into \eqref{eq:EOMbrane} leads to the solution (see e.g. \cite{Stelle:1998xg} for a derivation)
\begin{align}
	\dif s^{2}&=H^{-\frac{4\tilde{n}}{\Delta (d-2)}} \eta_{\mu\nu}\dif x^{\mu}\dif x^{\nu}+H^{\frac{4n}{\Delta (d-2)}}\left (\dif r^{2}+r^{2}\dif \Omega_{d-n-1}\right )\, ,
\end{align}
where we defined
\begin{equation}\label{eq:deltadef}
	\Delta = a_{p}^{2}+\dfrac{2n\tilde{n}}{d-2}\kom H(r) = 1+\dfrac{N\alpha}{r^{\tilde{n}}}\kom \alpha = \dfrac{g_s T_{p}}{\tilde{n}\Omega_{\tilde{n}+1}}\, \dfrac{\sqrt{\Delta}}{2}\kom \Omega_{m-1}=\dfrac{2\pi^{m/2}}{\Gamma(m/2)}\, .
\end{equation}
The profile of the dilaton is given by
\begin{equation}\label{dilatonlimitbrane}
    \mathrm{e}^{\phi} = H^{\frac{2a_{p}}{\zeta\Delta}}\kom \zeta=\begin{cases}
		+1 & \text{electric}\,,\\
		-1 & \text{magnetic}\,.
		\end{cases}
\end{equation}
The field strengths are
\begin{align}\label{eq:ElSolPBRANE} 
	F_{p+2}^{e}=(-1)^{pd+1}\dfrac{2N}{\sqrt{\Delta}} \alpha \tilde{n}\star\left (\mathrm{e}^{-a_{p}\phi}\epsilon_{S^{\tilde{n}+1}}\right )=\dfrac{2}{\sqrt{\Delta}}\dif H^{-1}\wedge\epsilon_{\bR^{1,n-1}}
\end{align}
for the electric solution and
\begin{equation}
	F_{p+2}^{m}= \dfrac{2N}{\sqrt{\Delta}} \alpha n\epsilon_{S^{p+2}}
\end{equation}
for the magnetic solution, where $N\in \mathbb{N}$ counts the number of coinciding branes.

These brane solutions are actually \emph{BPS states}. Let us compute the charge of the $p$-brane for the electric solution by computing
\begin{align}
\mu_{p} &= \dfrac{1}{N}\, \int_{S^{\tilde{n}+1}}\,\mathrm{e}^{a_{p}\phi}\star F_{p+2} = g_s T_{p}\, .
\end{align}
Thus, the tension $T_{p}$, corresponding to the mass, equals the electric charge $\mu_{p}$. This is the first hint of a BPS structure associated with these objects.  
To show that the solutions also preserve supersymmetry,
a bit more work is required. Notice however that the two functions $A(r), B(r)$ in our ansatz \eqref{eq:ansatzMetricPBrane} are given by
\begin{equation}
    A(r)=\dfrac{2\tilde{n}}{\Delta (d-2)}\log \left (H\right )\kom B(r)=-\dfrac{2n}{\Delta (d-2)}\log \left (H\right )
\end{equation}
and satisfy
\begin{equation}
    nA+\tilde{n}B=0\, .
\end{equation}
This condition implies that these brane solutions \emph{preserve half of the supersymmetry}, i.e., they are $1/2$-BPS states: see \cite{Stelle:1998xg,Blumenhagen:2013fgp} for details.

For the above system of $N$ coincident D$p$-branes, the radial dependence of the dilaton field is given by \eqref{dilatonlimitbrane}, which for $d=10$ reads
\begin{equation}\label{dilatonlimit}
    \mathrm{e}^{\phi} =  \left( 1 + \left( \frac{r_+}{r} \right)^{7-p} \right)^{\frac{1}{4}(3-p)} \, ,
\end{equation}
in terms of the characteristic curvature scale $r_+$, given by~\cite{Horowitz:1991cd} 
\begin{equation}\label{curvaturelimit}
    r_+^{7-p} = N\alpha  \, ,
\end{equation} 
where $\alpha$ was defined in \eqref{eq:deltadef}.

A classical supergravity description constitutes a valid approximation to the underlying string theory when the characteristic spacetime curvature is small in string units and the string coupling satisfies $ g_{s} \ll 1 $. Under these conditions, higher-derivative corrections proportional to powers of $\alpha'$, as well as string loop effects controlled by $g_{s}$, are parametrically suppressed, ensuring the consistency of the low-energy expansion. 

For $p < 7$, equation~\eqref{curvaturelimit} implies that the typical curvature scale becomes small whenever the combination $ g_{s} N $ is large. In this regime, the backreaction of $N$ coincident D$p$-branes produces a smooth and weakly curved geometry, rendering the classical supergravity description reliable. The dependence on $ g_{s} N $ thus governs the transition between the perturbative string regime and the strongly coupled gravitational regime. 

The case $ p = 3 $ is particularly special. For D3-branes, the dilaton field remains constant throughout spacetime, indicating that the effective string coupling does not vary across the background. Consequently, the D3-brane does not couple to the dilaton dynamics, and the corresponding supergravity solution is free from the strong-coupling regions that typically arise for $ p \neq 3 $. In this case, when the parameters satisfy 
\begin{equation}
    1 \ll g_{s} N \ll N\, ,
\end{equation}
both the higher-curvature and string-loop corrections are simultaneously suppressed. The resulting background therefore represents a well-controlled classical solution of type IIB supergravity, forming the foundational setting for the AdS/CFT correspondence \cite{Maldacena:1997re,Witten:1998qj}. 

For $ p \neq 3 $, however, the situation is qualitatively different. The dilaton profile acquires a non-trivial radial dependence, as expressed in equation~\eqref{dilatonlimit}. This behavior implies that the effective string coupling and the curvature cannot remain simultaneously small throughout the entire spacetime. Consequently, the classical supergravity approximation breaks down in certain regions, and a fully consistent description requires either string-theoretic corrections or a dual formulation valid in the strong-coupling domain. 

A systematic analysis of these extremal $p$-brane geometries, together with a detailed discussion of the conditions governing the validity of their supergravity limits, can be found in the comprehensive review \cite{Aharony:1999ti}.

\subsection{Orientifolds}\label{sec:orientifolds}
\index{Orientifolds}

To construct consistent string compactifications involving background fluxes, \emph{orientifolds} are essential. Orientifolds arise when one gauges a discrete symmetry that combines worldsheet parity with a spacetime involution acting on the target manifold. This operation introduces non-dynamical extended objects, so-called \emph{orientifold planes} or O$p$-planes, which carry negative R-R charge and tension. Their presence is essential for achieving tadpole cancellation and restoring overall charge neutrality in type II flux compactifications. Before describing the geometric action of orientifolds on Calabi-Yau threefolds, let us first recall the role of worldsheet parity in type II string theory.

\medskip

Worldsheet parity $\Omega_{\text{ws}}$ relates left-moving and right-moving modes on the  
worldsheet,\index{Worldsheet parity}
\begin{equation}
    \Omega_{\text{ws}}\,\alpha_{n}^{\mu}\, \Omega_{\text{ws}} = \tilde{\alpha}_{n}^{\mu}\,,
\end{equation} 
and exchanges $\text{R}$-$\text{NS} \Leftrightarrow \text{NS}$-$\text{R}$.
In the target spacetime description of type IIA string theory, this action exchanges a left-handed gravitino with a right-handed gravitino.  
Since type IIA is non-chiral, the two gravitini have opposite chiralities, and the theory is not invariant under worldsheet parity alone.  
In type IIB string theory, however, both gravitini possess the same chirality, and the R-NS sector coincides with the NS-R sector.  
As a result, $\Omega_{\text{ws}}$ is a genuine symmetry of the theory. 
One can therefore \emph{gauge} worldsheet parity, identifying states related by $\Omega_{\text{ws}}$ as physically equivalent.\footnote{In type IIA string theory, one can gauge a combination of worldsheet parity and a suitable spacetime symmetry, but that construction will not play a role in these lectures.}

Projecting onto states with $\Omega_{\text{ws}}=+1$ yields the unoriented type I theory, which preserves sixteen supercharges corresponding to $\mathcal{N}=1$ supersymmetry in ten dimensions.

Now suppose we have a Calabi-Yau threefold $X$ that admits a \emph{holomorphic isometric involution} $\sigma$, i.e.
\begin{enumerate}[\hphantom{V}a.]
    \item $\sigma$ acts holomorphically on the local complex coordinates $z_1$, $z_2$, $z_3$ of $X$.
    \item $\sigma$ leaves fixed the metric, in particular the complex structure and the K\"ahler form.
    \item $\sigma^2 =1$.
\end{enumerate}
Then the action of $\sigma$ on the $(3,0)$ form $\Omega$ is $\sigma(\Omega)=\pm\Omega$. 
To study the resulting fixed loci, we write $\Omega \propto \dif z_1 \wedge \dif z_2 \wedge \dif z_3$, and consider 
the action $\dif z_i \to \sigma(\dif z_i)$, with $\sigma(\dif z_i) = \pm \dif z_i$. 
At a fixed locus of type $\sigma(\Omega)=+\Omega$, zero or two of the $\dif z_i$ can be flipped, corresponding to all of $X$, or a curve, being fixed. At a fixed locus of type $\sigma(\Omega)=-\Omega$, one or three of the $\dif z_i$ can be flipped, corresponding to a four-cycle (divisor) or a point, respectively, in $X$ being fixed.

We will study involutions of type $\sigma(\Omega)=-\Omega$, whose fixed loci are four-cycles and points. One can check (see e.g.~\cite{Dabholkar:1996pc}) that to get a consistent, $\mathcal{N}=1$ supersymmetric spectrum, the projection onto $\Omega_{\text{ws}}\sigma = +1$ is not consistent, but the projection onto
\begin{equation}
    \mathcal{O} = (-1)^{F_L}\Omega_{\text{ws}}\sigma = +1\,,
\end{equation} 
with $F_L$ the spacetime fermion number of the left-movers,\footnote{That is, $F_L=0$ in the $(\text{NS},\bullet)$ sector, and $F_L=1$ in the $(\text{R},\bullet)$ sector.} is consistent.

The $(-1)^{F_L}$ action leaves the NS-NS fields $g$, $B$, and $\phi$ invariant, and acts as a minus sign on the R-R fields $C_0$, $C_2$, $C_4$.
The action of $\Omega_{\text{ws}}$ exchanges left-movers and right-movers, so in the NS-NS sector it acts as $+1$ on $g$ and $\phi$, and $-1$ on $B$.  In the R-R sector there is an additional sign from the action on the ground state, so  $\Omega_{\text{ws}}$ acts as $-1$ on $C_0$ and $C_4$, and as $+1$ on $C_2$.

The projection by $\mathcal{O}$ defines the type IIB orientifold theory. Locally, away from the fixed loci of $\sigma$, the theory coincides with the oriented type IIB background. At the fixed loci, however, non-dynamical extended objects appear, the \emph{orientifold planes} (O-planes)\index{O-planes}, whose dimensions are determined by the number of directions left invariant by the involution. For involutions of type~(ii), the fixed loci have total spacetime dimensions $3+1$ and $7+1$, respectively, and are called O3-planes and O7-planes. Given an $\{X,\mathcal{O}\}$, we call $X_{\mathcal{O}}\coloneqq X/\mathcal{O}$ a Calabi-Yau orientifold, specifically an \emph{O3/O7 orientifold}.\footnote{Case (i) leads to an O5/O9 orientifold.  See \cite{Grimm:2004uq} for a comprehensive treatment of both types of orientifolds.}

Orientifold planes are non-dynamical objects: the orientifold projection imposes boundary conditions that remove the worldvolume modes that would otherwise describe fluctuations in their positions. They carry negative R-R charge and tension, with magnitudes proportional to those of D-branes but with opposite sign.  
For an O$p$-plane, the charge and tension in string frame are given by
\begin{equation}\label{eq:QTOpDp}
    \mu_{\text{O}p} = -2^{p-5}\, \mu_{\text{D}p}\; , \quad T_{\text{O}p} = -2^{p-5} \, T_{\text{D}p}\,,
\end{equation}
where $\mu_{\text{D}p}$ and $T_{\text{D}p}$ denote the charge and tension of a D$p$-brane. In particular, an O3-plane carries $-\tfrac{1}{4}$ the charge and tension of a D3-brane, while an O7-plane carries $-8$ times that of a D7-brane. These negative contributions are essential for satisfying the tadpole cancellation condition \eqref{eq:D3Tadpole}, as we argue below.

The presence of orientifold planes has profound implications for the consistency of string compactifications. Because O-planes contribute negative R-R charge and tension, they must be accompanied by suitable combinations of D-branes and background fluxes to ensure global charge neutrality. In particular, in Calabi-Yau orientifolds of type IIB theory, the total D3-brane charge receives contributions from localized sources such as D3-branes, D7-branes, and O3/O7-planes, as well as from the background $3$-form fluxes $F_3$ and $H_3$. The delicate balance among these contributions is encoded in the D3-brane tadpole cancellation condition \eqref{eq:D3Tadpole} which we introduce in the next section.

Later, we will analyze in detail in \S\ref{sec:indD3ch} how D3-brane charge is \emph{induced} on D7-branes and O7-planes through curvature and worldvolume flux couplings in their Chern-Simons actions \eqref{eq:NCS}. We will then combine these local contributions to obtain the expression for the total D3-brane charge of a Calabi-Yau orientifold background, see Eq.~\eqref{eq:QOf}.

\section{Flux backgrounds in type IIB compactifications}\label{sec:EOM}

To write the ten-dimensional action of type IIB supergravity in a convenient form, we will define certain combinations of the bosonic fields that were introduced in \S\ref{sec:TypeIIspectrum}.
The \emph{axio-dilaton} is
\begin{equation}
    \tau \coloneqq  C_0 + \mathrm{i} \mathrm{e}^{-\phi}\,.
\end{equation}
In backgrounds where $\phi$ takes a constant value $\phi(x) \to \langle\phi\rangle$, we can define the string coupling by $g_s = \mathrm{e}^{\langle\phi\rangle}$.
In terms of the axio-dilaton, we define the    
complex flux
\begin{equation}
    G_3 \coloneqq F_3 - \tau H_3\,.
\end{equation}
Because $\tau \in \mathbb{C}$ does not obey a quantization condition, neither does $G_3$, even though its components $F_3$ and $H_3$ obey Dirac quantization.
Finally, we define
\begin{equation}
    \tilde{F}_5 \coloneqq  F_5 + \frac{1}{2} B_2 \wedge F_3 - \frac{1}{2} C_2 \wedge H_3\,,
\end{equation} where $F_5 = dC_4$.
The Bianchi identity for $\tilde{F}_5$ then reads
\begin{equation}\label{eq:D3tadpole_trivial}
    \dif \tilde{F}_5 = H_3\wedge F_3+\ldots\, ,
\end{equation}
where the ellipsis includes localized sources, cf.~Eq.~\eqref{eq:F5BianchiID}.
Recalling from \S\ref{sec:TypeIIspectrum} that $C_4$ has self-dual field strength, we have to impose $\tilde{F}_5 = \star_{10} \tilde{F}_5$ as a constraint on field configurations. Because implementing self-duality directly at the level of the action is technically challenging, we will, for the moment, work with a non-self-dual formulation of type IIB supergravity, following the approach of \cite{Bergshoeff:1995as,Bergshoeff:1995sq,Michelson:1996pn}. In this formulation, the $5$-form field strength $\tilde{F}_5$ is initially treated as an unconstrained field, and the self-duality condition $\tilde{F}_5 = \star_{10} \tilde{F}_5$ is imposed only at the level of the equations of motion rather than incorporated directly into the action. This strategy allows us to perform intermediate calculations more straightforwardly, while ensuring that the correct physical constraint is recovered at the end.

In terms of these variables, the type IIB supergravity action is  
\begin{equ}[Type IIB supergravity action]
\vspace{-0.5cm}
\begin{align}\label{eq:TypeIIBAxtion}
    S_{\mathrm{IIB}} &= \frac{1}{2\kappa_{10}^2}\int \, \Biggl(\mathcal{R}^{(10)}-\frac{|\nabla\tau|^2}{2\,(\mathrm{Im}(\tau))^2} -\frac{|G_3|^2}{12\,\mathrm{Im}(\tau)} -\frac{|F_5|^2}{4\cdot 5!} \Biggr) \, \sqrt{-g^{(10)}}\,\dif^{10}x\nonumber \\
    &+ \frac{1}{8\I\kappa_{10}^2}\int \frac{C_4 \wedge G_3 \wedge \overline{G}_3}{\mathrm{Im}(\tau)}+\text{local}\,.
\end{align} 
\end{equ}
In \eqref{eq:TypeIIBAxtion} we have shown explicitly   the closed string fields, 
and have given only a schematic expression for the couplings of local sources such as D-branes.

When we try to find a non-vacuum solution of this theory, in a Calabi-Yau compactification, we will find obstructions unless suitable local terms are included.  Moreover, it turns out that the necessary local terms are not 
possible in a Calabi-Yau threefold per se.
Instead, we will turn to an \emph{orientifold}\index{Orientifold} of a Calabi-Yau threefold, and there we will succeed.

Very heuristically, the argument for why orientifolds become necessary works as follows. If we integrate the Bianchi identity \eqref{eq:D3tadpole_trivial} over the Calabi-Yau, the left hand side vanishes due to Stokes's
theorem. However, the right hand side is generically non-vanishing for non-zero $H_3,F_3$. To allow for non-zero 3-form fluxes from $H_3,F_3$, we will need compensating local contributions. As it turns out, these contributions must carry negative D3-brane charge, which is exactly what occurs in orientifolds, as we will explain in \S\ref{sec:orientifolds}. Similar to \eqref{eq:ccvh}, constraints like Eq.~\eqref{eq:D3tadpole_trivial} on background configurations in string theory go by the name of \emph{tadpole cancellation conditions}\index{Tadpole cancellation conditions}\index{Tadpole}\index{D3-tadpole}. They arise as generalizations of Gauss’ law, ensuring that the net charge on a compact space $M$ necessarily vanishes (i.e., all field lines have to end somewhere on $M$). 

Though we are doomed to fail, we will first try to find a warped solution without local sources.  We make an ansatz for a $10$-dimensional metric 
\begin{equation}\label{eq:WarpedBackgroundAnsatzGKPSection} 
    \dif s^{2}=\mathrm{e}^{2 A(y)}\, \eta_{\mu\nu}\dif x^{\mu} \dif x^{\nu}+\mathrm{e}^{-2 A(y)}\,  \tilde{g}_{mn} \dif y^{m} \dif y^{n}\, .
\end{equation} 
The function $\mathrm{e}^{2 A(y)}$,
which depends only on the internal coordinates $y^{m}$, is called the \emph{warp factor}.
Preserving four-dimensional Poincar{\'e} invariance requires that the $3$-form flux has legs only along the internal directions, and that the self-dual 5-form $\tilde{ F}_{5}$ satisfies
\begin{equation}\label{eq:FiveFormAnsatzGKP} 
    \tilde{ F}_{5}=(1+\star_{10})\dif\alpha(y)\wedge\dif x^{0}\wedge\dif x^{1}\wedge\dif x^{2}\wedge\dif x^{3}\, .
\end{equation}

\subsection{Type IIB supergravity equations of motion}

Having specified a complete warped Calabi-Yau flux compactification ansatz, we can work out the classical equations of motion.  

Following the notation of \cite{Baumann:2008kq, Baumann:2010sx}, we introduce
\begin{equation}
    \label{eq:def}
    \Phi_\pm = \text{e}^{4A} \pm \alpha\,,\qquad G_\pm = (\star_6 \pm \mathrm{i}) G_3\,,\qquad \Lambda = \Phi_+ G_- + \Phi_- G_+\,,
\end{equation}
where $\star_6$ denotes the six-dimensional Hodge star operator with respect to the \emph{unwarped} metric $\tilde g_{ij}$.
We also use
\begin{equation}
    \overline{\Phi}=\Phi_++\Phi_-=2\mathrm{e}^{4A}\,.
\end{equation}
Below, we will raise and lower indices with the unwarped Calabi-Yau metric $\tilde{g}_{mn}$.

In terms of these fields, the equations of motion read \cite{Baumann:2008kq,Baumann:2010sx,Gandhi:2011id,McGuirk:2012sb} (extended by local terms, see also \cite{Giddings:2001yu} and \cite{Hebecker:2025} for a derivation):
\begin{itemize}
    \item equations of motion for the complex $3$-form $G_3$ together with the Bianchi identities for $F_3,H_3$:
        \begin{align}
            0 & = \mathrm{d}\Lambda +\frac{\mathrm{i}}{2\,\tau_2}\,\mathrm{d}\tau\wedge (\Lambda+\bar\Lambda)\;,\quad 0  = \mathrm{d}F_3= \mathrm{d}H_3\,,
        \end{align}
    where $\tau_2=\mathrm{Im}(\tau)$.
    \item Poisson-like equations for scalar quantities $\Phi_\pm$,
        \begin{align}\label{eq:eomPhi}
            \widetilde{\nabla}^2\Phi_\pm & =   \dfrac{\overline{\Phi}^2}{96\,\tau_2} \, |G_{\pm}|^2+ \frac{2}{\overline{\Phi}} |\tilde\partial\Phi_\pm|^2+ \widetilde{\mathcal{R}}^{(4)}+ \sqrt{2}\kappa_{10}^2\, \overline{\Phi}^{1/2} \left(\mathcal{J}_{\mathrm{loc}}\pm \mathcal{Q}_{\mathrm{loc}} \right)\,, 
        \end{align}
        where
        \begin{equation}
            |G_{\pm}|^2=(G_{\pm})_{mnp}\, (\overline{G}_{\pm})^{mnp}\, ,
        \end{equation}
        and for the axio-dilaton $\tau$
        \begin{align}\label{eq:eomtau}
            0 & = \widetilde{\nabla}^2\tau +\frac{\mathrm{i}}{\tau_2} (\tilde\partial\tau)^2 +\frac{\mathrm{i}}{48}\,\overline{\Phi} \, (G_+)_{mnp}\, (G_-)^{mnp} - \frac{4 \,\tau_2^2}{\sqrt{-g}} \frac{\delta S_\text{D7}}{\delta\bar\tau}\, .
        \end{align}
        Throughout, all internal indices are raised and lowered using the unwarped background metric $\tilde{g}_{mn}$. The equations governing the fields $\Phi_\pm$ arise from the Bianchi identity for the modified $5$-form flux $\tilde{F}_5$, which --- owing to the self-duality condition --- is equivalent to its equation of motion. These are supplemented by the trace of the ten-dimensional Einstein equations over the external (non-compact) directions, as given in Eq.~\eqref{eq:eomrmunu} below. 
        We denote the Ricci scalar of the external spacetime by $\widetilde{\mathcal{R}}^{(4)}$, which vanishes in the case of a four-dimensional Minkowski background. Both Eq.~\eqref{eq:eomPhi} and Eq.~\eqref{eq:eomtau} include contributions from \emph{localized sources}, which are charged, higher-dimensional objects present in the compactification. In particular, Eq.~\eqref{eq:eomPhi} includes source terms involving
        \begin{equation}
            \mathcal{J}_{\mathrm{loc}} \coloneqq \frac{1}{4}\left (T^{m}\,_{m}-T^{\mu}\,_{\mu}\right )^{\mathrm{loc}}\,,
        \end{equation} 
        which is the appropriately scaled tension of localized sources,
        and also involving the D3-brane charge density,
        \begin{equation}
            \mathcal{Q}_{\mathrm{loc}}\coloneqq  T_{3}\,\rho^{\mathrm{loc}}_{\mathrm{D3}}\,,
        \end{equation}
        with 
        \begin{equation}
        \rho_{\mathrm{D3}}^{\mathrm{loc}} = 2\kappa_{10}^2 \varrho_{\mathrm{D3}}^{\mathrm{loc}}\,,
        \end{equation}
        where $\varrho^{\mathrm{D3}}_{\mathrm{loc}}$ is the D3-brane \emph{charge} density of localized sources, normalized such that a D3-brane at $y_{\mathrm{D3}}$ has  $\varrho_{\mathrm{D3}}^{\mathrm{loc}} = \delta^6(y-y_{\mathrm{D3}})$.
        
        The axio-dilaton $\tau$ is famously sourced by 7-branes, as can be seen from the last term in Eq.~\eqref{eq:eomtau}.
        We will remain in the weakly-coupled type IIB theory in these lectures, but F-theory provides a vast generalization in which non-trivial axio-dilaton profiles sourced by seven-branes have a geometric meaning. 
    \item Einstein's equation for the internal components,
        \begin{align}\label{eq:eomR}
                \widetilde{R}_{mn} & =  \frac{\partial_{(m}\tau\partial_{n)}\bar\tau}{2\tau_2^2}  +\frac{2\partial_{(m}\Phi_+\partial_{n)}\Phi_-}{\overline{\Phi}^2} - \frac{\overline{\Phi}}{32\tau_2} \left( G_{+(m}^{~~~~~~~~~{pq}}\, \overline G_{-n)pq} + G_{-(m}^{~~~~~~~~~{pq}}\,\overline G_{+n)pq}\right) \nonumber \\
                 & \,\, + \frac{\tilde g_{mn}}{4\overline{\Phi}} \Biggl( -\tilde\nabla^2\Phi_+ + \frac{\overline{\Phi}^2}{96\tau_2}\, |G_+|^2 +\frac{2|\tilde\partial\Phi_+|^2}{\overline{\Phi}} + \sqrt{2}\kappa_{10}^2\, \overline{\Phi}^{1/2} \left(\mathcal{J}_{\mathrm{loc}}+ \mathcal{Q}_{\mathrm{loc}} \right) \\
                & \,\,-\tilde\nabla^2\Phi_- + \frac{\overline{\Phi}^2}{96\tau_2} |G_-|^2 +\frac{2|\tilde\partial\Phi_-|^2}{\overline{\Phi}} \Biggr)+ 
                \kappa_{10}^2\left( T^{\mathrm{loc}}_{mn} -\frac{g_{mn}}{4} ( T^p_p)^{\mathrm{loc}} \right)\,,\nonumber
        \end{align}
        where $\tilde R_{mn} = \tilde R^l_{~mln}$,
        and for the external components,
        \begin{align}\label{eq:eomrmunu}
            0 & =\widetilde{\mathcal{R}}^{(4)}_{\mu\nu} +\tilde g_{\mu\nu} \left( \frac{\text{e}^{8A}}{48 \,\tau_2} |G_3|^2 +\frac{\text{e}^{-4A}}{4} \left[ (\partial_m \alpha)\, (\partial^{m}\alpha) + (\partial_m \text{e}^{4A})\,(\partial^{m} \text{e}^{4A}) \right]\right) \nonumber\\[0.3em]
            & \qquad- \frac{1}{4}\tilde g_{\mu\nu} \widetilde\nabla^2 \text{e}^{4A} - \kappa_{10}^2 \left(T_{\mu\nu}^\text{loc}-\frac18 g_{\mu\nu} T^\text{loc}\right)\, .
        \end{align}
        Again, all internal indices are raised/lowered with the unwarped background metric $\tilde g_{mn}$.
        In Eq.~\eqref{eq:eomR}, $T_{\mu\nu}^{\mathrm{loc}}$ is the energy-momentum tensor of localized sources.
\end{itemize}
The full system appears somewhat complicated! However, we will ultimately work with a very special class of solutions in which \cite{Giddings:2005ff}
\begin{equation}\label{eq:BPSconditions}
    \Phi_- = G_- = \tilde R_{mn} = \widetilde{\mathcal{R}}^{(4)} = \partial_m \tau = 0 \,,
\end{equation}
whose physical justification will be explained at length below.
Taking \eqref{eq:BPSconditions} as a starting point, the equations of motion collapse to a Poisson-like equation for the inverse warp factor \cite{Giddings:2005ff},
\begin{equation}\label{eq:simplepoisson}
    - \tilde\nabla^2 \mathrm{e}^{-4A} = \frac{|G_{+}|^2}{12\, \mathrm{Im}(\tau)} +\rho_{\mathrm{D3}}^{\mathrm{loc}}\,.
\end{equation}

\subsection{Imaginary self-dual flux compactifications}\label{sec:GKP}

Let us now demonstrate how the conditions in Eq.~\eqref{eq:BPSconditions} arise directly from the equations of motion. Following the approach of \cite{Giddings:2001yu}, we will isolate and analyze a subset of the field equations, namely Eqs.~\eqref{eq:eomPhi} -- \eqref{eq:eomrmunu}, that capture the essential dynamical structure of the system.

\subsubsection{No-go theorem for flux compactifications}\label{sec:nogoMN}

Starting from the (trace reversed) Einstein's equations in $D$ dimensions,
\begin{equation}
R_{MN}=T_{MN}-\dfrac{1}{D-2}g_{MN}T^{K}\,_{K}\, ,
\end{equation}
the four-dimensional components can be written as
\begin{equation}\label{eq:EEFDGKP} 
R_{\mu\nu}=-g_{\mu\nu}\left (\dfrac{|G_{3}|^{2}}{48\mathrm{Im}(\tau)}+\dfrac{\mathrm{e}^{-8 A}}{4}(\p\alpha)^{2}\right )+\kappa_{10}^{2}\left (T_{\mu\nu}^{\mathrm{loc}}-\dfrac{1}{8}g_{\mu\nu}T^{\mathrm{loc}}\right )\, .
\end{equation}
Here, $T_{\mu\nu}^{\mathrm{loc}}$ is the energy-momentum tensor of localized sources.
After taking the trace, Eq.~\eqref{eq:EEFDGKP} yields
\begin{equation}\label{eq:EEFDGKP1} 
\tilde{\Delta}_{6}\,\mathrm{e}^{4 A}=\dfrac{\mathrm{e}^{2 A}}{12\mathrm{Im}(\tau)}\, |G_{3}|^{2}+\mathrm{e}^{-6 A}\Bigl((\p\alpha)^{2}+(\p\mathrm{e}^{4 A})^{2}\Bigr)+2\kappa_{10}^{2}\mathrm{e}^{2 A}\mathcal{J}_{\mathrm{loc}}\,.
\end{equation}

Since the left hand side integrates to zero on a compact manifold, but the right hand side is positive definite in the absence of \emph{negative-tension} localized sources, we find the
\begin{equ}[\hspace{-0.1cm}Maldacena-Nu{\~n}ez no-go theorem \cite{Maldacena:2000mw}]

In the absence of negative-tension localized sources, compactifications of type IIB supergravity in ten dimensions down to $\mathbb{R}^{1,3}$ require vanishing flux and constant warp factor.

\end{equ}
For a discussion of this no-go result, we refer the reader to \cite{Brennan:2017rbf,Danielsson:2018ztv,Obied:2018sgi,Akrami:2018ylq}.

\subsubsection{The GKP ansatz}

To circumvent this no-go result, we have to introduce \emph{negative tension} objects of some sort.  However, a further constraint comes from the $\tilde{ F}_{5}$ Bianchi identity,
\begin{equation}\label{eq:F5BianchiID}
    \mathrm{d}\tilde{F}_5 = H_3 \wedge F_3 + \rho_{\mathrm{loc}}^{\mathrm{D3}}\,,
\end{equation}
with D3-brane charge density $\mathcal{Q}_{\mathrm{loc}}\coloneqq  T_{3}\,\rho_{\mathrm{loc}}^{\mathrm{D3}}$.
Expressing the above in terms of $\alpha$ using \eqref{eq:FiveFormAnsatzGKP} yields
\begin{equation}\label{eq:FFBianchiIDGKP} 
\tilde{\Delta}_{6}\alpha=\dfrac{\I \mathrm{e}^{2 A}}{2\mathrm{Im}(\tau)}\,  G_{3}\cdot (\star_{6}\bar{ G}_{3})+2\mathrm{e}^{-6 A}\, (\p\alpha)\cdot(\p\mathrm{e}^{4 A})+2\kappa_{10}^{2}\mathrm{e}^{2 A}\, \mathcal{Q}_{\mathrm{loc}}\,,
\end{equation}
which is strikingly similar to \eqref{eq:EEFDGKP}.
Taking the difference of \eqref{eq:EEFDGKP1} and \eqref{eq:FFBianchiIDGKP} leads to
\begin{align}
\tilde{\Delta}_{6}(\mathrm{e}^{4 A}-\alpha)&=\dfrac{ \mathrm{e}^{2 A}}{4\,\mathrm{Im}(\tau)}\, \bigl |\I  G_{3}-\star_{6} G_{3}\bigl |^{2}+\mathrm{e}^{-6 A}\, \bigl |\p(\mathrm{e}^{4 A}-\alpha)\bigl |^{2}\nonumber\\
&\quad+2\kappa_{10}^{2}\mathrm{e}^{2 A}\, \Bigl(\mathcal{J}_{\mathrm{loc}}-\mathcal{Q}_{\mathrm{loc}}\Bigr)\, .
\end{align}
Once again integrating both sides, we see that a non-trivial and rather special solution is possible if $\mathcal{J}_{\mathrm{loc}}=\mathcal{Q}_{\mathrm{loc}}$, or more precisely if
$\int(\mathcal{J}_{\mathrm{loc}}-\mathcal{Q}_{\mathrm{loc}})=0$.
Since from \eqref{eq:EEFDGKP1} we need $\int \mathcal{J}_{\mathrm{loc}}<0$ to arrive at a non-trivially warped solution, the case of interest is
\begin{equation}
\sum \mathcal{J}_{\mathrm{loc}}=\sum \mathcal{Q}_{\mathrm{loc}}<0\,,
\end{equation}
i.e.~we seek localized sources with \emph{negative tension and equally negative charge}.  We have written sums rather than integrals because the sources are discrete and localized.

This leads to the
\begin{equ}[\hspace{-0.1cm}ISD solution of Giddings-Kachru-Polchinski (GKP) \cite{Giddings:2001yu}\index{GKP}]

Compactifications of type IIB supergravity to $\mathbb{R}^{1,3}$ with localized sources obeying
\begin{equation}\label{eq:GKPIEqu} 
    \sum \mathcal{J}_{\mathrm{loc}}=\sum \mathcal{Q}_{\mathrm{loc}}<0
\end{equation}
admit non-trivially warped solutions to the equations of motion, with the following properties:
\begin{enumerate}
\item \emph{Imaginary self-dual} (ISD) $3$-form flux\index{ISD flux}
\begin{equation}\label{eq:ISDGKP} 
    \star_{6} G_{3}=\I G_{3}\, ,
\end{equation}
\item The warp factor\footnotemark~and $\alpha$ are related as
\begin{equation}\label{eq:GKPFFWarpUpC} 
    \mathrm{e}^{4 A}=\alpha+\mathrm{const.}
\end{equation}
\item For a Ricci-flat metric $\tilde{g}_{mn}$, the equations of motion are satisfied for
\begin{equation}\label{eq:ConstADGKP} 
    \tau=\mathrm{const.}
\end{equation}
\end{enumerate}

\end{equ}

The above solution is \emph{conformally Calabi-Yau}, i.e.,~the internal part of the metric is $\mathrm{d}s^2 \supset \mathrm{e}^{-2A}\,\tilde{g}_{mn}\dif y^{m} \dif y^{n}$, with $\tilde{g}_{mn}$ the Ricci-flat metric. An advantage of conformally Calabi-Yau compactifications is that, up to effects of warping and the orientifold projection, which we will explain below, the moduli of the solution are those of the Calabi-Yau threefold, which we already understand. Moreover, the ISD condition is imposed in terms of the Hodge star $\star_{6}$ constructed from the \emph{unwarped} metric $\tilde{g}_{mn}$. Remarkably, however, the ISD condition is insensitive to the warp factor. This follows from the fact that on a complex three-dimensional manifold, the Hodge star acting on $3$-forms is invariant under conformal rescalings of the metric, see Eq.~\eqref{apA:eqnA.32}. As a result, solutions to the ISD condition are independent of the warp factor, and may be determined using only the unwarped Calabi-Yau geometry. In particular, one can reliably solve for the flux configuration satisfying the ISD condition without needing the explicit form of the warp factor, which would otherwise require detailed knowledge of the Calabi-Yau metric and a solution to the corresponding Poisson equation.

Much of the subject involves starting with a conformally Calabi-Yau solution and introducing perturbatively-small corrections, e.g.~$\mathcal{R}^{(4)} \neq 0$, $\nabla\tau \neq 0$.  That is what we will do in later chapters.

We observe that the warp factor appearing in Eq.~\eqref{eq:GKPFFWarpUpC} is determined only up to an additive constant. This ambiguity arises because the warp factor enters the Einstein equation \eqref{eq:EEFDGKP1} solely through its derivatives, rendering the equation invariant under the shift $\mathrm{e}^{-4A} \rightarrow \mathrm{e}^{-4A} + c$ for any constant $c \in \mathbb{R}$. Noting that the warped volume of the compactification manifold scales as $\mathcal{V} \sim c^{3/2}$, it is natural to associate this flat direction with the volume modulus of the Calabi-Yau threefold. At this stage, the overall volume remains unfixed and arbitrary reflecting the no-scale structure of the solution, see, e.g., \cite{Frey:2008xw}.

\subsubsection{Requirements on local sources}\label{sec:req}

In developing the GKP solution, we supposed that one can find a compactification in which the net effect of all the local sources obeys \eqref{eq:GKPIEqu}. In practice this is easiest to achieve if all local sources obey 
\begin{equation}\label{eq:GKPIEquMod} 
\mathcal{J}_{\mathrm{loc}}= \mathcal{Q}_{\mathrm{loc}}\,,
\end{equation} 
which corresponds to having equal amounts of D3-brane charge and tension.

In addition, localized sources and fluxes have to satisfy further consistency conditions. Integrating the Bianchi identity \eqref{eq:F5BianchiID} for $\tilde{F}_5$ over the Calabi-Yau threefold, we find the \emph{D3-brane tadpole cancellation condition}\index{D3-brane tadpole}
\begin{equation}\label{eq:D3Tadpole}
    0 = Q_{\mathrm{flux}}+\int_{X}\, \rho_{\mathrm{loc}}^{\mathrm{D3}}\, \sqrt{g}\,\dif^6 y\, ,
\end{equation}
where we defined the D3-brane charge $Q_{\mathrm{flux}}$ induced by fluxes as
\begin{equation}\label{eq:D3flux}
    Q_{\mathrm{flux}} \coloneqq \int_{X}\, H_3 \wedge F_3 = \dfrac{\I}{2}\int_{X}\, \dfrac{1}{\mathrm{Im}(\tau)}\, G_{3}\wedge\ov G_{3}\, .
\end{equation} 
Coming back to our earlier discussion at the beginning of \S\ref{sec:EOM}, we make the following crucial observation:
\begin{equ}[D3-brane tadpole for ISD fluxes]
    For ISD fluxes, the D3-brane charge $Q_{\mathrm{flux}}$ defined in Eq.~\eqref{eq:D3flux} is \emph{non-negative}. Such fluxes are possible only if suitable negative tension localized sources are present.
\end{equ}
\noindent Even so, it is not necessary that all local sources have negative D3-brane tension and negative D3-brane charge, only that the total is negative.  Writing $(\mathcal{J}_{\mathrm{loc}}, \mathcal{Q}_{\mathrm{loc}})$ for a D3-brane as $(1,1)$, an anti-D3-brane has $(1,-1)$ and violates \eqref{eq:GKPIEquMod}. But an O3-plane, as we will see, has $(\mathcal{J}_{\mathrm{loc}}, \mathcal{Q}_{\mathrm{loc}})=(-1/4,-1/4)$.  So a configuration of, for example, 64 O3-planes and 10 D3-branes would have $\sum \mathcal{J}_{\mathrm{loc}}=-6$ and $\sum \mathcal{J}_{\mathrm{loc}}-\sum \mathcal{Q}_{\mathrm{loc}}=0$, and so meets the conditions for \eqref{eq:GKPIEquMod}, even though the D3-branes have positive charge and positive tension. We will revisit this point in \S\ref{sec:indD3ch}, after introducing orientifolds and computing the induced D3-brane charge contributions from D7-branes and O7-planes.

\subsection{D-brane charges and tadpole cancellation}\label{sec:indD3ch}
\index{D-brane charges}\index{Tadpole cancellation}

On a curved D7-brane, an effective D3-brane charge is induced through higher-curvature and gauge-flux couplings in the D7-brane worldvolume action. These contributions arise from the Chern-Simons sector, where the R-R $4$-form $ C_4 $ couples to combinations of the worldvolume flux $\mathcal{F}$ and intrinsic curvature terms. As a result, even in the absence of explicit D3-branes, D7-branes act as distributed sources of D3-brane charge, modifying the total charge budget of the compactification.  

The induced charge typically contributes \emph{negatively} to the net D3-brane charge, analogous to the effect of orientifold planes. Both D7-branes and O7-planes therefore provide localized negative contributions to the total D3-brane tadpole, balancing the positive charge carried by background fluxes and mobile D3-branes. This interplay between flux-induced, curvature-induced, and localized contributions is a central ingredient in achieving global tadpole cancellation and ensuring the consistency of type IIB Calabi-Yau orientifold compactifications.  

In the following, we systematically derive these induced D3-brane charges, first by examining the relevant curvature couplings on D7-branes, and  
then by including the contributions from O7-planes. Together, these results will lead to the expression \eqref{eq:QOf} for the total D3-brane charge of an orientifold background.

\subsubsection*{D7-branes and O7-planes}

A D7-brane wrapped on a divisor $D$ can carry a worldvolume gauge flux $\mathcal{F}$: recall \eqref{eq:WVF}.
The Chern–Simons couplings \eqref{eq:NCS} for D7-branes imply that such fluxes source lower-dimensional charge.
The relevant piece for D3-brane charge schematically contains\footnote{We use units where $\ell_{s}^2 = (2\pi)^2\alpha^{\prime}=1$.}
\begin{equation}\label{eq:WVFCS} 
   S_{\text{CS}} \supset  \mu_7\,\int\, C_4 \wedge \Big(\tfrac12 \mathcal F\wedge\mathcal F \;+\; \ldots \Big)\,,
\end{equation}
so a D7-brane with non-trivial worldvolume flux $\mathcal{F}$ induces a D3-brane charge proportional to (see e.g. \cite{Crino:2022zjk})
\begin{equation}
    Q_{\mathrm{D3}}^{\mathrm{D7},\,\text{flux}} = -\dfrac{1}{2}\int_{D}\mathcal F\wedge\mathcal F \, .
\end{equation}

In addition to the D3-brane charge induced by worldvolume flux, curvature couplings appearing in Eq.~\eqref{eq:WVFCS} also generate an effective D3-brane charge even in the absence of flux, i.e., for $\mathcal{F}=0$. These curvature-dependent contributions originate from the gravitational sector of the D7-brane Chern-Simons action \eqref{eq:NCS}, which encapsulates the coupling of the brane to the ambient spacetime curvature. More concretely, they correspond to the expansion of the $A$-roof (or Todd) genus in the anomaly-canceling Chern-Simons term, see e.g. \cite{Junghans:2014zla}. Although we do not provide the full derivation of these curvature couplings here, their structure can be succinctly summarized by the modified Chern-Simons term
\begin{equation}\label{eq:WVFCSMod} 
   S_{\text{CS}} \supset  \mu_7\,\int\, C_4 \wedge \Big(\tfrac12 \mathcal F\wedge\mathcal F + \dfrac{1}{48}\Bigl[p_1(ND)-p_1(TD)\Bigr]\Bigr)\,,
\end{equation}
in terms of the Pontryagin classes $p_1(ND)$ and $p_1(TD)$ of the normal and tangent bundles, respectively. The second term in parentheses represents the purely gravitational correction to the induced D3-brane charge. Physically, it captures the contribution of curvature couplings that arise from the embedding of the D7-brane in a curved background. 

For a D7-brane wrapping a smooth complex divisor $D$ embedded in a Calabi-Yau threefold $X$, the normal bundle $ND$ and the tangent bundle $TD$ satisfy particularly simple topological relations. In this case, the normal bundle is trivial in the sense that its first Pontryagin class vanishes, while the tangent bundle inherits non-trivial curvature from the embedding. Explicitly, one finds
\begin{equation}
    p_1(ND)=0\; ,\quad p_1(TD) = c_1(TD)^2-2 c_2(TD)=-2 c_2(TD)\, .
\end{equation}
Substituting these relations into the curvature-dependent Chern-Simons coupling in Eq.~\eqref{eq:WVFCSMod}, the induced D3-brane charge due to purely curvature effects becomes 
\begin{equation}
     Q_{\mathrm{D3}}^{\mathrm{D7},\,\text{curv}}=-\dfrac{1}{24} \int_{D}\, c_2(TD)=-\dfrac{ \chi(D)}{24}
\end{equation}
where $\chi(D)$ denotes the Euler characteristic of the divisor $D$.\footnote{Explicit formulas for $\chi(D)$ are provided in \S\ref{sec:DivTops}: see Eq.~\eqref{eq:chiDexp}.} This relation expresses the correspondence between the curvature-induced D3-brane charge and the topology of the wrapped four-cycle. Combining the flux-induced and curvature-induced contributions, the total D3-brane charge sourced by a D7-brane wrapping $D$ is therefore
\begin{equation}\label{eq:D3CD7} 
    Q_{\mathrm{D3}}^{\mathrm{D7}} =  -\frac{1}{2}\int_{D}\mathcal F\wedge\mathcal F - \frac{\chi(D)}{24}\,.
\end{equation}

Orientifold planes wrapping cycles similarly contribute to the total D3-brane charge through curvature couplings in their effective Chern–Simons action. For O7-planes, the resulting contribution takes the form
\begin{equation}
	Q_{\mathrm{D3}}^{\mathrm{O7}}= - \frac{\chi(D)}{6}\,.
\end{equation} 
Curvature-induced charges, like orientifold planes, play an essential role in the global D3-brane tadpole cancellation condition,  
by providing negative contributions that can counterbalance  
the positive ones of background fluxes and D-branes.   

\subsubsection*{D7-brane tadpole}
\index{D7-brane tadpole}

In type IIB string theory, D7-branes act as sources for the R-R $8$-form potential $C_8$, whose field strength $F_9 = \mathrm{d}C_8$ is Hodge dual to the axion field strength $\mathrm{d}C_0$.  
The total D7-brane charge, defined as the integral of $F_9$ over a transverse two-sphere, must vanish in a compact space in order to satisfy Gauss’ law. If the net D7-brane charge were nonzero, the equations of motion \eqref{eq:eomtau} for $C_0$ would be inconsistent globally, and the ten-dimensional background could not be compactified smoothly. This requirement gives rise to the so-called \emph{D7-brane tadpole cancellation condition}. 

O7-planes contribute negative D7-brane charge and tension  
\cite{Sagnotti:1987tw,Gimon:1996rq}. To cancel these negative contributions, additional D7-branes must be introduced such that the combined configuration carries zero net D7-brane charge in each homology class of the compactification manifold $X$. In local terms, this condition can be written schematically as
\begin{equation}
    \sum_{\text{D7}} [\text{D7}] = 8\,[\text{O7}] \, ,
\end{equation}
where $[\text{D7}]$ and $[\text{O7}]$ denote the homology classes of the D7-branes and O7-planes, respectively.

A consistent and supersymmetric way to achieve this cancellation is to place four D7-branes directly on top of each O7-plane. The orientifold projection then identifies each brane with its mirror image, producing an $\mathfrak{so}(8)$ gauge algebra on the worldvolume of the stack (see e.g.~\cite{Blumenhagen:2002wn,Jockers:2004yj}). This configuration cancels both the D7-brane charge and the tension locally, ensuring that there is no tadpole obstruction to solving the R-R and NS-NS field equations. 
In what follows, such $\mathfrak{so}(8)$ stacks will serve as the basic building blocks for the computation of the induced D3-brane tadpole in orientifold compactifications.

Nonetheless, let us comment on a more general approach to D7-brane tadpole cancellation.
The D7-brane tadpole cancellation condition only requires that the \emph{total} D7-brane charge vanish globally in the compact space, 
\begin{equation}
    \sum_{\text{D7}} [\text{D7}] + [\text{O7}] = 0 \quad \in H_4(X,\mathbb{Z})\, ,
\end{equation}
which is a statement of charge conservation in cohomology. This global requirement does not necessarily imply that the cancellation must occur locally near each orientifold plane. Configurations in which the total D7-brane charge cancels globally but not locally are referred to as having non-local D7-brane tadpole cancellation.  

In such non-local configurations, the D7-branes are distributed away from the O7-planes in the compactification manifold, so that the local R-R flux sourced by each O7-plane is only partially compensated by nearby D7-branes, with the remainder canceled by distant ones. While the integrated tadpole condition remains satisfied, the local imbalance in D7-brane charge modifies the profile of the axio-dilaton $\tau = C_0 + \I \mathrm{e}^{-\phi}$: recall \eqref{eq:eomtau}. This generically leads to a non-trivial \emph{non-constant dilaton} background, with $\tau$ varying holomorphically over the internal space.

From the ten-dimensional viewpoint, such non-local D7/O7 configurations correspond to globally consistent but locally singular solutions of the IIB supergravity equations, where the deficit in local charge induces non-trivial monodromies of the axio-dilaton around the D7-brane loci. A natural geometric description of such backgrounds is provided by \emph{F-theory}\index{F-theory} \cite{Vafa:1996xn}, which encodes the varying axio-dilaton as the complex structure modulus of an auxiliary elliptic curve fibered over the compactification manifold. In this framework, D7-branes and O7-planes are represented by components of the discriminant locus $\Delta = 0$ of the elliptic fibration, and non-local D7-brane tadpole cancellation corresponds to the global requirement that the total vanishing order of $\Delta$ cancels across the compact space. This geometric formulation elegantly captures the interplay between localized sources, axio-dilaton monodromies, and the global consistency conditions required by charge conservation.

\subsubsection*{Total D3-brane charge of CY orientifolds}

Let us now assemble the different contributions to the D3-brane charge entering the tadpole cancellation condition \eqref{eq:D3Tadpole}. By isolating the localized sources that contribute to $\rho_{\mathrm{loc}}^{\mathrm{D3}}$, the condition \eqref{eq:D3Tadpole} can be written in the compact form  
\begin{equation}\label{eq:D3tadpole_compact}
    2(N_{\mathrm{D3}}-N_{\overline{\mathrm{D3}}})+Q_{\mathrm{flux}}-Q_{\text{O}} = 0
\end{equation}
where the flux and localized source contributions are defined as  
\begin{equation}
    Q_{\mathrm{flux}} = \int_{X}\, H_3\wedge F_3\; ,\quad -Q_{\text{O}} = \sum_{\mathrm{D7}}\, Q_{\mathrm{D3}}^{\mathrm{D7}} + \sum_{\mathrm{O7}}\, Q_{\mathrm{D3}}^{\mathrm{O7}} + \sum_{\mathrm{O3}}\, Q_{\mathrm{D3}}^{\mathrm{O3}}
\end{equation}
To evaluate the D3-brane charge induced by D7-branes and O7-planes, we employ the Lefschetz fixed-point theorem. This theorem connects the alternating sum of even and odd Betti numbers $b^{i}_{\pm}(X)$ to the Euler characteristic $\chi_{f}(X_{\mathcal{O}})$ of the corresponding fixed-point locus, thereby providing a topological measure of the D3-brane charge contribution from these objects.
Specifically, we have 
\begin{equation}\label{eq:LefTh}
	\sum_{i}\, (-1)^{i}\bigl(b_{+}^{i}(X)-b_{-}^{i}(X)\bigr)=\chi_{f}(X_{\mathcal{O}}) \kom b_{\pm}^{i}(X)=\sum_{p+q=i}\, h_{\pm}^{p,q}(X)\, .
\end{equation}
For Calabi-Yau threefolds, Eq.~\eqref{eq:LefTh} simplifies to  
\begin{equation}\label{eq:HodgeNumOrientCond} 
	\chi_{f}=2+2\left (h^{1,1}_{+}(X)-h^{1,1}_{-}(X)\right )-2\left (h^{1,2}_{+}(X)-h^{1,2}_{-}(X)-1\right )\, .
\end{equation}

In the absence of worldvolume fluxes $\mathcal{F}$ on the D7-branes and assuming configurations consisting solely of $\mathfrak{so}(8)$ stacks of seven-branes, 
the total contribution to the D3-brane tadpole is given by  
\begin{equation}
	\sum_{\mathrm{D7}}\, Q_{\mathrm{D3}}^{\mathrm{D7}} + \sum_{\mathrm{O7}}\, Q_{\mathrm{D3}}^{\mathrm{O7}} = - \dfrac{\chi_{f}-N_{\mathrm{O3}}}{2}\, .
\end{equation}
Including the O3-plane contributions yields  
\begin{equation}\label{eq:QOf}
	Q_{\text{O}} =  \dfrac{\chi_{f}}{2}\, .
\end{equation}
For the particular case of an orientifold with $h^{1,1}_{-}=h^{2,1}_{+}=0$,\footnote{Key examples for this type of orientifolds are the \emph{trilayer} orientifolds\index{Trilayer orientifolds} \cite{Moritz:2023jdb}; see also \S\ref{sec:toric}.} the Euler characteristic of the fixed-point locus \eqref{eq:HodgeNumOrientCond} becomes (see e.g. \cite{Demirtas:2021nlu,Crino:2022zjk,Moritz:2023jdb})
\begin{equation}\label{eq:ChiFtrilayer}
	\chi_{f}\bigl |_{h^{1,1}_{-}=h^{2,1}_{+}=0} = 4+2h^{1,1}+2h^{1,2}\, ,
\end{equation}
and consequently the total induced D3-brane charge reads  
\begin{equation}\label{eq:QOtrilayer}
	Q_{\text{O}}\bigl |_{h^{1,1}_{-}=h^{2,1}_{+}=0}=2+h^{1,1}+h^{1,2}\, .
\end{equation} 
This result provides a remarkably simple and universal expression for the total D3-brane charge induced by O-planes and D7-branes. 

Because the quantity $\chi_{f}$ can be expressed purely in terms of the Calabi-Yau Hodge numbers, the evaluation of the tadpole contribution becomes straightforward once a consistent orientifold satisfying the required projection conditions is identified. In practice, this means that the D3-brane tadpole can then be computed directly from the Hodge data of the Calabi-Yau manifold, without requiring detailed geometric information about the individual fixed loci.

\section{Recap of ISD compactifications}

We first sought a warped flux compactification on  a smooth Calabi-Yau: that is, a non-vacuum solution in which the internal space is a  Calabi-Yau threefold, the $3$-form flux $G_3$ and the $5$-form flux $\tilde{F}_5$ are nonzero, the metric is warped, and no local contributions are present.  We showed that no such solution is possible.

However, if the compactification is on an \emph{orientifold} of a  Calabi-Yau threefold, one finds an important class of non-vacuum solutions. These are type IIB compactifications with imaginary-self-dual (ISD) fluxes, on O3/O7 orientifolds of Calabi-Yau threefolds. These solutions are warped and are conformally Calabi-Yau.  

Recall that our interest in \emph{non}-vacuum solutions was the hope that the presence of stress-energy would create a potential for the moduli. This hope is realized in ISD compactifications: as we will show in the next chapter, for sufficiently general choices of quantized 3-form flux, the complex structure moduli and axio-dilaton have masses at tree level. However, the K\"ahler moduli remain light. 

To understand why the complex structure moduli and axio-dilaton become massive, we examine the flux kinetic term
\begin{equation}\label{eq:KinTG3}
    S_{10} \supset -\frac{1}{2\kappa_{10}^2}\int \frac{1}{2\,\mathrm{Im}\,\tau}\, G_3 \wedge \star_6 \overline{G}_3\,.
\end{equation}
The dependence on $\tau$ is manifest, and the dependence on the complex structure moduli occurs through the Hodge star, which depends on the metric.  Thus, it is clear that (at least some of) these fields will receive potential energy terms if $F_3$ and $H_3$ are nonzero in cohomology. Where in moduli space the moduli are stabilized is another matter.

On the other hand, the K\"ahler moduli are not affected: in particular, under the rescaling $g_{mn} \to \mathrm{e}^{2u}g_{mn}$ the potential \eqref{eq:KinTG3} is invariant.

To explore these structures in detail, we will now develop a four-dimensional theory of flux compactifications.

\chapter{Classical Theory of Flux Compactifications} \label{chap:classicalEFT}

In the previous chapter, we have analyzed flux compactifications primarily from a ten-dimensional viewpoint, emphasizing the role of fluxes, sources, and tadpole conditions in ensuring consistency of the higher-dimensional solutions. In this chapter, we now turn to the complementary and often more powerful perspective of the four-dimensional $\mathcal{N}=1$ effective supergravity that emerges after compactification on a Calabi-Yau orientifold. This low-energy theory captures the dynamics of the light moduli fields that arise from the geometry and fluxes of the internal space, and it provides a natural language for studying supersymmetry breaking, moduli stabilization, and vacuum structures. Our aim here is to construct this four-dimensional effective theory at the classical level, identifying the K\"ahler potential, superpotential, and resulting scalar potential that together encode the classical physics of flux compactifications. Corrections beyond this approximation, arising from quantum and stringy effects, will be addressed in Chapter~\ref{chap:quantumEFT}.

\section{Structure of $\mathcal{N}=1$ supergravity}

Chiral multiplets of $\mathcal{N}=1$ supersymmetry in four spacetime dimensions contain a complex scalar and a Weyl fermion. Suppose that $\Phi_A$, $A=1,\ldots,N$, are such scalars. The $\mathcal{N}=1$ supergravity action for the bosonic fields at the 2-derivative level  is
\begin{equation}\label{eq:4DSugra_classical}
	S = \int\dif^{4}x\,\sqrt{-g}\biggl \{\dfrac{M_{P}^{2}}{2}\mathcal{R}-\mathcal{K}_{A\bar{B}}(\partial_{\mu}\Phi^{A})(\partial_{\nu}\overline{\Phi}^{\bar{B}})\, g^{\mu\nu}-V_{F}(\Phi,\bar{\Phi})\biggl \}\, ,
\end{equation}
where $\mathcal{K}_{A\bar{B}}$ is the \emph{K\"ahler metric} on moduli space, obtained from a K\"ahler potential $\mathcal{K}$ via
\begin{equation}
	\mathcal{K}_{A\bar{B}} = \dfrac{\partial}{\partial \Phi^{A}}\dfrac{\partial}{\partial \bar{\Phi}^{\bar{B}}}\, \mathcal{K}(\Phi,\bar{\Phi})\, .
\end{equation}
Further, $V_{F}$ is the $F$-term scalar potential
\begin{equation}\label{eq:vfgen}
    V_{F}(\Phi,\bar{\Phi})=\mathrm{e}^{\mathcal{K}/M_{P}^{2}}\left (\mathcal{K}^{A\bar{B}} \, D_{A}W\, D_{\bar{B}}\overline{W}-\dfrac{3}{M_{P}^{2}}W\,\overline{W}\right )\,,
\end{equation}
where $W=W(\Phi)$ is the superpotential, a holomorphic function, and
\begin{equation}
    D_{A} W = \partial_A W+\dfrac{1}{M_{P}^{2}}(\p_{A}\mathcal{K})\, W
\end{equation}
is the K\"ahler-covariant derivative.

For $M_{P}\rightarrow\infty$, one has
\begin{equation}
	V_{F}\xrightarrow{\; M_{P}\rightarrow\infty\; }\mathcal{K}^{A\bar{B}} \, \partial_{A}W\, \partial_{\bar{B}}\overline{W}
\end{equation}
which is the familiar form of the $F$-term scalar potential for rigid supersymmetry, i.e., in the absence of gravity.

We will set $M_{P}=1$ henceforth, so
\begin{equation}
    V_{F}(\Phi,\overline{\Phi})=\mathrm{e}^{K}\left (\mathcal{K}^{A\bar{B}} \, D_{A}W\, D_{\bar{B}}\overline{W}-3W\,\overline{W}\right )\; , \quad D_{A} W = \partial_A W+(\p_{A}K)\, W\, .
\end{equation}

In general, in theories with vector fields in $\mathcal{N}=1$ vector multiplets, the corresponding kinetic terms would be obtained from holomorphic gauge kinetic functions, and the scalar potential would contain so-called $D$-terms. We have left out vector multiplets for simplicity of presentation.

In the above we have implicitly taken $W(\Phi)$, $\mathcal{K}(\Phi,\bar{\Phi})$, and even the K\"ahler coordinates $\Phi$ themselves, to be \emph{known} or computable data.  In practice one typically first computes these data of a supergravity theory at leading order in $g_s$ and $\alpha'$, and then incorporates corrections: we will say much more about this in Chapter \ref{chap:quantumEFT}.

\section{Effective theory of flux compactifications}

We will now examine the four-dimensional  $\mathcal{N}=1$ supergravity that is the low-energy effective theory of an ISD flux compactification on a Calabi-Yau orientifold. We have found a collection of scalars in Calabi-Yau compactifications, namely the geometric moduli and the scalars from $p$-forms: recall Table~\ref{tab:fieldcontentTypeII}. We note that, even when warping is included, the light fields of the four-dimensional effective theory~\eqref{eq:4DSugra_classical} are still captured, at leading order in $\alpha'$, by the familiar Calabi-Yau moduli: the axio-dilaton, the complex structure moduli, and the K\"ahler moduli. Although the presence of fluxes and warping partially lifts this moduli space by generating masses for some combinations of fields, it does not modify which fluctuations appear as genuine zero modes in the leading-order derivative expansion. Let us see how these fit into the data of an $\mathcal{N}=1$ theory.

\begin{table}[t!]
    \centering
    \begin{tabular}{c|c|c|c|c}
         &  &  & & \\[-1em]
         $\mathcal{N}=2$  multiplet & &  $\mathcal{N}=2$ &  $\mathcal{N}=1$ & name, symbol \\[0.3em]
         \hline
         \hline
          & &  & & \\[-1em]
         \multirow{ 2}{*}{\vspace{-0.25cm}vector} & number &  $h^{2,1}(X)$ & $h^{2,1}_{-}(X)$ & complex structure  \\[0.3em]\cline{2-4}
          & &  & & \\[-0.8em]
          & fields  & $\delta g_{ij}$, $\delta g_{\bar{\imath}\bar{\jmath}}$, $C_{\mu i j \bar{k}}$ & $\delta g_{ij}$, $\delta g_{\bar{\imath}\bar{\jmath}}$ &  moduli, $z^a$ \\[0.3em]
         \hline
         & &  & & \\[-1em]
         \multirow{ 2}{*}{\vspace{-0.25cm}hyper} & number &  $h^{1,1}(X)$ & $h^{1,1}_{+}(X)$ & \multirow{ 2}{*}{\vspace{-0.25cm} K\"ahler moduli, $T_i$}  \\[0.3em]\cline{2-4}
          & &  & & \\[-0.8em]
          & fields & $\delta g_{i\bar{\jmath}}$, $B_{i\bar{\jmath}}$, $C_{i\bar{\jmath}}$, $C_{\mu\nu i\bar{\jmath}}$ & $\delta g_{i\bar{\jmath}}$, $C_{\mu\nu i\bar{\jmath}}$& \\[0.3em]
          \hline
          & &  & & \\[-1em]
          universal hyper & fields & $\phi$, $C_0$, $B_{\mu\nu}$, $C_{\mu\nu}$ & $\tau = C_0+\mathrm{i}\mathrm{e}^{-\phi}$ & axio-dilaton, $\tau$\\[0.3em]
          \hline
    \end{tabular}
    \caption{Complex scalars residing in chiral multiplets in $\mathcal{N}=1$ compactifications of type IIB on Calabi-Yau orientifolds.  For simplicity we omit the 
    multiplets involving $C_{\mu i j\bar{k}}$, $B_{i\bar{\jmath}}$, $C_{i\bar{\jmath}}$, $B_{\mu\nu}$, and $C_{\mu\nu}$, which are present in general O3/O7 orientifolds, but are projected out in the particular ($h^{1,1}_{-}=h^{2,1}_{+}=0$) O3/O7 orientifolds appearing in these lectures.}
    \label{tab:fieldcontentTypeIIOrientifold}
\end{table}

\subsection{K\"ahler coordinates}\label{sec:kcoords}

Let $X$ be a Calabi-Yau threefold, and let $X_{\mathcal{O}}\coloneqq X/\mathcal{O}$ be an O3/O7 orientifold of $X$. We then decompose harmonic forms into eigenstates of $\mathcal{O}\coloneqq (-1)^{F_L}\,\Omega_{\mathrm{ws}}\,\sigma$:
\begin{align}
    h^{1,1}(X) = h^{1,1}_+(X_{\mathcal{O}}) + h_-^{1,1}(X_{\mathcal{O}})\,, \; 
    h^{2,1}(X) = h^{2,1}_+(X_{\mathcal{O}}) + h_-^{2,1}(X_{\mathcal{O}})\,.
\end{align}  
Because the orientifold-graded Hodge numbers $h^{p,q}_\pm(X_{\mathcal{O}})$ are always defined in terms of an orientifold
$X_{\mathcal{O}}$
rather than the Calabi-Yau threefold covering space $X$, we can dispense with writing $X_{\mathcal{O}}$ as an argument, and simply write $h^{p,q}_\pm$.
Now since the K\"ahler form on $X$ is given by
\begin{equation}\label{eq:defkahlerparam}
    J = t^{\hat{\imath}} \omega_{\hat{\imath}}\, \qquad \hat{\imath} = 1,\ldots, h^{1,1}(X)\,,
\end{equation}
and the complex structure deformations can be expressed as (recall Eq.~\eqref{eq:IDH1H21})
\begin{equation} 
    \delta g^{\hat{A}}_{\bar{\imath}\bar{\jmath}} = z^{\hat{A}}\,(\chi_{\hat{A}})_{k\ell\bar{\imath}}\,\bar{\Omega}^{k\ell}_{~~\bar{\jmath}}\, \qquad \hat{A} = 1,\ldots, h^{1,2}(X)\,,
\end{equation} 
and under the involution $\sigma$,
\begin{equation}
    J \to + J, \qquad \Omega \to -\Omega\,,
\end{equation}
we find that $X_{\mathcal{O}}$ has $h^{2,1}_-$ complex structure moduli $z^a$, $a=1,\ldots, h^{2,1}_-$ as well as $h^{1,1}_+$ K\"ahler moduli $T_i$ which are related to the K\"ahler parameters $t^i$ via\footnote{Quantum corrections to the relationship between the K\"ahler parameters $t^i$ and the K\"ahler moduli $T_i$ are discussed at length in Chapter \ref{chap:quantumEFT}.}
\begin{equation}\label{eq:KMcl}
    T_i =\dfrac{1}{g_s} \int_{X}\, J\wedge J\wedge \omega_i + \mathrm{i}\, \int_{X}\, C_4\wedge \omega_i\, \qquad i = 1,\ldots, h^{1,1}_+\, ,
\end{equation}
in terms of the R-R $4$-form $C_4$ and the K\"ahler form
\begin{equation}\label{eq:defkahlerparamdown}
    J = t^{i} \omega_{i}\, \qquad i = 1,\ldots, h^{1,1}_+\, .
\end{equation}

\begin{equ}[Geometric moduli]
In an O3/O7 orientifold there are
\begin{itemize}
    \item $h^{1,1}_+$ K\"ahler moduli $T_i$,
    $i=1,\ldots, h^{1,1}_+$, and
    \item $h^{2,1}_-$ complex structure moduli $z^a$, $a=1,\ldots, h^{2,1}_-$. 
\end{itemize}
In the remainder of these lectures, we will take $h^{1,1}_-=h^{2,1}_+=0$.
\end{equ}  
The conditions $h^{1,1}_-=h^{2,1}_+=0$ simplify the analysis, and will be fulfilled in all of the examples we discuss in Chapter \ref{chap:deSitter}.  Correspondingly, in the remainder we will use the symbol $h^{1,1}$ to mean $h^{1,1}_+(X_{\mathcal{O}})$, and $h^{2,1}$ to mean $h^{2,1}_-(X_{\mathcal{O}})$, i.e.,~we can omit the $\pm$ subscripts.

\subsection{Weil-Petersson metric on complex structure moduli space}\label{eq:WPcsm}
\index{Weil-Petersson metric}

To analyze the kinetic terms of the complex structure moduli $z^{i}$ in the four-dimensional effective action~\eqref{eq:4DSugra_classical}, we expand the internal metric as $g_{mn} = g^{(0)}_{mn} + \delta g_{mn}$ around the Calabi-Yau background $g^{(0)}_{mn}$, with the fluctuations $\delta g_{mn}$ expressed in terms of the moduli as described in \S\ref{sec:moduliofCY}. Reducing the ten-dimensional Einstein-Hilbert term of the type IIB supergravity action~\eqref{eq:TypeIIBAxtion} on this perturbed background then produces the corresponding kinetic terms for the moduli.\footnote{In general, deriving the full four-dimensional effective action~\eqref{eq:4DSugra_classical} from ten-dimensional Type~IIB supergravity is technically subtle because of kinetic terms and warping, see e.g.~\cite{Giddings:2005ff,Shiu:2008ry,Douglas:2008jx,Lust:2022xoq} for detailed discussions.} We will not present the details of this calculation here (see, e.g.,~\cite{Shiu:2008ry, Douglas:2008jx, Gandhi:2011id,Lust:2022xoq}), and instead simply state the resulting expressions below.

The standard \emph{Weil-Petersson metric}\index{Weil-Petersson metric} $G_{a\bar{b}}$ on the space of complex structure deformations $\delta_{\text{cs}}$ of a Calabi-Yau manifold is conventionally defined in terms of infinitesimal metric variations $\delta g_{mk}$, cf.~Eq.~\eqref{eq:delcs}. It takes the form (see, for example, \cite{Candelas:1990pi,Candelas:1989ug,Klemm:2005tw})\footnote{Note that the overall normalization and factors of 2 may vary slightly among conventions, depending on how one defines the complex structure variations and the metric fluctuations.}
\begin{equation}\label{eq:WPMdef}
	2G_{a\bar{b}}\, \delta z^{a}\delta\bar{z}^{\bar{b}} = \dfrac{1}{2\mathcal{V}}\int_{X}\, g^{m\bar{n}}g^{k\bar{l}}\delta g_{mk}\delta g_{\bar{n}\bar{l}}\, \sqrt{g}\, \dif^{6}y\, ,
\end{equation}
where $\mathcal{V} = \int_X \sqrt{g}\, \mathrm{d}^{6}y$ denotes the total volume of the internal space. In \S\ref{sec:moduliofCY} we showed that the complex structure deformations $\delta z^{a}$ can be identified with elements of the cohomology group $H^{1}(X, TX)$. More precisely, we established a one-to-one correspondence between infinitesimal deformations of the complex structure and harmonic $(2,1)$-form representatives $\chi_a \in H^{2,1}(X)$ through the relation given in Eq.~\eqref{eq:IDH1H21}.

This correspondence allows us to express the variations of the internal metric $\delta g_{mn}$ entirely in terms of the $(2,1)$-forms $\chi_a$ and their complex conjugates. Consequently, the metric on the moduli space, originally defined in terms of $\delta g_{mn}$ in Eq.~\eqref{eq:WPMdef}, can be rewritten in terms of the inner products of these forms as \cite{Candelas:1990pi}
\begin{equation}\label{eq:KMCSM}
    \mathcal{K}_{a\bar{b}}=-\dfrac{\int_{X}\, \chi_{a}\wedge\ov \chi_{b}}{\int_{X}\, \Omega\wedge\overline{\Omega}}\, .
\end{equation}
As proven rigorously by Tian \cite{tian1987smoothness} and further developed by Todorov \cite{todorov1989weil}, the metric $G_{a\bar{b}}$ defined in Eq.~\eqref{eq:WPMdef} coincides exactly with the expression in Eq.~\eqref{eq:KMCSM}, that is, $G_{a\bar{b}} = \mathcal{K}_{a\bar{b}}$.
We now present a more precise derivation of this equivalence, elaborating on the verbal argument given previously. As we have seen in \eqref{eq:delcs}, deformations of the complex structure can be expressed in terms of the metric fluctuation as \cite{Candelas:1990pi}
\begin{equation}\label{eq:deltaomega}
    (\delta_{a}  \Omega)_{ij\bar k} = \frac{1}{2} \Omega_{ijl}  g^{l\bar m} \p_{a}  g_{\bar m\bar k}\,.
\end{equation}
This relationship can be inverted to write the metric fluctuations in terms of the harmonic $(2,1)$-forms $\chi_{a}$ as
\begin{equation}\label{eq:deltag}
    \delta  g_{\bar m\bar k} = -\frac{\overline\Omega_{\bar m}^{~~ij}}{||\Omega||^2} \chi_{a,ij\bar k} \delta z^{a}\,.
\end{equation}
We can then compute \cite{tian1987smoothness,Candelas:1990pi}
\begin{align}
	2G_{a\bar{b}}\, \delta z^{a}\delta\bar{z}^{\bar{b}} &= \dfrac{1}{2\mathcal{V}}\int_{X}\, g^{m\bar{n}}g^{k\bar{l}}\delta g_{mk}\delta g_{\bar{n}\bar{l}}\, \sqrt{g}\, \dif^{6}y\nonumber\\ 
	&=\dfrac{1}{2\mathcal{V} (||\Omega||^2)^{2}} \delta z^{a}\delta\bar{z}^{\bar{b}}\int_{X}\,   \Omega^{\bar{n}\bar{\imath}\bar{\jmath}}  \bar{\chi}_{\bar{b},\bar{\imath}\bar{\jmath} k}\,  \overline\Omega_{\bar n}^{~~ij}  \chi_{a,ij}\,^{k} \sqrt{g}\, \dif^{6}y\nonumber\\
	&=-\dfrac{2}{\mathcal{V} ||\Omega||^2} \delta z^{a}\delta\bar{z}^{\bar{b}}\int_{X}\,    \chi_a\wedge \bar\chi_{\bar b}\,.
\end{align}
Making use of the relation 
\begin{equation}
    \mathcal{V} ||\Omega||^2 = ||\Omega||^2 \int \dif^6y\sqrt{ g^{(6)}}  = \int  \Omega\wedge\overline\Omega\, ,
\end{equation}
we find agreement with \eqref{eq:KMCSM}, and thus indeed $G_{a\bar{b}} = \mathcal{K}_{a\bar{b}}$ \cite{tian1987smoothness}. This equivalence is highly non-trivial as it establishes that the Weil-Petersson metric derived from variations of the Calabi-Yau metric is identical to the one obtained from the variation of the holomorphic $3$-form $\Omega$.

Let us now match this result with the underlying K\"ahler structure of complex structure moduli space. In particular, the metric defined in Eq.~\eqref{eq:KMCSM} should arise from a K\"ahler potential $\mathcal{K}_{\text{cs}}$ through
\begin{equation}
    \mathcal{K}_{a\bar{b}} = \partial_a \partial_{\bar{b}} \mathcal{K}_{\text{cs}} \, .
\end{equation}
We now show that the K\"ahler potential $\mathcal{K}_{\text{cs}}$ on the complex structure moduli space is
\begin{equation}\label{eq:Kcs}
	\mathcal{K}_{\text{cs}} = - \ln\left(-\I\int_{X}\, \Omega\wedge\overline{\Omega} \right) \, .
\end{equation}
As we will show now, differentiating this expression with respect to $z^{a}$ and $\bar{z}^{\bar{b}}$ then reproduces the Weil-Petersson metric Eq.~\eqref{eq:KMCSM}, confirming that $\mathcal{K}_{\text{cs}}$ correctly encodes the K\"ahler geometry of the complex structure moduli space. Indeed, the second derivative of the K\"ahler potential is given by 
\begin{equation}
    \p_a\p_{\bar b} \mathcal{K}_\text{cs} = -\frac{\int \p_{a} \Omega\wedge\bar{\p}_{\bar b}\overline\Omega}{\int  \Omega\wedge\overline\Omega}+ \frac{\left(\int  \p_{a} \Omega\wedge\overline\Omega\right)\left(\int  \Omega\wedge\bar\p_{\bar b}\overline\Omega\right)}{\left(\int  \Omega\wedge\overline\Omega\right)^2}\,. 
\end{equation}
This can be simplified by making use of the expression (see \cite{tian1987smoothness, Candelas:1990pi, Klemm:2005td}) 
\begin{equation}
    \partial_a \Omega = k_a \Omega + \chi_a \,,
\end{equation}
where $\chi_a$  are $(2,1)$ forms that are harmonic with respect to $g_{mn}$, and $k_a$ are independent of the internal coordinates. We thus have
\begin{equation}
    \int \p_{a} \Omega\wedge\bar{\p}_{\bar b}\overline\Omega = k_{a}\int  \Omega\wedge\bar{\p}_{\bar b}\overline\Omega +\int \chi_a \wedge\bar{\chi}_{\bar{b}}\, .
\end{equation}
Plugging this in, one notices that
\begin{align}
	 \p_{a} \p_{\bar b} \mathcal{K}_\text{cs} &= -\frac{\int \p_{a} \Omega\wedge\bar{\p}_{\bar b}\overline\Omega}{\int  \Omega\wedge\overline\Omega}+ \frac{\left(\int  \p_{a} \Omega\wedge\overline\Omega\right)\left(\int  \Omega\wedge\bar{\p}_{\bar b}\overline\Omega\right)}{\left(\int  \Omega\wedge\overline\Omega\right)^2}\nonumber\\
	 &=  -\frac{\int  \chi_a \wedge \bar{\chi}_{\bar{b}}}{\int  \Omega\wedge\overline\Omega}\,,
\end{align}
which matches \eqref{eq:KMCSM} as required.

To summarize, we have identified the K\"ahler potential $\mathcal{K}_{\mathrm{cs}}$ in \eqref{eq:Kcs}, which reproduces the K\"ahler metric \eqref{eq:KMCSM} on the complex structure moduli space. This, however, is only one component of the larger picture. Our ultimate goal, as outlined in Chapter~\ref{chap:gkp}, is to determine the effective potential generated for the complex structure moduli $z^a$ by the non-trivial $3$-form flux background. Establishing this potential will serve as the first step toward a concrete realization of moduli stabilization within the four-dimensional effective theory framework.

\subsection{Flux potential and superpotential}\label{sec:fluxpotential}

To determine the effective potential, we observe that the ten-dimensional kinetic term for $G_{3}$ in \eqref{eq:KinTG3} generates a contribution to the four-dimensional scalar potential. Early analyses of dimensional reduction with background fluxes include \cite{Polchinski:1995sm,Michelson:1996pn,Gukov:1999ya,Dasgupta:1999ss}. Adopting the approach developed in \cite{Giddings:2001yu,Grimm:2004uq,Grimm:2005fa}, we begin by rewriting the flux kinetic term \eqref{eq:KinTG3} as
\begin{equation}\label{eq:s10flux}
    S_{10} \supset \frac{1}{2\kappa_{10}^2}\int \biggl (-V_{\text{flux}}+\dfrac{1}{2} Q_{\mathrm{flux}}\biggl ) \,\sqrt{g}\,\mathrm{d}^4x\,,
\end{equation}
in terms of the D3-brane charge $Q_{\mathrm{flux}}$ induced by fluxes, as defined in Eq.~\eqref{eq:D3flux}, and the \emph{flux scalar potential} $V_{\text{flux}}$ defined as\index{Flux scalar potential}
\begin{equation}\label{eq:FC:39} 
    V_{\text{flux}}\equiv \dfrac{1}{4\,\mathrm{Im}(\tau)}\int_{X}\,   G_{-}\wedge\star_{6} \overline{ G}_{-}\, .
\end{equation}
Here, we choose normalizations of the (A)ISD components of $G_3$ as
\begin{equation} 
    G_{3}= \frac{1}{2\mathrm{i}}(G_{+}- G_{-})\kom G_{\pm} = (\star_6 \pm \mathrm{i}) G_3\kom \star_{6} G_{\pm}=\pm\I G_{\pm}\, .
\end{equation}
The Gauss's
law constraint \eqref{eq:D3Tadpole} enforces that the term in \eqref{eq:s10flux} proportional to $Q_{\mathrm{flux}}$ is canceled by local source terms upon integrating over the internal space.

For O3/O7 orientifolds, the imaginary anti-self-dual flux $G_{-}$ can be expanded terms of elements of $H^{3,0}_{-} \oplus H^{1,2}_{-}$, as \cite{DeWolfe:2002nn,Grimm:2004uq}
\begin{equation}\label{eq:decompGminus}
    G_{-}=\dfrac{2\mathrm{i}}{\int_{X}\, \Omega\wedge\overline{\Omega}}\left (\Omega \int_{X}\, \overline{\Omega}\wedge  G_{3}+\mathcal{K}^{a\bar{b}}\, \overline{\chi}_{b}\int_{X}\, \chi_{a}\wedge  G_{3}\right )\,,
\end{equation} 
where $\chi_a$ are a basis of $(2,1)$-forms, and $\mathcal{K}^{a\bar{b}}$ is the inverse of the Weil-Petersson metric on complex structure moduli space as defined in \eqref{eq:KMCSM}. We now plug \eqref{eq:decompGminus} into the expression \eqref{eq:FC:39} for the flux scalar potential. With $\star_{6} G_{-}=-\I G_{-}$, we obtain
\begin{equation}
    V_{\text{flux}}=\dfrac{\mathrm{e}^{\mathcal{K}_{\text{cs}}}}{\text{Im}(\tau)}\left (\int_{X}\, \Omega \wedge \ov{ G}_{3}\int_{X}\, \overline{\Omega}\wedge  G_{3}+\mathcal{K}^{a\bar{b}}\int_{X}\, \overline{\chi}_{b}\wedge\overline{ G}_{3}\, \int_{X}\, \chi_{a}\wedge  G_{3}\right )\, ,
\end{equation}
where we have used \eqref{eq:Kcs} to write
\begin{equation}
    \mathrm{e}^{\mathcal{K}_{\text{cs}}} = \dfrac{1}{-\I\int_{X}\, \Omega\wedge\overline{\Omega}}\, .
\end{equation}
We identify \cite{Candelas:1990pi,Grimm:2004uq}
\begin{equation}\label{eq:DefFTermsIntXT} 
    D_{\tau}W=\dfrac{1}{2\,\mathrm{Im}(\tau)}\int_{X}\Omega\wedge\overline{ G}_{3}\kom D_{a}W=\int_{X}\, \chi_{a}\wedge G_{3}\, .
\end{equation}
Hence, we obtain the flux scalar potential
\begin{equation}\label{eq:FC:33}
    V_{\text{flux}}=\dfrac{\mathrm{e}^{\mathcal{K}_{\text{cs}}}}{\mathrm{Im}(\tau)}\left (\mathcal{K}^{\tau\bar{\tau}}D_{\tau}W\,D_{\bar{\tau}}\overline{W}+\mathcal{K}^{a\bar{b}}D_{a}W\, D_{\bar{b}}\overline{W}\right )\,,
\end{equation}
where we have taken $\mathcal{K}^{\tau\bar{\tau}}=4\,\mathrm{Im}(\tau)^{2}$. This analysis shows that the induced superpotential $W$ is given by the \emph{Gukov-Vafa-Witten (GVW) superpotential}\index{Gukov-Vafa-Witten superpotential}\index{GVW superpotential} \cite{Gukov:1999ya}
\begin{equation}\label{eq:GVW} 
    W_{\text{GVW}}(z,\tau)\coloneqq  \sqrt{\tfrac{2}{\pi}}\int_{X}\,  G_{3}\wedge\Omega\,,
\end{equation} 
where the numerical factor is a convenient convention. This superpotential depends on the axio-dilaton $\tau$ through the complexified $3$-form flux $G_{3}$, as well as on the complex structure moduli $z^{a}$ due to the presence of $\Omega=\Omega(z)$.

We have just written the kinetic terms for the complex structure moduli and dilaton, and the kinetic terms for the K\"ahler moduli can be obtained by generalizing \eqref{eq:K3lg}.

As a shortcut, let us consider the Einstein-Hilbert term in ten dimensions, in addition to the kinetic term for $G_3$,
\begin{equation}
    S = \frac{1}{2\kappa_{10}^2}\int \,\biggl (\mathcal{R}-\dfrac{1}{12\,\text{Im}(\tau)}\, |G_3|^2\biggl )\, \sqrt{g}\dif^{10}x\, .
\end{equation}
Reducing this action on the background \eqref{eq:WarpedBackgroundAnsatzGKPSection} for arbitrary $h^{1,1}$ and ignoring the warp factor for simplicity, we find in four dimensions
\begin{equation}
    S = \frac{1}{2\kappa_{4}^2}\int \,\biggl (\mathcal{V}^{(0)}\,\mathcal{R}^{(4)}-V_{\text{flux}}+\ldots \biggl )\, \sqrt{{g}}\dif^{4}x\, ,
\end{equation}
where we ignore other couplings in four dimensions, such as the kinetic terms for the moduli. We have introduced the classical Calabi-Yau volume
\begin{equation}
    \mathcal{V}^{(0)} = \int_X\, \sqrt{{g}}\dif^6 y\, .
\end{equation}
Now we perform a Weyl rescaling of the four-dimensional metric,\footnote{For the rescaling to be dimensionless, we should use $\mathcal{V}^{(0)}/\langle \mathcal{V}^{(0)}\rangle$, see e.g. Appendix A of \cite{Conlon:2005ki}. For convenience, we are setting $\langle \mathcal{V}^{(0)}\rangle=1$ here.}
\begin{equation}
    \hat{g}_{\mu\nu} = \mathcal{V}^{(0)} g_{\mu\nu}
\end{equation}
so that the Einstein-Hilbert term becomes canonically normalized, leading to
\begin{equation}
    S = \frac{1}{2\kappa_{4}^2}\int \,\biggl (\hat{\mathcal{R}}^{(4)}-\dfrac{V_{\text{flux}}}{(\mathcal{V}^{(0)})^{2}}+\ldots \biggl )\, \sqrt{\hat{g}}\dif^{4}x\, .
\end{equation}
From the definition \eqref{eq:FC:33} of the flux scalar potential, we can then write
\begin{equation}
    \dfrac{V_{\text{flux}}}{(\mathcal{V}^{(0)})^{2}} = \mathrm{e}^{\mathcal{K}_{\text{tree}}}\left (\mathcal{K}^{\tau\bar{\tau}}D_{\tau}W\,D_{\bar{\tau}}\overline{W}+\mathcal{K}^{a\bar{b}}D_{a}W\, D_{\bar{b}}\overline{W}\right )
\end{equation}
where we identified the tree-level K\"ahler potential $\mathcal{K}_{\text{tree}}$ from the overall prefactor in front of the brackets as
\begin{equation}\label{eq:fullK}
    \mathcal{K}=\mathcal{K}_{\text{tree}} = -2\,\mathrm{log}\,\Bigl(\mathcal{V}^{(0)}(T_1,\ldots T_{h^{1,1}})\Bigr)-\mathrm{log}\,\Bigl(-i\int\Omega\wedge\bar{\Omega}\Bigr)-\mathrm{log}\,\Bigl(-i(\tau-\bar{\tau})\Bigr)+\mathcal{C} 
\end{equation} 
where the constant $\mathcal{C}$ depends on normalizations that we have not carefully fixed here.  In the conventions of \cite{McAllister:2024lnt}, one has $\mathcal{C}=-\mathrm{log}(8)$.

\section{Stabilization of complex structure moduli}\label{sec:classicalstab}
 
The chiral multiplets of the $\mathcal{N}=1$ theory from compactification on a Calabi-Yau orientifold are 
\begin{equation}
    T_i, z^a, \tau\,,
\end{equation}
while the classical superpotential is
\begin{equation}
    W = \sqrt{\tfrac{2}{\pi}}\,\int G_3 \wedge \Omega\,,
\end{equation}
and the tree-level K\"ahler potential is given in \eqref{eq:fullK}.

We would like to  seek supersymmetric solutions, i.e.,~solutions to the $F$-flatness equations
\begin{align}
    D_{z^a} W &= 0\; , \qquad D_{\tau} W = 0\; ,\qquad D_{T_i} W = 0\, .
\end{align}
Referring to
\eqref{eq:DefFTermsIntXT} 
and recalling that the ISD condition allows $G_3 = G_{2,1} + G_{0,3}$, we find:\footnote{The condition \eqref{eq:DTW1} as written is correct only at the classical level, for $W=W_{\text{GVW}}$: quantum effects lead to a crucial change in \eqref{eq:DTW1}, as we will explain in Chapter \ref{chap:quantumEFT}.}
\begin{equ}[ISD flux and $F$-terms]
The moduli $F$-terms and the Hodge type of the $3$-form flux are related as follows:
\begin{align}
    D_{z^a} W = 0 &\Leftrightarrow G_{1,2}=0\label{eq:DaW1}\, ,\\[0.3em]
    D_{\tau} W = 0 &\Leftrightarrow  G_{3,0}=0\label{eq:DtauW1}\, ,\\[0.3em]
    D_{T_i} W = 0&\Leftrightarrow G_{0,3}=0\label{eq:DTW1}\, ,\\[0.3em] 
    \mathrm{ISD}~\mathrm{flux} &\Leftrightarrow G_{1,2} =G_{3,0}=0
\end{align}
\end{equ} 

Evidently, $G_{2,1}$ flux is ISD and allowed by unbroken supersymmetry, while $G_{0,3}$ is ISD but breaks supersymmetry through $F$-terms for the K\"ahler moduli $T_i$.

Now suppose that we choose a set of flux quanta, in the sense of \eqref{eq:DiracQuantvec}.
Such a choice defines a nonnegative scalar potential 
$V_{\text{flux}}$, given in \eqref{eq:FC:39}, on the moduli space of the $z^a$ and $\tau$.
As one moves in this moduli space, the Hodge type of $G_3$ varies, even though the flux quanta are unchanged. 
At special loci where $G_{1,2}=G_{3,0}=0$, 
one has vanishing $F$-terms for $z^a$ and $\tau$, and as a result $V_{\text{flux}}=0$.

On the grounds of equation counting, one expects\footnote{We will substantiate this expectation in explicit examples in Chapter \ref{chap:modulistabilization}.} that if the choice of flux quanta is generic, then the solutions to 
$D_{a} W =D_\tau W = V_{\text{flux}} =0$ will be isolated points.  At such a point one has
\begin{equation}
     \sqrt{\tfrac{2}{\pi}}\,\int G_3 \wedge \Omega \to \Bigl\langle \sqrt{\tfrac{2}{\pi}}\,\int G_3 \wedge \Omega \Bigr\rangle \equiv W_0\,.
\end{equation}
Fluctuations of $z_a$, $\tau$ away from such a point are massive,
\begin{equation}\label{eq:massscale}
    m \sim \frac{\alpha'}{\sqrt{\mathcal{V}}}\,.
\end{equation}
Thus, the complex structure moduli $z^a$ and the axio-dilaton $\tau$ are \emph{stabilized} by a generic choice of $3$-form flux quanta.

\section{No-scale structure}\label{sec:noscale}
\index{No-scale structure}

Next we turn to the K\"ahler sector.  
For simplicity of presentation
we take $h^{1,1}_+=1$, so that there is a single K\"ahler modulus $T$.
The effective theory below the mass scale \eqref{eq:massscale} contains only $T$ (and perhaps D3-brane or D7-brane position fields, see later), with
\begin{equation}
    \mathcal{K} = -3\,\mathrm{log}\left(T+\bar{T}\right)\,, \qquad W = W_0 \,.
\end{equation}
The $F$-term for $T$ is
\begin{equation}
    D_T W = \partial_T W + W \partial_T K = - \frac{3 W_0 }{T+\bar{T}}\,.
\end{equation}
Supersymmetry is therefore broken, unless $W_0=0$ or $\mathcal{V} \to \infty$.  In other words, as noted above, $G_{0,3}$ flux breaks supersymmetry through a K\"ahler modulus $F$-term.

Nonetheless, the potential is actually zero at the order at which we are working!
The full potential is
\begin{equation}
    V_F = \mathrm{e}^{\mathcal{K}} \Bigl(\mathcal{K}^{A\bar{B}}D_A W \overline{D_B W} - 3 W \overline{W} \Bigr)\,,
\end{equation}
where the $A, B$ indices run over all the moduli (the K\"ahler modulus $T$, the complex structure moduli $z^a$, and the axio-dilaton).  
Since $D_a W = D_{\tau} W = 0$, we have 
\begin{equation}
    \mathrm{e}^{-\mathcal{K}}\,V_F = \mathcal{K}^{T\bar{T}}D_T W \overline{D_T W} - 3 W \overline{W}=\left(\mathcal{K}^{T\bar{T}}\mathcal{K}_T \overline{\mathcal{K}_T} -3\right)W \overline{W}\,,
\end{equation}
where we have used the fact that the Gukov-Vafa-Witten superpotential \eqref{eq:GVW} is independent of the K\"ahler modulus $T$.

Now we observe that
\begin{equation}\label{eq:onefieldnoscale}
    \mathcal{K}_{T\overline{T}}=\frac{3}{(T+\overline{T})^2} \Rightarrow \mathcal{K}^{T\bar{T}}\mathcal{K}_T \overline{\mathcal{K}_T}=3\,,
\end{equation} which is known as the \emph{no-scale} property.  We have exhibited no-scale structure in the example of a single K\"ahler modulus $T$ because \eqref{eq:onefieldnoscale} is trivially verified, but no-scale applies just as well when there are $h^{1,1}_+>1$ K\"ahler modulus $T_i$: from \eqref{eq:fullK} one finds
\begin{equation}\label{eq:NoScaleID}
    \mathcal{K}^{i\bar{\jmath}}\, \mathcal{K}_{i} \mathcal{K}_{\bar{\jmath}}-3=0\, .
\end{equation}

At the level of a no-scale theory, the K\"ahler moduli $T_i$ are unfixed, and the vacuum energy vanishes.  However, no-scale structure is only a property of the K\"ahler potential at leading order in $g_s$ (i.e., at the classical level) and in $\alpha'$.  Additional quantum corrections and sigma model corrections to the K\"ahler potential typically break the no-scale structure, as we discuss in \S\ref{sec:quantum_pert}.   Thus, no-scale supergravity theories do not, on their own, provide a solution to the cosmological constant problem.

\chapter{Quantum Theory of Flux Compactifications} \label{chap:quantumEFT}

So far, we have worked out the \emph{classical} theory of flux compactifications. At this level, the complex structure moduli and axio-dilaton are stabilized by 3-form flux, but the   complexified K\"ahler moduli remain unfixed.  Thus, the final vacuum structure of the type IIB landscape will be determined by \emph{quantum corrections} to the effective theory. Quantum effects, including higher-derivative $\alpha'$ corrections, string loop contributions, and non-perturbative effects, modify the four-dimensional effective theory. These corrections play a decisive role in lifting classical flat directions and generating potentials for the K\"ahler moduli. Our next task is to understand these corrections well enough, and compute them accurately enough, in order to find isolated/stabilized vacua.

\section{Overview of quantum corrections}\label{sec:scheme}

To orient ourselves to the problem of computing the structure of the scalar potential in a string compactification, we first consider the corresponding problem in a weakly-coupled quantum field theory with  a single scalar field $\Phi$ and a coupling constant $g \ll 1$.  The vacuum energy can be written as
\begin{equation}\label{eq:vgen}
    V(\Phi) = \sum_{n=0} g^n V_n(\Phi) + \mathcal{O}\bigl(\mathrm{e}^{-1/g}\bigr)\,,
\end{equation} and the terms $V_n$ can in principle be computed in perturbation theory.
Given a computation to some order, 
it is useful to write the schematic expression
\begin{equation}\label{eq:vco}
    V(\Phi) = V_{\mathrm{c.o.}}(\Phi) + \Delta V\,,
\end{equation} 
where the subscript ``c.o.'' denotes the \emph{computed order}, i.e.~the order in $g$ to which $V$ has been computed, while $\Delta V$ encodes an error term, whose parametric size may be known, but whose details are not.   

Let us extend the above logic to a flux compactification of type IIB string theory, where there are two fundamental expansions, in $g_s$ and in $\alpha'$, as well as a number of other approximation schemes. The symbol $\Phi^A$, $A=1,\ldots,N$ will now denote the moduli fields.

The task is to compute the moduli $\Phi^A$ and the scalar potential $V(\Phi^A)$ in terms of the underlying geometric data of the compactification.
We denote the geometric data by 
$\mathscr{G}$, and use the schematic decomposition
\begin{equation}\label{eq:Gsplit}
    \mathscr{G} = \mathscr{C} \cup \mathscr{D}_{\mathbb{Z}}\,,
\end{equation}
where $\mathscr{C}$ denotes continuous data, and $\mathscr{D}_{\mathbb{Z}}$ denotes discrete data.
For example, the triple intersection numbers $\kappa_{ijk} \in \mathbb{Z}$ are discrete data, whereas the K\"ahler form $J = t^{i} \omega_{i}$ given in \eqref{eq:defkahlerparamdown} involves the $h^{1,1}$ real-valued parameters $t^i$, and so can be viewed as continuous data.\footnote{Recall from \S\ref{sec:kcoords} that we are restricting to cases with $h^{1,1}_+=h^{1,1}$.}

As explained in \S\ref{sec:moduliofCY}, the moduli fields include the geometric moduli of the Ricci-flat metric, and so the continuous data $\mathscr{C}$ can be encoded by the expectation values of the $\Phi^A$.\footnote{The functional form of this encoding is an important piece of information: see the illustration in \S\ref{sec:kahlercoord}.}
Thus, we can think of the potential $V(\Phi^A;\mathscr{G})$ determined by a particular compactification as $V(\Phi^A;\mathscr{D}_{\mathbb{Z}})$: the data $\mathscr{D}_{\mathbb{Z}}$ encode what the moduli space is, and the $\Phi^A$ parameterize a location within that moduli space.

Given a compactification, specified initially at the level of classical geometry, the moduli fields can be written as
\begin{equation}\label{eq:phiofG}
    \Phi^A(\mathscr{G}) = \sum_{m,n}g_s^m (\alpha')^n\, \Phi^A_{m,n}(\mathscr{G}) + \mathcal{O}\bigl(\mathrm{e}^{-1/g_s}\bigr) +\mathcal{O}\bigl(\mathrm{e}^{-1/\alpha'}\bigr)\,,
\end{equation}
As in \eqref{eq:vgen}, the terms $\Phi^A_{m,n}(\mathscr{G})$ can \emph{in principle} be computed order by order in string perturbation theory, and the schematic exponential terms encode non-perturbative corrections, generated for example by D-instantons and worldsheet instantons. 

Next, the scalar potential in a theory that preserves $\mathcal{N}=1$ supersymmetry at the level of the Lagrangian --- but not necessarily in the state of interest --- can be written as
\begin{equation}\label{eq:vbreak}
    V = V_{\mathcal{N}=1}+V_{\mathrm{breaking}}\,,
\end{equation} 
in which $V_{\mathcal{N}=1}$ consists of terms encoding the potential in the supersymmetric theory, including the effects of spontaneous breaking of supersymmetry, while 
$V_{\mathrm{breaking}}$ contains all effects that do not have a manifest or convenient representation in terms of spontaneously broken supersymmetry.

In an $\mathcal{N}=1$ supersymmetric theory with (hypothetical) exactly known $W$ and $\mathcal{K}$, and with no vector multiplets yielding $D$-terms, 
we can write
\begin{equation}\label{eq:ffour}
    V_{\mathcal{N}=1} = V_{F} + c_4 \frac{F^4}{\Lambda^2} + \ldots\,,
\end{equation} 
where $\Lambda$ 
is a high cutoff scale, and where $V_F$ is given in terms of $W$ and $\mathcal{K}$ by \eqref{eq:vfgen}.  The quartic term in \eqref{eq:ffour} is not often encountered, because it is convenient, and usually sufficient, to work with the $\mathcal{N}=1$ supergravity Lagrangian up to two spacetime derivatives. The $F^4$ term, which involves zero spacetime derivatives, is paired by supersymmetry with higher-curvature corrections.  See e.g.~cite \cite{Koehn:2012ar} and \cite{Ciupke:2015msa} for more detailed discussion and additional references on higher $F$-terms.

In practice one does not know $W$ and $\mathcal{K}$ exactly: neither as functions of the exact $\Phi^A(\mathscr{G})$, nor as functions of the approximate $\Phi^A_{m,n}(\mathscr{G})$.\footnote{In very special cases $W$ might be exactly computable, but this has not yet been achieved in the examples relevant for de Sitter constructions.}  
Instead, both $W$ and $\mathcal{K}$ have double expansions
\begin{equation}\label{eq:woformalcorr}
    W(\Phi^A) = \sum_{m,n}g_s^m (\alpha')^n\, W_{m,n}(\Phi^A)\,,
\end{equation}
\begin{equation}\label{eq:kformalcorr}
    \mathcal{K}(\Phi^A,\bar{\Phi}^{\bar{A}}) = \sum_{m,n}g_s^m (\alpha')^n\, \mathcal{K}_{m,n}(\Phi^A,\bar{\Phi}^{\bar{A}})\,,
\end{equation} where now we emphasize that the dependence of $\mathcal{K}$ on the moduli fields $\Phi$, i.e.,~on the K\"ahler coordinates on moduli space, is not holomorphic.

A crucial requirement is that the fields $\Phi^A$ must furnish \emph{K\"ahler coordinates}\footnote{Such coordinates are often called `good K\"ahler coordinates' for emphasis, because so many objects are related to K\"ahler structures: every $\mathcal{N}=1$ supersymmetric moduli space is a K\"ahler manifold, whether or not it relates to 
deformations of the K\"ahler structure of the Ricci-flat metric.  Good K\"ahler coordinates are special, but because of the possibility of field redefinitions they are not uniquely defined. In particular, they provide a set of complex variables on which all holomorphic quantities of the theory depend holomorphically, most notably the superpotential $W(\Phi^A)$.} 
on the moduli space, i.e.~coordinates in terms of which $\mathcal{K}(\Phi^A,\bar{\Phi}^{\bar{A}})$ is a K\"ahler potential that yields a K\"ahler metric.
That is, the quantity
\begin{equation}
    \partial_{\Phi^A}\partial_{\bar{\Phi}^{\bar{B}}}\mathcal{K} \equiv \mathcal{K}_{A\bar{B}}
\end{equation}
must be a K\"ahler metric on the moduli space.

The form of the terms $W_{m,n}$ is far more constrained than that of the terms $\mathcal{K}_{m,n}$, because the superpotential is protected by non-renormalization theorems.  In \S\ref{sec:thew} below we will make extensive use of the associated structures, but for now it is more convenient to write  schematic expressions,
\begin{equation}
    W(\Phi)=W_{\mathrm{c.o.}}(\Phi)+ \Delta W\,,  
\end{equation} and 
\begin{equation}
    \mathcal{K}(\Phi,\bar{\Phi})=\mathcal{K}_{\mathrm{c.o.}}(\Phi,\bar{\Phi})+ \Delta \mathcal{K}\,.
\end{equation} 
As in \eqref{eq:vco}, the subscript ``c.o.'' denotes the order to which the quantities have been computed, while $\Delta W$ and $\Delta \mathcal{K}$ are error terms.

In fact we must be slightly more careful in order to properly compute the vacuum energy as a function of the geometric data $\mathscr{G}$.
In addition to perturbative corrections to the functional dependence of $W$ and $\mathcal{K}$ on the \emph{exact} $\Phi^A(\mathscr{G})$, we should account for perturbative corrections $\Phi^A_{m,n}(\mathscr{G})$ to the relation between the geometry $\mathscr{G}$ and the coordinates $\Phi^A$, as in \eqref{eq:phiofG}.

At the cost of a burdensome notation, we write
\begin{equation}
W(\Phi)=W_{\mathrm{c.o.}}\bigl(\Phi^A_{\mathrm{c.o.'}}(\mathscr{G})\bigr)+ \Delta W\,,  
\end{equation} and 
\begin{equation}
\mathcal{K}(\Phi^A,\bar{\Phi}^{\bar{A}})=\mathcal{K}_{\mathrm{c.o.}}\Bigl(\Phi^A_{\mathrm{c.o.'}}(\mathscr{G}),\bar{\Phi}^{\bar{A}}_{\mathrm{c.o.'}}(\mathscr{G})\Bigr)+ \Delta \mathcal{K}\,,
\end{equation} where we have allowed the possibility that the orders of computation of the $W-\Phi$ and $\Phi-\mathscr{G}$ relations may differ, and likewise for $\mathcal{K}$.

With these preliminaries, we can state the process of obtaining an approximation to the vacuum structure:
\begin{enumerate}
    \item First one explicitly computes the fundamental data: one calculates the quantities
$W_{\mathrm{c.o.}}(\Phi)$, 
$\mathcal{K}_{\mathrm{c.o.}}(\Phi,\bar{\Phi})$,
and $\Phi_{\mathrm{c.o.}}(\mathscr{G})$, each to some order.\footnote{The orders may well differ from one quantity to another, but as we have belabored this point sufficiently, we will suppress the distinction henceforth.}
\item Then one substitutes $W_{\mathrm{c.o.}}(\Phi)$, 
$\mathcal{K}_{\mathrm{c.o.}}(\Phi,\bar{\Phi})$,
and $\Phi_{\mathrm{c.o.}}(\mathscr{G})$ in $V_F$, arriving at
\begin{equation}
    V = V_{\mathrm{c.o.}}(\mathscr{G}) + \Delta V(\Delta W, \Delta\mathcal{K},\mathscr{G})\,,
\end{equation}
where the term $V_{\mathrm{c.o.}}(\mathscr{G})$ is fully specified and known, and $\Delta V$ is an error term.
\item Finally, to give evidence for a local minimum of the scalar potential, one computes, bounds, or estimates the error term.
\end{enumerate}
For example, establishing that $\Delta W$ is negligibly small can often be done by demonstrating that $\Delta W$ consists of sub-leading exponentials compared to $W_{\mathrm{c.o.}}$.  The corrections $\Delta \mathcal{K}$, in contrast, are generally only polynomially smaller, in control parameters such as $g_s$,
than $\mathcal{K}_{\mathrm{c.o.}}$.

The above process can be extended to incorporate explicit supersymmetry breaking by the term $V_{\mathrm{breaking}}$.  We will return to this point in Chapter \ref{chap:modulistabilization}, once we have explained how to compute the supersymmetry-breaking effects of anti-D3-branes.

\section{K\"ahler moduli space}

The above considerations were very general, so before proceeding we illustrate them in an example. In this section we revisit the K\"ahler moduli of a Calabi-Yau threefold, beginning with their classical description and then turning to quantum corrections.  At tree level, the K\"ahler class encodes the volumes of holomorphic cycles and determines the  
expansion
parameters in the $\alpha'$ series. 
Beyond tree level, the map between the geometric quantities and the coordinates on moduli space receives a variety of perturbative and non-perturbative corrections, which play a decisive role in moduli stabilization.  By first reviewing the classical structure and its relation to the topological data of the manifold, we will set the stage for understanding how this structure is deformed by quantum effects in the full flux compactification.

\subsection{K\"ahler moduli at tree level}\label{sec:kahlercoord}

We begin with the classical geometry of a Calabi-Yau threefold $X$. One important piece of the geometric data $\mathscr{G}$ of $X$ is the K\"ahler class $J$ of the Ricci-flat K\"ahler metric.  Let us see how this data is related to discrete data $\mathscr{D}_\mathbb{Z}$, and to complex K\"ahler coordinates $\Phi^A$ on moduli space.

Upon choosing a basis $\{\omega_i\}_{i=1}^{h^{1,1}(X)}$ of $H^2(X,\mathbb{Z})$, the K\"ahler class $J$ (in string frame) can be parametrized by $h^{1,1}$
real numbers $t^i$, the K\"ahler parameters, as in \eqref{eq:defkahlerparamdown}:
\begin{equation}\label{eq:jdef}
    J=\sum_i t^i\,\omega_i\,.
\end{equation}
The $t^i$ are constrained to lie in a cone, the \emph{K\"ahler cone}\index{K\"ahler cone} $\mathcal{K}_X\subset H^{1,1}(X,\mathbb{R})$ of $X$. The dual cone of $\mathcal{K}_X$ is the \emph{Mori cone}\index{Mori cone} $\mathcal{M}_{X} \subset H_{2}(X,\mathbb{R})$: for any $\mathbf{q}\in\mathcal{M}_{X}$, $q_it^i>0$ for all $\vec{t}\in\mathcal{K}_X$. The triple intersection numbers of $X$ are defined as
\begin{equation}
    \kappa_{ijk}\coloneqq \int_X\, \omega_i\wedge \omega_j\wedge \omega_k\, .
\end{equation}
The discrete data $\mathscr{D}_\mathbb{Z}$ at this stage are the $\kappa_{ijk}$ and the generators of the K\"ahler cone, and the (continuous) geometric data $\mathscr{G}$ consists of $J$ in \eqref{eq:jdef}.

Having organized the classical K\"ahler moduli in terms of their topological and geometric ingredients, we now turn to the corresponding metric on moduli space. This proceeds in close analogy with the analysis of \S\ref{eq:WPcsm} for the Weil-Petersson metric on complex structure moduli space: by studying the variations of the Ricci-flat metric induced by deformations of the K\"ahler class, we obtain the Weil-Petersson metric\index{Weil-Petersson metric} on K\"ahler moduli space. For the deformations defined in Eq.~\eqref{eq:kahlerdeformations}, this leads to
\begin{equation}\label{eq:WPMdefKM}
	2G_{ij}\, \delta t^{i}\delta t^{j} = \dfrac{1}{2\mathcal{V}^{(0)}}\int_{X}\, g^{m\bar{n}}g^{k\bar{l}}\delta g_{m\bar{l}}\delta g_{\bar{n}k}\, \sqrt{g}\, \dif^{6}y\, .
\end{equation}
We choose a basis $\omega_i\in H^{1,1}(X,\mathbb{R})$ so that
\begin{equation}
    \delta g_{m\bar{n}} = (\omega_i)_{m\bar{n}}\, \delta t^{i}\,,
\end{equation}
which implies that \eqref{eq:WPMdefKM} can be written as
\begin{equation}\label{eq:WPKMS}
	G_{ij} = \dfrac{1}{4\mathcal{V}^{(0)}}\int_{X}\, \omega_i\wedge\star_6\omega_j\, ,
\end{equation}
which agrees with \cite{Strominger:1985ks} --- see also \cite{Candelas:1990pi,Grimm:2004uq} and recall our discussion around Eq.~\eqref{eq:zeroform}. One can then show that the action of the Hodge star $\star_6$ on $\omega_j\in H^{1,1}(X,\mathbb{R})$ is given by
\begin{equation}
    \star_6\, \omega_j = -J\wedge \omega_j +\dfrac{\int_{X}\, J\wedge J\wedge \omega_j}{4\mathcal{V}^{(0)}}\, J\wedge J\, .
\end{equation}
Let us introduce the quantities
\begin{equation}\label{eq:Kahlercoordinatestree}
    \mathcal{T}_j^{\text{tree}} = \dfrac{1}{2}\int_{X}\, J\wedge J\wedge \omega_j = \dfrac{1}{2}\kappa_{jkl}t^kt^l\, ,
\end{equation}
corresponding to the (classical) string frame divisor volumes.
We can then write \eqref{eq:WPKMS} as \cite{Candelas:1990pi}
\begin{align}\label{eq:WPKMSfinal}
    G_{ij} &= \dfrac{1}{4\mathcal{V}^{(0)}}\biggl [-\int_{X}\, J\wedge\omega_i\wedge \omega_j+\dfrac{\mathcal{T}_i^{\text{tree}}}{2\mathcal{V}^{(0)}}\, \int_{X}\, \omega_i\wedge J\wedge J\biggl ]\nonumber\\
    &=\dfrac{1}{4\mathcal{V}^{(0)}}\biggl [-\kappa_{ijk}t^k+\dfrac{\mathcal{T}_i^{\text{tree}}\,\mathcal{T}_j^{\text{tree}}}{\mathcal{V}^{(0)}}\biggl ] \, .
\end{align} 
The resulting Weil-Petersson metric $G_{ij}$ on K\"ahler moduli space is completely fixed by the intersection numbers of $X$, together with a choice of K\"ahler parameters $t^i$ and the associated classical divisor volumes $\mathcal{T}^{\text{tree}}_{i}$ defined in Eq.~\eqref{eq:Kahlercoordinatestree}.
The classical result \eqref{eq:WPKMSfinal} for $G_{ij}$  
will provide the reference point for incorporating perturbative and non-perturbative corrections in the next sections.

With the moduli-space metric written purely in terms of the classical geometric data $(t^{i}, \kappa_{ijk})$, the next step is to identify the holomorphic coordinates of the four-dimensional $\mathcal{N}=1$ effective theory. This amounts to complexifying the real K\"ahler parameters by combining them with the corresponding R-R axions. Equivalently, we seek an explicit relation between the real variables $t^{i}$ and the complex K\"ahler coordinates $\Phi^{A}$ defined above. Guided by the analysis of \S\ref{sec:kintermex}, we introduce the complex coordinates
\begin{equation}\label{eq:Ttree}
    T_i^{\text{tree}} \coloneqq \frac{1}{2g_s}\int_{X}\, J \wedge J\wedge\omega_i + \mathrm{i}\int_{X}\, C_4\wedge\omega_i \equiv \dfrac{1}{g_s}\mathcal{T}_i^{\text{tree}} + \mathrm{i} \theta_i
\end{equation}
where the classical string frame divisor volumes $\mathcal{T}_i^{\text{tree}}$ are given by \eqref{eq:Kahlercoordinatestree}. The $T_i^{\text{tree}}$ defined by \eqref{eq:Ttree} are good K\"ahler coordinates on the K\"ahler moduli space, at leading order in $g_s$ and $\alpha'$, i.e.~corresponding to the leading term in the expansion of \eqref{eq:phiofG}. The corresponding K\"ahler potential is  
\begin{equation}\label{eq:ktreeex}
    \mathcal{K}_{\text{tree}} = - 2\,\mathrm{log}\bigl(\mathcal{V}^{(0)} \bigr)\,,
\end{equation}
where the classical Calabi-Yau volume in string frame is given by  
\begin{equation}\label{eq:VolClassical}
   \mathcal{V}^{(0)} \coloneqq \dfrac{1}{6}\int_X  \, J\wedge J\wedge J= \frac{1}{6}\kappa_{ijk}t^it^jt^k\, .
\end{equation}
The classical superpotential $W^{\text{tree}}$ is independent of $T_i^{\text{tree}}$, and so is not shown here.
 
Let us show that the tree-level K\"ahler potential \eqref{eq:ktreeex} leads to the appropriate field space metric $\mathcal{K}_{i\bar{\jmath}}$, which is directly related to the Weil-Petersson metric $G_{ij}$ in \eqref{eq:WPKMSfinal}, and corresponds to a K\"ahler metric defined as $\mathcal{K}_{i\bar{\jmath}}=\partial_{T_i}\partial_{\overline{T}_{\bar{\jmath}}}\mathcal{K}_{\text{tree}}$.  In the 
tree-level theory,
\begin{equation}
    \dfrac{\partial T_i^{\text{tree}}}{\partial t^j} = \dfrac{1}{g_s} \kappa_{ijk}t^k =\dfrac{1}{g_s}  \kappa_{ij}\; , \quad \partial_{t^i}\partial_{t^j}\mathcal{K}_{\text{tree}} = 8G_{ij}
\end{equation}
and so we obtain the expression\footnote{This expression matches the results of \cite{Bobkov:2004cy,AbdusSalam:2020ywo,Cicoli:2021tzt} at leading order in the $\alpha'$ expansion.}
\begin{align}
    \partial_{T_i}\partial_{\overline{T}_{\bar{\jmath}}}\mathcal{K}_{\text{tree}} &= \dfrac{g_s^{2}}{8(\mathcal{V}^{(0)})^{2}}\biggl [-2 \mathcal{V}^{(0)} \kappa^{ij}+ t^it^j \biggl ] 
\end{align}
where $\kappa^{ij}$ denotes the inverse of $\kappa_{ij}$, and the additional factors of $g_s$ arise because we are working with string frame volumes.

\subsection{K\"ahler moduli beyond tree level}

Beyond tree level, the functional dependence $T_i^{\text{tree}}(t^i)$ is corrected compared to \eqref{eq:Kahlercoordinatestree}, and likewise $\mathcal{K}$ is corrected to a more complicated function of the $t^i$ (or $T_i$) than that given in \eqref{eq:ktreeex}: that is,
\begin{equation}
    T_i = T_i^{\text{tree}}+ \delta T_i\; ,\quad \mathcal{K} =  - 2\,\mathrm{log}\bigl(\mathcal{V} \bigr) \; ,\quad \mathcal{V} = \mathcal{V}^{(0)}+\delta \mathcal{V}\, .
\end{equation}
The origin and structure of these corrections can be intricate. Already at string tree level, $\alpha'$ corrections induce shifts in the relation between the $t^{i}$ and the complex coordinates $T_{i}$, as in \eqref{eq:firstdefcalT}. At higher order in the string loop expansion, further corrections appear: see \eqref{eq:TWSILoop} for the loop-corrected definition of $T_i$, and \eqref{eq:Valphap3Loop} and \eqref{eq:VWSILoop} for the corresponding corrections to the K\"ahler potential. 

\newpage

It is therefore useful to view the quantum-corrected moduli space as receiving corrections at two intertwined levels: 
\begin{enumerate}
    \item corrections to the metric for the \emph{real} moduli $t^i$ and the axions, and
    \item corrections requiring \emph{field redefinitions} to obtain the holomorphic coordinates $T_i$. 
\end{enumerate}
This distinction becomes important when interpreting the corrected kinetic terms, especially since axionic shift symmetries restrict the possible corrections to the imaginary parts of the $T_i$ but do not prevent mixing in the real parts --- see in particular \cite{Haack:2018ufg}. 

From the standpoint of the underlying (super)symmetry, different types of corrections arise for different reasons: axionic shift symmetries restrict the dependence of the effective action on the R-R fields, holomorphy constrains the possible quantum corrections to the superpotential, and the pattern of $\mathcal{N}=2 \rightarrow \mathcal{N}=1$ breaking limits which combinations of fields may mix at loop level. Thus, rather than treating individual corrections in isolation, it is useful to organize them according to the symmetries they respect or break, and to understand how these symmetries shape the quantum-corrected moduli space.
To this end, we now provide a more systematic perspective on the origin and structure of quantum corrections. 

\section{Supersymmetry and the structure of corrections}\label{sec:QCorPrel}

In \S\ref{sec:scheme} we gave a very general description of corrections to the K\"ahler coordinates $\Phi^A$, the superpotential $W$, and the K\"ahler potential $\mathcal{K}$, writing the formal expansions \eqref{eq:phiofG}, \eqref{eq:woformalcorr}, and \eqref{eq:kformalcorr} to include perturbative corrections at all orders, as well as non-perturbative corrections. In so doing we did not incorporate any non-renormalization results, nor did we specify a parameter regime and an associated truncation of the series in \eqref{eq:phiofG}, \eqref{eq:woformalcorr}, and \eqref{eq:kformalcorr}: the treatment of \S\ref{sec:scheme} could apply to essentially any $\mathcal{N}=1$ supergravity theory. However, one can expect different phenomena depending on the relative rates at which the weak-coupling ($g_s \to 0$) and weak-curvature ($\alpha' \to 0$) limits are approached. Moreover, weak breaking of $\mathcal{N}=2$ to $\mathcal{N}=1$ supersymmetry, as well as holomorphy of the superpotential, further restrict the form of the effective theory.

We will now specialize to a particular class of theories, and to a particular parameter regime, for which we can be much more specific about the form of the quantum corrections to the effective theory.

First of all, the structure of quantum corrections is far more constrained in the $\mathcal{N}=2$ supersymmetric theory resulting from a type IIB vacuum solution on a Calabi-Yau threefold $X$ than in the $\mathcal{N}=1$ supersymmetric theory resulting from a flux compactification on an orientifold of $X$.  The moduli space $\mathcal{M}(X)$ of the $\mathcal{N}=2$ type IIB theory is locally a product
\begin{equation}\label{eq:factor}
    \mathcal{M}(X) = \mathcal{M}_{\text{vector}}(X)\times \mathcal{M}_{\text{hyper}}(X)\,,
\end{equation}
with the axio-dilaton $\tau$ and the K\"ahler moduli $T_i$ living in hypermultiplets, and the complex structure moduli residing in vector multiplets.  Because the sizes of corrections in the $g_s$ and  $\alpha'$ expansions are parameterized by the dilaton and the K\"ahler moduli, respectively, the vector multiplet moduli space receives no corrections in $g_{s}$ or in $\alpha'$, while the hypermultiplet moduli space suffers corrections in both expansions.

In a general $\mathcal{N}=1$ theory, on the other hand, the K\"ahler potential receives corrections at all orders in perturbation theory, and also non-perturbatively.  
In the spirit of Eq.~\eqref{eq:kformalcorr}, we can write
\begin{equation}
    \mathcal{K}(\Phi^A,\bar{\Phi}^{\bar{A}}) = \mathcal{K}^{\mathcal{N}=2}+\mathcal{K}^{\mathcal{N}=1}\,,
\end{equation}
where the term $\mathcal{K}^{\mathcal{N}=2}$ encodes the K\"ahler potential\footnote{Strictly speaking, a K\"ahler potential is an $\mathcal{N}=1$ quantity, and the hypermultiplet moduli space is a quaternionic K\"ahler manifold, and in particular is not K\"ahler \cite{Strominger:1985ks,Ferrara:1988ff,Ferrara:1989ik,Candelas:1990pi}.  The precise statement is that $\mathcal{N}=2$ corrections entering $\mathcal{K}^{\mathcal{N}=2}$ depend only on data of the Calabi-Yau threefold (oftentimes topological), while the genuinely $\mathcal{N}=1$ corrections in $\mathcal{K}^{\mathcal{N}=1}$ arise from additional sources such as fluxes, D-branes, and orientifold planes that modify the background and break some of the supersymmetry.} for the metric on moduli space that arises in the theory with unbroken $\mathcal{N}=2$ supersymmetry, and the term $\mathcal{K}^{\mathcal{N}=1}$ contains all corrections from breaking of $\mathcal{N}=2$ to $\mathcal{N}=1$ by the presence of fluxes, orientifold planes, and D-branes in the compactification background. 

In particular, the K\"ahler potential for the complex structure moduli can be written
\begin{equation}
    \mathcal{K}_{\text{cs}} = \mathcal{K}_{\text{cs}}^{\mathcal{N}=2}+\delta\mathcal{K}_{\text{cs}}^{\mathcal{N}=1}=-\log\left (-\mathrm{i}\int_{X}\, \Omega\wedge\overline{\Omega}\right )+\delta\mathcal{K}_{\text{cs}}^{\mathcal{N}=1}\,,
\end{equation}
and the factorization \eqref{eq:factor} implies that the $\mathcal{N}=2$ term is exact in both $g_s$ and $\alpha'$.

In contrast, the K\"ahler potential for the K\"ahler moduli enjoys no such protection, and both terms in
\begin{equation}\label{eq:KKtmp}
    \mathcal{K}_{\text{K\"ahler}} = \mathcal{K}^{\mathcal{N}=2}_{\text{K\"ahler}} + \delta\mathcal{K}^{\mathcal{N}=1}_{\text{K\"ahler}}\,,
\end{equation}
are corrected order by order, and non-perturbatively, in the string loop and $\alpha'$ expansions.\footnote{We have denoted the $\mathcal{N}=1$ term as a correction for reasons that will be explained more fully below.}

To characterize these corrections, we first write $\mathcal{K}_{\text{K\"ahler}}^{\mathcal{N}=2}$ as a double sum in the $g_s$ and $\alpha'$ expansions, indexed by $m$ and $n$ respectively,
\begin{equation}
    \mathcal{K}_{\text{K\"ahler}}^{\mathcal{N}=2} = \sum_{m=0}^{\infty}\sum_{n=0}^{\infty}g_s^{m-2}\delta\mathcal{K}_{m,n} = \sum_{m=0}^{\infty}g_s^{m-2}\delta\mathcal{K}_{m,\bullet} \,,
\end{equation}
where we use $\bullet$ to denote effects at all perturbative orders, combined with non-perturbative effects, in the $\alpha'$ expansion. We note that the $m=0$ terms in the sums scale as $g_s^{-2}$, and represent closed string tree-level effects.

Next, we decompose $\mathcal{K}_{\text{K\"ahler}}^{\mathcal{N}=2}$ as
\begin{equation}\label{eq:kformalcorrforKahler1pt5}
    \mathcal{K}_{\text{K\"ahler}}^{\mathcal{N}=2}  = 
    \underbrace{g_s^{-2}\,\delta\mathcal{K}^{\mathcal{N}=2}_{0,\,0}}_{\mathcal{K}_{\text{tree}}}
    +\underbrace{g_s^{-2} \Bigl(\sum_{n=1}^{\infty} \delta\mathcal{K}_{m=0,\,n}^{\mathcal{N}=2}
    + \delta\mathcal{K}_{m=0,\,\text{n.p.}}^{\mathcal{N}=2}\Bigr)}_{\delta\mathcal{K}^{\mathcal{N}=2}_{\text{sphere}}\, \equiv\, \delta\mathcal{K}_{m=0,\,\bullet}^{\mathcal{N}=2}}
    +\underbrace{\sum_{m=1}^{\infty}g_s^{m-2} \delta\mathcal{K}_{m,\,\bullet}^{\mathcal{N}=2}}_{\delta\mathcal{K}^{\mathcal{N}=2}_{(g_s)}}\,,
\end{equation}
where the term $\delta\mathcal{K}_{m=0,\,\text{n.p.}}^{\mathcal{N}=2}$ is non-perturbative in $\alpha'$ but tree-level in $g_s$.

We likewise decompose $\mathcal{K}_{\text{K\"ahler}}^{\mathcal{N}=1}$ as 
\begin{equation}\label{eq:kformalcorrforKahlerN}
    \mathcal{K}_{\text{K\"ahler}}^{\mathcal{N}=1}  = 
    \underbrace{g_s^{-2}\Bigl(\sum_{n=0}^{\infty}\delta\mathcal{K}_{m=0,\,n}^{\mathcal{N}=1}
    + \delta\mathcal{K}_{m=0,\,\text{n.p.}}^{\mathcal{N}=1} \Bigr)}_{\delta\mathcal{K}^{\mathcal{N}=1}_{\text{sphere}}}
    +\underbrace{\sum_{m=1}^{\infty}g_s^{m-2} \delta\mathcal{K}_{m,\,\bullet}^{\mathcal{N}=1}}_{\delta\mathcal{K}^{\mathcal{N}=1}_{(g_s)}}\,.
\end{equation}

Combining the above expansions, we arrive at
\begin{equation}\label{eq:kformalcorrforKahler3P1}
    \mathcal{K}_{\text{K\"ahler}}  = \mathcal{K}_{\text{tree}}+
    \delta\mathcal{K}^{\mathcal{N}=2}_{\text{sphere}}+
    \delta\mathcal{K}^{\mathcal{N}=2}_{(g_s)}+
    \delta\mathcal{K}^{\mathcal{N}=1}_{\text{sphere}} +
    \delta\mathcal{K}^{\mathcal{N}=1}_{(g_s)}\,.
\end{equation}
The term $\delta \mathcal{K}^{\mathcal{N}=1}_{\text{sphere}}$ captures the backreaction of fluxes at the classical level, as well as axio-dilaton gradients that can be sourced by seven-branes (see e.g.~\cite{Minasian:2015bxa}). In our compactifications the flux is imaginary self-dual and all seven-branes coincide in $\mathfrak{so}(8)$ stacks.

We now make a very important observation, following \cite{Demirtas:2021nlu}. 
In compactifications with imaginary self-dual fluxes and vanishing axio-dilaton gradients, all effects of $\mathcal{N}=2 \to \mathcal{N}=1$ breaking carry \emph{at least one power of $g_s$}, relative to the tree-level theory \cite{Demirtas:2021nlu,Cho:2023mhw}: in other words, $\delta \mathcal{K}^{\mathcal{N}=1}_{\text{sphere}}$ is suppressed by a factor of $g_s$ compared to $\mathcal{K}_{\text{tree}}$ and $\delta\mathcal{K}^{\mathcal{N}=2}_{\text{sphere}}$:
\begin{equation}\label{eq:nomzero}
    \delta\mathcal{K}^{\mathcal{N}=1}_{\text{sphere}}  \lesssim \mathcal{O}(g_s^{-1})\,.
\end{equation}
The reason is as follows: in the absence of orientifolding, fluxes, and D-branes, the compactification preserves $\mathcal{N}=2$ supersymmetry.  The strength of the breaking can be quantified by examining how the sources of breaking enter the ten-dimensional Lagrangian, in comparison to the tree-level terms.  D-branes and orientifold planes famously have tension $\propto \frac{1}{g_s}$, which is suppressed by a factor $g_s^1$ compared to the $\frac{1}{g_s^2}$ tree-level terms.
The stress-energy of NS-NS flux appears to enter at tree level, because the string-frame action
\eqref{eq:TypeIIAct} includes the term
\begin{equation}
S_{\text{IIB}} \supset -\frac{1}{4\kappa_{10}^2}\int \mathrm{e}^{-2\phi}\,|H_3|^2\,,\end{equation}
which scales as $g_s^{-2}$.
However, in an ISD background, and taking vanishing axio-dilaton for simplicity, the condition \eqref{eq:ISDGKP} implies (in the conventions of \cite{Giddings:2001yu})
\begin{equation}
    \star_6 H_3 = - g_s F_3\,,
\end{equation} 
and so the flux backreaction is proportional to $\frac{1}{g_s^2}\times g_s \int H_3 \wedge F_3$, 
which, using \eqref{eq:D3flux}, is a relative correction  by a factor  of $g_s Q_{\text{D3}}^{\text{flux}}$.   
We stress that \eqref{eq:nomzero}
does not mean that the K\"ahler potential for a type IIB Calabi-Yau orientifold compactification receives no $\mathcal{N}=1$ corrections at the level of sphere worldsheets.
Rather, \eqref{eq:nomzero} records the fact that 
$\delta\mathcal{K}_{m=0,\,\bullet}^{\mathcal{N}=1}$ computes the backreaction of sources --- fluxes, D-branes, O-planes, and axio-dilaton gradients --- whose effects are suppressed, in the class of backgrounds studied here, by a factor of $g_s$; see, however, the discussion in \S\ref{sec:sft}.

At weak string coupling, we can therefore write 
\begin{equation}\label{eq:kformalcorrforKahler3}
g_s^2\,\mathcal{K}_{\text{K\"ahler}}  = \underbrace{g_s^2\,\Bigl(\mathcal{K}_{\text{tree}}+
\delta\mathcal{K}^{\mathcal{N}=2}_{\text{sphere}}\Bigr)}_{\mathcal{O}(1)}+
\underbrace{g_s^2\Bigl(\delta\mathcal{K}^{\mathcal{N}=2}_{(g_s)}+
\delta\mathcal{K}^{\mathcal{N}=1}_{\text{sphere}} +
\delta\mathcal{K}^{\mathcal{N}=1}_{(g_s)}\Bigr)}_{\mathcal{O}(g_s)}\,.
\end{equation}
The point is that the order-unity terms in \eqref{eq:kformalcorrforKahler3}  
are inherited from the $\mathcal{N}=2$ supersymmetric theory prior to orientifolding, and can be computed exactly using the c-map \cite{Cecotti:1988qn},\footnote{The corrections
$\mathcal{K}_{m>0,\,\bullet}^{\mathcal{N}=2}$, 
are known to all perturbative orders in $g_s$ \cite{Robles-Llana:2006hby,Robles-Llana:2007bbv}, as we will explain in
\S\ref{sec:N2loop},
and
turn out to be
safely 
negligible in our analysis.
However, as these terms are comparable to the unknown terms
$\delta\mathcal{K}^{\mathcal{N}=1}$, 
evaluating the $\mathcal{K}_{m>0,\,\bullet}^{\mathcal{N}=2}$ is of limited practical benefit.}
while the remainder  
is suppressed by at least one power of $g_s$ in comparison.

\begin{table}
    \centering
    \resizebox{.5\textwidth}{!}{
    \begin{tabular}{|c||c|c||c|c|}
        \hline
        &  &  &  &   \\ [-0.9em]
        & $\mathcal{N}=2$ &  Eq.  & $\mathcal{N}=1$ & Eq. \\ [0.2em]
        \hline
        \hline
         &  & &  &    \\ [-0.9em]
       $g_s^0$ &  $\mathcal{K}_{m=0,\,\bullet}^{\mathcal{N}=2}$ & \eqref{eq:Kahlerpotential}  
       &  $0^*$  & \eqref{eq:nomzero}
       \\[0.3em]
       \hline
       &  &  &  &   \\ [-0.9em]
       $g_s^1$ & $\mathcal{K}_{m=1,\,\bullet}^{\mathcal{N}=2}$  & \eqref{eq:dVloop}
       & $\mathcal{K}_{m=1,\,\bullet}^{\mathcal{N}=1}$ & \eqref{equ:KKgs}
       \\[0.3em]
       \hline
    \end{tabular}
    }
    \caption{Schematic form of corrections in the string loop expansion.  Powers of $g_s$ shown are given relative to closed string tree level, so $g_s^0$ means tree level and $g_s^1$ means closed string one-loop. The $\bullet$ subscripts indicate that $\alpha'$ corrections at all perturbative orders, as well as worldsheet instantons, are included.  The $0^*$ indicates that in our class of flux backgrounds, there are no $\mathcal{N}=1$ corrections with this $g_s$ scaling: see the discussion following \eqref{eq:nomzero}, but also note the remarks in \S\ref{sec:sft}.}\label{tab:overview_corr}
\end{table}

Usually, corrections to the K\"ahler potential $\mathcal{K}_{\text{K\"ahler}}$ can be parametrized as additional contributions $\delta\mathcal{V}$ to the corrected Calabi-Yau volume $\mathcal{V}$
\begin{equation}\label{eq:CorrCYvol}
    \mathcal{V}  = \mathcal{V}^{(0)} + \delta\mathcal{V}\, ,
\end{equation}
where $\mathcal{V}^{(0)}=\frac{1}{6}\kappa_{ijk}t^it^jt^k$ was introduced in \eqref{eq:VolClassical}, 
and we decompose $\delta\mathcal{V}$ as
\begin{equation}\label{eq:vdecomp}
    \delta\mathcal{V}=
    \delta\mathcal{V}^{\mathcal{N}=2}_{\text{sphere}}+
    \delta\mathcal{V}^{\mathcal{N}=2}_{(g_s)}+
    \delta\mathcal{V}^{\mathcal{N}=1}_{\text{sphere}} +
    \delta\mathcal{V}^{\mathcal{N}=1}_{(g_s)}\, .
\end{equation}
Considering the leading terms in more detail, we have
\begin{equation}\label{eq:deln2}
    \delta\mathcal{V}^{\mathcal{N}=2}_{\text{sphere}} = \mathcal{V}_{(\alpha')^3} + \mathcal{V}_{\text{WSI}}\,,
\end{equation}
which consists of a perturbative term at order $(\alpha')^3$ \cite{Grisaru:1986kw,Gross:1986iv,Antoniadis:1997eg,Becker:2002nn}, and a series of worldsheet instanton contributions $\mathcal{K}_{\text{WSI}}$ \cite{Dine:1986zy,Dine:1987bq}.
The explicit forms of these corrections are given in \eqref{eq:Valphap3} and \eqref{eq:VWSI} below.

The intrinsically $\mathcal{N}=1$ corrections are not as well-characterized. The explicit form of $\delta\mathcal{K}_{\mathcal{N}=1}^{(g_s)}$ is not known, but in \S\ref{sec:N1loop} we collect the available results and present representative expressions that capture the expected parametric dependence of these corrections based on naive dimensional analysis and comparisons to well-understood toroidal orientifold examples. These formulas are not derived from first principles in full generality, but they are expected to be sufficient to estimate the size of $\mathcal{N}=1$ loop effects. In Chapter~\ref{chap:deSitter}, we use these expressions to assess the magnitude of $\delta\mathcal{K}_{\mathcal{N}=1}^{(g_s)}$ and to verify that such corrections remain under control in the explicit de Sitter constructions considered there.
 
\medskip

Thus far we have discussed corrections to $\mathcal{K}$, and now we turn to corrections to the K\"ahler coordinates.  The decomposition into various classes of corrections is precisely analogous, so we will be brief.
In \eqref{eq:Ttree} we defined the
tree-level K\"ahler coordinates $T^{\text{tree}}_i(t)$
in terms of their real parts,
\begin{equation}
    \mathcal{T}_i^{\text{tree}} =\frac{1}{2}\kappa_{ijk}t^jt^k\,,
\end{equation}
as
\begin{equation}
    T_i^{\text{tree}} =\dfrac{1}{g_s} \mathcal{T}_i^{\text{tree}} +\mathrm{i}\,\int_X C_4\wedge \omega_i =\dfrac{1}{g_s} \mathcal{T}_i^{\text{tree}} + \mathrm{i}\,\theta_i\, ,
\end{equation}
thus furnishing a relation between the coordinates on K\"ahler moduli space --- namely, $T^{\text{tree}}_i(t)$ --- and the volumes $t^i$ of two-cycles.
This relationship
receives perturbative and non-perturbative corrections, 
as laid out in the general expression \eqref{eq:phiofG}.
Just as in \eqref{eq:kformalcorrforKahler3P1},
we can decompose the corrections into terms arising at (open string) tree level, versus loop corrections:
\begin{equation}\label{eq:kformalcorrforKahler3calT}
    \mathcal{T}_i  = \mathcal{T}^{\text{tree}}_i+
\delta \mathcal{T}^{\mathcal{N}=2,\text{tree}}_i+
\delta \mathcal{T}^{\mathcal{N}=2,(g_s)}_i+
\delta \mathcal{T}^{\mathcal{N}=1,\text{tree}}_i +
\delta \mathcal{T}^{\mathcal{N}=1,(g_s)}_i\,.
\end{equation}
The imaginary part is unchanged, i.e.,  
\begin{equation}
T_i = \mathcal{T}_i + \mathrm{i}\,\theta_i\,,
\end{equation}
so the complexified K\"ahler coordinates can be written
\begin{equation}\label{eq:kformalcorrforKahler3T}
    T_i  = T^{\text{tree}}_i+
\delta T^{\mathcal{N}=2,\text{tree}}_i+
\delta T^{\mathcal{N}=2,(g_s)}_i+
\delta T^{\mathcal{N}=1,\text{tree}}_i +
\delta T^{\mathcal{N}=1,(g_s)}_i\,.
\end{equation}
 
The corrections $\delta T^{\mathcal{N}=2,\text{tree}}_i$ 
can be obtained from the c-map \cite{Cecotti:1988qn} (cf.~\cite{Baume:2019sry,Marchesano:2019ifh}) as
\begin{equation}
\label{eq:Kahlercoordinatessplitpreview}
    \delta T^{\mathcal{N}=2,\text{tree}}_i = \delta T_i^{(\alpha')^2} +\delta T_i^{\text{WSI}} \,,
\end{equation}
where $T_i^{\text{tree}}$,
$\delta T_i^{(\alpha')^2}$, and $\delta T_i^{\text{WSI}}$ are given 
in \eqref{eq:Ttree}, \eqref{eq:KahlercoordinatesBBHL}, and \eqref{eq:KahlercoordinatesWSI}.
Like
$\delta\mathcal{K}^{\mathcal{N}=2}_{(g_s)}$,
$\delta T^{\mathcal{N}=2,(g_s)}_i$ follows from the results of \cite{Robles-Llana:2006hby,Robles-Llana:2007bbv}, as discussed in \S\ref{sec:N2loop}.  
As with \eqref{eq:nomzero}, the correction
$\delta T^{\mathcal{N}=1,\text{tree}}_i$ is suppressed at weak string coupling. 
The intrinsically $\mathcal{N}=1$ correction $\delta T^{\mathcal{N}=1,(g_s)}_i$, on the other hand, is not completely characterized (but see e.g.~\cite{Witten:2012bh,Haack:2018ufg,Sen:2024nfd}).  As with $\delta\mathcal{K}_{\mathcal{N}=1}^{(g_s)}$,
in Chapter 8 we provide estimates of the size of the corrections $\delta\mathcal{T}_i^{\mathcal{N}=1,(g_s)}$, and argue that these effects are under control in the candidate de Sitter vacua presented there.\\

In contrast to the K\"ahler potential $\mathcal{K}$, the superpotential $W$ is protected by non-renormalization theorems and receives no corrections in perturbation theory --- a result we explain in detail in \S\ref{sec:non-renormalization}. Nevertheless, $W$ does acquire non-perturbative contributions from Euclidean D3-branes wrapping cycles of the internal geometry, 
as well as from gaugino condensation on stacks of seven-branes wrapping such cycles,  as discussed in \S\ref{sec:Wnp}.

Looking ahead, our strategy will be to rely on the non-renormalization
of $W$, and also on weak breaking of $\mathcal{N}=2$ to $\mathcal{N}=1$ supersymmetry, as in
\eqref{eq:kformalcorrforKahler3P1}. 
We will compute the full non-perturbative superpotential --- whose structure is topological and, in principle, accessible --- and search for candidate vacua based on this  
$W$. We then verify that the resulting solutions remain robust under reasonable estimates of corrections to the K\"ahler potential
and the K\"ahler coordinates, using naive dimensional analysis (NDA) or more detailed inputs where available, as laid out in general in \S\ref{sec:scheme}.\\

In the remainder of this chapter we will examine in great detail the functional form of $\Phi^A(\mathscr{G})$, $W(\Phi)$, and $\mathcal{K}(\Phi,\bar{\Phi})$, 
for type IIB flux compactifications on Calabi-Yau orientifolds, in a particular parameter regime:
\begin{equ}[Control regime] 
\begin{fleqn}[0.75\parindent]
\begin{subequations}\label{eq:control_regime}

    \vspace*{-0.65cm}

    \begin{align}
        &\text{1. weak string coupling: $g_s \ll 1$.} \label{eq:control_regime_cond1} \\[1em]
        &\text{2. weak breaking of $\mathcal{N}=2 \to \mathcal{N}=1$, controlled by $g_s \ll 1$: cf.~\eqref{eq:kformalcorrforKahler3}.} \label{eq:control_regime_cond2}\\[1em]
        &\text{3. large complex structure, $\mathrm{Im}(z^a) \gg 1$, cf.~\eqref{eq:LCS}}. \label{eq:control_regime_cond3}\\[1em]
        &\text{4. large Einstein-frame volumes of four-cycles, $2\pi\,\mathcal{T}_i \gg 1$, 
        cf.~\eqref{eq:kformalcorrforKahler3calT}}.
        \label{eq:control_regime_cond4}  
    \end{align}
    
    \vspace*{-0.01cm}
    
\end{subequations}
\end{fleqn}
\end{equ} 

The choices that lead to conditions \eqref{eq:control_regime_cond1} -- \eqref{eq:control_regime_cond4} will be explained in due course.

\section{Quantum corrections to K\"ahler coordinates and K\"ahler potential}\label{sec:quantum_pert}

We have just explained in \S\ref{sec:QCorPrel}
that in the string loop expansion and in the $\alpha'$ expansion, there are non-vanishing perturbative corrections to the K\"ahler potential, viewed as a function of the K\"ahler coordinates, and there are also corresponding corrections to the K\"ahler coordinates, viewed as functions of the underlying geometric data. 
We will now present the form of these corrections in more detail.

Throughout we make use of the control expansion laid out in \S\ref{sec:QCorPrel}. In particular,
we will repeatedly use the decomposition of the 
K\"ahler coordinates
and the 
K\"ahler potential
into $\mathcal{N}=2$ and $\mathcal{N}=1$
pieces, followed by truncation to the terms
$\mathcal{K}_{m=0,\,\bullet}^{\mathcal{N}=2}$ arising
at string tree level but at all orders in $\alpha'$, as in
\eqref{eq:deln2}.

\subsection{$\mathcal{N}=2$ corrections}\label{sec:N2Cor}
\index{$\mathcal{N}=2$ corrections}

In this section, we provide detailed expressions for the $\mathcal{N}=2$ corrections $\delta\mathcal{K}_{\mathcal{N}=2}$ and $\delta T^{\mathcal{N}=2}_i$ to the K\"ahler potential and K\"ahler coordinates, respectively. First, 
as an instructive exercise, 
in \S\ref{sec:BBHL} we compute the tree-level $(\alpha')^3$ correction
$\mathcal{K}_{(\alpha')^3}$ explicitly  by performing a dimensional reduction of higher derivative corrections on a Calabi-Yau manifold. Subsequently, we comment on the tree level corrections $\delta\mathcal{K}_{\mathcal{N}=2}^{\text{sphere}}$ and $\delta T^{\mathcal{N}=2,\text{tree}}_i$ in \S\ref{sec:N2tree}. We conclude with general $\mathcal{N}=2$ string loop corrections $\delta\mathcal{K}_{\mathcal{N}=2}^{(g_s)}$ and $\delta T^{\mathcal{N}=2,(g_s)}_i$ in \S\ref{sec:N2loop}.

\subsubsection{Derivation of the tree level $(\alpha')^3$ corrections}\label{sec:BBHL}
 
String theory reduces, in its low-energy limit, to an effective supergravity theory, but one must move beyond this approximation to capture genuinely stringy behavior. Such corrections may arise both in the $\alpha'$ expansion and in the genus expansion: the former introduces higher-derivative terms to the supergravity action, while the latter corresponds to quantum corrections in spacetime arising from string loops. In type II superstring theory, the former corrections begin at the eight-derivative level and were first derived at tree level from four-graviton scattering amplitudes \cite{Schwarz:1982jn,Gross:1986iv},\footnote{The corresponding expression at one loop was obtained in \cite{Sakai:1986bi}.} as well as via the $\cN=2$ non-linear $\sigma$-model at four loops \cite{Callan:1985ia,Grisaru:1986px,Grisaru:1986dk,Freeman:1986br,Freeman:1986zh}.

In this section we will lay out the computation of the best-known example of such a correction arising from curvature corrections in the ten-dimensional theory. Because the calculation will be lengthy, we will first sketch the process and the result. In the ten-dimensional supergravity action \eqref{eq:TypeIIBAxtion} for the massless fields, the relevant curvature correction amounts to adding a term involving a contraction of four Riemann tensors, given below in Eq.~\eqref{eq:SuperInvRF}. Upon carrying out the dimensional reduction, we will find that the K\"ahler potential \eqref{eq:KKtmp} is corrected as
\begin{equation}\label{eq:KPBBHL}
    \mathcal{K}_{\text{K\"ahler}}\bigl(T_i,\overline{T}_i\bigr) \to -2\log\Bigl(\mathcal{V}(T_i,\overline{T}_i)+ \xi \,g_s^{-3/2}\Bigr)\,,
\end{equation} 
where $\xi$ is a constant to be obtained in Eq.~\eqref{eq:ValueXiBBHL},
\begin{equation}\label{eq:ValueXiBBHL1} 
    \xi = -\frac{\zeta(3)\, \chi(X)}{2\,(2\pi)^3}\,.
\end{equation}
Here $\chi(X)$ is the Euler characteristic of the Calabi-Yau manifold $X$,
which appears because 
$\int_X R^3 \sim -\chi(X)$. 
 
With these preliminaries, we will begin the computation.
Including the stringy $(\alpha^{\prime})^{3}$ corrections, the supergravity action reads
\begin{equation}\label{eq:R4action}
    S \supset \dfrac{1}{2\kappa_{10}^{2}}\int\,\mathrm{e}^{-2\phi}\biggl \{R+\dfrac{\zeta(3)\, (\alpha^{\prime})^{3}}{3\cdot 2^{11}}\,\mathcal{J}_{0}\biggl \}\sqrt{-g}\dif^{10}x\, .
\end{equation}
The quantity $\mathcal{J}_{0}$ encodes the detailed structure of the corrections mentioned in \S\ref{sec:QCorPrel}, and is given by \cite{Schwarz:1982jn,Gross:1986iv,Sakai:1986bi}
\begin{equation}\label{eq:SuperInvRF} 
\mathcal{J}_{0}=\left (t_{8}t_{8}-\dfrac{1}{4}\epsilon_{8}\epsilon_{8}\right )R^{4}\, ,
\end{equation}
in terms of the totally anti-symmetric Levi-Civita symbol $\epsilon_{D}$ in $D$ dimensions and an 8-index tensor $t_{8}$ whose action on an anti-symmetric matrix $A$ is defined as \cite{Freeman:1986zh,Freeman:1986br}
\begin{equation}\label{eq:t8}
    t_{8}A^{4}=24\left (\text{tr}(A^{4})-\dfrac{1}{4}\text{tr}(A^{2})^{2}\right )\, .
\end{equation}
By computing \eqref{eq:SuperInvRF} explicitly, one finds that $\mathcal{J}_{0}$ can be expressed in terms of the Weyl tensor $C_{MNPQ}$ as \cite{Gross:1986iv,Green:2005qr}\footnote{The fact that $\mathcal{J}_{0}$ can be written only in terms of the Weyl tensor is most easily proven by studying a superspace approach to type IIB supergravity \cite{Howe:1983sra}, see also \cite{deHaro:2002vk,Green:2003an,Paulos:2008tn,Liu:2022bfg}.}
\begin{equation}\label{eq:R4WeylTensor} 
    \dfrac{\mathcal{J}_{0}}{3\cdot 2^{8}}=C^{MNPQ}\left (-\dfrac{1}{4}C_{MN}\,^{RS}C_{PR}\,^{TU}C_{QSTU}+C_{M}\,^{R}\,_{P}\,^{S}C_{R}\,^{T}\,_{N}\,^{U}C_{STQU}\right )\, .
\end{equation}

\

Next, let us sketch the derivation of the $(\alpha^{\prime})^{3}$-corrected K\"ahler potential \eqref{eq:KPBBHL} by explicitly reducing the above ten-dimensional higher-derivative corrections in Eq.~\eqref{eq:R4action} in a one-modulus example \cite{Antoniadis:1997eg,Becker:2002nn}.
We begin with the following parametrization of the  
type IIB supergravity action in string frame
\begin{align}\label{eq:BBHLStringFrameAction} 
S&=\dfrac{1}{2\kappa_{10}^{2}}\int\,\mathrm{e}^{-2\phi}\biggl \{R+4|\, \mathrm{d} \phi|^{2}+\alpha a_{T}\left [\mathcal{J}_{0}+  3\cdot 2^{8}\, \gamma (\nabla^{2}\phi)\,  Q\right ] \biggl \}\sqrt{-g}\dif^{10}x\, ,
\end{align}
where we introduced the constants
\begin{equation}
    \alpha=\dfrac{(\alpha^{\prime})^{3}}{3\cdot 2^{12}}\kom a_{T}=2\zeta(3)\, .
\end{equation} 
For reasons of consistency  
we have added 
a higher-derivative operator of the form $(\nabla^{2}\phi)\,  Q$, 
where $Q$ is the six-dimensional Euler density, defined as 
\begin{align}\label{eq:SixDEulerNLSM} 
Q&\coloneqq \dfrac{1}{24}\left (R_{MN}\,^{RS}R^{MNOP}R_{OPRS}-4R_{M}\,^{R}\,_{O}\,^{S}R^{MNOP}R_{NSPR}\right )\, .
\end{align}
When integrated over $X$, this object has the useful property
\begin{equation}\label{eq:IntQChi} 
\int_{X}\, Q\sqrt{g}\dif^{6}y=(2\pi)^{3}\chi(X)\, .
\end{equation}

For a  compactification with a single K\"ahler modulus $\mathcal{V}=\mathrm{e}^{6u}$, and omitting fluxes for simplicity,
we make the metric ansatz
\begin{equation}\label{eq:backgroundBBHL}
\mathrm{d} s^{2}=\eta_{\mu\nu}\, \mathrm{d} x^{\mu}\, \mathrm{d} x^{\nu}+\mathrm{e}^{2u}\tilde{g}_{mn}\, \mathrm{d} y^{m}\, \mathrm{d} y^{n}\, .
\end{equation}
We aim to carry out the reduction in such a way that the K\"ahler metric for the modulus $u$ can be extracted directly from its kinetic terms by comparison with Eq.~\eqref{eq:4DSugra_classical}. This, in turn, allows us to infer the corresponding correction to the K\"ahler potential. To achieve this, we restrict our attention to couplings involving precisely two spacetime derivatives in four dimensions.
In the background defined by Eq.~\eqref{eq:backgroundBBHL}, one verifies that the components of the Riemann tensor, Ricci tensor and Ricci scalar are given by
\begin{align}\label{eq:RTRTRSBBHL}
R^{m}\,_{\mu n\nu}&=-\delta^{m}_{~n}\left (\p_{\mu}u\p_{\nu}u+\p_{\mu}\p_{\nu}u\right )\kom R^{\mu}\,_{m\nu n}=-g_{mn}\left (\p_{\nu}u\p^{\mu}u+\p_{\nu}\p^{\mu}u\right )\, ,\nonumber\\[0.3em]
R^{k}\,_{mnp}&=(\p u)^{2}\left (\delta^{k}_{~p}g_{mn}-\delta^{k}_{~n}g_{pm}\right )\kom R_{\mu\nu}=-6(\p_{\mu}u\p_{\nu}u+\p_{\mu}\p_{\nu}u)\, ,\nonumber\\[0.3em]
R_{mn}&=-g_{mn}\Bigl(6(\p u)^{2}+\Delta u\Bigr)\kom R=-42(\p u)^{2}-12\Delta u\, .
\end{align}

First, we have to solve the equations of motion as in 
our discussion in \S\ref{sec:GKP}, but this time including $(\alpha^{\prime})^{3}$ corrections as parametrized in \eqref{eq:BBHLStringFrameAction}.
The $\alpha'$-corrected equations of motion for the metric and the dilaton are given by\footnote{Similar observations can be made by computing $\beta$-functions in the 
two-dimensional $\cN=2$ non-linear $\sigma$-model at four loops \cite{Callan:1985ia,Grisaru:1986px,Grisaru:1986dk}.}
\begin{equation}\label{eq:EOMCorBBHLDIL} 
R_{MP}+\dfrac{(\alpha')^{3}\zeta(3)}{2} \nabla_{M}\nabla_{P}Q = 0\kom R+4\Delta^{2}\phi-4(\nabla\phi)^{2}+\alpha a_{T}\mathcal{J}_{0}=0\, .
\end{equation}
Solving these equations for the corrected dilaton, one finds \cite{Becker:2002nn}
\begin{equation}\label{eq:CorrectedDialtonBBHL} 
 \phi=\phi_{0}+c\, Q \kom c=\dfrac{\alpha'^3 a_{T}}{2^{5}}\,,
\end{equation}
with $\phi_{0}=\text{const.}$ on the internal manifold $X$. 
Then, reducing the  
two-derivative 
action to four dimensions, using $\mathcal{V}=\mathrm{e}^{6u}$ and \eqref{eq:IntQChi}, and writing $\widetilde{\chi}=(2\pi)^3 \chi$,
we obtain
\begin{align}\label{eq:BBHLStringFrameActionCl}
    2\kappa_{10}^{2}\, S &\supset\int_{X}\,\mathrm{e}^{-2\phi}\, \bigl(R+4|\, \mathrm{d} \phi|^{2}\bigr)\sqrt{g}\dif^{6}x \nonumber\\[0.5em]
    &=\mathrm{e}^{-2\phi_{0}}\, \biggl \{\left (\mathcal{V}-2c\widetilde{\chi}\right ) R^{(4)}+4\left (\mathcal{V}-2c\widetilde{\chi}\right ) (\p\phi_{0})^{2}\nonumber\\[0.3em]
    &\hphantom{=\mathrm{e}^{-2\phi_{0}}\, \bigl \{} +6\left (5\mathcal{V}+14c\widetilde{\chi}\right )(\p u)^{2}-24 \mathcal{V} (\p_{\mu}\phi_{0})(\p^{\mu}u) \biggr \}\,,
\end{align}
in terms of the four-dimensional Ricci scalar $R^{(4)}$.

Second, we also have to reduce the higher derivative terms like $\mathcal{J}_0$ on the classical background.
That is, plugging the components of the Riemann tensor \eqref{eq:RTRTRSBBHL} into $\mathcal{J}_{0}$ in \eqref{eq:R4WeylTensor} and keeping only terms up to the two-derivative level,
one verifies that
\begin{align}\label{eq:ExpansionJ0TwoDerivatives} 
\alpha a_{T}\mathcal{J}_{0}&\supset 12\lambda\, Q\left (-6(\p u)^{2}-R^{(4)} \right )\kom \lambda=\dfrac{\alpha'^3 a_{T}}{3\cdot 2^{6}}\, .
\end{align}
Then, the reduction of higher derivative terms up to two derivatives in four dimensions reads
\begin{align}\label{eq:BBHLStringFrameActionHDT} 
2\kappa_{10}^{2} S&\supset \int_{X}\,\mathrm{e}^{-2\phi}\, \alpha a_{T}\,\biggl \{\mathcal{J}_{0}+  3\cdot 2^{8}\, \gamma (\nabla^{2}\phi)\,  Q \biggl \}\sqrt{g}\dif^{6}x \nonumber\\[0.5em]
&=\mathrm{e}^{-2\phi_{0}}\, \biggl \{-12\lambda\widetilde{\chi}\, R^{(4)}-24 \gamma\lambda \widetilde{\chi}\, (\p\phi_{0})^{2}-72\lambda\widetilde{\chi}\,(\p u)^{2}\biggr \} \, .
\end{align}

Putting everything together,
we find by adding the contributions from \eqref{eq:BBHLStringFrameActionCl} and \eqref{eq:BBHLStringFrameActionHDT} that
\begin{align}\label{eq:BBHLReductionAction4} 
S&=\dfrac{M_{\mathrm{pl}}^2}{2}\int\, \mathrm{e}^{-2\phi_{0}}\, \biggl \{\left (\mathcal{V}-\left [2c+12\lambda\right ]\widetilde{\chi}\right ) R^{(4)}+4\left (\mathcal{V}-\left [2c-6 \gamma\lambda\right ]\widetilde{\chi}\right ) (\p\phi_{0})^{2}\nonumber\\[0.3em]
&\hphantom{=\dfrac{1}{2\kappa_{4}^{2}}\int\,\mathrm{e}^{-2\phi_{0}}\, \bigl \{}+6\left (5\mathcal{V}+\left [14c-12\lambda\right ]\widetilde{\chi}\right )(\p u)^{2}-24 \mathcal{V} (\p_{\mu}\phi_{0})(\p^{\mu}u)  \biggr \}\sqrt{-g}\dif^{4}x\, ,
\end{align}
in terms of the four-dimensional reduced Planck mass $M_{\mathrm{pl}}$.
Let us note that both the kinetic term for $\phi_{0}$ and the four-dimensional Einstein-Hilbert term are corrected by two sources:
\begin{enumerate}
    \item terms like $\sim 2c\, \widetilde{\chi}$ originating from the corrected dilaton background~\eqref{eq:CorrectedDialtonBBHL}, and
    \item terms like $\sim \lambda\, \widetilde{\chi}$ or $\sim \gamma\, \widetilde{\chi}$ coming from dimensionally reducing the higher-derivative terms $R^{4}$ and $(\nabla^{2}\phi)R^{3}$, respectively.
\end{enumerate}
Four-dimensional supersymmetry requires that the numerical prefactors of the four-dimensional Ricci scalar $R^{(4)}$ and the dilaton kinetic term $(\p \phi_{0})^{2}$ match, and thus \cite{Becker:2002nn}
\begin{equation}\label{eq:DetCoeffDOFSFD} 
2c+12\lambda=2c-6 \gamma\lambda\quad\Rightarrow\quad \gamma=-2\, .
\end{equation}
While this is an indirect argument used in \cite{Becker:2002nn} (see also \cite{Bonetti:2016dqh}) to fix the a priori unknown coefficient $\gamma$, a proper string amplitude computation performed in \cite{Garousi:2020lof,Liu:2025uqu} precisely reproduced this expected value. This suggests a close relationship between the coefficient $\gamma$ in front of $(\nabla^{2}\phi_{0})R^{3}$ in \eqref{eq:BBHLStringFrameAction} and $\alpha a_{T}$ in front of $\mathcal{J}_{0}\sim R^{4}$. At last, using $c=6\lambda$, we get
\begin{align}\label{eq:BBHLReductionAction5} 
S&=\dfrac{M_{\mathrm{pl}}^2}{2}\int\, \mathrm{e}^{-2\phi_{0}}\, \biggl \{\left (\mathcal{V}-4c\widetilde{\chi}\right ) R^{(4)}+4\left (\mathcal{V}-4c\widetilde{\chi}\right ) (\p\phi_{0})^{2}\nonumber\\
&\hphantom{=\dfrac{1}{2\kappa_{4}^{2}}\int\,\mathrm{e}^{-2\phi_{0}}\, \bigl \{}+6\left (5\mathcal{V}+12c\widetilde{\chi}\right )(\p u)^{2}-24\mathcal{V} (\p_{\mu}\phi_{0})(\p^{\mu}u) \biggl \}\sqrt{-g}\dif^{4}x\, .
\end{align}

By introducing the four-dimensional dilaton as
\begin{equation}\label{eq:FourDimDilOrig} 
\mathrm{e}^{-2\phi_{4}}=\mathrm{e}^{-2\phi_{0}}\,\left (\mathcal{V}-4c\chi\right )\, , 
\end{equation}
the Einstein-Hilbert term is canonically normalized, and so
\begin{align}\label{eq:BBHLReductionAction7} 
S&=\dfrac{M_{\mathrm{pl}}^2}{2}\int\,  \biggl \{ R^{(4)}-2 (\p\phi_{4})^{2}-6\left (1+\dfrac{16c\chi}{\mathcal{V}}\mathrm{e}^{-\frac{3\phi_{0}}{2}}\right )(\p u)^{2}
\biggl \}\sqrt{-g}\dif^{4}x\, .
\end{align}
The K\"ahler potential consistent with this K\"ahler metric can be written,
in string units $\ell_{s}=1$,
as
\begin{equation}\label{eq:KaehlerTestTreeBBHL} 
    \mathcal{K}=-2\log\left (\mathcal{V}+\dfrac{\xi}{2}\,\mathrm{Im}(\tau)^{3/2}\right )-\log\Bigl(-\I(\tau-\bar{\tau})\Bigr)\, ,
\end{equation}
where we have introduced $\tau=C_0+\mathrm{i}\mathrm{e}^{-\phi_0}$, and we have defined \cite{Antoniadis:1997eg,Becker:2002nn}
\begin{equation}\label{eq:ValueXiBBHL} 
    \xi = -\frac{\zeta(3)\, \chi(X)}{2\,(2\pi)^3}\, .
\end{equation}
This analysis has been generalized in the literature 
by including arbitrary numbers of K\"ahler moduli \cite{Bonetti:2016dqh,Grimm:2017okk,Liu:2022bfg}, by including additional 10-dimensional higher-derivative terms in the reduction \cite{Liu:2019ses,Liu:2022bfg}, and by computing higher $F$-term contributions in four dimensions \cite{Ciupke:2015msa}.
Beyond the closed-string tree level, this analysis extends straightforwardly \cite{Antoniadis:1997eg,Liu:2022bfg}, see \S\ref{sec:N2loop}. At each closed string loop order, additional contributions arise, each suppressed by an extra factor of $g_s^2$. Moreover, D$(-1)$-instantons contribute non-perturbatively, with effects suppressed by positive powers of $\mathrm{e}^{-2\pi/g_s}$ \cite{Sen:2021tpp,Sen:2021jbr}, see Eq.~\eqref{eq:Valphap3Loop}.

Thus far, we have neglected background fluxes and inferred the K\"ahler potential by computing the kinetic terms for the dilaton and moduli fields. Upon reintroducing fluxes, the resulting $F$-term scalar potential receives additional contributions stemming from corrections to the K\"ahler potential. In principle, such terms in the scalar potential can be derived from a direct dimensional reduction of higher-derivative operators in ten dimensions, such as those involving $|G_3|^2 R^3$. However, this analysis lies beyond the scope of these lectures; for a detailed treatment, see \cite{Liu:2022bfg}. 
Instead, we now demonstrate that the corrected K\"ahler potential given in Eq.~\eqref{eq:KaehlerTestTreeBBHL} explicitly breaks the no-scale structure of the scalar potential derived from the Gukov-Vafa-Witten superpotential in Eq.~\eqref{eq:GVW}. 

Specifically, we find that the $F$-term scalar potential acquires a new term due to the $\alpha'$ correction \eqref{eq:ValueXiBBHL}:\footnote{The correction $V_\xi$ in Eq.~\eqref{eq:BBHLCorSPTreeLevelTREE} can be seen in ten dimensions as arising from five-point couplings of the schematic form $|G_3|^2 R^3$, as shown in \cite{Liu:2019ses,Liu:2022bfg}.}
\begin{equation}\label{eq:BBHLCorSPTreeLevelTREE} 
V_{F}=\dfrac{V_{\text{flux}}}{\mathcal{V}^{2}}-\dfrac{\xi\,\mathrm{e}^{-\frac{3\phi_{0}}{2}}}{\mathcal{V}^{3}}\, V_{\text{flux}}+V_{\xi}+\ldots\, ,
\end{equation}
where
\begin{equation}
    V_{\xi}=\dfrac{3\xi\,\mathrm{e}^{-\frac{\phi_{0}}{2}}\, \mathrm{e}^{\mathcal{K}_{\text{cs}}}}{4\mathcal{V}^{3}}\, \bigl |W+(\tau-\bar{\tau})\,  D^{(0)}_{\tau}W  \bigl |^{2}
\end{equation}
in terms of 
\begin{equation}\label{eq:KCVDer} 
D^{(0)}_{\tau}W=\dfrac{1}{\bar{\tau}-\tau} \int_{X}\, \overline{G}_{3}\wedge\Omega\, .
\end{equation}
It is now manifest that the classical no-scale structure is broken by the inclusion of the $\alpha'$ correction \eqref{eq:ValueXiBBHL}. Unlike the tree-level potential, which exhibits an exact cancellation among terms involving the K\"ahler moduli (cf.~Eq.~\eqref{eq:NoScaleID}), the corrected scalar potential develops a non-trivial dependence on the overall Calabi–Yau volume $\mathcal{V}$. Specifically, the identity Eq.~\eqref{eq:NoScaleID} is modified as \cite{Becker:2002nn}
\begin{equation}\label{eq:NoScaleIDBBHL}
    \mathcal{K}^{i\bar{\jmath}}\, \mathcal{K}_{i} \mathcal{K}_{\bar{\jmath}}-3=\dfrac{3\hat{\xi}(\mathcal{V}^2+7\hat{\xi}\mathcal{V}+\hat{\xi}^2)}{(\mathcal{V}-\hat{\xi})(2\mathcal{V}+\hat{\xi})^2}\; ,\quad  \hat{\xi} = \xi\,\mathrm{e}^{-\frac{3\phi_{0}}{2}}\, .
\end{equation}
This dependence on $\mathcal{V}$ introduces a potential for the K\"ahler moduli at leading order in $g_s$, lifting the flat directions associated with the classical no-scale symmetry. Consequently, the volume modulus is no longer unfixed, and its dynamics must now be taken into account when analyzing vacuum solutions.

\subsubsection{$\mathcal{N}=2$ string tree level corrections}\label{sec:N2tree}

We now turn to the full structure of the string tree level corrections, to all orders in $\alpha'$, that are   
inherited from the $\mathcal{N}=2$ parent theory:
that is, the terms denoted
$\delta\mathcal{K}^{\mathcal{N}=2}_{m=0,\,\bullet}
\equiv \delta\mathcal{K}^{\mathcal{N}=2}_{\text{sphere}}$ in
\eqref{eq:kformalcorrforKahler1pt5},
and the terms denoted
$\delta T^{\mathcal{N}=2,\text{tree}}_i = T_i^{(\alpha')^2} + T_i^{\text{WSI}}$
in \eqref{eq:Kahlercoordinatessplitpreview}.

Recalling \eqref{eq:vdecomp},
the closed string tree level contribution to the K\"ahler potential
can be written 
\begin{align}\label{eq:Kahlerpotential}
\delta\mathcal{V}^{\mathcal{N}=2}_{\text{sphere}}=\delta\mathcal{V}_{(\alpha')^3} +\delta\mathcal{V}_{\text{WSI}}\,.
\end{align}
Here $\mathcal{V}^{(0)}$ is
the string-frame volume defined in Eq.~\eqref{eq:VolClassical}, 
the $(\alpha')^3$ correction at string tree level is \index{BBHL} \cite{Grisaru:1986kw,Gross:1986iv,Antoniadis:1997eg,Becker:2002nn}
\begin{align}\label{eq:Valphap3}
    \delta\mathcal{V}_{(\alpha')^3} &= -\frac{\zeta(3)\chi(X)}{4(2\pi)^3}\, , 
\end{align}
and the worldsheet instanton corrections\index{Worldsheet instanton corrections}\index{WSI corrections} take the form \cite{Dine:1986zy,Dine:1987bq,Grimm:2007xm}
\begin{equation}\label{eq:VWSI}
     \delta\mathcal{V}_{\text{WSI}} = \frac{1}{2(2\pi)^3}
\sum_{\mathbf{q}\in \mathcal{M}_X}\, \mathscr{N}_{\mathbf{q}}\,\Biggl( \text{Li}_3\Bigl((-1)^{\mathbf{\gamma}\cdot \mathbf{q}}\mathrm{e}^{-2\pi \mathbf{q}\cdot \mathbf{t}}\Bigr) + 2\pi \mathbf{q}\cdot \mathbf{t}\,\,\text{Li}_2\Bigl((-1)^{\mathbf{\gamma}\cdot \mathbf{q}}\mathrm{e}^{-2\pi \mathbf{q} \cdot \mathbf{t}}\Bigr)\Biggr)\,.
\end{equation}
We have introduced
\begin{equation}
    \gamma^i\coloneqq \int_X [\text{O7}]\wedge \omega^i
\end{equation}
to denote twice the class of the B-field, and  $\mathscr{N}_{\mathbf{q}}$ denote the genus-zero Gopakumar-Vafa (GV) invariants \cite{Gopakumar:1998ii,Gopakumar:1998jq} of $X$.
The polylogarithms used above are defined as \begin{equation}\label{eq:polylog}
    \text{Li}_k(z)=\sum_{n=1}^\infty \frac{z^n}{n^k}\, .
\end{equation} 
The Einstein-frame volume including $\alpha'$ corrections is written
\begin{equation}\label{eq:veinsteindef}
    \mathcal{V}_E \coloneqq \frac{\mathcal{V}}{g_s^{3/2}}\,.
\end{equation}

At string tree level 
the K\"ahler moduli are corrected as in \eqref{eq:Kahlercoordinatessplitpreview}, 
\begin{equation}\label{eq:firstdefcalT}
    T_i \rightarrow \frac{1}{g_s} \bigl (\mathcal{T}_i^{\text{tree}} + \delta\mathcal{T}_i^{(\alpha')^2} +  \delta\mathcal{T}_i^{\text{WSI}}\bigl )+i\int_X C_4\wedge \omega_i\, ,
\end{equation}
in terms of $\mathcal{T}_i^{\text{tree}}$ as defined in Eq.~\eqref{eq:Kahlercoordinatestree}, and corrections at order $(\alpha')^2$  (cf.~\cite{Cecotti:1988qn,Grimm:2007xm})
\begin{align}
    \delta\mathcal{T}_i^{(\alpha')^2} &= -\frac{\chi(D_i)}{24}\, ,\label{eq:KahlercoordinatesBBHL} 
\end{align}
as well as the worldsheet instanton contributions
\begin{align}
    \delta\mathcal{T}_i^{\text{WSI}} &= \frac{1}{(2\pi)^2}\sum_{\mathbf{q}\in \mathcal{M}_X}q_i\, \mathscr{N}_{\mathbf{q}} \,\text{Li}_2\Bigl((-1)^{\mathbf{\gamma}\cdot \mathbf{q}}\mathrm{e}^{-2\pi \mathbf{q}\cdot \mathbf{t}}\Bigr)\, .\label{eq:KahlercoordinatesWSI} 
\end{align}
The coordinates $\tau$ and $z^a$, however, are not corrected at string tree level.

\subsubsection{$\mathcal{N}=2$ string loop corrections}\label{sec:N2loop}

We now turn to the string loop corrections   
$\mathcal{K}^{\mathcal{N}=2}_{(g_s)}$
and 
$\delta T^{\mathcal{N}=2,(g_s)}_i$ 
that were introduced in  
\eqref{eq:kformalcorrforKahler1pt5}
and
\eqref{eq:kformalcorrforKahler3T}, respectively.

As we have seen above, at string tree level, the K\"ahler potential for the hypermultiplet sector receives perturbative $\alpha'$ corrections as well as worldsheet instanton contributions, which are well understood. Beyond tree level, the situation becomes more intricate.
String loop corrections 
appear in the terms 
$\delta T^{\mathcal{N}=2,(g_s)}_i$ in \eqref{eq:kformalcorrforKahler3T} and $\delta\mathcal{K}^{\mathcal{N}=2}_{(g_s)}$ in \eqref{eq:kformalcorrforKahler1pt5}.
Their structure is tightly constrained by the underlying $\mathrm{SL}(2,\mathbb{Z})$ symmetry of type IIB superstring theory, as we now explain.

The string loop corrections $\mathcal{K}^{\mathcal{N}=2}_{(g_s)}$ can be parametrized as an additional contribution $\delta\mathcal{V}^{\mathcal{N}=2}_{(g_s)}$ to the corrected Calabi-Yau volume \eqref{eq:CorrCYvol} given by
\begin{equation}\label{eq:dVloop}
   \delta \mathcal{V}^{\mathcal{N}=2}_{(g_s)} = \delta\mathcal{V}_{1\text{-loop}}
    +\delta\mathcal{V}_{\text{ED}(-1)}+\delta\mathcal{V}_{\text{ED1}}\, .
\end{equation}
Here $\delta\mathcal{V}_{1\text{-loop}}$ and $\delta\mathcal{V}_{\text{ED}(-1)}$ will be defined in Eq.~\eqref{eq:Valphap3Loop} and $\delta\mathcal{V}_{\text{ED1}}$ in Eq.~\eqref{eq:VWSILoop}. The corrections to the K\"ahler coordinates are given by
\begin{equation}\label{eq:dTloop}
    \delta T^{\mathcal{N}=2,(g_s)}_i  = \dfrac{1}{g_s}\biggl (\delta \mathcal{T}^{\text{ED}(-1)}_i+\delta \mathcal{T}^{\text{ED}1}_i\biggl )\, ,
\end{equation} 
whose origins are explained below.

\paragraph{Corrections from closed string loops and ED$(-1)$-branes.}
We begin with the loop corrections $\delta\mathcal{K}^{\mathcal{N}=2}_{(g_s)}$ to the K\"ahler potential, as in~\eqref{eq:kformalcorrforKahler1pt5}.

Let us briefly explain the ten-dimensional origin of these corrections.
By computing four graviton scattering at the closed string one-loop level,
the effective action \eqref{eq:R4action} for the ten-dimensional metric is modified as \cite{Sakai:1986bi} 
\begin{equation}\label{eq:R4actionLoop}
    S \supset \dfrac{1}{2\kappa_{10}^{2}}\int\,\mathrm{e}^{-2\phi}\biggl \{R+\dfrac{(\alpha^{\prime})^{3}}{3\cdot 2^{12}}\,\left (2\zeta(3)+\dfrac{2\pi^{2}}{3}\mathrm{e}^{2\phi}\right )\mathcal{J}_{0}\biggl \}\sqrt{-g}\dif^{10}x\, .
\end{equation}
Here, the term $\propto {2\pi^{2}}/{3}\mathrm{e}^{2\phi}$ is the new one-loop contribution, which looks kinematically like the tree level contribution --- it is also given by $\mathcal{J}_{0}\sim R^{4}$ as defined in \eqref{eq:SuperInvRF}.
This observation has its root in an additional $\mathrm{SL}(2,\mathbb{Z})$ symmetry, together with the  
chirality  
of the type IIB superstring (see e.g.~\cite{Liu:2013dna} for a discussion).
Taking into account the contributions from D$(-1)$-instantons, the full effective action can be written in the form \cite{Green:1997tv,Green:1997di,Green:1997as}\footnote{This is most easily seen via duality to M-theory \cite{Green:1997as}. In string theory, such corrections have been derived explicitly via D-instanton calculus in \cite{Sen:2021tpp,Sen:2021jbr}.}
\begin{equation}\label{eq:R4Sl2Z}
    S \supset \dfrac{1}{2\kappa_{10}^{2}}\int\,\tau_{2}^{2}\biggl \{R+\dfrac{(\alpha^{\prime})^{3}}{3\cdot 2^{12}\,\tau_{2}^{3/2}}\, f_{0}(\tau,\bar{\tau})\mathcal{J}_{0}\biggl \}\sqrt{-g}\dif^{10}x\,,
\end{equation}
where $f_0(\tau,\bar{\tau})$ is the non-holomorphic Eisenstein series of weight $3/2$ defined as
\begin{equation}
    f_0(\tau,\bar{\tau}) = \sum_{(0,0)\neq (m,n)\in \mathbb{Z}^2}\, \dfrac{\tau_2^{3/2}}{|n+\tau m|^{3}}\kom \tau=\tau_{1}+\mathrm{i}\tau_{2}\kom \tau_{2}=\mathrm{e}^{-\phi}\, .
\end{equation}
Under $\mathrm{SL}(2,\mathbb{Z})$ transformations, the metric is invariant, but the axio-dilaton $\tau$ transforms as
\begin{equation}
    \tau\rightarrow\dfrac{a\tau+b}{c\tau+d} \kom ad-bc=1\kom a,b,c,d\in\mathbb{Z}\,,
\end{equation}
under which $f_{0}(\tau,\bar{\tau})$ is invariant, that is,
\begin{equation}
    f_{0}\left (\dfrac{a\tau+b}{c\tau+d},\dfrac{a\bar{\tau}+b}{c\bar{\tau}+d}\right )=f_{0}(\tau,\bar{\tau})\, .
\end{equation}
Thus, the effective action \eqref{eq:R4Sl2Z} is also invariant under such transformations.

Dimensionally reducing this action on a Calabi-Yau threefold works analogously\footnote{Additional higher derivative corrections involving the dilaton have to be taken into account: see e.g. \cite{Liu:2022bfg}.} to the discussion in \S\ref{sec:BBHL}.\footnote{Explicit dimensional reductions to four dimensions involving the modified action \eqref{eq:R4Sl2Z} were studied in \cite{Antoniadis:1997eg,Liu:2022bfg}.}
Together with the tree-level term $\mathcal{V}_{(\alpha')^3}$, the total correction to the K\"ahler potential enters the corrected string frame volume \eqref{eq:Kahlerpotential} and takes the compact form\footnote{The explicit factor of $g_s^{3/2}=\mathrm{Im}(\tau)^{-3/2}$ shows up because we are reporting these terms in string frame.}
\begin{align}\label{eq:Valphap3Loop}
    \delta\mathcal{V}_{(\alpha')^3}+\delta\mathcal{V}_{1\text{-loop}}
    +\delta\mathcal{V}_{\text{ED}(-1)} &= -\frac{\chi(X)g_s^{3/2}}{8(2\pi)^3}\, f_0(\tau,\bar{\tau})\, .
\end{align}
At large $\tau_{2}=\mathrm{Im}(\tau)$, this modular function $f_0(\tau,\bar{\tau})$ enjoys an expansion
\begin{align}\label{eq:EisensteinExpForDiscussionSP} 
    f_{0}(\tau,\bar{\tau})&=2\zeta(3)\, \tau_2^{3/2} +\dfrac{2\pi^{2}}{3} \tau_2^{-1/2}\nonumber\\
    &\quad+8\pi \,  \tau_2^{-1/2}\sum_{n\neq 0,\,  m\geq 1}\, \left |\frac{n}{m}\right |   K_{1}(2\pi\, |nm|\,  \tau_2) \, \mathrm{e}^{2\pi\I nm\, \tau_1}\, ,
\end{align}
in terms of the modified Bessel function $K_{1}$ of the second kind.
The first term, proportional to $2\zeta(3)$, corresponds to the well-known tree-level correction discussed in \S\ref{sec:BBHL}. The second term represents the perturbative one-loop correction, suppressed by an additional factor of $\tau_2^{-2} = g_s^{2}$. 
The final line captures non-perturbative contributions from a dilute gas of D$(-1)$-instantons with charge $nm$. Using the asymptotic form $K_1(x) \sim \mathrm{e}^{-x}/\sqrt{x}$ for $x \gg 1$, these terms are exponentially suppressed by $\mathrm{e}^{-2\pi/g_s}$ relative to the tree-level result.

Similarly, the tree-level correction \eqref{eq:KahlercoordinatesBBHL} to the holomorphic K\"ahler coordinates is modified through loop effects.
As discussed in \S\ref{sec:indD3ch}, the Chern-Simons action of D-branes receives higher-order curvature corrections in the $\alpha'$ expansion, which induce D3-brane charge through terms of the schematic form $C_4 \wedge R^2$.  
Analogous higher-derivative corrections appear in the DBI action \eqref{equ:DBIXX}, as first pointed out in \cite{Bachas:1999um}.
Such corrections are essential in ensuring modular covariance of the effective action under the $SL(2,\mathbb{Z})$ duality of type IIB string theory, and lead to a
modification of the definition of the  
K\"ahler moduli $\mathcal{T}_i$, as we now explain.  

At order $(\alpha')^2$, the DBI action for a D$p$-brane acquires curvature-squared terms of the form \cite{Bachas:1999um}
\begin{equation}\label{eq:DBIR2}
    S_{\mathrm{DBI}}^{(\alpha')^2} \sim \int \, E_1(\tau,\bar{\tau}) \, t_8 R^2+\ldots\, ,
\end{equation}
where the tensor $t_8$ was defined earlier in Eq.~\eqref{eq:t8}.
The coefficient $E_1(\tau,\bar{\tau})$ is a non-holomorphic modular function which can be expressed as \cite{Bachas:1999um,Basu:2008gt,Garousi:2011fc}
\begin{equation}\label{eq:E1modular}
    E_1(\tau,\bar{\tau}) = 2\zeta(2)\tau_2 - \pi\ln(\tau_2)+2\pi\,\sqrt{\tau_2}\sum_{n\neq 0,\,  m\geq 1}\, \biggl |\dfrac{m}{n}\biggl |^{1/2}\, K_{1/2}(2\pi\, |nm|\,  \tau_2)\, \mathrm{e}^{2\pi\I mn\tau_1}\, ,
\end{equation}
in terms of the modified Bessel function $K_{1/2}$.
The first term in~\eqref{eq:E1modular} corresponds to the open string tree-level contribution, the logarithmic term arises from one-loop (annulus) corrections in the open string sector \cite{Basu:2008gt}, and the exponentially suppressed terms encode an infinite tower of non-perturbative D-instanton contributions proportional to $\mathrm{e}^{-2\pi m n \tau_2} = \mathrm{e}^{-2\pi m n / g_s}$.  
These latter terms can again be identified with the effects of ED$(-1)$-branes in the compactification space.

When evaluated on a D3-brane wrapping a smooth divisor $D_i \subset X$, the curvature-squared correction~\eqref{eq:DBIR2} integrates to a topological invariant proportional to the Euler characteristic $\chi(D_i)$.  
The resulting correction to the complexified K\"ahler coordinate $\mathcal{T}_i$ is therefore of the schematic form
\begin{equation}\label{eq:KCLoop}
    \delta\mathcal{T}_i^{(\alpha')^2}+ \delta \mathcal{T}^{\text{ED}(-1)}_i = -g_s E_1(\tau,\bar{\tau})\,\dfrac{\chi(D_i)}{12}\, ,
\end{equation}
where we have normalized the prefactor following~\cite{Bachas:1999um,Basu:2008gt,Garousi:2011fc}.  
The logarithmic dependence in~\eqref{eq:E1modular} implies that the correction \eqref{eq:KCLoop} to the K\"ahler coordinate introduces a term proportional to $\ln(g_s^{-1})$, which can be traced to the annulus diagram in the one-particle-irreducible effective action~\cite{Basu:2008gt}.  
Meanwhile, the non-perturbative pieces, proportional to $\chi(D_i)\,\mathrm{e}^{2\pi\I\tau}$, correspond to the effects of ED$(-1)$-branes attached to the D3-brane.

\paragraph{Corrections from Euclidean D1-branes.}
Next, we turn to Euclidean D1-brane corrections inherited from the $\mathcal{N}=2$ parent model. 
The full $\mathrm{SL}(2,\mathbb{Z})$-invariant expressions for the K\"ahler potential and K\"ahler coordinates can be written in the form 
\begin{equation}\label{eq:VWSILoop}
     \delta\mathcal{V}_{\text{WSI}}+\delta\mathcal{V}_{\text{ED1}} = \frac{1}{4(2\pi)^3}
    \sum_{\mathbf{q}\in \mathcal{M}_X}\,  \sum_{(0,0)\neq (m,n)\in \mathbb{Z}^2 }\frac{\mathscr{N}_{\mathbf{q}}(1 + S_{m,n})}{|m\tau +n|^3} \, (-1)^{n\mathbf{\gamma}\cdot \mathbf{q}}\, \mathrm{e}^{-S_{m,n}} \,,
\end{equation}
and
\begin{equation}\label{eq:TWSILoop}
    \delta\mathcal{T}_i^{\text{WSI}}+\delta\mathcal{T}_i^{\text{ED1}} = \frac{1}{2(2\pi)^2}\sum_{\mathbf{q}\in \mathcal{M}_X}\,\sum_{(0,0)\neq (m,n)\in \mathbb{Z}^2 }\frac{q_i\, \mathscr{N}_{\mathbf{q}} }{|m\tau +n|^2} \, (-1)^{n\mathbf{\gamma}\cdot \mathbf{q}}\, \mathrm{e}^{-S_{m,n}}\, ,
\end{equation}
where we have defined
\begin{equation}
    S_{m,n} = 2\pi |m\tau +n|(\mathbf{q}\cdot \mathbf{t})\, .
\end{equation}

To illustrate the basic behavior of these corrections, we consider a single curve with string-frame volume $t$ and B-field $b$. This simplified setting will later serve as a reference point in evaluating the smallest curves in the candidate de Sitter vacua of Chapter \ref{chap:deSitter}. In this case, the corrections can be schematically expressed as 
\begin{equation}\label{eq:N2ED1gen}
    I^{(p)}= \sum_{(0,0)\neq (m,n)\in \mathbb{Z}^2 }\frac{\mathrm{e}^{2\pi \mathrm{i}\, z_{m,n}}}{|m\tau +n|^p} \, ,\quad z_{m,n}\coloneqq nb+\mathrm{i} \, |m\tau+n|t\, ,
\end{equation}
for some $p\in \mathbb{Z}$, $p\leq 3$. 
Worldsheet instanton effects are encoded in the $m=0$ terms,
\begin{equation}
    I_{m=0}^{(p)}=2\,\text{Li}_p\Bigl((-1)^{2b} \mathrm{e}^{-2\pi t}\Bigr)\, ,
\end{equation}
which are associated with the tree level corrections in Eq.~\eqref{eq:VWSI} to the K\"ahler potential (for $p=3$ and $p=2$) and in Eq.~\eqref{eq:KahlercoordinatesWSI} to the K\"ahler coordinates (for $p=2$).
The remainder for $m\neq 0$ can be Poisson-resummed, see e.g. \cite{Robles-Llana:2006hby}.

Let us show this explicitly for the corrections to the K\"ahler potential in Eq.~\eqref{eq:VWSILoop} from a single curve.
In this case, we have
\begin{equation}
	\delta\mathcal{V}^{(t)} = I^{(3)}+(2\pi t) I^{(2)}\,.
\end{equation}
For the contributions with $m\neq 0$, we write
\begin{equation}\label{eq:N2ED1genPS1}
    I^{(p)}_{m\neq 0}=\sum_{m\neq 0}\,\sum_{n\in\mathbb{Z}} \,\frac{\mathrm{e}^{-2\pi \,  [(m\tau_1)^2 +(m\tau_2+n)^2]^{1/2}t}}{\bigl((m\tau_1)^2 +(m\tau_2+n)^2\bigr)^{p/2}} \, \mathrm{e}^{2\pi \mathrm{i}\,  nb}\,,
\end{equation}
where we introduced
\begin{equation}
	\tau=\tau_{1}+\mathrm{i}\tau_{2}\, .
\end{equation}
To perform the Poisson resummation, we use
\begin{equation}
	\sum_{n\in\mathbb{Z}}\,  g(x+na)=\dfrac{1}{a} \sum_{n\in\mathbb{Z}}\, \tilde{g}(2\pi n/a)\,\mathrm{e}^{2\pi\mathrm{i}\, nx/a}
\end{equation}
for some function $g:\,\mathbb{R}\rightarrow \mathbb{R}$ in terms of the Fourier transform
\begin{equation}
	\tilde{g}(k)=\int_{-\infty}^{\infty}\, g(x)\mathrm{e}^{-\mathrm{i}kx}\, \dif x\, .
\end{equation}
Comparing with \eqref{eq:N2ED1genPS1}, we identify
\begin{equation}\label{eq:N2ED1genPS2}
	\tilde{g}(2\pi n) = \frac{(2\pi)^{p}}{\bigl(\alpha^2 +(\gamma+2\pi n)^2\bigr)^{p/2}}\, \mathrm{e}^{- [\alpha^2 +(\gamma+2\pi n)^2]^{1/2}t}\,,
\end{equation}
in terms of
\begin{equation}
	a=1\kom x=b\kom \alpha=2\pi m\tau_2\kom \gamma=2\pi m\tau_1\, .
\end{equation}
We can then use
\begin{equation}
\int_{0}^{\infty}\, \dfrac{1}{\sqrt{x^{2}+\alpha^{2}}}\, \mathrm{e}^{-\beta\sqrt{x^{2}+\alpha^{2}}}\cos(xy)\dif x = K_{0}\left (\alpha\sqrt{\beta^{2}+y^{2}}\right )\,,
\end{equation}
and differentiate by $\alpha$ to find
\begin{equation}
\int_{0}^{\infty}\, \dfrac{1+\beta \sqrt{x^{2}+\alpha^{2}}}{(x^{2}+\alpha^{2})^{3/2}}\, \mathrm{e}^{-\beta\sqrt{x^{2}+\alpha^{2}}}\cos(xy)\dif x =\dfrac{1}{\alpha}\sqrt{\beta^{2}+y^{2}} K_{1}\left (\alpha\sqrt{\beta^{2}+y^{2}}\right )\, ,
\end{equation}
with $K_1$ the modified Bessel function of the second kind.
Combining all the pieces, we find 
\begin{align}
	\delta\mathcal{V}^{(t)} &= 2\,\text{Li}_3\Bigl((-1)^{2b} \mathrm{e}^{-2\pi t}\Bigr) +2\,(2\pi t) \text{Li}_2\Bigl((-1)^{2b} \mathrm{e}^{-2\pi t}\Bigr) +\delta\mathcal{V}_{m\neq 0}^{(t)} \, ,
\end{align}
where
\begin{equation}
	\delta\mathcal{V}_{m\neq 0}^{(t)} = \dfrac{1}{\pi\tau_{2}} \sum_{m\neq 0}\,\sum_{n\in\mathbb{Z}} \,\dfrac{\sqrt{t^{2}+(n+b)^{2}}}{|m|} K_{1}\left (2\pi |m\tau_{2}|\,  \sqrt{t^{2}+(n+b)^{2}}\right )\mathrm{e}^{-2\pi \mathrm{i}\, m \tau_{1}(b+n)}\,.
\end{equation} 
The terms encoded in $K_1$ 
are  
Euclidean D1-brane corrections,
with a sum over worldvolume fluxes. 
For all examples in these lectures, $b\in \{0,1/2\}$, 
for which the corresponding leading terms are
\begin{equation}\label{eq:LoopCorN2}
    \delta\mathcal{V}_{m\neq 0}^{(t)} \simeq \begin{cases}
        \frac{2}{\pi \tau_{2}}K_1\left(2\pi\tau_{2} t\right) & b=0\\[0.3em]
        \frac{4}{\pi\tau_{2}}\sqrt{1+\frac{1}{4t^2}}\cos(\pi \tau_{1})K_1\left( 2\pi\tau_{2}\sqrt{t^2+1/4}\right) & b=\frac{1}{2}\, .
    \end{cases}
\end{equation}

\subsection{$\mathcal{N}=1$ corrections}\label{sec:N1Cor} 
\index{$\mathcal{N}=1$ corrections}
 
While $\mathcal{N}=2$ corrections to the K\"ahler potential and K\"ahler coordinates are well understood and under good perturbative control, much less is known about genuinely $\mathcal{N}=1$ corrections arising after the orientifold projection. 
From the perspective of effective field theory, these $\mathcal{N}=1$ effects are more subtle
than their $\mathcal{N}=2$ counterparts, 
but can nevertheless exert a decisive influence on the structure of the scalar potential, the stabilization of moduli, and therefore the viability of vacua. In this section, we review what is currently known about these corrections and outline the assumptions under which they may be neglected. We will mostly focus on corrections to the K\"ahler \emph{coordinates}, omitting $\mathcal{N}=1$ corrections to the K\"ahler potential, as the former have greater impact on the computation of the moduli potential: as discussed in \S\ref{sec:stabKah}, corrections to $T_i$ modify the holomorphic variables in which the four-dimensional supergravity is expressed, and therefore have a more direct impact on the structure of the resulting vacua than do small additive corrections to $\mathcal{K}$.

In terms of the decomposition \eqref{eq:kformalcorrforKahler3T}, reproduced here for convenience,
\begin{equation}
    T_i  = T^{\text{tree}}_i
        +\delta T^{\mathcal{N}=2,\text{tree}}_i
        +\delta T^{\mathcal{N}=2,(g_s)}_i
        +\delta T^{\mathcal{N}=1,\text{tree}}_i 
        +\delta T^{\mathcal{N}=1,(g_s)}_i\,,
\end{equation}
the terms $T^{\text{tree}}_i$, $\delta T^{\mathcal{N}=2,\text{tree}}_i$, and $\delta T^{\mathcal{N}=2,(g_s)}_i$ were obtained in \S\ref{sec:N2Cor} and originate entirely from the pre-orientifold $\mathcal{N}=2$ theory. The last two terms encode the genuinely $\mathcal{N}=1$ corrections, which include:  
\begin{itemize}
    \item \emph{tree-level $\mathcal{N}=1$ effects}, denoted $\delta T_{i}^{\mathcal{N}=1,\text{tree}}$, principally the effects of warping, sourced by fluxes, D-branes, and O-planes. 
    \item \emph{string loop $\mathcal{N}=1$ corrections}, denoted $\delta T_{i}^{\mathcal{N}=1,(g_s)}$, which arise at the loop level and are therefore suppressed by explicit powers of the string coupling. 
\end{itemize}
In both cases, their structure is tied to the details of the orientifold projection and typically cannot be inferred from the simpler $\mathcal{N}=2$ parent theory.
Unlike the $\mathcal{N}=2$ corrections, which are largely fixed by hypermultiplet geometry or underlying symmetries of the parent theory like $\mathrm{SL}(2,\mathbb{Z})$, the $\mathcal{N}=1$ contributions lack a universal form and remain an active area of research. 
The limited understanding of these $\mathcal{N}=1$ corrections is one of the main obstacles to obtaining fully controlled moduli stabilization scenarios, particularly in settings that rely on delicate cancellations or hierarchical tunings in special regions of the K\"ahler cone. Developing a more complete description of such effects, and clarifying the regimes in which they can be neglected, remains an important open problem in the quantum theory of flux compactifications.

\subsubsection{$\mathcal{N}=1$ string tree level corrections}\label{sec:10deffects}

We begin with the terms $\delta T^{\mathcal{N}=1,\text{tree}}_i$. The definition \eqref{eq:Ttree} of the tree-level K\"ahler coordinates is valid in an unwarped Calabi-Yau compactification. In the presence of warping, instead of $\mathcal{T}_i^{\text{tree}} \coloneqq \frac{1}{2 }\int_{\Sigma_4^{i}}\, J \wedge J$, one has \cite{DeWolfe:2002nn,Giddings:2005ff,Frey:2013bha,Martucci:2014ska,Martucci:2016pzt} 
\begin{equation}\label{eq:Ttreewarped}
    \mathcal{T}_i^{\text{tree}} + \delta \mathcal{T}^{\mathcal{N}=1,\text{tree}}_i = \frac{1}{2 }\int_{\Sigma_4^{i}}\mathrm{e}^{-4A}\, J \wedge J\,,
\end{equation}
where $\mathrm{e}^{A}$ is the warp factor defined in \eqref{eq:WarpedBackgroundAnsatzGKPSection}.
The appearance of the warp factor in \eqref{eq:Ttreewarped} can be easily understood by computing the classical gauge coupling on D7-branes in a warped background (see e.g.~\cite{Baumann:2006th}), and the proper definitions were systematically treated in 
\cite{Martucci:2016pzt}.

In an ISD background that fulfills the conditions \eqref{eq:BPSconditions},
the warp factor obeys the Poisson equation \eqref{eq:simplepoisson} in the internal space. Thus, evaluating $\delta \mathcal{T}^{\mathcal{N}=1,\text{tree}}_i$ requires solving \eqref{eq:simplepoisson} and integrating the resulting warp factor over a basis $\{\Sigma_i\}$ of divisors.  In a non-compact Calabi-Yau cone where the Ricci-flat metric is known, one can sometimes compute the necessary Green's function, and then perform the integral in \eqref{eq:Ttreewarped} over a locally-defined four-cycle.
This computation was carried out for the conifold, and for cones over the Sasaki-Einstein manifolds $Y^{p,q}$, in \cite{Baumann:2006th}, and the resulting corrections play a key role in D-brane inflation (see \cite{Baumann:2014nda}).

To perform the corresponding computation in a \emph{compact} Calabi-Yau, one would first need to obtain a numerical approximation to the Ricci-flat metric, and then solve the Poisson equation.  For threefolds with large numbers of K\"ahler moduli, including all the examples of Chapter \ref{chap:deSitter}, computing the metric numerically is not currently possible.  An advance in this direction would be important, but for the present, we will necessarily resort to estimates.

As a proxy for estimating the size of $\mathcal{N}=1$ corrections induced by warping and localized sources, it is useful to introduce the dimensionless measures proposed in~\cite{Demirtas:2021nlu}; see also \cite{Hebecker:2025} for a recent analysis of warping effects.\footnote{The literature on warped flux backgrounds contains analyses of issues such as moduli dynamics, backreaction, localized sources, and the interplay between warping and supersymmetry breaking \cite{DeWolfe:2002nn, Giddings:2005ff, Burgess:2006mn, Frey:2006wv, Douglas:2007tu, Koerber:2007xk, Shiu:2008ry, Douglas:2008jx, Frey:2008xw, Marchesano:2008rg, Martucci:2009sf, Douglas:2009zn, Underwood:2010pm, Marchesano:2010bs, Grimm:2012rg, Frey:2013bha, Martucci:2014ska, Grimm:2014efa, Grimm:2015mua, Martucci:2016pzt, Lust:2022xoq, Frey:2025rvf, Agarwal:2025rqd}. Together, these works have significantly
clarified the role of warping in the four-dimensional effective theory, but several aspects remain unresolved.}  
In an O3/O7 orientifold, each four-cycle $D_i$ acquires an induced D3-brane charge through worldvolume flux and curvature couplings in the D7/O7 Chern-Simons action, cf. \S\ref{sec:indD3ch}. Denoting this induced charge by $N_{\mathrm{D3}}(\omega_i)$, we define the dimensionless measures
\begin{equation}\label{eq:Demirtas:2021nlu417}
    g_{\mathcal{N}=1}^i = \frac{|N_{\text{D3}}(\omega_i)|}{\mathcal{T}_{i,E}^{(0)}}\, ,
\end{equation}
where $\mathcal{T}^{(0)}_{i,E}$ is the leading (Einstein-frame) tree-level volume of the divisor $D_i$.
A complementary global measure compares the total orientifold-induced D3-brane charge $Q_{\mathrm{O}}$, arising from O3-planes and O7-planes as discussed around Eq.~\eqref{eq:QOf}, to the Einstein-frame volume $\mathcal{V}_E$ of the Calabi-Yau orientifold:
\begin{equation}\label{eq:Demirtas:2021nlu418}
    g_{\mathcal{N}=1}^X = \frac{Q_{\text{O}}}{\mathcal{V}_{E}^{2/3}}\, .
\end{equation}
Because $Q_{\mathrm{O}}$ encodes the negative induced D3-brane charge of the orientifold planes, the ratio $ g_{\mathcal{N}=1}^X$ provides a coarse diagnostic of the possible size of warping and localized curvature effects in the $\mathcal{N}=1$ effective theory.

In the examples in Chapter \ref{chap:deSitter}, the proxy quantities $g_{\mathcal{N}=1}^i$ in \eqref{eq:Demirtas:2021nlu417} can be as large as $30\%$, while $g_{\mathcal{N}=1}^X$ in \eqref{eq:Demirtas:2021nlu418} is around $10\%$.  Thus, evaluating the effects of warping in an explicit supergravity solution is an urgent problem.
 
A particularly serious concern involving warp factor corrections is known as the \emph{singular bulk problem}: negative D3-brane charge on seven-brane stacks can source a singular warp factor \cite{Carta:2019rhx,Gao:2020xqh,Carta:2021lqg}.  A proper quantitative assessment of this issue will require a numerical supergravity solution, which is not presently available. 
We note that in \cite{Moritz:2025bsi} it was found that the singular bulk problem, and more generally the warping corrections to K\"ahler coordinates, become more severe at large complex structure.

\subsubsection{$\mathcal{N}=1$ string loop corrections}\label{sec:N1loop}

Next, we turn to the terms $\delta T^{\mathcal{N}=1,(g_s)}_i$, which encode genuine loop-level corrections to the K\"ahler coordinates arising within the four-dimensional $\mathcal{N}=1$ theory.

To describe these effects it is convenient to work with the real parts of the K\"ahler coordinates, including all string-tree-level $\alpha'$ corrections but excluding loop effects,
\begin{equation}
    \mathcal{T}_i^{\text{l.o.}} \coloneqq \text{Re}\, T_i^{\text{l.o.}}\,,
\end{equation} 
where $T_i^{\mathrm{l.o.}}$ was defined in Eq.~\eqref{eq:firstdefcalT}.
String loop corrections enter through powers of the \emph{string-frame} divisor volumes.  
Let $\mathcal{T}_{i,s}^{\mathrm{l.o.}}$ denote the corresponding string-frame volume of a divisor $D_i$, again including all tree-level $\alpha'$ corrections. 
This is related to the Einstein frame volume $\mathcal{T}_i^{\text{l.o.}}$  via $1/\mathcal{T}_i^{\text{l.o.}} = g_s/\mathcal{T}_{i,s}^{\text{l.o.}}$.

The most relevant corrections arise from small divisors, and from small curves:
\begin{equation}\label{eq:ModelForCorrectionsToTscheme}
   \delta T^{\mathcal{N}=1,(g_s)}_i =  \delta T^{\mathcal{N}=1,(g_s)}_{i,\text{divisor}} + \delta T^{\mathcal{N}=1,(g_s)}_{i,\text{curve}} \, .
\end{equation}
We begin with divisor volumes.
For a gauge theory with $N$ colors and with gauge coupling $g_i$,
on the grounds of naive dimensional analysis the loop counting parameter should be
\begin{equation}
    \lambda = \frac{g_i^2 N}{16\pi^2}\,.
\end{equation}
Converting to our parameters, we have
\begin{equation}
    \lambda = \frac{c_{D_i}}{4\pi\mathcal{T}_i^{\text{l.o.}}}\,,
\end{equation}
where $c_{D_i}$ is the dual Coxeter number of the seven-brane gauge group on divisor $D_i$,
and so
\begin{equation}\label{eq:ModelForCorrectionsToTdiv}
   \delta T^{\mathcal{N}=1,(g_s)}_{i,\text{divisor}} = \frac{c_{D_i}}{4\pi}\sum_{n=0}^\infty k^{i}_{\text{self},n} \,\Biggl(\frac{c_{D_i}}{4\pi} \cdot \frac{g_s}{\mathcal{T}_{i,s}^{\text{l.o.}}}\Biggr)^n+\ldots  \,,
\end{equation} 
in terms of unknown coefficients $k^i_{\text{self}}$ that one would expect to be of order unity (see for comparison the explicit computation in toroidal orientifolds carried out in \cite{Berg:2004ek}). The $n=0$ term in \eqref{eq:ModelForCorrectionsToTdiv} can be absorbed into the definition of the Pfaffian $\mathcal{A}_D$, over which we marginalize later, see the discussion in \S\ref{sec:twee}. Hence, the first correction we will need evaluate in our examples is $n=1$, which is at $\mathcal{O}(g_s)$.

An additional correction entering $\delta T^{\mathcal{N}=1,(g_s)}_{i,\text{divisor}}$ in \eqref{eq:ModelForCorrectionsToTdiv} is a term proportional to $\log(\mathcal{V}_E)$, namely
\begin{align}\label{eq:ReTlog}
    \text{Re}(T_i)\supset \text{Re}(T_i^{\mathrm{tree}})+ \beta_{i}  \ln( \mathcal{V}_E^{(0)})\, .
\end{align}
Corrections of this form are expected on general field-theoretic grounds \cite{Conlon:2010ji}, and also have been argued to arise from specific higher-derivative couplings in M-theory compactifications \cite{Grimm:2017pid,Weissenbacher:2019mef}. The most direct and unambiguous evidence for their presence, however, comes from the heterotic/F-theory duality analysis of \cite{Klaewer:2020lfg}, where such logarithmic terms were derived explicitly. The variation of a $\log(\mathcal{V}_E^{(0)})$ term with respect to the K\"ahler parameters is suppressed by a factor of $1/\mathcal{V}_E^{(0)}$, so its effect on the K\"ahler metric and on the scalar potential is sub-leading at large volume. To leading order in the $1/\mathcal{V}_E^{(0)}$ expansion, this logarithmic contribution can be absorbed into a rescaling of the Pfaffians $\mathcal{A}_D$ appearing in the non-perturbative superpotential. Since in our analysis we marginalize over the $\mathcal{A}_D$ across a broad range, we do not include the logarithmic term in our model \eqref{eq:ModelForCorrectionsToTdiv}.

We now consider the corrections $\delta T^{\mathcal{N}=1,(g_s)}_{i,\text{curve}}$. The primary risk comes from small curves, and in \cite{Berg:2005ja,Berg:2007wt} it was argued that in fact the dominant effect arises from winding strings that wrap a curve where two stacks of D7-branes intersect. In our examples, there are no such intersections.  A more conservative model is that a comparable correction could also arise from winding strings on a curve where a Euclidean D3-brane intersects a stack of D7-branes. But in our compactifications all such curves of intersection have genus $g=0$, and so support no winding states.
       
An even more conservative model assigns contributions
\begin{equation}\label{eq:potloopest}
    \delta T^{\mathcal{N}=1,(g_s)}_{i,\text{curve}} =  k_{\mathcal{C}}\,\frac{g_s}{\vol_s{\mathcal{C}}}\,, 
\end{equation}
supported even on curves $\mathcal{C}$ that are \emph{not} the intersection loci of Euclidean D3-branes and D7-branes (see~\cite{Gao:2022uop}), provided that $\mathcal{C}$ has $g>0$ and thus can host winding modes.  As usual $k_{\mathcal{C}}$ is an undetermined coefficient.

In the examples in Chapter \ref{chap:deSitter} we therefore evaluate the string-frame volume $\vol_s{\mathcal{C}}$ of the smallest potent\footnote{We call a curve $\mathcal{C}$ potent if the series $\mathscr{N}(\mathcal{C}), \mathscr{N}(2\mathcal{C}), \mathscr{N}(3\mathcal{C}), \ldots$, with $\mathscr{N}(\bullet)$ the genus-zero Gopakumar-Vafa invariant, has infinitely many nonzero terms.  Such a curve  cannot be flopped: as it shrinks to zero volume, an infinite tower of states becomes light.} (and hence, potentially nonzero-genus) curve $\mathcal{C}$, and use this value to estimate curve corrections to the coordinates $\mathcal{T}_{i}$.

\subsubsection*{A model of string loop corrections to $\mathcal{K}$}

Finally, we examine string loop corrections $\delta \mathcal{K}_{(g_{s})}^{\mathcal{N}=1}$ to the K\"ahler potential, cf.~\eqref{eq:kformalcorrforKahler3P1}. 
As we noted at the beginning of 
\S\ref{sec:N1Cor}, the corresponding effects are sub-leading compared to those of $\delta T^{\mathcal{N}=1,\text{tree}}_i$ 
and $\delta T^{\mathcal{N}=1,(g_s)}_i$, and we discuss
$\delta \mathcal{K}_{(g_{s})}^{\mathcal{N}=1}$ here only for the sake of completeness.\footnote{In particular, the status of the candidate de Sitter vacua presented in chapter \ref{chap:deSitter} does not rely on the specific forms given below.}
 
The no-scale property \eqref{eq:NoScaleID} of the classical K\"ahler metric is already broken at the level of $\mathcal{N}=2$ corrections, among others by the tree level $(\alpha')^3$ correction \eqref{eq:Valphap3} of \cite{Becker:2002nn}, as we verified in \eqref{eq:NoScaleIDBBHL}. Beyond that, perturbative corrections from higher-genus string worldsheets also tend to break the tree-level no-scale property \eqref{eq:NoScaleIDBBHL}. Unfortunately, explicit computations of these corrections are scarce. At present, the only fully known results come from the $\mathcal{N}=1$ orientifold compactification on $T^{6}/(\mathbb{Z}_{2}\times\mathbb{Z}_{2})$ studied in \cite{Berg:2005ja, Berg:2005yu}, with subsequent work extending these insights to general Calabi-Yau threefolds~\cite{Berg:2007wt, Cicoli:2007xp}. A useful organizing principle for loop corrections is provided by the Berg-Haack-Pajer (BHP) conjecture \cite{Berg:2007wt} (see also~\cite{vonGersdorff:2005bf}), which proposes two characteristic scalings for the string loop contribution to the K\"ahler potential. These effects fall into two classes,
\begin{equation}
    \delta \mathcal{K}_{(g_{s})}^{\mathcal{N}=1} \supset \delta \mathcal{K}_{(g_{s})}^{\rm KK}  + \delta \mathcal{K}_{(g_{s})}^{\rm W}\, ,
    \label{equ:KKgs}
\end{equation}
associated with
\begin{enumerate}
    \item corrections from exchange of closed strings carrying Kaluza-Klein momentum, and  
    \item corrections from wound strings associated with non-contractible one-cycles on the intersection curves of D7-branes.\footnote{These one-cycles arise only on the intersection curves of D7-branes because the Calabi-Yau itself has no non-trivial (non-torsion) one-cycles.}
\end{enumerate}
More recently, Ref.~\cite{Gao:2022uop} sharpened the understanding of loop effects by identifying three distinct types of corrections: genuinely non-local quantum loops from KK modes, local higher-derivative $\alpha'$ corrections arising in the ten-dimensional bulk or on branes, and additional localized contributions tied to brane and orientifold sectors. This classification clarifies which terms in the loop-corrected K\"ahler potential originate from true quantum effects and which stem from local higher-curvature operators.

Following 
\cite{Gao:2022uop}, the loop corrections of \cite{Berg:2007wt} can be defined as follows. The conjectured KK and winding contributions take the form
\begin{equation}\label{equ:KGS}
    \delta \mathcal{K}_{(g_{s})}^{\rm KK}
      =  g_{s}\sum_{\alpha}\, \frac{\mathcal{C}_{\alpha}^{\rm KK}(z,\bar{z})\, \mathcal{J}_\alpha(t^i)}{\mathcal{V}} \; , \quad \delta \mathcal{K}_{(g_{s})}^{\rm W} = \sum_{\alpha} \frac{\mathcal{C}_{\alpha}^{\rm W}(z,\bar{z})}{\mathcal{I}_\alpha(t^i)\,\mathcal{V}}\, ,
\end{equation}
where $\mathcal{J}_{\alpha}$ and $\mathcal{I}_{\alpha}$ were argued in the original BHP proposal to be linear in the K\"ahler parameters $t^{i}$, although \cite{Gao:2022uop} emphasized that $\mathcal{I}_{\alpha}$ can more generally depend on ratios of the $t^{i}$ and need not be strictly linear.  
A key feature of these corrections is the appearance of the functions $\mathcal{C}^{\rm KK}_{\alpha}(z,\bar{z})$ and $\mathcal{C}^{\rm W}_{\alpha}(z,\bar{z})$, which encode non-trivial dependence on the complex structure moduli $z^a$.  As a result, the moduli space metric no longer factorizes into independent K\"ahler and complex structure sectors, and hence loop corrections
are believed to unavoidably mix the two. This mixing is characteristic of genuine $\mathcal{N}=1$ corrections.

Counting dimensions in \eqref{equ:KGS}, we see that $\delta \mathcal{K}_{(g_{s})}^{\rm KK}$ enters at order $g_s(\alpha')^2$, whereas $\delta \mathcal{K}_{(g_{s})}^{\rm W}$ enters at order $g_s^2(\alpha')^4$. In particular, $\delta \mathcal{K}_{(g_{s})}^{\rm KK}$ is the dominant contribution which, naively, also dominates over the tree level corrections $\delta K_{(\alpha')^3}$ in \eqref{eq:Valphap3}.
The situation is in fact more subtle than a naive analysis would suggest, as first emphasized in~\cite{vonGersdorff:2005bf} and later clarified and made more explicit in~\cite{Berg:2007wt, Cicoli:2007xp}: although loop corrections dominate at the level of the K\"ahler potential, a partial cancellation occurs in the scalar potential. As a result, the leading contribution to the scalar potential is still controlled by the tree level $(\alpha')^3$ term in \eqref{eq:Valphap3}. This cancellation, known as \emph{extended no-scale structure}~\cite{Cicoli:2007xp}, arises from the fact that $\delta \mathcal{K}_{(g_{s})}^{\rm KK}$ is homogeneous of degree $-2$ in the two-cycle volumes $t^{i}$.

A simple example with one K\"ahler modulus $\tau$~\cite{vonGersdorff:2005bf,Cicoli:2011zz} illustrates this
phenomenon. The K\"ahler potential may be written schematically as
\begin{equation}
    \mathcal{K} = -2\ln(\mathcal{V}) + \frac{\mathcal{C}^{\rm KK}\sqrt{\tau}}{\mathcal{V}} - \frac{\hat{\xi}}{\mathcal{V}}\; ,\quad \mathcal{V}=\tau^{3/2}\, . \label{equ:KKK}
\end{equation}
For simplicity, we assume that the complex structure moduli $z^a$ have been integrated out so that we can treat $\mathcal{C}^{\rm KK}$ as a constant.
For a constant superpotential $W=W_{0}$, the resulting scalar potential becomes
\begin{equation}
    V = \frac{|W_{0}|^{2}}{\mathcal{V}^{3}}
        \left[
            0 + \hat{\xi} + 0\cdot \mathcal{C}^{\rm KK}\sqrt{\tau}
            + \frac{\mathcal{C}^{\rm KK}}{\sqrt{\tau}} + \ldots
        \right]\,  .
    \label{equ:Vgs}
\end{equation}
The first zero reflects the standard no-scale structure \eqref{eq:NoScaleID}. The vanishing of the term proportional to $\sqrt{\tau}$ is the hallmark of extended no-scale structure. The dominant loop correction scales as $1/\sqrt{\tau}$ and is therefore \emph{sub-leading} compared to the tree level $(\alpha')^3$ term, which breaks no-scale structure due to \eqref{eq:NoScaleIDBBHL}.

\newpage

\subsubsection{Towards $\mathcal{N}=1$ corrections from F-theory and string field theory}\label{sec:sft}

In order to understand the structure of quantum corrections to the K\"ahler moduli in genuine $\mathcal{N}=1$ compactifications, it is essential to go beyond the classical type IIB description and identify the corrections that survive the $\mathcal{N}=2 \rightarrow \mathcal{N}=1$ truncation. Concrete progress has nevertheless been made: for example, several works have shown that higher-derivative $(\alpha')^3$ terms induce shifts of $\chi$ in Eq.~\eqref{eq:KaehlerTestTreeBBHL}, arising both at string tree level~\cite{Minasian:2015bxa} and at string loop level~\cite{Berg:2014ama,Haack:2015pbv,Haack:2018ufg,Antoniadis:2018hqy}. A powerful method for deriving these corrections makes use of duality: rather than computing $\alpha'$ effects directly in the orientifold limit, one can lift the problem to M-theory on a Calabi-Yau fourfold $Y_4$ and then pass to F-theory\index{F-theory}. In this setting, the relevant perturbative $\alpha'$ corrections arise from higher-derivative terms in the eleven-dimensional M-theory effective action \cite{Grimm:2013bha,Grimm:2013gma,Junghans:2014zla,Grimm:2017pid,Weissenbacher:2019mef}. After taking the limit to F-theory, these corrections descend to the four-dimensional $\mathcal{N}=1$ effective theory, modifying both the K\"ahler potential and K\"ahler coordinates. Alternatively, one can use the duality of F-theory to compactifications of heterotic string theory to determine additional corrections \cite{Klaewer:2020lfg}. In the long term, to advance our understanding of $\mathcal{N}=1$ corrections, it will be essential to combine insights from these dual perspectives to obtain a consistent picture of the corrected K\"ahler coordinates $T_i$.

Another promising avenue for understanding quantum corrections is \emph{string field theory}\index{String field theory}. String scattering amplitudes for states that undergo renormalization exhibit infrared ambiguities, which stem from the fact that the corresponding external states require a consistent prescription for wavefunction renormalization \cite{Witten:2012bh}. The appearance of non-trivial loop corrections to the K\"ahler metric is a direct manifestation of this issue: the moduli fields of a Calabi-Yau compactification are not protected and therefore acquire wavefunction renormalization in the full quantum theory. String field theory provides a natural and systematic framework for addressing this issue, because it supplies an off-shell formulation of the theory in which the field basis of string states and hence their renormalization is fixed consistently.

At first sight, it might appear that such a sophisticated formalism is unnecessary. One could attempt to work order by order in the worldsheet genus and carry out perturbative computations in the $\alpha'$ expansion. Given that gravitons are not renormalized in ten-dimensional flat space, one might hope to compute graviton loop amplitudes in $\mathbb{R}^{9,1}$ and then dimensionally reduce the result to extract corrections to the four-dimensional K\"ahler potential. This approach has the advantage of relying only on well-understood perturbative string techniques.

However, a  perturbative expansion in $\alpha'$   misses a  class of finite-coupling effects that are crucial in compactifications with fluxes: worldsheet instantons, D-instantons, and related non-perturbative contributions are invisible in any finite-order perturbative expansion. Thus, it would be valuable to develop fully string field theoretic tools to compute the loop-corrected K\"ahler potential directly. Such methods have the potential to unify perturbative and non-perturbative corrections within a single framework and to clarify the structure of quantum corrections in flux compactifications. Recent progress in this direction \cite{Sen:2014pia,Cho:2018nfn,Cho:2023mhw,Kim:2024dnw,Mazel:2025fxj,Frenkel:2025wko}\footnote{For a review, see \cite{Sen:2024nfd}.} demonstrates that string field theory can be used to compute properties of the four-dimensional effective action, including D-instanton corrections to the metric on hypermultiplet moduli space  
in type IIB compactifications \cite{Alexandrov:2021shf,Alexandrov:2021dyl}. Techniques inspired by string field theory have been employed to derive the one-loop renormalization of the dilaton and graviton kinetic terms for general Calabi-Yau orientifolds \cite{Kim:2023sfs,Kim:2023eut}.

\section{The superpotential}\label{sec:thew}

In type IIB flux compactifications, the superpotential enjoys powerful non-renormalization properties: it receives no corrections in perturbation theory, either in the string coupling $g_s$ or in the $\alpha'$ expansion. As we explain in \S\ref{sec:non-renormalization}, this is a consequence of holomorphy, $\mathcal{N}=1$ supersymmetry, and underlying shift symmetries of the axionic fields. However, the superpotential does receive non-perturbative corrections, arising from Euclidean D-branes or from strong gauge dynamics on D7-branes via gaugino condensation. These effects, introduced in \S\ref{sec:Wnp}, are exponentially suppressed in the large-volume and weak-coupling limits, and take the schematic form $W_{\mathrm{np}} \sim \mathcal{A}_D\, \mathrm{e}^{-a T_D}$, where $T_D$ is a K\"ahler modulus associated to the divisor $D$ wrapped by the instanton, and $\mathcal{A}_D$ is a prefactor potentially depending on complex structure moduli. 
Although these contributions are non-perturbative, they induce the leading dependence of the superpotential on the K\"ahler moduli --- 
recall that the classical flux superpotential $W_{\mathrm{flux}}$ depends only on the complex structure moduli and axio-dilaton ---
and so they can 
play a central role in stabilizing K\"ahler moduli, and hence in constructing fully stabilized vacua.

\subsection{Non-renormalization}\label{sec:non-renormalization}
\index{Non-renormalization}

As demonstrated in \S\ref{sec:fluxpotential}, the scalar potential $\propto \int G_3 \cdot \overline{G}_3$ can be obtained from the K\"ahler potential \eqref{eq:fullK}, and from the flux superpotential
\begin{equation}
    W = \sqrt{\tfrac{2}{\pi}} \int G_3 \wedge \Omega\, .
\end{equation} 
This expression defines the \emph{classical} flux superpotential in type IIB compactifications. A key feature of this superpotential is that it is not renormalized: it is exact to all orders in $g_s$ and $\alpha'$. Here is the argument, for the $\alpha'$ expansion --- for the $g_s$ expansion, see \cite{Burgess:2005jx}.

The non-renormalization in $\alpha'$ follows from a shift symmetry of the ten-dimensional action. At leading order in $\alpha'$, the classical type IIB action is invariant under the continuous shift
\begin{equation}
    \mathscr{S}_\mathbb{R}: \qquad C_4 \to C_4 + \Lambda, \qquad  \Lambda \in H^4(X,\mathbb{R})\,.
\end{equation}
This descends in four dimensions to a shift symmetry of the real pseudoscalar fields $\theta_i \coloneqq \int_{\Sigma^i_4} C_4$. This implies that the fields $\theta_i$ have purely derivative couplings in the classical, leading order in $\alpha'$ action in four dimensions. Hence all Feynman rules for $\theta_i$ involve momenta, and hence all couplings generated by loop corrections do as well: these couplings all vanish at zero momentum. It follows that perturbative quantum corrections to the effective action cannot generate a potential for $\theta_i$ at zero momentum, and thus the continuous shift symmetry $\mathscr{S}_\mathbb{R}$ is preserved to all orders in $\alpha'$. It is broken only by instantons that carry charge under the Ramond-Ramond 4-form $C_4$, i.e., only by Euclidean D-branes, or by strong gauge dynamics on seven-branes.  Such effects break $\mathscr{S}_\mathbb{R}$ down to the discrete shift symmetry
\begin{equation}
    \mathscr{S}_\mathbb{Z}: \qquad C_4 \to C_4 + \Lambda, \qquad  \Lambda \in H^4(X,\mathbb{Z})\,.
\end{equation}

The K\"ahler coordinates on K\"ahler moduli space are given by
\begin{equation}
    T_i \coloneqq \frac{1}{2g_s} \int_{\Sigma_4^{i}}J \wedge J + i \int_{\Sigma_4^{i}} C_4
\end{equation}
and supersymmetry requires that the superpotential $W$ be a holomorphic function of these coordinates. Invariance under the discrete shift symmetry $\mathscr{S}_\mathbb{Z}$ then constrains the superpotential to have the general form
\begin{equation}
    W= \mathcal{C}_0+ \sum_{n=1}^{\infty} \, \mathcal{C}^{(i)}_n\, \mathrm{e}^{-2\pi n T_i}
\end{equation} 
with the prefactors $\mathcal{C}^{(i)}_n$ depending holomorphically on moduli other than K\"ahler moduli.

Putting all of this together, the exact superpotential in type IIB flux compactifications takes the form
\begin{equation}\label{eq:Wexact}
    W(z^a,\tau,T_i) = W_{\text{flux}}(z^a,\tau) + W_{\text{np}}(z^a,\tau,T_i)\,,
\end{equation}
where $W_{\mathrm{flux}}$ is determined by background $3$-form fluxes \eqref{eq:GVW}, and $W_{\mathrm{np}}$ encodes non-perturbative effects such as Euclidean D3-instantons or gaugino condensation. In the remainder of this chapter, we describe how each contribution to the superpotential can be computed explicitly: the flux term from period integrals, and the non-perturbative term from instanton calculus.

\subsection{The flux superpotential}\label{sec:flux_superpotential}

In type IIB Calabi–Yau orientifold compactifications, background $3$-form fluxes play a crucial role in stabilizing moduli and generating a non-trivial scalar potential as discussed in \S\ref{sec:classicalstab}. These fluxes enter the four-dimensional effective theory through the GVW superpotential \eqref{eq:GVW}, which depends holomorphically on the axio-dilaton $\tau$ and the complex structure moduli $z^a$. In this section, we describe how the flux superpotential can be computed explicitly by evaluating suitable complex integrals.\index{Flux superpotential}

\subsubsection{Formal structure}

Our goal is to evaluate the GVW superpotential generated by background $3$-form fluxes in type IIB compactifications,
\begin{equation}
    W_{\mathrm{flux}} = \sqrt{\tfrac{2}{\pi}} \, \int_X\, G_3 \wedge\Omega\,,
\end{equation}
where $G_3 = F_3 - \tau H_3$ is the complexified $3$-form flux, and $\Omega$ is the holomorphic $3$-form of the Calabi–Yau threefold $X$. To evaluate this expression, we first introduce a symplectic basis $\{\alpha_A, \beta^A\}$, with $A = 0, \ldots, h^{2,1}$, for $H^3(X, \mathbb{Z})$, satisfying
\begin{equation}
    \int_{X}\,\alpha_A \wedge \beta^B = \delta_{A}^{~B}\, .
\end{equation}
In this basis, $\Omega$ admits the standard expansion
\begin{equation}
    \Omega = X^A\, \alpha_A-\mathcal{F}_B\,\beta^B\,,
\end{equation}
where $X^A,\mathcal{F}_B$ are called \emph{periods}. These can be assembled into \emph{period vector} $\vec{\Pi}$ defined by
\begin{equation}\label{eq:PvecFirstDef}
    \vec{\Pi}\coloneqq
        \begin{pmatrix}
            \int_X\,\Omega\wedge \beta_A \\
            \phantom{\Bigl.}\int_X\,\Omega\wedge \alpha^A
        \end{pmatrix}
        =
        \begin{pmatrix}
            \mathcal{F}_A\\
            X^A
        \end{pmatrix}\, .
\end{equation}
Similarly, the background fluxes can be expanded in the same basis. We define the flux vectors
\begin{equation} 
    \vec{f} \coloneqq \biggl(\int_X\, F_3 \wedge \alpha_A, \int_X\, F_3 \wedge \alpha^A\biggr)\,, \qquad  \vec{h} \coloneqq \biggl(\int_X\, H_3 \wedge \alpha_A, \int_X\, H_3 \wedge \alpha^A\biggr)\,.
\end{equation}
Due to the Dirac quantization condition \eqref{eq:DiracQuant}, the fluxes are integers: $\vec{f}, \vec{h} \in \mathbb{Z}^{2h^{2,1}+2}$. 
We can then write
\begin{equation}\label{eq:WfluxPeriods}
    W_{\mathrm{flux}} = \sqrt{\tfrac{2}{\pi}} \, \vec{\Pi} \cdot \Sigma \cdot \left(\vec{f}-\tau\vec{h}\right)\, ,
\end{equation}
where $\Sigma$ is the symplectic pairing matrix,
\begin{equation}  
    \Sigma\coloneqq
    \begin{pmatrix} 
        0 & {\mathbb{I}}\\
        -{\mathbb{I}} & 0
    \end{pmatrix}\, .
\end{equation}

In a suitable patch, the complex structure moduli space of $X$
can be parametrized by the periods $X^A$ corresponding to  
\textit{homogeneous} complex coordinates.
By normalizing $\Omega$ such that $X^0=1$, which can always be achieved away from the locus $z^0=0$, the remaining coordinates $z^a=X^a/X^0$, $a=1,\ldots,h^{2,1}(X)$, serve as complex structure moduli. 

The dual periods $\mathcal{F}_A$ are determined in terms of the $z^a$ by the \textit{prepotential} $\mathcal{F}(z)$ via 
\begin{equation}\label{eq:relper}
    \mathcal{F}_a(z)=\p_{z^a} \mathcal{F}(z)\, ,\; \mathcal{F}_0 =2\mathcal{F}-z^a\p_{z^a}\mathcal{F}\, .
\end{equation}
In general the K\"ahler potential $\mathcal{K}_{\text{cs}}$ for the complex structure moduli and the prepotential $\mathcal{F}$ are related by 
\begin{equation}
    \mathcal{K}_{\text{cs}} = \mathrm{log}\Bigl(\mathrm{i}\,\overline{X}^A\mathcal{F}_A- \mathrm{i}\,X^A\overline{\mathcal{F}}_A\Bigr) = \mathrm{log}\Bigl(2\,\mathrm{Im}\bigl (X^A\overline{\mathcal{F}}_A\bigl )\Bigr)\,,
\end{equation}
which can be simplified using \eqref{eq:relper} as
\begin{equation}
    \mathcal{K}_{\text{cs}} = \mathrm{log}\Bigl(2\,\mathrm{Im}\bigl (2\overline{\mathcal{F}}+(z^a-\bar{z}^a)\overline{\mathcal{F}}_a\bigl )\Bigr)\, .
\end{equation}
In summary, the flux superpotential $W_{\mathrm{flux}}$ and the K\"ahler potential $\mathcal{K}_{\text{cs}}$ can be explicitly computed once the periods $\mathcal{F}_A$ and $X^A$ are known or, alternatively, a prepotential $\mathcal{F}(z)$ is available.
In the next subsection, we discuss how the periods are computed from complex integrals and how a large family of prepotentials can be computed using mirror symmetry.

\subsubsection{Period integrals from computational mirror symmetry}\label{sec:CMS}

To compute the prepotential, we consider a mirror pair
$X$, $\widetilde{X}$ of Calabi-Yau threefolds, and compactify type IIB string theory on $X$.  In the following, tilded quantities are those of $\widetilde{X}$.

In type IIB compactification on ${X}$, the vector multiplet moduli space parameterizes complex structures ${X}$, and is classically \emph{exact}: the metric receives no corrections beyond leading order in $g_s$ or $\alpha'$.
In the symplectic basis $\{\alpha_A, \beta^A\}$, $A = 0, \ldots, h^{2,1}$, for $H^3(X, \mathbb{Z})$ from above,
the periods of ${\Omega}$ are given by (cf.~Eq.~\eqref{eq:PvecFirstDef})
\begin{equation}\label{eq:iibperdef}
    \Vec{\Pi}_{\text{IIB}}\coloneqq \begin{pmatrix}
            \int_X\,\Omega\wedge \beta_A \\
            \phantom{\Bigl.}\int_X\,\Omega\wedge \alpha^A
        \end{pmatrix}\, .
\end{equation}
In type IIA compactification on the mirror $\widetilde{X}$, one likewise writes a prepotential for the vector multiplet moduli space in terms of a vector $\Vec{\Pi}_{\text{IIA}}$.

In this setting, the content of mirror symmetry is that
\begin{equation}\label{eq:mirrorsays}
\Vec{\Pi}_{\text{IIA}}=\Vec{\Pi}_{\text{IIB}}\equiv \Vec{\Pi}\, ,
\end{equation}
provided we choose an appropriate basis and normalization.

To compute the prepotential on the type IIB side, we need to compute the periods $\vec{\Pi}$ of the holomorphic $(3,0)$ form ${\Omega}$ in an integral symplectic basis.  This is difficult in general.  
However, in a particular limit, and for $X$ a hypersurface in a toric variety,
one can obtain the periods following a procedure laid out by Hosono, Klemm, Theisen, and Yau \cite{Hosono:1993qy}.

The limit in question is \emph{large complex structure} (LCS), which is mirror to large volume.  Expanding $\Vec{\Pi}_{\text{IIB}}$ at LCS is equivalent to expanding
$\Vec{\Pi}_{\text{IIA}}$ at large volume.  Around the LCS point, the periods undergo monodromy: specifically, the  monodromy is maximally unipotent \cite{MUM}, and corresponds to the (known) large-volume monodromies of type IIA string theory on $\widetilde{X}$.  Thus, if one can obtain the periods expanded around LCS in a general $\mathbb{C}$-basis, they can be expressed in an integral symplectic basis by matching the monodromies required by mirror symmetry.

\paragraph{Fundamental period}

The trick used in \cite{Hosono:1993qy} to compute the LCS periods in a $\mathbb{C}$-basis is as follows: when $X$ is a Calabi-Yau hypersurface, 
one can compute a \textit{fundamental period} directly, by the residue theorem, and from this obtain all the periods.  Let us briefly outline this computation.

We suppose that 
$(\widetilde{X},X)$ is
a mirror pair of
Calabi-Yau hypersurfaces in toric varieties $(\widetilde{V},V)$
defined by fine, regular, star triangulations
$(\widetilde{\mathcal{T}},\mathcal{T})$ of a dual pair of reflexive polytopes $(\Delta,\Delta^\circ)$.
Specifically, we take 
${X}$ to be the vanishing locus of a generic anticanonical polynomial $f$ in ${V}$.

In ${V}$, we consider toric coordinates $(\mathbbm{t}^1,\ldots,\mathbbm{t}^4)$ that parameterize a dense algebraic torus $(\mathbb{C}^*)^{4}$, and we write for the generic anticanonical polynomial $f$
\begin{equation}\label{eq:DefFX}
    f(\mathbbm{t})=\Psi^0 S_0(\mathbbm{t})-\sum_{I=1}^m \Psi^{I}S_{I}(\mathbbm{t})\, ,
\end{equation}
specified in terms of $m+1$ complex parameters $\Psi^{I}$.
$S_{0}(\mathbbm{t})$ and $S_{I}(\mathbbm{t})$ are suitable monomials of the toric coordinates $(\mathbbm{t}^1,\ldots,\mathbbm{t}^4)$.
For later use we introduce the gauged linear sigma model (GLSM) matrix with entries $Q^a_{~I}$, $a=1,\ldots h^{1,1}$, $I=1,\ldots h^{1,1}+4$, as well as the charge of the
anti-canonical divisor ${Q^a}_0\coloneqq\sum_I {Q^a}_I$. The $Q^a_{~I}$ record the coefficients of the prime toric divisor homology classes $[\widehat{D}_I]$ in terms of those of a basis $\{\widehat{H}_a\}$ of $H_4(V,\mathbb{Z})$, via 
$[\widehat{D}_I] = \sum_a Q^a_{~I}\widehat{H}_a$.

We  write the holomorphic $3$-form as
\begin{equation}
    {\Omega}=\oint_{f=0}\frac{\dif\mathbbm{t}^1\wedge \dif\mathbbm{t}^2\wedge \dif\mathbbm{t}^3\wedge \dif\mathbbm{t}^4}{(2\pi\mathrm{i})^4\cdot f(\mathbbm{t})}\, .
\end{equation}
We now carry out this contour integral.
In terms of a constant $\epsilon\ll 1$, we write $\mathbbm{t}^i=\epsilon\cdot \mathrm{e}^{2\pi\mathrm{i}\,  \phi^i}$ for $i=1,2,3$, with phases $\phi^i\in [0,1)$.
By a choice of $f$ we arrange that $f=0$ admits a solution with $\mathbbm{t}^4(\mathbbm{t}^1,\mathbbm{t}^2,\mathbbm{t}^3)=\mathcal{O}(\epsilon)$, which defines a $T^3\subset {X}$, which is the SYZ fiber \cite{Strominger:1996it}.

The \emph{fundamental period} $\varpi_0(\psi)$ is then defined by
\begin{equation}\label{eq:fundef}
\varpi_0(\psi)\coloneqq \Psi^0\int_{T^3}\widetilde{\Omega}=\Psi^0\oint_{|\mathbbm{t}^1|=\epsilon}\frac{\dif\mathbbm{t}^1}{2\pi\mathrm{i}\, }\cdots \oint_{|\mathbbm{t}^4|=\epsilon} \frac{\dif\mathbbm{t}^4}{2\pi\mathrm{i}\, }\frac{1}{f(\mathbbm{t})}\, .
\end{equation}

We next apply the residue theorem, arriving at
\begin{equation}
\varpi_0(\psi)=\oint\frac{\dif\mathbbm{t}^1}{2\pi\mathrm{i}\,  \mathbbm{t}^1}\cdots \oint \frac{\dif\mathbbm{t}^4}{2\pi\mathrm{i}\,  \mathbbm{t}^4}\frac{1}{1-\sum_i \tfrac{\Psi^I}{\Psi^0} S_I} = \sum_{k=0}^\infty \left.\left(\sum_I \tfrac{\Psi^I}{\Psi^0} S_I\right)^k\right|_{\mathbbm{t}\text{-independent}}\, ,
\end{equation}
where only the constant terms are retained in the sum on the right.  Using the multinomial theorem
we find\footnote{One also needs a fact from the toric setting: $\prod_I S_I^{k_I}=1$ iff $\sum_I k_Iq^I=0$, with $q^I$ the points of $\Delta$.}
\begin{align}\label{eq:fundper}
\varpi_0(\psi)= \sum_{k_1=0}^\infty\cdots \sum_{k_{h^{2,1}+4}=0}^\infty \frac{(\sum_I k_I)!}{\prod_I k_I!}\begin{cases}
\prod_{I=1}^{h^{2,1}+4}\left(\tfrac{\Psi^I}{\Psi^0}\right)^{k_I} & \text{if}\quad k_I q^I=0\\
    0 & \text{otherwise}
\end{cases}\, .\end{align}
Recasting \eqref{eq:fundper} as a sum over curves in the Mori cone $\mathcal{M}_{\widetilde{V}}$ of the mirror toric variety $\widetilde{V}$ (see \cite{Demirtas:2023als} for an explanation of this step)
we arrive at our final expression for the fundamental period:
\begin{equation}\label{eq:fundamental_period_final}
\varpi_0(\psi)=\sum_{\tilde{\mathbf{q}}\in \mathcal{M}_{\widetilde{V}}\cap H^2(\widetilde{V},\mathbb{Z})}\frac{\Gamma(1+\tilde{q}_a {Q^a}_0)}{\prod_I \Gamma(1+\tilde{q}_a {Q^a}_I)}\psi^{\tilde{\mathbf{q}}}=: \sum_{\tilde{\mathbf{q}}}c_{\tilde{\mathbf{q}}}\psi^{\tilde{\mathbf{q}}}\,,
\end{equation}
in terms of 
the anti-canonical divisor
${Q^a}_0\coloneqq \sum_I {Q^a}_I$ and
\begin{equation}
\psi^{\tilde{\mathbf{q}}} \coloneqq  \prod_{a=1}^{h^{1,1}(\widetilde{V})} (\psi^a)^{\tilde{q}_a}\,.
\end{equation}
In expansion around LCS, all the remaining periods are determined by the fundamental period,
\begin{equation}\label{eq:fundamental_period_final2}
\varpi_0(\psi)=\sum_{\tilde{\mathbf{q}}}c_{\tilde{\mathbf{q}}}\psi^{\tilde{\mathbf{q}}}\,,
\end{equation}
as follows \cite{Hosono:1993qy}:
\begin{align} \label{eq:pfrelations}
\varpi^a(\psi)=\sum_{\tilde{\mathbf{q}}\in \mathcal{M}_{\widetilde{V}}}\left. \frac{\partial_{\rho_a}}{2\pi\mathrm{i}\, }\left(c_{\tilde{\mathbf{q}}+\Vec{\rho}}\,\psi^{\tilde{\mathbf{q}}+\Vec{\rho}}\right)\right|_{\Vec{\rho}=0}&\, ,\quad \varpi^{ab}(\psi)=\sum_{\tilde{\mathbf{q}}\in \mathcal{M}_{\widetilde{V}}}\left.\frac{\partial_{\rho_a}\partial_{\rho_b}}{(2\pi\mathrm{i})^2}\left(c_{\tilde{\mathbf{q}}+\Vec{\rho}}\,\psi^{\tilde{\mathbf{q}}+\Vec{\rho}}\right)\right|_{\Vec{\rho}=0}\,,\nonumber\\
\varpi^{abc}(\psi)=\sum_{\tilde{\mathbf{q}}\in \mathcal{M}_{\widetilde{V}}}&\left. \frac{\partial_{\rho_a}\partial_{\rho_b}\partial_{\rho_c}}{(2\pi\mathrm{i})^3}\left(c_{\tilde{\mathbf{q}}+\Vec{\rho}}\,\psi^{\tilde{\mathbf{q}}+\Vec{\rho}}\right)\right|_{\Vec{\rho}=0}\, .
\end{align}
At zeroth order in $\psi$ we have
\begin{align}
&\varpi_0\simeq 1\, ,\quad  \varpi^a\simeq \frac{\log(\psi^a)}{2\pi\mathrm{i}}\, ,\quad
\frac{1}{2}\widetilde\kappa_{abc}\varpi^{ab}\simeq \frac{1}{2}\widetilde\kappa_{abc}\varpi^a \varpi^b-\frac{1}{24}\tilde{c}_a\nonumber\\
&\frac{1}{3!}\widetilde\kappa_{abc}\varpi^{abc}\simeq \frac{1}{3!}\widetilde\kappa_{abc}\varpi^a \varpi^b \varpi^c-\frac{1}{24}\tilde{c}_a \varpi^a+\frac{\zeta(3)}{(2\pi\mathrm{i})^3}\chi(\widetilde{X})\, ,
\end{align}
where $\tilde{c}_a$ and $\chi(\widetilde{X})$ are defined as
\begin{equation}
\tilde{c}_a=\int_{\widetilde{X}}\, c_2(\widetilde{X})\wedge \tilde{\beta}_a\, ,\quad
 \chi(\widetilde{X})=\int_{\widetilde{X}}\,  c_3(\widetilde{X})\, ,
\end{equation}
for a basis $\{\tilde{\beta}_a\}_{a=1}^{h^{1,1}(\widetilde{X})}$ of $H^2(\widetilde{X},\mathbb{Z})$,
and with $c_n(\widetilde{X})$ denoting the $n$-th Chern class of $\widetilde{X}$.

Mirror symmetry implies
that the LCS
monodromies in type IIB on $X$ are equal to the large volume monodromies of type IIA on $\widetilde{X}$, and the latter are determined entirely by the intersection form of $\widetilde{X}$.
Using the large volume monodromies to fix all integration constants, and adopting a suitable normalization, we at last arrive at the periods in an integral symplectic basis: 
\begin{equation}\label{eq:period2}
\Pi(\psi)=\frac{1}{\varpi_0}\begin{pmatrix}
\frac{1}{3!}\widetilde\kappa_{abc}\varpi^{abc}+\frac{1}{12}\tilde{c}_a \varpi^a\\
-\frac{1}{2}\widetilde\kappa_{abc}\varpi^{bc}+\tilde{a}_{ab}\varpi^b\\
\varpi_0\\
\varpi^a
\end{pmatrix}\, ,
\end{equation}
where $\tilde{a}_{ab}$ is defined as
\begin{equation}
\tilde{a}_{ab}\equiv \frac{1}{2}\begin{cases}
\widetilde{\kappa}_{aab} & a\geq b\\
\widetilde{\kappa}_{abb} & a<b
\end{cases} \, .
\end{equation}

\paragraph{Gopakumar-Vafa invariants from mirror symmetry} \label{subsec:enumerative_invariants} We now write flat coordinates $z^a$ that are related to the $\psi^a$ via
\begin{equation}
    z^a=\frac{\varpi^a}{\varpi_0}=\frac{\log(\psi^a)}{2\pi\mathrm{i}\, }+\frac{1}{2\pi\mathrm{i}\,  }  \frac{c^a(\psi)}{\varpi_0}\, ,
\end{equation}
where the $c^a(\psi)$ are given by 
\begin{equation}\label{eq:cadef}
    c^a(\psi)\coloneqq \sum_{\tilde{\mathbf{q}}\in \mathcal{M}_V}  c^a_{\tilde{\mathbf{q}}}\,\psi^{\tilde{\mathbf{q}}}\, ,\quad \text{with} \quad c^a_{\tilde{\mathbf{q}}}\coloneqq \partial_{\rho_a}c_{\tilde{\mathbf{q}}+\Vec{\rho}}\Bigl|_{\vec{\rho}=0}\, .
\end{equation}
A practical definition of the LCS limit can be given in terms of the flat coordinates as
\begin{equation}\label{eq:LCS}
    \mathrm{e}^{2\pi\mathrm{i}z^a} \ll 1\,.
\end{equation}
Next, we introduce
\begin{equation}\label{eq:cabdef}
    c^{ab}(\psi)\coloneqq \sum_{\tilde{\mathbf{q}}\in \mathcal{M}_V} c^{ab}_{\tilde{\mathbf{q}}} \,\psi^{\tilde{\mathbf{q}}}\, ,\quad \text{with} \quad c^{ab}_{\tilde{\mathbf{q}}}\coloneqq \partial_{\rho_a}\partial_{\rho_b}c_{\tilde{\mathbf{q}}+\Vec{\rho}}\Bigr|_{\vec{\rho}=0}\, .
\end{equation}
The periods can then be written
\begin{equation}\label{eq:periodf}
    \mathcal{F}_a=-\frac{1}{2}\widetilde\kappa_{abc}\frac{\varpi^{bc}}{\varpi_0}+\tilde{a}_{ab}\frac{\varpi^b}{\varpi_0}=-\frac{1}{2}\widetilde\kappa_{abc}z^bz^c+\tilde{a}_{ab}z^b+\frac{\tilde{c}_a}{24}-\frac{1}{2(2\pi\mathrm{i})^2}\widetilde\kappa_{abc}\frac{\hat{c}^{bc}-c^bc^c}{\varpi_0}\, ,
\end{equation}
in terms of
\begin{align}\label{eq:chat}
    \hat{c}^{ab}(\psi)&\coloneqq \sum_{\tilde{\mathbf{q}}\in \mathcal{M}_{\widetilde{V}}}\hat{c}^{ab}_{\tilde{\mathbf{q}}}\psi^{\tilde{\mathbf{q}}}\coloneqq \sum_{\tilde{\mathbf{q}}\in \mathcal{M}_{\widetilde{V}}}\left(c^{ab}_{\tilde{\mathbf{q}}}-\frac{\pi^2}{6}\Bigl[{Q^a}_0{Q^b}_0-\sum_I {Q^a}_I{Q^b}_I\Bigr]c_{\tilde{\mathbf{q}}}\right)\psi^{\tilde{\mathbf{q}}}\nonumber\\
    &\equiv c^{ab}(\psi)-\frac{\pi^2}{6}\Bigl[{Q^a}_0{Q^b}_0-\sum_I {Q^a}_I{Q^b}_I\Bigr]\varpi_0\, .
\end{align}
The key point is that one can also write the non-perturbative terms in the type IIA prepotential in terms of genus-zero Gopakumar-Vafa invariants $\mathscr{N}_{\tilde{\mathbf{q}}}$:
\begin{equation}\label{eq:IIAprep}
	\mathcal{F}_{\mathrm{inst}}(z)=-\frac{1}{(2\pi \mathrm{i})^3}\sum_{\tilde{\mathbf{q}}\in \mathcal{M}(\widetilde{X})}\, \mathscr{N}_{\tilde{\mathbf{q}}}\,\mathrm{Li}_3\Bigl(\mathrm{e}^{2\pi \mathrm{i}\,\tilde{\mathbf{q}}\cdot \mathbf{z}}\Bigr)\,,
\end{equation}
where the sum runs over curves $\tilde{\mathbf{q}}$ in the Mori cone $ \mathcal{M}_{\widetilde{X}}$ of $\widetilde{X}$.

Comparing \eqref{eq:IIAprep} and \eqref{eq:periodf}, we learn that
\begin{equation}
\label{eq:perturbative_GV_formula}
    \sum_{\tilde{\mathbf{q}}\in \mathcal{M}_{\widetilde{X}}} \, \tilde{q}_a \, \mathscr{N}_{\tilde{\mathbf{q}}}\,\text{Li}_2\biggl(\psi^{\tilde{\mathbf{q}}}\,\mathrm{exp} \Bigl({ \tilde{q}_b \tfrac{c^b(\psi)}{\varpi_0(\psi)}}\Bigr)\biggr)=\phantom{\Biggl(}\frac{1}{2}\widetilde\kappa_{abc}\frac{\hat{c}^{bc}(\psi)-c^b(\psi)c^c(\psi)}{\varpi_0(\psi)}\,.
\end{equation}
By expanding both sides of \eqref{eq:perturbative_GV_formula} in powers of the $\psi^a$, one can read off the GV invariants.  To accomplish this in practice in models with many moduli, specialized algorithms are necessary, and were implemented in \cite{Demirtas:2023als}.

\paragraph{Summary of LCS formulas}

To summarize, the prepotential $\mathcal{F}$ in the large complex structure (LCS) patch can be written as
\begin{equation}\label{eq:prepotential}
\mathcal{F}(z)=\mathcal{F}_{\mathrm{poly}}(z)+\mathcal{F}_{\mathrm{inst}}(z)\,,
\end{equation}
with
\begin{align}
    \label{eq:fpoly} \mathcal{F}_{\mathrm{poly}}(z)&=-\frac{1}{3!}\widetilde{\kappa}_{abc}z^az^bz^c+\frac{1}{2}\tilde{a}_{ab}z^az^b+\frac{1}{24}\tilde{c}_a z^a+\frac{\zeta(3)\chi(\widetilde{X})}{2(2\pi \mathrm{i})^3}\, ,\\[0.5em]
\label{eq:finst} \mathcal{F}_{\mathrm{inst}}(z)&=-\frac{1}{(2\pi \mathrm{i})^3}\sum_{\tilde{\mathbf{q}}\in \mathcal{M}(\widetilde{X})}\mathscr{N}_{\tilde{\mathbf{q}}}\,\mathrm{Li}_3\Bigl(\mathrm{e}^{2\pi \mathrm{i}\,\tilde{\mathbf{q}}\cdot \mathbf{z}}\Bigr)\, ,
\end{align}
in terms of the $k$-th polylogarithm $\mathrm{Li}_k(x)$ defined as
\begin{equation}
    \mathrm{Li}_k(x)\coloneqq\sum_{n=1}^\infty \, \dfrac{x^n}{n^k}\, .
\end{equation}
The polynomial terms in \eqref{eq:fpoly} are computed from the triple intersection numbers $\widetilde{\kappa}_{abc}$ of the mirror threefold $\widetilde{X}$, and
\begin{equation}
\tilde{c}_a=\int_{\widetilde{X}}\, c_2(\widetilde{X})\wedge \tilde{\beta}_a\, ,\quad
\tilde{a}_{ab}\equiv \frac{1}{2}\begin{cases}
\widetilde{\kappa}_{aab} & a\geq b\\
\widetilde{\kappa}_{abb} & a<b
\end{cases}\, , \quad \mathrm{and} \quad \chi(\widetilde{X})=\int_{\widetilde{X}} c_3(\widetilde{X})\, ,
\end{equation}
for a basis $\{\tilde{\beta}_a\}_{a=1}^{h^{2,1}(X)}$ of $H^2(\widetilde{X},\mathbb{Z})$. Further, $c_n(\widetilde{X})$ corresponds to the $n$-th Chern class of $\widetilde{X}$. 
The contribution $\mathcal{F}_{\mathrm{inst}}(z)$ accounts for type IIA worldsheet instanton corrections, which are given in terms of effective curve classes $\tilde{\mathbf{q}}$ and the associated genus-zero Gopakumar-Vafa invariants $\mathscr{N}_{\tilde{\mathbf{q}}}\in \mathbb{Z}$ of $\widetilde{X}$.

\subsection{Non-perturbative superpotential}\label{sec:Wnp}
\index{Non-perturbative superpotential}

Coming back to \eqref{eq:Wexact}, we want to compute the non-perturbative superpotential $W_{\text{np}}=W_{\text{np}}(z^a,\tau,T_i)$,  
which can be written in the form
\begin{equation}\label{eq:Wnp}
     W_{\text{np}} = W_{\text{ED}3}+W_{\text{ED}(-1)}+W_{\lambda\lambda}\, .
\end{equation}
The first and second terms account for contributions from Euclidean D3-branes wrapping holomorphic four-cycles \cite{Witten:1996bn}, and from Euclidean D$(-1)$-branes, respectively.  The third term is induced by gaugino condensation on seven-branes\footnote{Gaugino condensation can also occur on a stack of coincident spacetime-filling D3-branes, but this possibility will not be relevant in these lectures.} wrapping holomorphic four-cycles, an effect that we explain in \S\ref{sec:WnpGaugino}. 

The contributions from Euclidean D(-1)-branes can be written schematically as\footnote{For related work, see e.g.~\cite{Grimm:2007xm,Sen:2020cef,Sen:2021tpp,Alexandrov:2022mmy,Agmon:2022vdj}.}  
\begin{equation}
    W_{\text{ED}(-1)} \supset \mathcal{A}(\phi^i) \, \mathrm{e}^{\pi\mathrm{i}\, n\tau}
\end{equation}
for some $n\in\mathbb{N}$.
Here, $\phi^i$ collectively denote other moduli fields (e.g. complex structure moduli, or open-string moduli). In all explicit configurations studied later on, pure Euclidean D(-1) corrections are actually absent because all the D7-brane gauge algebras are $\mathfrak{so}(8)$ \cite{Kim:2022jvv}. Hence, we will ignore $W_{\text{ED}(-1)}$ in the subsequent discussion.

\subsubsection{Euclidean D3-branes}
\index{Euclidean D3-branes}

A Euclidean D3-brane wrapping a four-cycle $D \in H_4(X, \mathbb{Z})$ in the Calabi–Yau threefold $X$ appears as an instanton --- i.e.,  
a saddle point of the Euclidean path integral --- from the four-dimensional spacetime perspective. Such instantons can, under suitable conditions that we now summarize, generate non-perturbative corrections to the $\mathcal{N}=1$ superpotential $W_{\mathrm{np}}$. 

\begin{table}[t!]
\centering
\bgroup
\def\arraystretch{1.5}
\begin{tabular}{c|c|c|c}
zero modes & statistics & cohomology IIB & cohomology F-theory \\ 
\hline 
\hline 
$(X_{\mu},\theta_{\alpha})$ & (Bose, Fermi) & $H_{+}^{0,0}(D)$& $H^{0,0}(\widehat{D})$ \\ 
\hline 
$\bar{\tau}_{\dot{\alpha}}$ & Fermi & $H_{-}^{0,0}(D)$ & \multirow{2}{*}{$H^{0,1}(\widehat{D})$} \\ 
$\gamma_{\alpha}$ & Fermi & $H^{0,1}_{+}(D)$&  \\ 
\hline 
$(\omega,\bar{\gamma}_{\dot{\alpha}})$ & (Bose, Fermi) & $H_{-}^{0,1}(D)$& \multirow{2}{*}{$H^{0,2}(\widehat{D})$} \\ 
$\chi_{\alpha}$ & Fermi & $H_{+}^{0,2}(D)$&  \\ 
\hline 
$(c,\bar{\chi}_{\dot{\alpha}})$ & (Bose, Fermi) & $H_{-}^{0,2}(D)$& $H^{0,3}(\widehat{D})$ \\ 
\end{tabular} 
\egroup
\caption{Zero mode structure for D3/M5-instantons.}\label{tab:FtheoryLiftCohomDiv} 
\end{table}

Superpotential terms arise from  path integral
contributions of the schematic form
\begin{equation}\label{eq:PathIntWNP}
\int\, \mathcal{D} X\, \mathcal{D} \vartheta\, \mathrm{e}^{-S_{\text{D3}}[X,\vartheta]}\sim \int\dif^{4}x\dif^{2}\Theta\, W_{\text{np}}
\end{equation}
where $X$ and $\vartheta$ denote the bosonic and fermionic worldvolume fields, respectively.
Just as in the case of Yang–Mills instantons,
a non-vanishing contribution occurs only if there are zero modes that saturate the Grassmann integral $\mathrm{d}^2 \Theta$,
and so the structure of the non-perturbative superpotential is controlled by the spectrum of fermionic zero modes associated with the brane worldvolume theory.

The zero modes appearing in \eqref{eq:PathIntWNP} are summarized in Table~\ref{tab:FtheoryLiftCohomDiv}. Among these, the modes $X^\mu$, $\theta_\alpha$, and $\bar{\tau}_{\dot{\alpha}}$ are always present and do not depend on the geometry of the wrapped divisor $D$; they are referred to as \emph{universal zero modes}.  The universal moduli  correspond to Goldstone bosons and Goldstinos associated with the spontaneous breaking of symmetries by the branes.
The remaining zero modes arise from Wilson line and geometric deformation moduli of the brane counted by the Hodge numbers $h^{0,p}(D)$.

In the M-theory lift, Euclidean D3-branes correspond to Euclidean M5-branes wrapping six-dimensional divisors $\widehat{D} \subset Y_4$ of an elliptically fibered Calabi–Yau fourfold $Y_4$ \cite{Becker:1995kb,Robbins:2004hx}; see also \cite{Blumenhagen:2010ja,Bianchi:2011qh,Baumann:2014nda} for detailed treatments in the F-theory context.
To discuss necessary and sufficient conditions for
for Euclidean M5-branes wrapping a divisor $\widehat{D}$
to contribute to the superpotential, we introduce the
arithmetic genus of $\widehat{D}$,
\begin{equation}\label{eq:6cycleCondEuler} 
    \chi(\widehat{D},\mathcal{O}_{\widehat{D}}) \coloneqq \sum_{i=0}^{3}\, (-1)^{i}h^{0,i}(\widehat{D})\,,
\end{equation}
with $\mathcal{O}_{\widehat{D}}$ denoting the structure sheaf.
For a \emph{smooth} divisor $\widehat{D}$, and in the absence of flux, 
$\chi(\widehat{D},\mathcal{O}_{\widehat{D}})=1$ is a necessary condition for a superpotential contribution \cite{Witten:1996bn}.
Next, a sufficient but not necessary condition is that $\widehat{D}$ is \emph{rigid}, meaning that
\begin{equation}\label{eq:Rigid6Cycle} 
    h^{0,1}(\widehat{D})=h^{0,2}(\widehat{D})=h^{0,3}(\widehat{D})=0\, .
\end{equation}
The condition \eqref{eq:Rigid6Cycle}  guarantees the absence of additional fermionic zero modes, which would otherwise lead to a vanishing superpotential.

In the type IIB orientifold limit, Euclidean D3-branes wrapping divisors $D \subset X$ can be classified according to their behavior under the orientifold involution $\sigma$. If $D$ is mapped to a distinct image under $\sigma$, the instanton comes in a pair and carries a unitary Chan-Paton factor; such objects are called \emph{U(1)-instantons}.  
If instead the divisor is fixed under $\sigma$, the brane is its own image, and the orientifold projection acts non-trivially on its zero modes; such configurations are known as 
\emph{O(1)-instantons}. 
The zero modes for both cases are depicted in Table~\ref{tab:OUOneInstantonDivCohomZM}, where we define the orientifold-graded Hodge numbers $h_{\pm}^{p,q}(D)\coloneqq\mathrm{dim}(H_{\pm}^{p,q}(D))$.

\begin{table}[t!]
\centering
\resizebox{\columnwidth}{!}{
\bgroup
\def\arraystretch{1.3}
\begin{tabular}{c|c|c}
zero modes & statistics & number \\ 
\hline 
\hline 
$X_{\mu}$ & Bose & 1 \\ 
\hline 
$\theta_{\alpha}$ & Fermi & 1 \\ 
\hline 
$\bar{\tau}_{\dot{\alpha}}$ & Fermi & 1 \\ 
\hline 
$(\omega,\gamma_{\alpha},\bar{\gamma}_{\dot{\alpha}})$ & (Bose, Fermi) & $h^{0,1}(D)$ \\ 
\hline 
$(c,\chi_{\alpha},\bar{\chi}_{\dot{\alpha}})$ & (Bose, Fermi) & $h^{0,2}(D)$ \\ 
\end{tabular} 
\egroup
\hspace{0.5cm}
\bgroup
\def\arraystretch{1.3}
\begin{tabular}{c|c|c}
zero modes & statistics & number \\ 
\hline 
\hline 
$(X_{\mu},\theta_{\alpha})$ & (Bose, Fermi) & $1=h_{+}^{0,0}(D)$ \\ 
\hline 
$\bar{\tau}_{\dot{\alpha}}$ & Fermi & $0=h_{-}^{0,0}(D)$ \\ 
\hline 
$\gamma_{\alpha}$ & Fermi & $h^{0,1}_{+}(D)$ \\ 
\hline 
$(\omega,\bar{\gamma}_{\dot{\alpha}})$ & (Bose, Fermi) & $h_{-}^{0,1}(D)$ \\ 
\hline 
$\chi_{\alpha}$ & Fermi & $h_{+}^{0,2}(D)$ \\ 
\hline 
$(c,\bar{\chi}_{\dot{\alpha}})$ & (Bose, Fermi) & $h_{-}^{0,2}(D)$ \\ 
\end{tabular} 
\egroup
}
\caption{\emph{Left:} Zero modes for $\mathrm{U}(1)$-instantons. \emph{Right:} Zero modes for $\mathrm{O}(1)$-instantons.}\label{tab:OUOneInstantonDivCohomZM} 
\end{table}

As it stands, only $\mathrm{O}(1)$-instantons can contribute to the
superpotential.  
The reason is that the orientifold projection removes certain zero modes that would otherwise violate the necessary selection rules. In particular, the unwanted fermionic mode $\bar{\tau}_{\dot{\alpha}}$, associated with $H^{0,0}_-(D)$, is projected out in the $\mathrm{O}(1)$ case but generically present for $\mathrm{U}(1)$-instantons. Matching the uplift to F-theory via Table~\ref{tab:FtheoryLiftCohomDiv}, the absence of these modes requires
\begin{equation}
h_{-}^{0,0}=h_{+}^{0,1}=h_{-}^{0,1}=h_{+}^{0,2}=h_{-}^{0,2}=0\, .
\end{equation}
Under these conditions, and assuming rigidity of the divisor $D$, a superpotential correction from an $\mathrm{O}(1)$-instanton is allowed, in accord with Witten's criterion \cite{Witten:1996bn}.

The corresponding conditions for the divisor $D\subset X$ to be rigid can be written more cleanly as
\begin{equation}\label{eq:drigid}
    h^{\bullet}_+(D,\mathcal{O}_D)=(1,0,0)\,, \qquad h^{\bullet}_-(D,\mathcal{O}_D)=0\,.
\end{equation}
We note that, while Witten's original argument assumed smoothness of $D$, this may not always be necessary. It was recently demonstrated in \cite{Gendler:2022qof} that, under certain conditions, singular divisors can contribute to the superpotential as well. The roles of non-trivial worldvolume fluxes, $2\pi\alpha'\mathcal{F}=2\pi\alpha' F-\iota^{*}B_{2}\neq 0$, and of background bulk fluxes, on the zero modes of a Euclidean D3-brane, have been examined in \cite{Bianchi:2011qh,Bianchi:2012pn,Bianchi:2012kt}.  In the examples of Chapter \ref{chap:deSitter}, the rigid divisors' topology does not allow magnetization by worldvolume flux.

To summarize, a Euclidean D3-brane wrapping $D\in H_4(X,\mathbb{Z})$ contributes the non-perturbative superpotential term
\begin{equation}
    W_{\text{ED}3} \supset \mathcal{A}_{D}(\phi^i)\,\mathrm{e}^{-2\pi T_D}\, ,
\end{equation}
with $T_D$ denoting the Einstein-frame volume of the divisor. Here, $\phi^i$ denotes other moduli fields on which the Pfaffian prefactors $\mathcal{A}_D$ depend, namely the complex structure moduli, the axio-dilaton, and the positions of D3-branes and D7-branes. The Pfaffian $\mathcal{A}_D$ is identically zero for divisors $D$ that support more than two fermion zero modes.

\subsubsection{Gaugino condensation}\label{sec:WnpGaugino}
\index{Gaugino condensation}

In addition to Euclidean D3-branes, the non-perturbative superpotential \eqref{eq:Wnp} receives corrections from strongly coupled gauge dynamics in the worldvolume theory of seven-branes wrapping four-cycles $D\subset X$ in the Calabi-Yau threefold $X$. Let us describe this process in more detail.

Consider a stack of $N>1$ D7-branes on a divisor $D\subset X$ in the Calabi-Yau threefold $X$. The resulting four-dimensional gauge theory is $\mathcal{N}=1$ super-Yang-Mills (SYM) with some gauge group $G$, and some matter content. Deformations of the divisor $D$, in particular those counted by $h^{0,2}(D)=h^2(\mathcal{O}_D)$, give rise to adjoint matter fields. If $D$ is rigid, i.e., if $h^{0,2}(D)=h^{0,1}(D)=0$ (recall Eq.~\eqref{eq:drigid}), there are no such deformations and the resulting worldvolume theory on the stack of D7-branes is pure glue SYM.
 
At low energies, this theory generates a gaugino condensate that gives rise to a superpotential contribution of the form
\begin{equation}
	W_{\lambda\lambda} \supset M_{\text{UV}}^{3}\, \exp\left (-\dfrac{8\pi^{2}}{g_{\text{YM}}^{2}}\, \dfrac{1}{c_D}\right )\times\text{ phase}\,,
\end{equation}
where $c_D$ is the dual Coxeter number of $G$. The gauge coupling can be written in terms of geometric quantities as
\begin{equation}
	\dfrac{8\pi^{2}}{g_{\text{YM}}^{2}} = \dfrac{1}{(2\pi)^{3}g_{^^s}}\, \dfrac{1}{(\alpha')^{2}}\, \int_{D}\dif^{4}\xi\sqrt{g}\,\mathrm{e}^{-4A} = T_{3}V_{D}^{w}\,,
\end{equation}
with $V_{D}^{w}$ the warped volume of the divisor $D$. Expressed in terms of the K\"ahler coordinates, we have
\begin{equation}
	\dfrac{8\pi^{2}}{g_{\text{YM}}^{2}} = 2\pi\,\mathrm{Re}(T_D)\,,
\end{equation}
although we note that warping corrections to the K\"ahler coordinates $T_i$ have not been written explicitly elsewhere in our analysis.

We conclude that 
\begin{equation}
	W_{\lambda\lambda} \supset \mathcal{A}_D(z^{a},\tau)\,\exp\left (-\tfrac{2\pi}{c_D}\, T_D\right )
\end{equation}
where the Pfaffian $\mathcal{A}_D(z^{a},\tau)$ encodes moduli-dependent threshold corrections to the classical gauge coupling.

\subsubsection{Pfaffian prefactors and pure rigid divisors}\label{sec:pure_rigid}

To summarize, the non-perturbative superpotential in \eqref{eq:Wexact}, arising from Euclidean D3-branes and gaugino condensation on stacks of D7-branes, takes the general form
\begin{equation}\label{eq:Wnpexact}
    W_{\text{np}}(z^a,\tau,T_i) = \sum_{D}\, \mathcal{A}_D\,\exp\left (-\tfrac{2\pi}{c_D}\, T_D\right )\, ,
\end{equation}
where $T_D$ denotes the K\"ahler modulus associated with the divisor $D$, and $c_D$ is the dual Coxeter number of the gauge group supported on the brane wrapping $D$. For divisors supporting Euclidean D3-branes, one has $c_D = 1$, while for gaugino condensation on stacks of D7-branes with gauge algebra $\mathfrak{so}(8)$, one has 
$c_D = 6$.

The final ingredient required to specify the non-perturbative superpotential \eqref{eq:Wnpexact} is the set of Pfaffian prefactors $\mathcal{A}_D$. These arise from one-loop determinants of worldvolume fluctuations around the Euclidean D3-brane instanton background. In general, the Pfaffians can depend non-trivially on complex structure moduli and open-string fields, making their explicit form difficult to compute. To simplify the analysis and avoid complications associated with such moduli dependence, it is useful to impose a stronger condition\footnote{In the examples discussed in Chapter~\ref{chap:deSitter}, such conditions are imposed by rejection sampling, i.e.,~by selecting Calabi-Yau orientifolds in which the desired conditions are demonstrably fulfilled.} than just rigidity of $D$.

Type IIB flux compactifications of the type considered here are dual, via F-theory, to compactifications on elliptically-fibered Calabi–Yau fourfolds. Let $\widehat{X}$ denote the F-theory fourfold that is dual to type IIB on $X / \mathcal{I}$, where $\mathcal{I}$ is the orientifold involution. A Euclidean D3-brane wrapping a divisor $D \subset X$ uplifts in F-theory to a Euclidean M5-brane wrapping a vertical divisor $\widehat{D} \subset \widehat{X}$. The Pfaffian prefactor $\mathcal{A}_D$ can then be interpreted as a section of a line bundle over the complex structure moduli space of $\widehat{D}$, determined by its intermediate Jacobian \cite{Witten:1996hc}.

Following \cite{Demirtas:2021nlu}, we define a rigid divisor $D$ to be \emph{pure rigid} if its F-theory uplift $\widehat{D}$ satisfies
\begin{equation}\label{eq:dpurerigid}
    h^{2,1}\big(\widehat{D}\big)=0\, .
\end{equation}
For pure rigid divisors, the intermediate Jacobian is trivial, and the Pfaffian $\mathcal{A}_D$ reduces to a number, independent of moduli. This statement holds up to potential corrections from warping sourced by D3-brane charge, which modify the effective action of the instanton. These effects may be captured as part of $\delta T_i^{\mathcal{N}=1,\text{tree}}$ in Eq.~\eqref{eq:kformalcorrforKahler3T} --- cf.~\S\ref{sec:10deffects} ---
and have been discussed in \cite{Ganor:1996pe,Berg:2004ek,Giddings:2005ff,Baumann:2006th,Shiu:2008ry,Frey:2013bha,Martucci:2016pzt}.

Although the precise values of these prefactors remain unknown, recent progress has shed light on their structure \cite{Kim:2022jvv,Alexandrov:2022mmy,Kim:2022uni,Kim:2023cbh}. In line with this literature, we adopt the normalization
\begin{equation}\label{eq:K0}
    \mathcal{A}_D = \sqrt{\frac{2}{\pi}}\frac{n_D}{4\pi^2}\,,
\end{equation}
where $n_D$ is a numerical constant capturing the one-loop determinant over worldvolume fields. While a complete computation of $n_D$ has not yet appeared, the analysis of \cite{Kim:2023cbh} suggests that $n_D$ is generically an order-one quantity.

Moreover, as first argued in \cite{McAllister:2024lnt} and later corroborated in \cite{Moritz:2025bsi}, there are  
indications that $n_D \in \mathbb{Z}$. The reasoning is based on mirror symmetry and duality to M-theory: O3/O7 orientifolds in type IIB are mirror to O6 orientifolds in type IIA, which in turn lift to compactifications of M-theory on $G_2$-holonomy manifolds. In this dual frame, both the flux-induced and instanton-induced contributions to the superpotential arise from Euclidean M2-branes wrapping three-cycles. Since the exponential flux corrections $W_{\mathrm{flux}}$ appear with integer coefficients in our normalization, it is natural to expect that $n_D$ is also integer-valued. This provides further support for the normalization in Eq.~\eqref{eq:K0}.

\subsubsection{Freed-Witten anomaly cancellation}

In compactifications with D-branes, consistency of the low-energy effective theory imposes additional constraints beyond e.g. tadpole cancellation. One such requirement arises from the cancellation of the \emph{Freed-Witten anomaly}\index{Freed-Witten anomaly} \cite{Freed:1999vc}, which ensures that the worldvolume theory of a D-brane wrapping a divisor $D$ is well-defined from the perspective of the open string worldsheet theory.  
More specifically, consistency of the 
the path integral for worldsheet fermions  
imposes a quantization constraint that relates the NS-NS $B$-field to the topology of the cycle. In type IIB string theory, a D-brane wrapping a divisor $D$ must satisfy the Freed-Witten anomaly cancellation condition\index{Freed-Witten anomaly cancellation} \cite{Freed:1999vc}
\begin{equation}\label{eq:FreedWittenAnCan} 
	\tfrac{1}{2\pi} F_{2}+\tfrac{1}{2}c_{1}(D)\in H^{2}(D,\mathbb{Z})\, ,
\end{equation}
where $F_2$ is the field strength of the worldvolume gauge theory, and  $\iota^{*}:\, H^2(X)\rightarrow H^2(D)$ is the pull-back map on $D$.\footnote{For a Calabi-Yau threefold, one can show that $c_{1}(D)=-\iota^{*}D$.}
Crucially, if $D$ is non-spin\footnote{A non-spin divisor $D$ has non-trivial second Stiefel-Whitney class $w_{2}(D)\in H^{2}(D, \bZ_{2})$, and thus $c_{1}(D)$ cannot be even.} and hence $c_{1}(D)$ is odd, one needs to turn on gauge fluxes $F_{2}$ to cancel the anomaly. The Freed-Witten anomaly cancellation condition \eqref{eq:FreedWittenAnCan} thus constrains the allowed background fluxes and D-brane configurations.

The constraint discussed above plays a critical role when D-branes wrapping a divisor $D$ are expected to generate a contribution to the superpotential, such as from an O(1) Euclidean D3-instanton or from gaugino condensation on a stack of D7-branes. In these cases, the relevant gauge-invariant worldvolume field strength is given by
\begin{equation}\label{eq:GIfluxBrane} 
	2\pi\alpha'\mathcal{F}=2\pi\alpha' F_{2}-\iota^{*}B_{2}\, ,
\end{equation}
where $F_2$ is the worldvolume gauge field and $\iota^* B_2$ is the pullback of the background NS-NS 2-form, recall the discussion in \S\ref{sec:Dbranes}. A non-trivial gauge flux $\mathcal{F} \neq 0$ typically induces charged chiral matter in the brane worldvolume theory, which obstructs the generation of a superpotential term by lifting necessary fermionic zero modes. To avoid this, one typically tries to ensure that $\mathcal{F} = 0$, which can in principle be achieved by an appropriate choice of the background value of $B_2$, provided the Freed-Witten quantization condition \eqref{eq:FreedWittenAnCan} is also satisfied.

However, it is important to note that while the $2$-form field $F_2$ can be chosen independently on each D-brane worldvolume, $B_2$ is a globally-defined bulk field. Its pullback $\iota^* B_2$ contributes locally to the gauge-invariant flux via Eq.~\eqref{eq:GIfluxBrane}, but the background value of $B_2$ must be chosen globally and consistently across all divisors. This global constraint can impose significant limitations in generic compactifications, particularly when attempting to ensure $\mathcal{F} = 0$ simultaneously on multiple divisors, as discussed in \cite{Cicoli:2012vw,Cicoli:2017shd}.

In the explicit constructions presented in Chapter~\ref{chap:deSitter}, the Freed-Witten anomaly cancellation condition \eqref{eq:FreedWittenAnCan} can be satisfied 
by choosing
\begin{equation}\label{eq:fwb}
	B_{2} = \dfrac{1}{2}\sum_{A \in \{\text{O}7\}}\, [D_{A}]\, \in\, H^{2}(X,\mathbb{Z}/2)\, ,
\end{equation} 
where the sum runs over the divisors supporting O7-planes.
The  choice \eqref{eq:fwb}  ensures that the pullback of $B_2$ precisely cancels the half-integer Freed-Witten shift, avoiding worldvolume flux (i.e., keeping $\mathcal{F} = 0$), and enabling consistent superpotential contributions from the wrapped branes.

A noteworthy fact about the compactifications studied in 
Chapter~\ref{chap:deSitter} is that 
even though the choice \eqref{eq:fwb} only involves the O7-planes,  
imposing \eqref{eq:fwb}
\emph{also} ensures the 
cancellation of Freed-Witten anomalies on all the Euclidean D3-branes wrapping prime toric divisors.
This full cancellation can be verified explicitly in each example, but a systematic understanding of this phenomenon has not been presented.

\section{Summary of the leading-order EFT}\label{sec:leadingEFT}
\index{Leading-order EFT}

Let us now recapitulate the main results of this chapter, and collect the leading terms of the effective theory. In \S\ref{sec:scheme}, we described how the moduli fields $\Phi^A$, together with the superpotential $W(\Phi^A)$ and K\"ahler potential $\mathcal{K}(\Phi^A,\bar{\Phi}^A)$, can be expressed in terms of the geometric data $\mathscr{G}$ of a compactification. Within $\mathcal{N}=1$ supergravity, this information fully determines the two-derivative effective Lagrangian $\mathscr{L}(\Phi^A)$.

As noted in Eq.~\eqref{eq:Gsplit}, the geometric data $\mathscr{G}$ can be divided into discrete parameters, denoted by $\mathscr{D}_{\mathbb{Z}}$, and continuous parameters $\mathscr{C}$. The continuous parameters correspond e.g. to expectation values of the moduli fields $\Phi^A$.
More precisely, 
the $\Phi^A$ are good K\"ahler coordinates on the moduli space of the physical theory, while the data $\mathscr{G}$ might be given in a form that is adapted to differential geometry, without the additional $p$-form fields of string theory.  For example, the \emph{complexified} K\"ahler moduli $T_i$ are related to the K\"ahler form $J$ by \eqref{eq:Ttree}.

The fundamental data specifying a moduli space of theories is thus the discrete information $\mathscr{D}_{\mathbb{Z}}$. As a result, the effective theories of interest can be expressed in a schematic way by writing the four-dimensional Lagrangian as  
\begin{equation}\label{eq:lol}
    \mathscr{L}=\mathscr{L}(\Phi_A;\mathscr{D}_{\mathbb{Z}})\,.
\end{equation}
In sum, the functional form of $\mathscr{L}$ is determined by the discrete parameters $\mathscr{D}_{\mathbb{Z}}$, which include the topology of a Calabi-Yau threefold, a choice of an orientifold involution, and a choice of quantized $3$-form fluxes.  
A complete list of discrete parameters will be given below in Eq.~\eqref{eq:discrete_data}.

As discussed in \S\ref{sec:scheme}, the computation of $\mathscr{L}$ is necessarily carried out to some order in one or more expansion schemes. 
In \S\ref{sec:QCorPrel}
we laid out the particular regime of control used in \cite{Demirtas:2021nlu} and \cite{McAllister:2024lnt}:
see \eqref{eq:control_regime}.
Working in this scheme, we 
recorded in \S\ref{sec:quantum_pert} and
\S\ref{sec:thew}
the leading terms of the resulting approximate theory, which we term the \emph{leading-order effective theory}, and denote by
\begin{equation}
    \mathscr{L} \approx \mathscr{L}_{\text{l.o.}}\,.
\end{equation}
In the remainder of this section, we collect these terms.

\newpage
 
\subsection{Data of the leading-order supersymmetric EFT}\label{sec:leadingEFTSUSY}

We recall from \S\ref{sec:QCorPrel} 
that, at leading order in the string coupling $g_s$, the K\"ahler potential is known to \emph{all} orders in $\alpha'$, both perturbative and non-perturbative.  Using \eqref{eq:Kahlerpotential},
it can be written as
\begin{align}\label{eq:detailedform2}
\mathcal{K} \approx \mathcal{K}_{\text{l.o.}} \coloneqq  &\,\,\mathcal{K}_{\text{tree}} + \mathcal{K}_{(\alpha')^3} + \mathcal{K}_{\text{WSI}}\, \\
= & -2\log\Bigl(2^{3/2}g_s^{-3/2}\mathcal{V} \Bigr) -\log\bigl(-\I\left(\tau-\bar{\tau}\right)\bigr) - \log\Bigl(-\I\int_{X}\, \Omega(z^a)\wedge\overline{\Omega}({\bar z}^a )\Bigr)\,, \nonumber
\end{align}
in terms of the $\alpha'$-corrected, string-frame volume $\mathcal{V}$ at  string tree level given in \eqref{eq:Kahlerpotential},
\begin{align}
\mathcal{V} & =  \frac{1}{6}\kappa_{ijk}t^it^jt^k-\frac{\zeta(3)\chi(X)}{4(2\pi)^3}\nonumber\\[0.3em]
&+\frac{1}{2(2\pi)^3}\sum_{\mathbf{q}\in \mathcal{M}_X}\, \mathscr{N}_{\mathbf{q}}\,\Biggl( \text{Li}_3\Bigl((-1)^{\mathbf{\gamma}\cdot \mathbf{q}}\mathrm{e}^{-2\pi \mathbf{q}\cdot \mathbf{t}}\Bigr)+ 2\pi \mathbf{q}\cdot \mathbf{t}\,\,\text{Li}_2\Bigl((-1)^{\mathbf{\gamma}\cdot \mathbf{q}}\mathrm{e}^{-2\pi \mathbf{q} \cdot \mathbf{t}}\Bigr)\Biggr)\,. \label{eq:detailedform3}
\end{align}
Furthermore, the K\"ahler coordinates at leading order are given by
\begin{align}
T_i \approx T^{\text{l.o.}}_i
\coloneqq \frac{1}{g_s}\Bigl(\mathcal{T}^{\text{tree}}_i+ 
\delta\mathcal{T}_i^{(\alpha')^2}    
+
\delta\mathcal{T}_i^{\text{WSI}}\Bigr) +i\int_X C_4\wedge \omega_i\,,
\end{align}
\vspace{-0.5cm} 
\begin{align}
\text{with}\qquad\mathcal{T}^{\text{tree}}_i+ 
\delta\mathcal{T}_i^{(\alpha')^2}   
+
\delta\mathcal{T}_i^{\text{WSI}} 
&=\frac{1}{2}\kappa_{ijk}t^jt^k-\frac{\chi(D_i)}{24}\nonumber\\
&\hphantom{=}+\frac{1}{(2\pi)^2}\sum_{\mathbf{q}\in \mathcal{M}_X}q_i\, \mathscr{N}_{\mathbf{q}} \,\text{Li}_2\Bigl((-1)^{\mathbf{\gamma}\cdot \mathbf{q}}\mathrm{e}^{-2\pi \mathbf{q}\cdot \mathbf{t}}\Bigr)\,. \label{eq:detailedform4}
\end{align}

Adopting the normalization \eqref{eq:K0} for the Pfaffian prefactors,
the superpotential is 
\begin{align} 
W = W_{\text{flux}} + W_{\text{ED}3} + W_{\lambda\lambda}\,  
=\sqrt{\tfrac{2}{\pi}} \, \, \vec{\Pi}^\top \,{\cdot}\,\Sigma \,{\cdot}\,\big(\vec{f}-\tau\vec{h}\big) + \sqrt{\frac{2}{\pi}}\frac{1}{4\pi^2}\sum_D\, n_D\, \mathrm{e}^{-\frac{2\pi}{c_D}T_D}\,,
\label{eq:wlo3}    
\end{align}
where the sum runs over pure rigid prime toric divisors.
The period vector $\Pi$ in \eqref{eq:wlo3}, obtained from the 
prepotential \eqref{eq:prepotential} using the definition \eqref{eq:PvecFirstDef}, is 
\begin{equation}\label{eq:PiLCS} 
\Pi=\left (\begin{array}{c}
\frac{1}{6} \widetilde{\kappa}_{abc}\, z^a z^b z^c + \tfrac{\tilde{c}_a}{24} z^a- \tfrac{\mathrm{i} \,\zeta(3)\, \chi(\widetilde{X})}{(2\pi)^3}  + (2-z^a \partial_{z^a})F_{\text{inst}}\\[0.75em]
-\frac{1}{2}\widetilde{\kappa}_{abc} z^b z^c + \tilde{a}_{ab} z^b + \tfrac{\tilde{c}_a}{24}  +\partial_{z^a}F_{\text{inst}}  \\[0.3em]
1 \\ 
z^{a} 
\end{array}\right )\, ,
\end{equation} 
\begin{equation}
\text{where}\qquad \qquad \qquad    \partial_{z^a}F_{\text{inst}} =-\frac{1}{(2\pi \mathrm{i})^2}\sum_{\tilde{\mathbf{q}}\in \mathcal{M}(\widetilde{X})}\mathscr{N}_{\tilde{\mathbf{q}}}\, \tilde{q}_a\, \mathrm{Li}_2\Bigl(\mathrm{e}^{2\pi \mathrm{i}\,\tilde{\mathbf{q}}\cdot \mathbf{z}}\Bigr)\, , 
\end{equation}
and
\begin{align}
    (2-z^a \partial_{z^a})F_{\text{inst}} &= -\frac{1}{(2\pi \mathrm{i})^3}\sum_{\tilde{\mathbf{q}}\in \mathcal{M}(\widetilde{X})}\mathscr{N}_{\tilde{\mathbf{q}}}\, \biggl [2\,\mathrm{Li}_3\Bigl(\mathrm{e}^{2\pi \mathrm{i}\,\tilde{\mathbf{q}}\cdot \mathbf{z}}\Bigr)-2\pi \mathrm{i}(\tilde{\mathbf{q}}\cdot \mathbf{z})\, \mathrm{Li}_2\Bigl(\mathrm{e}^{2\pi \mathrm{i}\,\tilde{\mathbf{q}}\cdot \mathbf{z}}\Bigr) \biggl ]\, .
\end{align}

\newpage

We have now fully specified an $\mathcal{N}=1$ supersymmetric supergravity theory:
\begin{equ}[The leading-order supersymmetric EFT]
The leading-order Lagrangian density is
\begin{equation}\label{eq:loldef}
    \mathscr{L}_{\text{l.o.}} = \dfrac{1}{2}\mathcal{R} - \mathcal{K}_{A\bar{B}}^{\text{l.o.}}\, (\partial_{\mu}\Phi^A_{\text{l.o.}})(\partial^{\mu}\bar{\Phi}^{\bar{B}}_{\text{l.o.}}) - V_F^{\text{l.o.}}\, ,
\end{equation}
where the K\"ahler coordinates  $\Phi^A_{\text{l.o.}}$ at leading order are given by
\begin{equation}
    \Phi^A_{\text{l.o.}} = \lbrace \tau, z^a, T^{\text{l.o.}}_i\rbrace\,,
\end{equation} 
as defined in \eqref{eq:detailedform4},
and the leading-order $F$-term potential in \eqref{eq:loldef} is 
\begin{equation}\label{eq:vfsum}
    V_F^{\text{l.o.}} = V_F(W;\mathcal{K}_{\text{l.o}};\tau, z^a, T^{\text{l.o.}}_i;\mathscr{D}_{\mathbb{Z}})\,,
\end{equation} 
with $W$  given in \eqref{eq:wlo3}, and $\mathcal{K}_{\text{l.o.}}$ given in \eqref{eq:detailedform2} and 
\eqref{eq:detailedform3}. 
\end{equ}

The potential $V_F^{\text{l.o.}}$ depends on all moduli and can be computed \emph{explicitly} in a given compactification, in terms of discrete data $\mathscr{D}_{\mathbb{Z}}$: the topology of a Calabi-Yau threefold $X$ and its mirror $\widetilde{X}$, an associated orientifold involution  $\mathcal{I}$ of $X$, and a choice of flux quanta.
Specifically, we have
\begin{equation}
    \mathscr{D}_{\mathbb{Z}} = \mathscr{D}_{\mathbb{Z}}^{X}\cup \mathscr{D}_{\mathbb{Z}}^{\widetilde{X}}\cup \mathscr{D}_{\mathbb{Z}}^{\text{sources}}\, ,
\end{equation}
where $\mathscr{D}_{\mathbb{Z}}^{X}$, $\mathscr{D}_{\mathbb{Z}}^{\widetilde{X}}$ and $\mathscr{D}_{\mathbb{Z}}^{\text{sources}}$ are defined as
\begin{subequations}\label{eq:discrete_data}
    \begin{align}
    \mathscr{D}_{\mathbb{Z}}^{X} &\coloneqq \lbrace \, \kappa_{ijk} \, , \,  \mathscr{N}_{\mathbf{q}} \, , \, c_2(X)\, , \, \chi(X)\, ,\, \mathcal{I}\, ,\, h^{\bullet}(D)\, ,\,  n_D \rbrace\, ,\\
    \mathscr{D}_{\mathbb{Z}}^{\widetilde{X}} &\coloneqq \lbrace \widetilde{\kappa}_{abc}  \, , \, \mathscr{N}_{\tilde{\mathbf{q}}} \, , \, c_2(\widetilde{X})\, , \chi(\widetilde{X})\rbrace\, ,\\
    \mathscr{D}_{\mathbb{Z}}^{\text{sources}} &\coloneqq \lbrace \vec{f}\, ,\, \vec{h}\,  , \, \text{O}p\, , \, \text{D}p\, , \, \overline{\text{D}3}\rbrace\, .
\end{align}
\end{subequations}
In $\mathscr{D}_{\mathbb{Z}}^{X}$ we collect the topological data of the Calabi-Yau threefold $X$ and its orientifold, including the triple intersection numbers, GV invariants, and related quantities. The corresponding data for the mirror threefold $\widetilde{X}$ are summarized in $\mathscr{D}_{\mathbb{Z}}^{\widetilde{X}}$. Finally, $\mathscr{D}_{\mathbb{Z}}^{\text{sources}}$ encodes the remaining ingredients from sources, such as the choice of $3$-form fluxes $\vec{f}, \vec{h}$, the locations of O$p$-planes and D$p$-branes, and the locations of anti-D3-branes, if present.

\subsection{Corrections to the leading-order theory}\label{sec:leadingEFTcorrections}

In arriving at the leading-order theory \eqref{eq:loldef}, we have omitted sub-leading corrections. Here we assemble the effects that have been omitted, and we recall the justification.

\medskip

\noindent \textbf{Corrections within the supersymmetric theory.} In writing the K\"ahler potential and K\"ahler coordinates of the supersymmetric theory, we have retained all terms at \emph{leading order in $g_s$}, and omitted all terms at sub-leading orders in $g_s$, according to the decompositions of \eqref{eq:kformalcorrforKahler3} and \eqref{eq:kformalcorrforKahler3T},
which we reproduce here for reference:
\begin{equation}\label{eq:kformalcorrforKahler3bis}
g_s^2\,\mathcal{K}_{\text{K\"ahler}}  = \underbrace{g_s^2\,\Bigl(\mathcal{K}_{\text{tree}}+
\delta\mathcal{K}^{\mathcal{N}=2}_{\text{sphere}}\Bigr)}_{\text{leading order}}+
\underbrace{g_s^2\Bigl(\delta\mathcal{K}^{\mathcal{N}=2}_{(g_s)}+
\delta\mathcal{K}^{\mathcal{N}=1}_{\text{sphere}} +
\delta\mathcal{K}^{\mathcal{N}=1}_{(g_s)}\Bigr)}_{\mathcal{O}(g_s),~\text{omitted}}\,,
\end{equation}
\begin{equation}
    T_i  = \underbrace{T^{\text{tree}}_i+
\delta T^{\mathcal{N}=2,\text{tree}}_i}_{\text{leading order}}+
\underbrace{\delta T^{\mathcal{N}=2,(g_s)}_i+
\delta T^{\mathcal{N}=1,\text{tree}}_i +
\delta T^{\mathcal{N}=1,(g_s)}_i}_{\mathcal{O}(g_s),~\text{omitted}}\,.
\end{equation}
\begin{itemize}
    \item Because the $\mathcal{N}=2$ string loop corrections $\delta \mathcal{K}^{\mathcal{N}=2}_{(g_s)}$ and $\delta T^{\mathcal{N}=2,(g_s)}_i$ were obtained explicitly in \S\ref{sec:N2loop}, one can evaluate these terms in the candidate vacua of Chapter \ref{chap:deSitter} and verify that their effects are negligibly small, and so omitting them is well-justified.
    \item The $\mathcal{N}=1$ string loop corrections   $\delta\mathcal{K}^{\mathcal{N}=1}_{(g_s)}$ and $\delta T^{\mathcal{N}=1,(g_s)}_i$ are not fully known, so the argument for their omission is the smallness of $g_s$.
    \item The terms $\delta\mathcal{K}^{\mathcal{N}=1}_{\text{sphere}}$ and $\delta T^{\mathcal{N}=1,\text{tree}}_i$ formally arise at closed-string tree level and open-string tree level, respectively, but as explained in \S\ref{sec:QCorPrel}, their effects are suppressed by one power of $g_s$ because the breaking of $\mathcal{N}=2$ to $\mathcal{N}=1$ supersymmetry is correspondingly weak.  Warping corrections in $\delta T^{\mathcal{N}=1,\text{tree}}_i$ are important, but cannot be computed explicitly without knowing the warp factor: see \cite{Frey:2013bha,Martucci:2016pzt}.
\end{itemize}
In writing the superpotential \eqref{eq:wlo3}, 
\begin{itemize}
    \item We have omitted $W_{\text{ED}(-1)}$, which is justified because all the seven-brane gauge algebras are $\mathfrak{so}(8)$ \cite{Kim:2022jvv}.  
    \item We have truncated the sum of Euclidean D3-brane terms to include only prime toric divisors: that is, we have omitted autochthonous divisors, as well as positive sums of prime toric divisors.  The consistency of this truncation is examined in \S\ref{sec:conv}.  
    \item We have taken the Pfaffian numbers $\mathcal{A}_D$ to be given by \eqref{eq:K0} with $n_D=1$.  
    \item We have explicitly computed the periods $\vec{\Pi}$, and hence the flux superpotential, including mirror worldsheet instanton effects to very high degrees.  The truncation of this series is considered in \S\ref{sec:conv}.
\end{itemize}

Of all the corrections to the supersymmetric theory that we have omitted,
the most dangerous ones, in our opinion, are the contributions to $\delta T^{\mathcal{N}=1,\text{tree}}_i$ from warping: see
Eq.~\eqref{eq:Ttreewarped}.

\medskip

\noindent \textbf{Supersymmetry-breaking corrections.}
Moving beyond the two-derivative supersymmetric theory, we have omitted:
\begin{itemize}
\item Higher $F$-terms, denoted $c_4 \frac{F^4}{\Lambda^2} + \ldots$ in \eqref{eq:ffour}.  These are demonstrably small for $F \ll \Lambda$, which can be verified from the data given in Chapter \ref{chap:deSitter}. 
\item Anti-D3-brane supersymmetry-breaking effects, denoted $V_{\mathrm{breaking}}$ in \eqref{eq:vbreak}. These effects are not sub-leading, but it is convenient to postpone considering them until Chapter \ref{chap:modulistabilization}.  
We will compute the supersymmetry-breaking effects of anti-D3-branes at leading order in $\alpha'$ and $g_s$ --- see the detailed discussion in \S\ref{sec:KPV} --- and omit corrections beyond leading order.
\end{itemize}

Of all the corrections to the non-supersymmetric theory that we have omitted,
the most dangerous ones, in our opinion, are $\alpha'$ corrections to the anti-D3-brane action: see Eq.~\eqref{eq:antibranecurvecorr}.

\chapter{Exponential Hierarchies}\label{chap:modulistabilization}

Thus far we have focused on deriving the four-dimensional quantum effective theories of flux compactifications. The eventual goal is to exhibit local minima of these theories, i.e.~vacua with stabilized moduli.  

As we explained in \S\ref{sec:hier}, controlled moduli stabilization relies on mechanisms for generating hierarchies. The KKLT scenario \cite{Kachru:2003aw} is a scheme for constructing (A)dS$_4$ vacua by combining three sorts of hierarchical effects: a small classical flux superpotential $W_0$; non-perturbative contributions to the superpotential from Euclidean D3-branes or gaugino condensation on D7-branes (\S\ref{sec:Wnp}); and, in order to break supersymmetry, the tension of a small number of anti-D3-branes \cite{Kachru:2002gs} at the tip of an exponentially warped throat.

Together, these elements provide the foundation for the explicit construction of supersymmetric AdS$_4$ vacua and of metastable de Sitter vacua in Chapter~\ref{chap:deSitter}. In this chapter we will explain how each 
of
these components of the scenario can be constructed.

\section{A first glance at the KKLT scenario}\label{sec:KKLTintro}
\index{KKLT}
 
We briefly remarked in \S\ref{sec:classicalstab}
that the complex structure moduli and axio-dilaton are generically stabilized at the classical level once one accounts for the effects of $3$-form flux, as captured in the 
Gukov-Vafa-Witten superpotential \eqref{eq:GVW}. 
However, the K\"ahler moduli remain unfixed at this level due to the no-scale structure 
\eqref{eq:NoScaleID} of the classical scalar potential.  Their stabilization requires sub-leading effects --- most importantly, non-perturbative contributions to the superpotential from Euclidean D3-branes or gaugino condensation on D7-branes, as introduced in \S\ref{sec:Wnp}.  These corrections introduce exponential terms that generate a potential for the K\"ahler moduli.

\begin{figure}[t!]
\begin{center} 
\includegraphics[width=\linewidth]{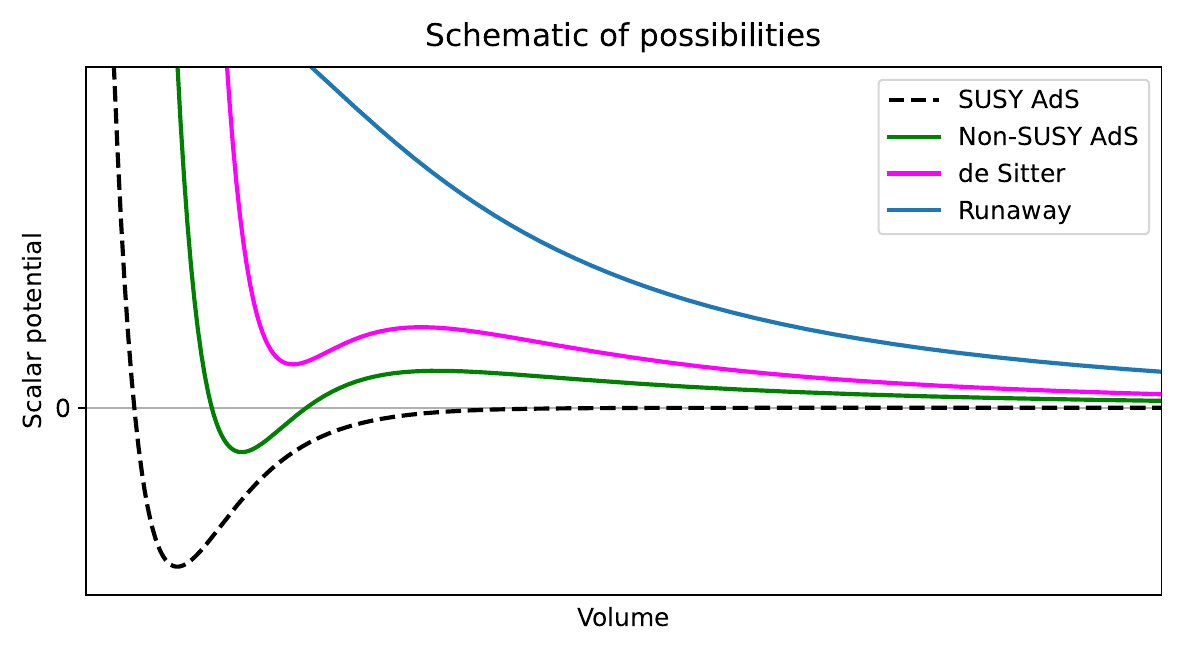}
\caption{A schematic of the possible results of uplift from a supersymmetric AdS vacuum.
The horizontal axis is the Calabi-Yau volume $\mathcal{V}=(T+\overline{T})^{3/2}$, and the vertical axis is the scalar potential $V$.  
}\label{fig:ExamplePlot}
\end{center}
\end{figure}

Here, we aim to illustrate the underlying stabilization mechanism in the simplest possible setting, in order to motivate the essential ingredients that will be developed in the following sections. For an orientifold of a Calabi-Yau threefold $X$ with $h^{1,1}=1$, we have
\begin{equation}
    W = \sqrt{\tfrac{2}{\pi}}\, \int_X\, G_3\wedge\Omega+\mathcal{A}\,\mathrm{e}^{-2\pi T}\, .
\end{equation}
Upon turning on ISD fluxes, the complex structure moduli and the axio-dilaton can be integrated out. Then, below the complex structure mass scale,\footnote{Care is required when integrating out moduli that are not much heavier than the scales of interest.  We give a simplified treatment here, and describe the complete process in \S\ref{sec:stabKah}.} the K\"ahler potential and superpotential are given by
\begin{equation}\label{kkltonemod}
    W = W_0+\mathcal{A}\,\mathrm{e}^{-2\pi T}\; ,\quad \mathcal{K} = -3\log(T+\overline{T})
\end{equation} 
The resulting potential for the K\"ahler modulus,
\begin{equation}\label{eq:Vsch}
    V = \mathrm{e}^{\mathcal{K}}\biggl (\mathcal{K}^{T\overline{T}}\, D_T W\, \overline{D_T W}-3|W|^2\biggl )\,,
\end{equation}
is shown as a function of the Calabi-Yau volume $\mathcal{V}=(T+\overline{T})^{3/2}$ as the black dashed line in Fig.~\ref{fig:ExamplePlot}.
The no-scale structure is clearly broken due to the presence of the non-perturbative effects. In particular, $\partial_T W\neq 0$ now, even though $\partial_T W_0 = 0$.

The toy theory specified in \eqref{kkltonemod} has an $\mathcal{N}=1$ supersymmetric AdS$_4$ vacuum satisfying $D_T W=0$.
The solution is obtained by first computing
\begin{align}
    D_T W &= \partial_T W + W \partial_T \mathcal{K} \nonumber\\
    &=-2\pi \mathcal{A}\,\mathrm{e}^{-2\pi T} - \tfrac{3}{T+\overline{T}}\bigl (W_0+\mathcal{A}\,\mathrm{e}^{-2\pi T}\bigr )\,,
\end{align}
so that 
\begin{equation}
    W_0 = -\mathcal{A}\,\mathrm{e}^{-2\pi T}\biggl [ \dfrac{2\pi}{3}(T+\overline{T})+1\biggl ]\,.
\end{equation}
Thus, to leading order,
\begin{equation}\label{eq:twapprox}
    2\pi T=\log(|W_0|^{-1})+\ldots \, .
\end{equation}
The approximation \eqref{eq:twapprox} holds good provided that $|W_0|\ll 1$. In this case, the minimum arises at a point $T=T_\star$ with $\mathrm{Re}(T_\star)$ large enough so that higher instanton corrections to $W$, and perturbative and non-perturbative corrections to $\mathcal{K}$, can arguably be neglected.

The claim of \cite{Kachru:2003aw} is that supersymmetric AdS$_4$ vacua arise in this way, with AdS vacuum energy
\begin{equation}\label{eq:VAdSsch}
    V_{\text{AdS}} = -3 \mathrm{e}^{\mathcal{K}} |W|^2\, .
\end{equation}

Having stabilized all moduli in a supersymmetric AdS$_4$ vacuum, the next step is to uplift the vacuum energy to achieve a metastable de Sitter minimum. This requires introducing a source of positive energy that breaks supersymmetry, while preserving control over the effective theory. In the KKLT scenario, the supersymmetry-breaking source is a collection of one or more anti-D3-branes \cite{Kachru:2002gs} at the tip of a Klebanov-Strassler warped throat geometry \cite{Klebanov:2000hb}. As we will explain in \S\ref{sec:KS}, such a region can arise in a flux compactification in which the complex structure moduli are stabilized near a conifold point in moduli space. The anti-D3-branes break all remaining supersymmetry, and their tension contributes a positive ``uplift'' term $V_{\overline{\text{D3}}}$ to the scalar potential \eqref{eq:Vsch}, as we will show in \S\ref{sec:KPV}. The resulting energy scale is exponentially suppressed by the warp factor, allowing in principle for precise tuning of the vacuum energy. 

The qualitative behavior of the resulting vacuum then depends on the relative size of this contribution compared to the negative vacuum energy $V_{\text{AdS}}$ of the supersymmetric AdS solution \eqref{eq:VAdSsch}, yielding three distinct possibilities:
\begin{enumerate}[1)]
    \item If $V_{\text{AdS}} + V_{\overline{\text{D3}}} < 0$, the minimum remains AdS, but supersymmetry is broken.
    \item If $V_{\overline{\text{D3}}} \gg |V_{\text{AdS}}|$, the anti-D3-brane destabilizes the vacuum and leads to a runaway direction corresponding to decompactification, where the volume increases indefinitely.
    \item If $V_{\overline{\text{D3}}} \gtrsim |V_{\text{AdS}}|$ but not parametrically larger, a metastable de Sitter vacuum can arise, with the anti-D3-brane effectively uplifting the vacuum energy to a positive value.
\end{enumerate}
These three scenarios are illustrated schematically in Figure~\ref{fig:ExamplePlot}.

\bigskip
In summary, constructing a KKLT-type de Sitter vacuum requires the \emph{simultaneous} realization of the following ingredients in a Calabi-Yau orientifold compactification,
in a region of moduli space where quantum corrections to the effective theory are under control:
\begin{enumerate}[(a)]
    \item \label{item:AdSVacuum} a supersymmetric AdS vacuum with exponentially small negative vacuum energy,
    \item \label{item:KSthroat} a Klebanov-Strassler throat, generated by fluxes near a conifold point, providing the necessary warping to suppress the uplift energy, and
    \item \label{item:antiD3} an anti-D3-brane at the bottom of the throat whose redshifted tension is sufficient to uplift the vacuum energy to a small, positive value.
\end{enumerate}
In particular, (a) requires an exponentially small value of the flux superpotential $W_0$, as well as a suitable structure of terms in $W_{\text{np}}$, such that there exists a supersymmetric vacuum inside the moduli space.

In this chapter, we will describe in more detail each of these necessary ingredients of the KKLT construction, emphasizing   the mechanisms that lead to hierarchies. In \S\ref{sec:pfv} we introduce a method of selecting quantized fluxes that yields exponentially small values of the flux superpotential: this is the \emph{perturbatively flat vacuum} (PFV) mechanism \cite{Demirtas:2019sip}. In \S\ref{sec:KS} we provide background on the Klebanov-Strassler solution. Then, in \S\ref{sec:conipfv}, we show how appropriately-designed PFVs lead to complex structure stabilization near a conifold, and with small $|W_0|$, allowing for a Klebanov-Strassler throat region and KKLT-type K\"ahler moduli stabilization in the same compactification.  Finally, in \S\ref{sec:KPV} we give the details of anti-D3-brane supersymmetry breaking and uplift.

\section{Perturbatively flat vacua (PFVs)}\label{sec:pfv}

To achieve small values of the flux superpotential,
\begin{equation}
    |W_0|\coloneqq \bigl |\langle W_{\mathrm{flux}}\rangle \bigl | = \biggl |\biggl\langle \sqrt{\tfrac{2}{\pi}}\int_X\, G_3\wedge\Omega\biggl\rangle \biggl |\ll 1\, ,
\end{equation}
let us consider the following idea:
we split the period vector into two contributions
\begin{equation}
    \vec{\Pi} = \vec{\Pi}_{\text{poly}} + \vec{\Pi}_{\text{exp}}\, ,
\end{equation}
where $\vec{\Pi}_{\text{poly}}$ and 
$\vec{\Pi}_{\text{exp}}$ encode polynomial and exponential contributions, respectively.
Then, we can write
\begin{equation}
    \int_X\, G_3\wedge\Omega = \vec{\Pi}^{T}\,\cdot\Sigma\cdot (\vec{f}-\tau\vec{h})
\end{equation}
as
\begin{equation}\label{eq:PFVWsplit}
    \sqrt{\tfrac{\pi}{2}}\, W_{\mathrm{flux}} = W_{\mathrm{poly}}  +W_{\mathrm{exp}}\, ,
\end{equation}
where
\begin{equation}
    W_{\mathrm{poly}}=\vec{\Pi}_{\text{poly}}^{T}\,\cdot\Sigma\cdot (\vec{f}-\tau\vec{h})
    \; ,\quad 
    W_{\mathrm{exp}}=\vec{\Pi}_{\text{exp}}^{T}\,\cdot\Sigma\cdot (\vec{f}-\tau\vec{h})\, .
\end{equation}
By choosing the fluxes $\vec{f},\vec{h}$ appropriately, we would like to engineer a situation where in the vacuum
\begin{equation}
    \langle W_{\text{poly}}\rangle = 0\, .
\end{equation}
Then we would naturally obtain
\begin{equation}
    |W_0|\sim \bigl |\langle W_{\mathrm{exp}}\rangle \bigl | \ll 1\, .
\end{equation}

We now want to implement the above idea in a concrete setup. 
To do so, we recall the LCS prepotential in Eq.~\eqref{eq:prepotential}, for which we compute the flux superpotential $W_{\mathrm{flux}}$ as follows.
First we make an ansatz for the fluxes,
\begin{equation}\label{eq:PFVflux}
    \vec{f} = \left(R_0,R_a,0,M^a\right)^\top \, ,\; \vec{h} = \left(0,K_a,0,0^a\right)^\top\,,
\end{equation} 
where the entries are all integers, and $a=1,\ldots, h^{2,1}$. 
The flux superpotential can be written in the form \eqref{eq:PFVWsplit}, where
\begin{align}\label{eq:WfluxPoly}
    W_{\mathrm{poly}}(z^a,\tau)=\frac{1}{2}M^a \widetilde{\kappa}_{a bc}z^b z^c-\tau K_a z^a+\left(R_a- \tilde{a}_{ab} \, M^b\right)z^a+\left(R_0-\frac{M^a \tilde{c}_a}{24}\right)  \,  ,
\end{align}
and
\begin{align}\label{eq:WfluxInst}
    W_{\mathrm{exp}}(z^a)=-\frac{1}{(2\pi)^2}\sum_{\tilde{\mathbf{q}}\in\mathcal{M}_{\widetilde{X}}} \mathscr{N}_{\tilde{\mathbf{q}}}\,\tilde{\mathbf{q}}_a M^a\,\mathrm{Li}_2(\mathrm{e}^{2\pi \I \tilde{\mathbf{q}}_a z^a})\,  .
\end{align}
The flux choice \eqref{eq:PFVflux} makes $W_{\mathrm{poly}}(z^a,\tau)$ a quadratic function of the fields $z^a,\tau$.
Looking at $W_{\mathrm{poly}}(z^a,\tau)$, we can cancel the constant and linear term in $z^a$
by choosing $R_0$ and $R_a$ such that
\begin{equation}\label{eq:PFVPconst}
    R_a= \tilde{a}_{ab} \, M^b\in \bZ\kom R_0=\frac{M^a \tilde{c}_a}{24}\in \bZ\, .
\end{equation}
Of course, because the flux vectors in \eqref{eq:PFVflux} have to be quantized, the condition \eqref{eq:PFVPconst} can only be achieved for certain choices of $M^a$ for which the right hand sides in \eqref{eq:PFVPconst} take values in $\bZ$.
If that is the case, the polynomial superpotential $W_{\mathrm{poly}}(z^\alpha,\tau)$ simplifies to
\begin{align}\label{eq:WfluxPoly2}
    W_{\mathrm{poly}}(z^a,\tau)=&\frac{1}{2}N_{ab}\, z^a z^b-\tau K_a z^a \,  ,
\end{align}
where we have introduced the quantity
\begin{equation}
	N_{ab}\coloneqq  \kappa_{abc}\, M^c\, .
\end{equation}

Next, we want to find  conditions under which $\partial_{z^a} W_{\mathrm{poly}}=0$.
We first compute
\begin{align}\label{eq:dWfluxPoly}
    \partial_{z^a} W_{\mathrm{poly}}= N_{ab}\,  z^b-\tau K_a \,  ,
\end{align}
which, assuming $N_{ab}$ is invertible, vanishes along the locus
\begin{equation}\label{eq:znp}
    z^a = N^{ab} K_b\,\tau = p^a \tau\, ,
\end{equation}
where we have defined
\begin{equation}\label{eq:PvecDef}
    p^a\coloneqq  N^{ab} K_b\, .
\end{equation} 
Clearly, we have to ensure $\vec{p}\in \mathcal{K}_{\widetilde{X}}$ in order for the complex structure moduli $z^a$ to take values in the complexified K\"ahler cone $\mathcal{K}_{\widetilde{X}}$ of the mirror $\widetilde{X}$.

Plugging \eqref{eq:znp} back into \eqref{eq:WfluxPoly2}, we obtain
\begin{align}\label{eq:WfluxPoly3}
    W_{\mathrm{poly}}(z^a,\tau)=-\frac{1}{2}\tau K_a z^a= -\frac{1}{2}\tau^2 K_a p^a\,  ,
\end{align}
and so, if we further impose
\begin{equation}\label{eqref:pfvcond}
    K_a p^a = K_a N^{ab} K_b = 0\, ,
\end{equation}
we see that
\begin{equation}
    W_{\mathrm{poly}} = \partial_{z^a} W_{\mathrm{poly}} = \partial_{\tau} W_{\mathrm{poly}}=0
\end{equation}
are satisfied along the one-complex-dimensional locus $z^a=p^a \tau$.
This locus defines a so-called \emph{perturbatively flat vacuum} (PFV) \cite{Demirtas:2019sip}.

To summarize, given a choice of $M^a,K_a\in \bZ$, we have to require
\begin{align}
    \label{eq:PFV}
    \det N\neq 0\, \kom\; \vec{p}\in \mathcal{K}_{\widetilde{X}}\, \kom\; K_{a}\, p^a=0\, \kom\; \tilde{a}_{ab}\, M^b\in \mathbb{Z}\, \kom\;  \tilde{c}_a\,  M^a\in 24\mathbb{Z}\, .
\end{align}
Then the full flux vectors $\vec{f},\vec{h}$ are determined by \eqref{eq:PFVflux}, with $R_0$ and $R_a$ fixed by \eqref{eq:PFVPconst}, and one has $W_{\mathrm{poly}}=0$ along the one-dimensional locus $z^a=p^a \tau$.

Along this direction, the flux superpotential is given by $W_{\mathrm{exp}}$ in Eq.~\eqref{eq:WfluxInst}, and so
\begin{equation}
    \sqrt{\tfrac{\pi}{2}}\, W_{\mathrm{flux}}(z^a=p^a \tau) =-\frac{1}{(2\pi)^2}\sum_{\tilde{\mathbf{q}}\in\mathcal{M}_{\widetilde{X}}} \mathscr{N}_{\tilde{\mathbf{q}}}\,\tilde{\mathbf{q}}_a M^a\,\mathrm{Li}_2(\mathrm{e}^{2\pi \I \tilde{\mathbf{q}}_a\, p^a\, \tau})\,  .
\end{equation}
This expression is explicitly computable and purely exponential. A minimum for $\tau$ is generically generated by the \emph{racetrack mechanism}, see e.g.~\cite{Denef:2004dm}. Specifically, in the presence of multiple non-perturbative contributions with different powers, here given by the dot product $\tilde{\mathbf{q}}_a\, p^a$, the competing exponential terms can generate a local minimum of the scalar potential, thereby stabilizing the remaining modulus $\tau$.

Let us illustrate the PFV racetrack mechanism in an explicit example, with $h^{2,1}(X)=2$.
For the degree 18 hypersurface in weighted projective space $\mathbb{CP}_{[1,1,1,6,9]}$ studied in \cite{Candelas:1994hw}, we consider an orientifold involution described in \cite{Louis:2012nb}, with two O7-planes, each with four D7-branes, and in which $Q_{\mathrm{D3}}=138$.\footnote{To be more precise, the Calabi-Yau manifold under consideration possesses $h^{2,1} = 272$ complex structure moduli. However, as observed in \cite{Giryavets:2003vd}, it is advantageous to restrict attention to a particular locus in moduli space where the geometry is symmetric under the action of a discrete group $G = \mathbb{Z}_6 \times \mathbb{Z}_{18}$. By including only those fluxes that are invariant under $G$, one ensures that the resulting $F$-term equations admit solutions that lie entirely within the $G$-invariant subspace. In such a setting, it suffices to solve only the reduced system of $F$-term conditions along the directions tangent to this invariant locus, thereby significantly simplifying the analysis. The periods along these directions coincide with those of the mirror Calabi-Yau threefold \cite{Greene:1990ud} and are encoded in an effective two-modulus prepotential of the form \eqref{eq:prepotential}, with geometric and instanton data specified in \eqref{eq:geodata} and \eqref{eq:theinst}, respectively.  On the other hand, the flux quantization conditions can be subtle in Greene-Plesser mirror pairs \cite{MoritzUnpublished}.}
To compute the prepotential in Eq.~\eqref{eq:prepotential}, we need the following data:
\begin{align}\label{eq:geodata}
\widetilde{\kappa}_{111}=\,\,9\, ,\quad \widetilde{\kappa}_{112}=3\, ,\quad \widetilde{\kappa}_{122}=1\, ,\quad  \tilde{a}=\,\,\frac{1}{2}\begin{pmatrix}
9 & 3 \\
3 & 0
\end{pmatrix}\, ,\quad \tilde{c}=\begin{pmatrix}
102\\
36
\end{pmatrix}\, .
\end{align}
The instanton corrections entering \eqref{eq:finst} take the form \cite{Candelas:1994hw}
\begin{align}\label{eq:theinst}
& (2\pi \I )^3\mathcal{F}_{\text{inst}}=\mathcal{F}_1 + \mathcal{F}_2 + \cdots\, , 
\end{align}
where
\begin{align}
& \mathcal{F}_1 =  - 540 \mathrm{e}^{2\pi \mathrm{i} z^1} - 3\mathrm{e}^{2\pi \mathrm{i} z^2}\, ,\label{eqf1} \\[0.3em]
& \mathcal{F}_2 = -\frac{1215}{2}\mathrm{e}^{4\pi \mathrm{i} z^1} +1080 \mathrm{e}^{2\pi \mathrm{i} (z^1+z^2)} + \frac{45}{8} \mathrm{e}^{4\pi \mathrm{i} z^2}\, .\label{eqf2}
\end{align}
The hierarchy in the coefficients of the one-instanton terms   will turn out to be crucial below.

From \eqref{eq:geodata}, the condition \eqref{eqref:pfvcond} defining a perturbatively flat vacuum
reads 
\begin{equation}\label{eq:lemmaeq}
	M_1 = \frac{M_2 K_2 (2K_1 - 3K_2)}{(K_1 - 3K_2)^2}\,.
\end{equation}
We find the flat direction 
\begin{equation}
	\vec{z} = \tau \begin{pmatrix}
	p^1\\
	p^2
	\end{pmatrix} = \frac{\tau (K_1 - 3K_2)}{M_2} \begin{pmatrix}
	- K_2/K_1\\
	1
	\end{pmatrix}\,.
\end{equation}

The full effective superpotential along the flat direction, including the worldsheet instantons \eqref{eq:theinst}, is
\begin{equation}\label{weffis}
	W_{\text{eff}}(\tau)=c\,\Bigl(\mathrm{e}^{2\pi \mathrm{i} p^1 \tau}+A\mathrm{e}^{2\pi \mathrm{i} p^2 \tau}\Bigr)\, + \ldots\, ,
\end{equation}
in terms of factors $c$ and $A$ that depend on $\vec{M}$ and  $\vec{K}$, but do not depend on $\tau$.
When  $|p^1-p^2| \ll p^2$, the 
two terms in \eqref{weffis} can have comparable magnitude, even for $\mathrm{i}\tau \ll -1$.  
Specifically, using
\begin{equation}
\p_{\tau} W_{\text{eff}}(\tau)=2\pi \mathrm{i} c\,\Bigl(p^{1} \mathrm{e}^{2\pi \mathrm{i} p^1 \tau}+p^{2}A\mathrm{e}^{2\pi \mathrm{i} p^2 \tau}\Bigr)\,,
\end{equation}
and solving
\begin{equation}
\p_{\tau} W_{\text{eff}}=0 \quad \Rightarrow\quad \mathrm{e}^{2\pi \mathrm{i} (p^1-p^{2}) \langle\tau\rangle}=-A\, \dfrac{p^{2}}{p^{1}}\, ,
\end{equation}
we obtain
\begin{equation}
\langle\tau\rangle = \dfrac{1}{2\pi\mathrm{i}}\,  \dfrac{1}{(p^1-p^{2})}\, \ln\left (-A\, \dfrac{p^{2}}{p^{1}}\right )\, .
\end{equation}
Plugging this back into $W_{\text{eff}}(\tau)$ in Eq.~\eqref{weffis},
we find
\begin{equation}
W_{\text{eff}}(\langle \tau\rangle) = c\, \dfrac{p^{2}-p^{1}}{p^{2}}\,\left (-A\, \dfrac{p^{2}}{p^{1}}\right )^{\frac{p^1}{p^1-p^{2}}}\,,
\end{equation}
which is small precisely when $|p^1-p^2| \ll p^2$.

\begin{figure}[t!]
	\centering
	\includegraphics[keepaspectratio,scale=0.28]{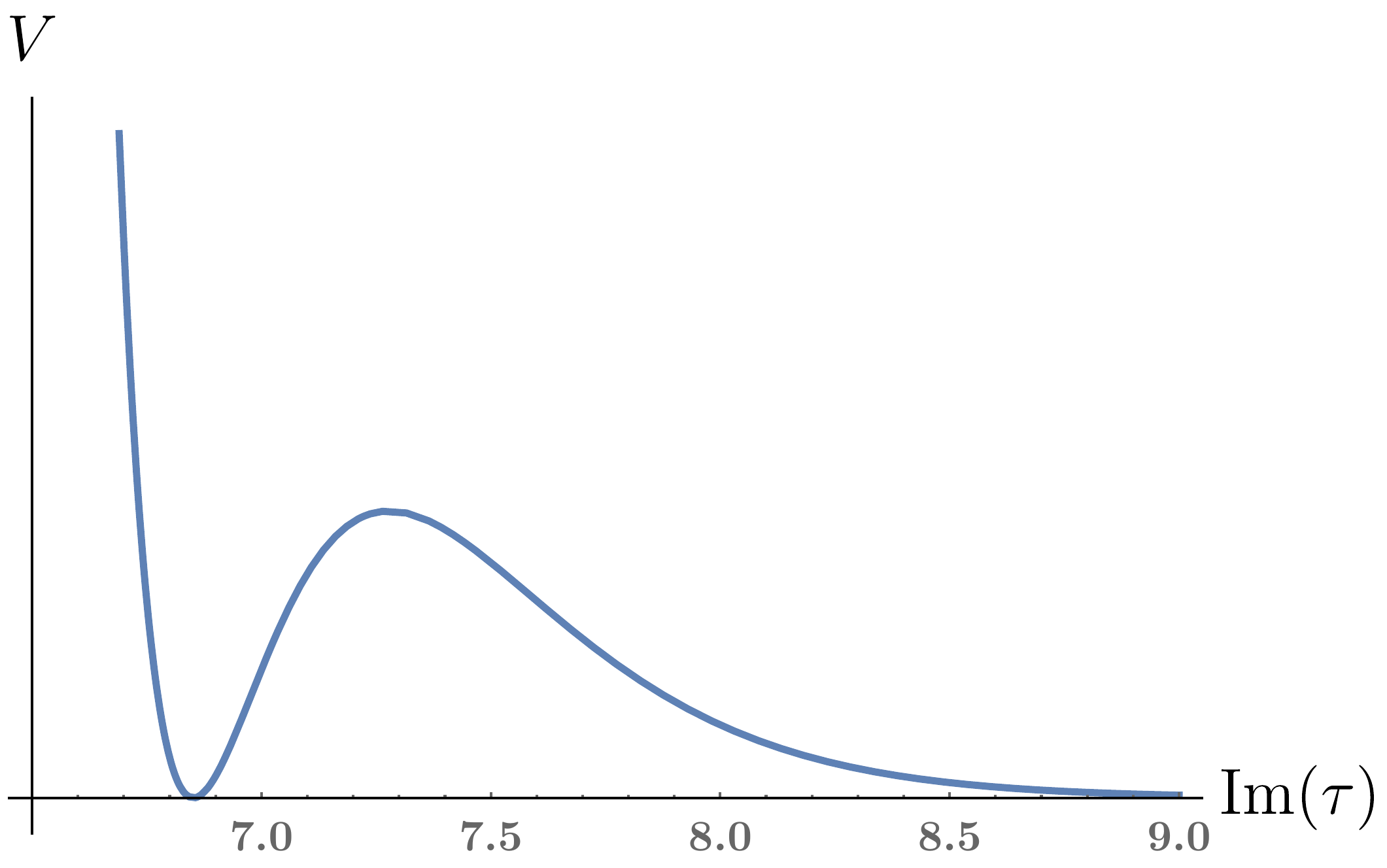}
	\caption{The scalar potential on the positive $\mathrm{Im}(\tau)$ axis.}
	\label{fig:racetrackpot}
\end{figure}

To realize a racetrack potential, we therefore seek 
$\vec{M}$ and $ \vec{K}$ obeying \eqref{eq:lemmaeq}
and $|K_1+K_2| \ll |K_2|$, and fulfilling the tadpole constraint
$Q_{\mathrm{flux}} \equiv - \frac{1}{2} \vec{M} \cdot \vec{K} \leq 138$.  One such choice is
\begin{equation}
\vec{M}=\begin{pmatrix*}
-16\\
\phantom{-}50
\end{pmatrix*}\, ,\quad \vec{K}=\begin{pmatrix*}
\phantom{-}3\\
-4
\end{pmatrix*}\, ,
\end{equation}
for which we find $A=-\frac{5}{288}$, $c=-\sqrt{\tfrac{2}{\pi}}\frac{8640}{(2\pi \I )^3}$, and $Q_{\mathrm{flux}}=124$. 
Figure \ref{fig:racetrackpot} shows the resulting racetrack potential.  
The $F$-terms for the moduli vanish at
\begin{equation}
\langle \tau\rangle =6.856 \mathrm{i}\, , \quad
\langle z^1\rangle =2.742 \mathrm{i}\, , \quad
\langle z^2\rangle = 2.057\mathrm{i}\, ,
\end{equation}
and at these stabilized vevs we have
\begin{equation}\label{theans}
|W_0|=2.037\times 10^{-8}\, .
\end{equation}
The racetrack potential used above was obtained from the one-instanton terms \eqref{eqf1}, but the two-instanton terms \eqref{eqf2} are sub-leading by a factor $\mathcal{O}(10^{-5})$, and so can be safely neglected.

\section{Warping and the Klebanov-Strassler solution}\label{sec:KS}

One of the central challenges in string compactifications --- and indeed, in contemporary physics --- is to account for the large disparity between the fundamental Planck scale and observable physical scales, such as the electroweak scale or the cosmological constant. 
Warped geometries, in which energy scales vary along the internal dimensions, offer a natural mechanism for generating such hierarchies dynamically.  Incarnations of warped hierarchies occur in type IIB flux compactifications on orientifolds of Calabi–Yau threefolds. As described in \S\ref{sec:EOM}, background $3$-form fluxes backreact on the ten-dimensional metric \eqref{eq:WarpedBackgroundAnsatzGKPSection}, leading to a non-trivial warp factor that varies over the internal Calabi-Yau manifold $X$.

A particularly important warped geometry is the \emph{Klebanov-Strassler (KS) solution}\index{Klebanov-Strassler solution}\index{KS solution} \cite{Klebanov:2000hb},
a regular, $\mathcal{N}=1$ supersymmetric solution of type IIB supergravity  
that describes the backreaction of NS-NS and R-R $3$-form fluxes on a deformed conifold geometry.  
This solution is holographically dual to a confining $\mathcal{N}=1$ gauge theory, and it provides an explicit realization of the Randall-Sundrum mechanism \cite{Randall:1999ee} within string theory: states localized at the tip of the throat experience an exponentially suppressed energy scale relative to the bulk.

\medskip

In order to explain the Klebanov-Strassler solution, we will build up a series of related but simpler warped geometries,
\begin{equation}
     \mathrm{AdS}_5 \times S^5 \rightarrow \mathrm{AdS}_5 \times T^{1,1} \rightarrow \text{Klebanov-Tseytlin} \rightarrow \text{Klebanov-Strassler}\,,
\end{equation}
which we now introduce in turn.

We begin with the geometry sourced by $N\gg 1$ coincident D3-branes in flat space.  
Placing the branes at the origin of $\mathbb{R}^{6}$, with metric
\begin{equation}
    \mathrm{d}s_6^{2} = \mathrm{d}r^{2} + r^{2} \mathrm{d}s_{S^{5}}^{2} \,,
\end{equation}
gives rise to the well-known $\mathrm{AdS}_{5}\times S^{5}$ solution supported by RR $5$-form flux \cite{Aharony:1999ti}.  
The D3-branes serve as electric and magnetic sources for the RR $4$-form $C_{4}$ (recall the discussion in \S\ref{sec:DbraneSources}), inducing $N$ units of quantized $5$-form flux,
\begin{equation}
    N = \frac{1}{(4\pi^{2}\alpha')^{2}} \int_{S^{5}} F_{5}\, .
\end{equation}
A solution consistent with self-duality of $F_{5}$ is
\begin{equation}
    F_{5} = (1+\star)\, \frac{(4\pi^{2}\alpha')^{2} N}{\mathrm{Vol}(S^{5})}\,\Omega_{5},
\end{equation}
where $\Omega_{5}$ is the volume form on the round $S^{5}$.  
Solving the equations of motion yields the warp factor
\begin{equation}
    \mathrm{e}^{-4A} = 1 + \frac{r_{+}^{4}}{r^{4}}\;,
    \qquad r_{+}^{4} = 4\pi g_{s} N (\alpha')^{2}\, .
\end{equation}
In the near-horizon region $r\ll r_{+}$, it is convenient to introduce a new radial coordinate
\begin{equation}
    y = r_{+}\,\ln\left(\frac{r}{r_{0}}\right) \, .
\end{equation}
The constant $r_{0}$ is simply an arbitrary reference scale introduced to make the logarithm dimensionless.
It has no physical significance and corresponds to the freedom to shift the AdS radial coordinate by a constant: $y \;\rightarrow\; y + c\;\Leftrightarrow\; r_{0} \;\rightarrow\; r_{0}\, \mathrm{e}^{-c/r_{+}}$.
This casts the ten-dimensional metric into the form
\begin{equation}\label{eq:AdS5S5}
    \mathrm{e}^{2y/r_{+}}\,\mathrm{d}x_{4}^{2}+ \mathrm{d}y^{2}+ r_{+}^{2}\,\mathrm{d}s_{S^{5}}^{2}\, .
\end{equation}
This is manifestly $\mathrm{AdS}_{5}\times S^{5}$ with sphere radius $r_{+}$.  
For large 't Hooft coupling $g_{s}N\gg 1$, curvatures are small and the supergravity description is valid.
In this limit, the D3-branes sit infinitely deep in the throat.

The $\mathrm{AdS}_{5}\times S^{5}$ solution is the foundational example of the \emph{AdS/CFT correspondence}\index{AdS/CFT correspondence} \cite{Maldacena:1997re}, which equates the near-horizon limit of type IIB string theory on this background to $\mathcal{N}=4$ $\mathrm{SU}(N)$ super-Yang-Mills theory.  
The complexified gauge coupling is identified with the axio-dilaton,
\begin{equation}
	\tau = \frac{\theta}{2\pi} + \mathrm{i}\,\frac{4\pi}{g^{2}_{\text{YM}}}\, .
\end{equation}
The dual descriptions apply in complementary regimes:  
the gauge theory is weakly coupled for $g_s N\ll 1$, while the supergravity picture is reliable when the 't~Hooft coupling is large, $g_s N\gg 1$.

A natural generalization arises when $N \gg 1$ D3-branes are instead placed at the tip of the \emph{singular conifold}, 
which is defined as the subvariety of $\mathbb{C}^4$ satisfying
\begin{equation}\label{eq:conifold_sing}
    \sum_{i=1}^{4}\, z_i^2 = 0\; , \quad z_i \in \mathbb{C}\, .
\end{equation} 
The singular conifold has a conical singularity at the origin.
The Ricci-flat metric on the singular conifold is that of a cone over the five-manifold $T^{1,1}$, i.e.,
\begin{equation}\label{eq:singconising} 
    \mathrm{d}s_6^2 = \mathrm{d}r^2 +r^2 \mathrm{d}s_{T^{1,1}}^2\,,
\end{equation}
where $T^{1,1}$, defined as the coset space  
\begin{equation}
    T^{1,1} \coloneqq  \bigl(\mathrm{SU}(2)\times \mathrm{SU}(2)\bigr)/\mathrm{U}(1)
\end{equation}
is topologically $S^2 \times S^3$, with the metric
\begin{equation}
   \mathrm{d} s_{T^{1,1}}^2 = \frac{1}{9}\Bigl(\mathrm{d}\psi + \sum_{i=1}^{2} \mathrm{cos}(\theta_i) \mathrm{d}\phi_i\Bigr)^2 + \frac{1}{6}\sum_{i=1}^{2} \Bigl(\mathrm{d}\theta_i^2 + \mathrm{sin}^2(\theta_i) \,\mathrm{d}\phi_i^2 \Bigr)\,.
\end{equation} 

\medskip

The long-distance supergravity solution that results when N D3-branes are placed at the tip of the singular conifold is $\mathrm{AdS}_5 \times T^{1,1}$, described by a warped metric of the form
\begin{equation}\label{eq:AdST11} 
    \mathrm{d}s^{2} = \mathrm{e}^{2A(r)}\,\mathrm{d}x_{4}^{2}
    + \mathrm{e}^{-2A(r)}\,\mathrm{d}s_{6}^{2}\,,
\end{equation}
where the unwarped internal metric is that of the singular conifold, given in \eqref{eq:singconising}, 
and the warp factor for a stack of $N$ D3-branes takes the same functional form as in the flat case,
\begin{equation}
    \mathrm{e}^{-4A(r)} \equiv h(r) =1+ \frac{27\pi g_s N \alpha'^2}{4 r^4}\,.
\end{equation}
The dual $\mathcal{N}=1$ superconformal field theory,
known as the \emph{Klebanov-Witten theory}, \index{Klebanov-Witten}
has gauge group $\mathrm{SU}(N) \times \mathrm{SU}(N)$ with bifundamental matter fields $A_1$, $A_2$, $B_1$, $B_2$ \cite{Klebanov:1998hh}. 

A further deformation of this setup is obtained by adding $M$ D5-branes wrapping the shrinking $S^{2}$ of the conifold. These wrapped D5-branes appear as \emph{fractional} D3-branes, and the gauge theory is modified to $\mathrm{SU}(N+M)\times \mathrm{SU}(N)$. The theory is no longer conformal: the two gauge couplings now run logarithmically. On the gravity side, this flow is encoded in a warped throat geometry supported by non-trivial $3$-form fluxes $F_{3}$ and $H_{3}$. The effective D3-brane charge becomes a slowly varying function of the radial coordinate,
\begin{equation}
    N_{\text{eff}}(r) \;\sim\; N + \frac{3}{2\pi}\,(g_s M)^{2}\,\ln\left(\frac{r}{r_{0}}\right),
\end{equation}
and diminishes towards the infrared. The warp factor acquires a corresponding logarithmic correction, and takes the form, for sufficiently large $r$,
\begin{equation}\label{eq:kswarplog}
    \mathrm{e}^{-4A(r)}  =1+ \frac{27\pi \alpha'^2}{4 r^4}\Bigl[g_s N + \frac{3}{2\pi}(g_s M)^2\,\mathrm{log}\biggl (\frac{r}{r_0}\biggl) + \frac{3}{8\pi}(g_s M)^2\Bigr]\,,
\end{equation}
where $r_0$ is an exponentially small infrared scale: see \eqref{eq:approxwarping} and \S\ref{sec:alignment}. 

The resulting ten-dimensional supergravity background \eqref{eq:AdST11}, corresponding asymptotically to $\mathrm{AdS}_5 \times T^{1,1}$ with the warp factor given by \eqref{eq:kswarplog}, is the \emph{Klebanov-Tseytlin (KT) solution}\index{Klebanov-Tseytlin solution} \cite{Klebanov:2000nc}. The logarithmic running in \eqref{eq:kswarplog} reflects the slow depletion of the effective D3-brane charge as one moves towards the infrared along the throat.

While the KT solution correctly captures the ultraviolet behavior of the warped throat sourced by $N$ regular and $M$ fractional D3-branes on the singular conifold, it becomes singular in the deep infrared. The origin of this pathology is geometric: in the KT background the entire $T^{1,1}$ collapses at $r=0$, and the fluxes threading the shrinking cycles generate divergent curvatures. A smooth completion requires a modification of the conifold geometry itself: the singularity is deformed and replaced by a finite three-sphere. The singular conifold \eqref{eq:singconising} is replaced by its \emph{deformed} counterpart,
\begin{equation}\label{eq:defcondef}
    \sum_{i=1}^{4} z_i^{\,2}=\varepsilon^{2}\,,
\end{equation}
where the complex parameter $\varepsilon$ controls the size of the three-sphere at the tip. The deformation smooths out the conical singularity by keeping an $S^{3}$ of finite radius,
\begin{equation}
    R_{S^{3}} \sim \sqrt{g_s M \alpha'}\,,
\end{equation}
so that the total geometry remains non-singular. For $g_s M \gg 1$, the resulting geometry is weakly curved and therefore well-described by supergravity.

Klebanov and Strassler \cite{Klebanov:2000hb} constructed the full ten-dimensional background describing this situation. Their solution includes non-trivial NS-NS and R-R $3$-form fluxes $H_{3}$ and $F_{3}$ threading the $A$- and $B$-cycles of the deformed conifold, together with an induced self-dual $5$-form field strength. These fluxes generate a strongly warped metric that is smooth at the tip and asymptotes to the KT solution at large radius. Thus the \emph{Klebanov-Strassler (KS) geometry} provides the complete and non-singular supergravity dual of the cascading $\mathrm{SU}(N+M)\times\mathrm{SU}(N)$ gauge theory, capturing both the cascade of Seiberg dualities \cite{Seiberg:1994pq} beginning in the ultraviolet, and the eventual  chiral symmetry breaking 
in the 
deep 
infrared.\footnote{Although the KT and KS solutions agree in the ultraviolet, the ansatz for the KT solution respects a $U(1)$ isometry that is interpreted as an R-symmetry in the dual CFT, while the KS ansatz preserves only a $\mathbb{Z}_2 \subset U(1)$.  Comparing to \eqref{eq:defcondef}, which is invariant under $z_i \to \pm z_i$ but not under phase rotation of the $z_i$, one sees that the less-restrictive KS ansatz allows the possibility of deformation.}

\medskip

The ten-dimensional KS metric takes the warped form,
\begin{equation}
    \mathrm{d}s_{10}^{2}= h^{-1/2}(\tau)\,\mathrm{d}x_{4}^{2} + h^{1/2}(\tau)\,\mathrm{d}s^{2}_{CY}\,,
\end{equation}
where $\tau$ is a radial coordinate on the deformed conifold and $\mathrm{d}s_{CY}^{2}$ is the Ricci-flat metric on the internal space (see \cite{Herzog:2002ih} for details), with the warp factor now being determined by the integral expression
\begin{equation}\label{eq:KSwarpItau}
    h(\tau) = (g_s M)^2 2^{2/3}\varepsilon^{-8/3} I(\tau)\,,
    \qquad 
    I(\tau)=\int_{\tau}^{\infty} \mathrm{d}x\,
    \frac{x\coth x-1}{\sinh^{2}x}\,
    \bigl(\sinh(2x)-2x\bigr)^{1/3}.
\end{equation}
The warp factor $h(\tau)$ approaches the logarithmically-running KT form \eqref{eq:kswarplog} at large $\tau$, while remaining finite and smooth at the tip $\tau=0$.

\medskip

The KS throat plays a central role in flux compactifications of type IIB string theory. Warped throats arise naturally in compact Calabi-Yau orientifolds because conifold points are ubiquitous in complex structure moduli space: as a three-cycle collapses, its dual cycle can support background R-R flux, thereby generating a KS-like region. In a compact setting the throat does not extend to arbitrarily large radius, but is smoothly glued to the bulk Calabi-Yau at a finite matching scale, providing the analogue of a UV brane in the Randall-Sundrum picture, while the KS tip serves as the IR brane. The resulting matching conditions and their impact on the normalization of fields and fluxes will be analyzed in detail in \S\ref{sec:alignment}.

\begin{figure}[t!]
	\centering
	\includegraphics[width=0.5\linewidth]{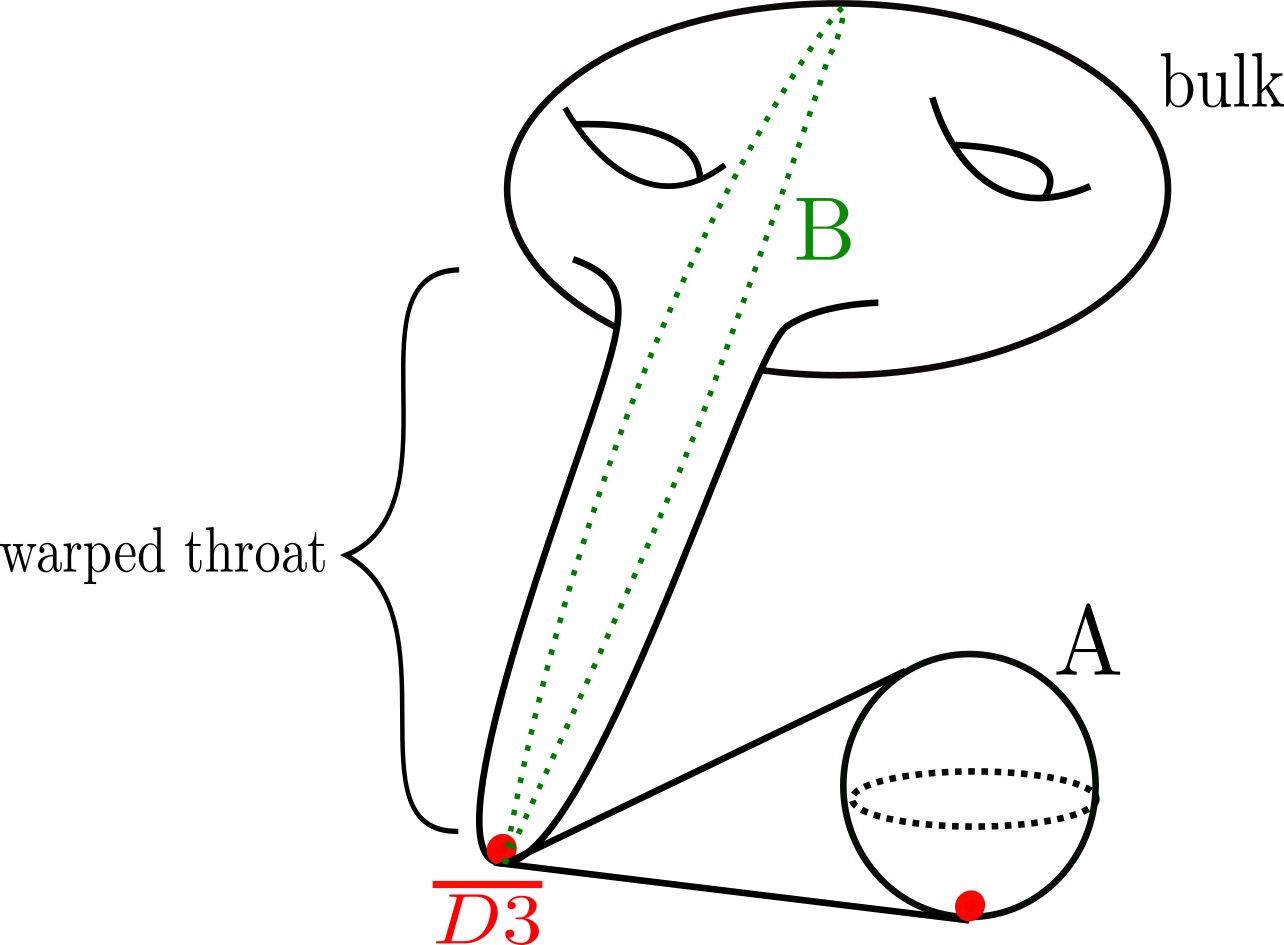}
	\caption{Sketch of the KS throat. The warping is generated by $F_3$ and $H_3$ fluxes threading the $A$- and $B$-cycles of the deformed conifold, producing a smooth geometry that caps off at a finite three-sphere. The warping grows stronger near the tip, yielding a highly redshifted region suitable for localized supersymmetry breaking. Figure courtesy of S.~Schreyer.}\label{fig:KSsolution} 
\end{figure}

In practice, KS throats can be embedded into a compact Calabi-Yau manifold  by stabilizing the complex structure moduli near a conifold singularity, where a three-cycle shrinks to zero size and can be threaded by flux. As we now demonstrate a simple toy example, the resulting flux-induced warping depends on the ratio of flux quanta and controls the hierarchy generated in the throat. 

Let us suppose we have a compact Calabi-Yau orientifold admitting a conifold locus.  In such a compact setting, the collapsing two-cycle at the conifold tip is part of a compact three-cycle, denoted the \emph{$A$-cycle}, which is Poincar\'{e} dual to the non-compact \emph{$B$-cycle} that carries the NS-NS $3$-form flux of the KS background, cf.~Fig.~\ref{fig:KSsolution}. This geometric structure allows the $3$-form fluxes to be interpreted in terms of periods over $(A,B)$ cycles, and these fluxes naturally generate a flux superpotential. We denote by $K$ the number of units of NS-NS $3$-form $H_{3}$ flux on the $B$-cycle,
\begin{equation}\label{eq:coniKA}
    K = -\frac{1}{(2\pi)^{2}\alpha'} \int_{B\text{-cycle}} H_{3}\, ,
\end{equation}
while we use $M$ to denote the number of units of RR flux $F_{3}$ threading the $A$-cycle,
\begin{equation}\label{eq:coniMB}
    M = \frac{1}{(2\pi)^{2}\alpha'} \int_{A\text{-cycle}} F_{3}\, .
\end{equation}
As emphasized already by GKP \cite{Giddings:2001yu}, this flux configuration leads directly to
the GVW superpotential \eqref{eq:GVW}
\begin{equation}\label{eq:GVWconi}
    W_{\text{GVW}} = \sqrt{\tfrac{2}{\pi}}\int_{X} G_{3} \wedge \Omega = \sqrt{\tfrac{2}{\pi}} (2\pi)^{2}\alpha' \, ( M \, \Omega_{B} - \tau\, K \,\Omega_{A} )\, ,
\end{equation}
where $\Omega$ is the holomorphic $3$-form and $\Omega_{A,B}$ are its periods.
For the conifold, the periods take a particularly simple form.  The vanishing three-cycle, the $A$-cycle, has period
\begin{equation}
    \Omega_{A} \equiv z_{\mathrm{cf}} \, ,
\end{equation}
while the period over the dual $B$-cycle has the logarithmic behavior characteristic of the conifold,
\begin{equation}
    \Omega_{B}  = \frac{z_{\mathrm{cf}}}{2\pi \mathrm{i}} \ln(z_{\mathrm{cf}}) + \text{holomorphic}\, .
\end{equation}
Inserting these expressions into \eqref{eq:GVWconi} yields the well-known stabilization of the conifold modulus $z_{\mathrm{cf}}$ at an exponentially small value,
\begin{equation}\label{eq:conifoldEXP}
    z_{\mathrm{cf}} \sim \exp\left( -\frac{2\pi K}{g_{s} M} \right)\, .
\end{equation}
This exponentially small modulus in turn controls the hierarchy of warp factors between
the KS tip (IR) and the compact bulk (UV).  In particular,
\begin{equation}\label{eq:approxwarping}
    \mathrm{e}^{A_{\rm IR} - A_{\rm UV}} \sim z^{1/3}_{\mathrm{cf}}\, ,
\end{equation}
demonstrating explicitly how fluxes supported on conifold cycles generate large warping.
A more precisely normalized version of \eqref{eq:approxwarping} is given in \eqref{eq:warpz} below.
 
We now turn to examining in detail how the Klebanov-Strassler geometry arises in concrete compact models.  In particular, we will show how conifold singularities and their associated warped throats emerge directly from the period structure analyzed in \S\ref{sec:flux_superpotential}.

\section{Conifolds in flux compactifications}\label{sec:conipfv}

Let us next show how a warped throat region of Klebanov-Strassler type can be engineered in an explicit flux compactification. We achieve this by stabilizing the complex structure moduli near a \emph{conifold singularity}. Such singularities arise at specific loci in the complex structure moduli space $\mathcal{M}_{\mathrm{cs}}(X)$ of a Calabi–Yau threefold $X$, where a collection of three-cycles $\{\mathcal{C}_i\}$, all in the same homology class $[\mathcal{C}] \in H_3(X,\mathbb{Z})$, shrink to zero volume. 

Mirror symmetry identifies the LCS patch of $\mathcal{M}_{\mathrm{cs}}(X)$ with the complexified K\"ahler cone $\mathcal{K}_{\widetilde{X}}$ of the mirror manifold $\widetilde{X}$. In this dual description, the conifold locus corresponds to a facet $\mathcal{K}_{\mathrm{cf}} \subset \partial \mathcal{K}_{\widetilde{X}}$ along which a nilpotent\footnote{We call a curve class $[\mathcal{C}_{\mathrm{cf}}]$ nilpotent 
if the GV invariants of curve classes $k[\mathcal{C}_{\mathrm{cf}}]$ vanish for all but finitely many $k>1$.  For the present discussion, we suppose that $[\mathcal{C}_{\mathrm{cf}}]$ is nilpotent of order one, i.e.~that the GV invariants of curve classes $k[\mathcal{C}_{\mathrm{cf}}]$ vanish for all $k>1$.} curve in $\widetilde{X}$ shrinks to zero volume.\footnote{Strictly speaking,  $\mathcal{K}_{\mathrm{cf}}$ denotes the \emph{interior} of the facet where the relevant curve shrinks.}

The shrinking curves $\mathcal{C}_{\mathrm{cf}}$ belong to an effective curve class $\tilde{\mathbf{q}}_{\mathrm{cf}} \in H_2(\widetilde{X}, \mathbb{Z}) \cap \mathcal{M}_{\widetilde{X}}$, which we refer to as the \emph{conifold class} \cite{Demirtas:2020ffz}. The modulus controlling the size of the shrinking cycle is given by the complex parameter $z_{\mathrm{cf}} = \tilde{\mathbf{q}}_{\mathrm{cf}} \cdot \mathbf{z}$, where $\mathbf{z}$ denotes the vector of complex structure moduli $z^a$. For convenience, we choose a basis such that $\tilde{\mathbf{q}}_{\mathrm{cf}} = (1,0,\ldots,0)$, and identify $z_{\mathrm{cf}} \equiv z^1$ as the \emph{conifold modulus}, with the remaining \emph{bulk} complex structure moduli denoted $z^\alpha$, $\alpha = 1,\ldots,h^{2,1}-1$.
 
In \S\ref{sec:CMS} we explained how to use computational mirror symmetry to evaluate the period vector \eqref{eq:PvecFirstDef} and the corresponding flux superpotential \eqref{eq:WfluxPeriods} in the LCS patches of the moduli space $\mathcal{M}_{\mathrm{cs}}(X)$. With only modest additional work, these techniques can be extended to determine the periods in the vicinity of a conifold locus, where a holomorphic curve $\mathcal{C}_{\mathrm{cf}}$ collapses and the corresponding complex structure modulus approaches $z_{\mathrm{cf}}\to 0$. Polynomial terms in the prepotential, $\mathcal{F}_{\mathrm{poly}}$ in \eqref{eq:fpoly}, remain regular in this limit. By contrast, the instanton-corrected part $\mathcal{F}_{\mathrm{inst}}$ in \eqref{eq:finst} involves polylogarithms whose convergence properties change as one moves away from the LCS regime. To obtain the correct behavior near the conifold point, we must therefore analytically continue $\mathcal{F}_{\mathrm{inst}}$ to the region $z_{\mathrm{cf}}\to 0$.
 
Because $\mathcal{C}_{\mathrm{cf}}$ is nilpotent of order one, 
only a single $z_{\mathrm{cf}}$-dependent instanton term survives in $\mathcal{F}_{\mathrm{inst}}$ in \eqref{eq:finst}, and the relevant piece of the prepotential reduces to
\begin{equation}
   \mathcal{F}_{\mathrm{inst}} \;\supset\; -\frac{1}{(2\pi \mathrm{i})^3}\,\mathscr{N}_{\tilde{\mathbf{q}}_{\mathrm{cf}}}\, \mathrm{Li}_3\bigl(\mathrm{e}^{2\pi \I \, z_{\mathrm{cf}}}\bigr)\, ,
\end{equation}
which must be analytically continued to obtain the correct period behavior near the conifold. This continuation produces the characteristic logarithmic monodromy in $z_{\mathrm{cf}}$ discussed above, and thereby captures the full structure of the periods in the conifold regime.  
 
Using Euler’s reflection formula, one finds the following expansion of the trilogarithm term near the conifold point \cite{Demirtas:2020ffz}:
\begin{equation}
	-\dfrac{1}{(2\pi \I)^{3}} \mathrm{Li}_3\Bigl(\mathrm{e}^{2\pi \I\, z_{\mathrm{cf}}}\Bigr)=  \dfrac{z_{\mathrm{cf}}^{2}}{4\pi \I}\,\ln(-2\pi\I \, z_{\mathrm{cf}}) - \dfrac{1}{(2\pi \I)^{3}}\sum_{n=0}^{\infty}\, \dfrac{\hat{\zeta}(n-3)}{n!}\,  (2\pi \I\, z_{\mathrm{cf}})^{n}\,,
\end{equation}
where $\hat{\zeta}(x) = \zeta(x)$ for $x \neq 1$ and $\hat{\zeta}(1) = 3/2$. Inserting this into the full prepotential \eqref{eq:prepotential}, we obtain a controlled expansion around small $|z_{\mathrm{cf}}|\ll 1$,
\begin{equation}\label{eq:FconiLCS} 
	\mathcal{F}(z_{\mathrm{cf}},z^{\alpha}) = n_{\mathrm{cf}}\, \dfrac{z_{\mathrm{cf}}^{2}}{4\pi \I}\,\ln(-2\pi\I \,z_{\mathrm{cf}})+\sum_{n=0}^{\infty}\, \dfrac{\mathcal{F}^{(n)}(z^\alpha)}{n!}\,  z_{\mathrm{cf}}^{n}\,,
\end{equation}
where $n_{\mathrm{cf}} = \mathscr{N}_{\tilde{\mathbf{q}}_{\mathrm{cf}}}$ is the GV invariant of the conifold class (corresponding to the number of conifolds), and the coefficients $\mathcal{F}^{(n)}(z^\alpha)$ are given by
\begin{equation}
\mathcal{F}^{(n)}(z^\alpha) = (\partial_{z_{\mathrm{cf}}}^n \mathcal{F}_{\mathrm{poly}})\bigl |_{z_{\mathrm{cf}}=0} - n_{\mathrm{cf}}\, \dfrac{\hat{\zeta}(3-n)}{(2\pi\mathrm{i})^{3-n}}\,- \dfrac{1}{(2\pi\mathrm{i})^{3-n}}\, \sum_{\tilde{\mathbf{q}}\neq \tilde{\mathbf{q}}_{\mathrm{cf}}}\,  \mathscr{N}_{\tilde{\mathbf{q}}}\, (\tilde{q}_{1})^n\, \mathrm{Li}_{3-n}\Bigl(\mathrm{e}^{2\pi \I\,\tilde{\mathbf{q}}\cdot \mathbf{z}}\Bigr)\bigl |_{z_{\mathrm{cf}}=0}
\end{equation}
in terms of the polynomial prepotential $\mathcal{F}_{\mathrm{poly}}$ defined in Eq.~\eqref{eq:fpoly}.

The above expansion provides a controlled way to compute the period vector near the conifold locus, to any desired order in $|z_{\mathrm{cf}}|$. In many cases, keeping only the leading terms in the expansion --- e.g., up to linear order in $z_{\mathrm{cf}}$ --- already yields a sufficiently accurate approximation for computing the flux superpotential and its derivatives.

\medskip

We now want to replicate the idea of perturbatively flat vacua introduced in \S\ref{sec:pfv}, but now in the presence of a conifold singularity.
In spirit, this analysis proceeds as in \S\ref{sec:pfv}, but will lead to slightly different choices of fluxes, which have to be carefully selected to achieve $|z_{\text{cf}}|\ll 1$ at the minimum.

Let us choose quantized fluxes \cite{Demirtas:2020ffz}
\begin{equation}\label{eq:coniFH}
\vec{f} = \left(P_0,P_a,0,M^a\right)^\top \, ,\; \vec{h} = \left(0,K_a,0,0^a\right)^\top\,.
\end{equation}
 At leading order in $z_{\mathrm{cf}}$,
 the superpotential can be written as
\begin{equation}\label{eq:WExpConi} 
    \sqrt{\tfrac{\pi}{2}} \cdot W(z^\alpha,z_{\mathrm{cf}},\tau)=W_{\mathrm{bulk}}(z^\alpha,\tau)+z_{\mathrm{cf}} W^{(1)}(z^\alpha,z_{\mathrm{cf}},\tau)+\mathcal{O}(z_{\mathrm{cf}}^2)\, ,
\end{equation}
in terms of
\begin{align}\label{eq:WfluxBulk}
    W_{\mathrm{bulk}}(z^\alpha,\tau)=&\frac{1}{2}M^a \widetilde{\kappa}_{a \beta\gamma}z^\beta z^\gamma-\tau K_\alpha z^\alpha+\left(P_\beta- M^a \tilde{a}_{a\beta}\right)z^\beta+\left(P_0-\frac{1}{24}M^a \tilde{c}'_a\right)  \nonumber \\
    &-\frac{1}{(2\pi)^2}\sum_{\tilde{\mathbf{q}}\neq \tilde{\mathbf{q}}_{\mathrm{cf}}} \mathscr{N}_{\tilde{\mathbf{q}}}\,\tilde{\mathbf{q}}_a M^a\,\mathrm{Li}_2(\mathrm{e}^{2\pi \I \tilde{\mathbf{q}}_\alpha z^\alpha})\,  ,
\end{align}
where in the constant term we defined
\begin{equation}
    \tilde{c}'_a\coloneqq \tilde{c}_a+n_{\mathrm{cf}} \delta_{a,1}\, .
\end{equation}
That is, compared to standard PFVs with the equivalent expression \eqref{eq:WfluxPoly}, the constant term $M^a \tilde{c}'_a$ is shifted by $n_{\mathrm{cf}} \delta_{a,1}$ due to taking the conifold limit.
Similarly, the superpotential contribution in \eqref{eq:WExpConi} at linear order in the conifold modulus $z_{\mathrm{cf}}$
is given by
\begin{align}\label{eq:W1coniLCS}
    W^{(1)}(z^\alpha,z_{\mathrm{cf}},\tau)=&-M\frac{n_{\mathrm{cf}}}{2\pi \I }\Bigl(\log(-2\pi \I  z_{\mathrm{cf}})-1\Bigr)-\tau K+ \widetilde{\kappa}_{1a\gamma} M^az^\gamma+P_1-\tilde{a}_{1b}M^b \nonumber\\
    &+\frac{1}{2\pi \I  }\sum_{\mathbf{q}\neq \tilde{\mathbf{q}}_{\mathrm{cf}}}
     \tilde{\mathbf{q}}_1 (\tilde{\mathbf{q}}_a\,M^a)\, \mathscr{N}_{\tilde{\mathbf{q}}}\,\mathrm{Li}_1(\mathrm{e}^{2\pi \I  \tilde{\mathbf{q}}_\alpha z^\alpha})\, ,
\end{align}
where we introduced
\begin{equation}\label{eq:coniMdef}    
    M\coloneqq  M^1\, ,\; K\coloneqq K_1\, .
\end{equation}

The $F$-flatness condition for the conifold modulus $z_{\mathrm{cf}}$ is satisfied for
\begin{equation}\label{eq:conifold_vev}
    \langle| z_{\mathrm{cf}}|\rangle= \frac{1}{2\pi}\exp\Biggl(-\frac{2\pi K'}{(g_s M) \, n_{\mathrm{cf}}}\Biggr)\, ,
\end{equation}
where we introduced
\begin{equation}\label{eq:Kprime} 
    K' = K-g_{s}\widetilde{\kappa}_{1a\beta}M^{a} \mathrm{Im}(z^{\beta}) \, ,
\end{equation}
neglecting terms of the order $\mathrm{e}^{2\pi \I  \tilde{\mathbf{q}}_\alpha z^\alpha}$. The expression \eqref{eq:conifold_vev} generalizes
the result \eqref{eq:conifoldEXP} obtained in the case of a single conifold modulus, and manifests the same qualitative behavior as \eqref{eq:conifoldEXP}: 
provided that $K'/M>0$, $z_{\mathrm{cf}}$ is stabilized at an exponentially small value, giving rise to a warped throat region.  
Note that 
\begin{equation}\label{eq:D3_charge_in_throats}
    Q_{\mathrm{flux}}^{\mathrm{throat}}=K'\,M>0
\end{equation}
is a measure for the D3-brane charge carried by fluxes residing in local conifold regions.
Importantly, \eqref{eq:D3_charge_in_throats} departs from the naive expectation laid out in  
\S\ref{sec:KS}: the D3-brane charge stored in the throat is not simply given by the product $M \cdot K$ of the integer fluxes on the $A$-cycle and $B$-cycle, 
as defined in \eqref{eq:coniKA} and \eqref{eq:coniMB}. When the throat is embedded into a compact Calabi-Yau threefold, the presence of additional bulk moduli stabilized near the LCS point alters the effective D3-brane charge through the mechanism encoded in \eqref{eq:Kprime}. This shift arises from additional  contributions to the superpotential coefficient $W^{(1)}$ given in \eqref{eq:W1coniLCS}: these new terms depend on the bulk moduli $z^{\alpha}$,
and are therefore absent in the 
non-compact KS solution. This modification is substantial, and we will analyze its implications in detail below.

Provided that the conifold modulus $ \langle|z_{\mathrm{cf}}|\rangle$
as computed from \eqref{eq:conifold_vev} is very small,
the $F$-term conditions for the remaining fields $(z^\alpha,\tau)$ can be studied independently of $z_{\mathrm{cf}}$
by working with $W_{\mathrm{bulk}}(z^\alpha,\tau)$ in \eqref{eq:WfluxBulk} \cite{Demirtas:2020ffz} (see also \cite{Alvarez-Garcia:2020pxd}).
Canceling the polynomial terms in \eqref{eq:WfluxBulk} requires
\begin{equation}
    P_\beta= M^a \tilde{a}_{a\beta}\kom P_0=\frac{1}{24}M^a \tilde{c}'_a\, .
\end{equation}
Then, we introduce the quantities
\begin{equation}
	N_{\alpha\beta}\coloneqq M^a \kappa_{a\alpha\beta}\kom p^\alpha\coloneqq  N^{\alpha\beta} K_{\beta}\, ,
\end{equation}
which leads to
\begin{align}\label{eq:WfluxBulk2}
    W_{\mathrm{bulk}}(z^\alpha,\tau)=&\frac{1}{2}N_{\alpha\beta}\, z^\alpha\, z^\beta -\tau K_\alpha z^\alpha -\frac{1}{(2\pi)^2}\sum_{\tilde{\mathbf{q}}\neq \tilde{\mathbf{q}}_{\mathrm{cf}}} \mathscr{N}_{\tilde{\mathbf{q}}}\,\tilde{\mathbf{q}}_a M^a\,\mathrm{Li}_2(\mathrm{e}^{2\pi \I \tilde{\mathbf{q}}_\alpha z^\alpha})\,  .
\end{align}
As proposed in \cite{Demirtas:2019sip,Demirtas:2020ffz},
by imposing
\begin{align}
    \label{eq:coniPFV}
    \det N\neq 0\, \kom\; \vec{p}\in \mathcal{K}_{\mathrm{cf}}\, \kom\; K_{\alpha}p^\alpha=0\, \kom\; \tilde{a}_{\alpha b}M^b\in \mathbb{Z}\, \kom\;  \tilde{c}'_a M^a\in 24\mathbb{Z}\, ,
\end{align} 
and by momentarily neglecting the exponential terms in $W_{\mathrm{bulk}}$ in \eqref{eq:WfluxBulk2},
the remaining $F$-term conditions $D_{\alpha}W=D_{\tau}W=0$ are satisfied along a one-dimensional locus given by
\begin{equation}
    z^\alpha=p^\alpha \tau\, .
\end{equation}
This locus defines a generalized PFV in the presence of a conifold, which we refer to as a \emph{conifold PFV}.\index{conifold PFV}
As described in \S\ref{sec:pfv}, the remaining flat direction is lifted by mirror worldsheet instanton corrections and stabilized by the racetrack mechanism using the effective superpotential
\begin{align}\label{eq:WfluxBulk3}
    W_{\mathrm{bulk}}^{\text{eff}}(z^\alpha=p^\alpha \tau,\tau)=& -\frac{1}{(2\pi)^2}\sum_{\tilde{\mathbf{q}}\neq \tilde{\mathbf{q}}_{\mathrm{cf}}} \mathscr{N}_{\tilde{\mathbf{q}}}\,\tilde{\mathbf{q}}_a M^a\,\mathrm{Li}_2(\mathrm{e}^{2\pi \I \tilde{\mathbf{q}}_\alpha p^\alpha \tau})\,  .
\end{align}
This can be computed to any desired degree using the methods of \cite{Demirtas:2023als}. 
We also define
\begin{equation}
    W^{\text{eff}}_{\text{bulk}}(\tau) = \sum_{N=1}^{\infty}\, W_N(\tau)\,,
\end{equation}
where the individual terms are given by
\begin{equation} 
   W_N(\tau)  \coloneqq -\frac{1}{(2\pi)^2} 
\sum_{\mathbf{p}_{\text{int}}\cdot\tilde{\mathbf{q}}=N}\mathscr{N}_{\tilde{\mathbf{q}}}\,\tilde{\mathbf{q}}_a M^a\,\text{Li}_2\Bigl(\mathrm{e}^{\frac{2\pi \I}{\mathtt{r}} N \tau}\Bigr). \label{eq:Widef}
\end{equation}
In this expression we use the integer vector $\mathbf{p}_{\mathrm{int}} \equiv \mathtt{r}\,\mathbf{p}$ obtained by rescaling the generally rational vector $\mathbf{p}$ by a positive integer $\mathtt{r}$ so that all of its components become integral. 
 
The effective superpotential \eqref{eq:WfluxBulk3}
generally receives corrections of order $\left(W^{\text{eff}}_{\text{bulk}}(\tau)\right)^2$. As such, care must be taken when the vacuum expectation value of the effective superpotential is not parametrically small. In the examples considered later, the flux superpotential is moderately small, but not exponentially suppressed, and so a controlled, multi-step approach is necessary to reliably identify true solutions to the full $F$-term conditions.
We describe this elaborate process in \S\ref{sec:csms}.

\section{Supersymmetry breaking from anti-D3-branes}\label{sec:KPV}

In this section we review the mechanism proposed by Kachru, Pearson, and Verlinde (KPV) \cite{Kachru:2002gs} for uplifting supersymmetric $\mathrm{AdS}_4$ vacua to metastable de Sitter solutions in type IIB flux compactifications by including anti-D3-branes in the internal space. 

In a generic compactification, the anti-D3-brane energy is parametrically too large to yield a controlled uplift. This is because $V_{\overline{\text{D3}}}$ is determined by the string scale, while $V_{\text{AdS}}$ is set by the much lower scales in the four-dimensional effective theory below the Kaluza-Klein threshold.  As a result, the system exhibits runaway behavior --- rapid decompactification --- rather than a stable uplifted vacuum (blue curve in Fig.~\ref{fig:ExamplePlot}). 

To achieve a controlled de Sitter vacuum, one must suppress the anti-D3-brane contribution, and this is where warping plays a crucial role. The idea is that anti-D3-branes placed at the tip of a strongly warped KS throat (see Fig.~\ref{fig:KSsolution}) as introduced in \S\ref{sec:KS} experience a large gravitational redshift, which suppresses their positive energy contribution to the four-dimensional potential. This redshift allows the uplift term to be comparable to the negative cosmological constant of the AdS vacuum without destabilizing the moduli. The KPV construction thus provides a concrete 
framework for realizing metastable de Sitter vacua from fully stabilized supersymmetric AdS minima.

\subsection{The Kachru-Pearson-Verlinde model}
\index{KPV}
\index{Supersymmetry breaking}

\begin{figure}[t!]
    \centering
    \includegraphics[width=0.4\textwidth]{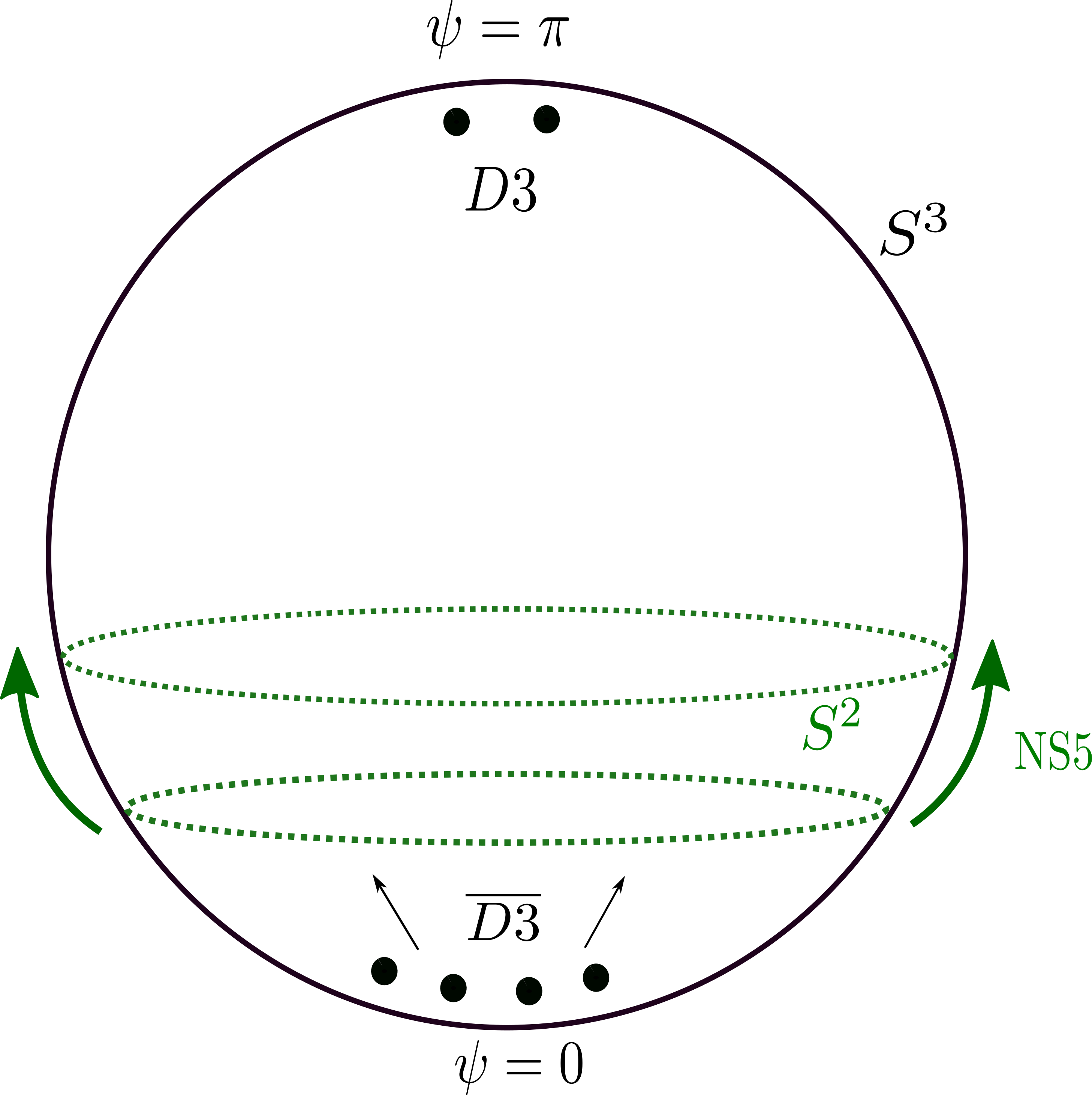}
    \caption{Illustration of brane–flux annihilation in the KPV mechanism. The anti-D3-branes polarize into an NS5-brane that wraps a contractible $S^2$ inside the $S^3$ at the tip of the KS throat. As the NS5-brane 
    moves across the $S^3$, its worldvolume flux decreases, annihilating anti-D3 charge against background three-form flux and ultimately producing $(M-p)$ D3-branes. Figure courtesy of S.~Schreyer.}
    \label{fig:KPV}
\end{figure}

When $p$ anti-D3-branes are localized at the apex of a warped throat, their dynamics are governed by strong background fluxes. A well-known phenomenon, first identified in \cite{Kachru:2002gs}, is that $p>1$ anti-D3-branes may polarize into an NS5-brane through the \emph{Myers effect} \cite{Myers:1999ps}\index{Myers effect}. In this configuration, the NS5-brane carries $p$ units of worldvolume flux and wraps an $S^2$ inside the finite $S^3$ residing at the tip of the deformed conifold: see Fig.~\ref{fig:KPV}. The metric on this $S^3$ can be expressed in standard polar coordinates as  
\begin{equation}
    R_{S^3}^2 \dif\Omega_3^2 
    = R_{S^3}^2\left( \dif\psi^2 + \sin^2(\psi)  \dif\Omega_2^2 \right)\, ,
\end{equation}
where\footnote{In this section, we set $\alpha'=1$.} 
\begin{equation}
    R_{S^3}= b_0 \sqrt{g_s M}\, , \qquad 
    b_0^2=\frac{2\, a_0^{1/2}}{6^{1/3}} \approx 0.93266\, , \qquad 
    a_0 = I(0) \approx 0.71805 \, ,
\end{equation}
and where the KS warp factor $I(\tau)$ was defined in \eqref{eq:KSwarpItau}. The coordinate $\psi$ parametrizes the latitude on the three-sphere, with $\psi=0$ located at the north
pole (the position of the anti-D3-branes) and $\psi=\pi$ at the south 
pole.

From this viewpoint, the system admits a non-trivial decay channel. The NS5-brane, initially located near $\psi=0$, can move across the $S^3$, sweeping out the two-sphere it wraps. During this process its induced anti-D3-brane charge diminishes through brane-flux annihilation, as the worldvolume flux on the NS5-brane is coupled to the background three-form flux. If the NS5-brane 
reaches the opposite pole, the configuration transitions to one containing $(M-p)$ D3-branes, and supersymmetry is restored. Whether the NS5-brane remains trapped in a metastable configuration or proceeds to the supersymmetric state is governed by the ratio $p/M$: only sufficiently small values of this ratio admit a metastable minimum, see below.

\begin{figure}[t!]
    \centering
    \includegraphics[width=0.8\textwidth]{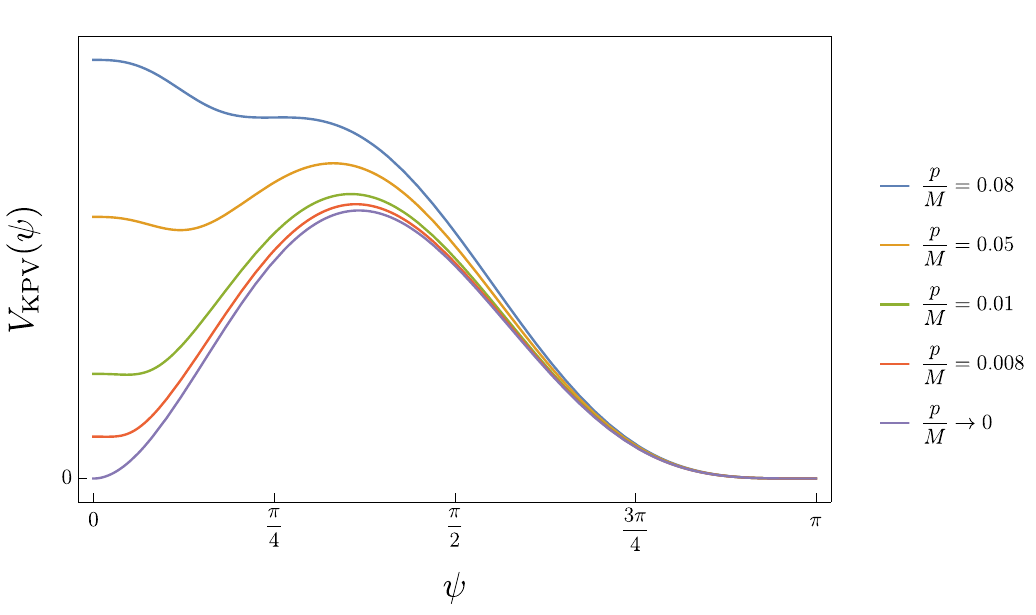}
    \caption{The NS5-brane polarization potential $V_{\text{KPV}}(\psi)$ \eqref{eq:vns5} (shown here in a suitably normalized form) for several representative values of the ratio $p/M$. As $p/M$ increases, the metastable minimum near $\psi=0$ becomes progressively shallower and eventually disappears when the minimum and the nearby maximum merge into an inflection point. This illustrates the loss of metastability at the critical value $(p/M)_\ast \approx 0.08$ identified by KPV. Figure from \cite{Hebecker:2022zme}, courtesy of S.~Schreyer.}
    \label{fig:KPVpot_lo}
\end{figure}

Kachru, Pearson, and Verlinde quantified this behavior by computing the effective potential $V(\psi)$ for the NS5-brane to leading order in $\alpha'$, obtaining \cite{Kachru:2002gs}
\begin{equation}\label{eq:vns5}
    V_{\mathrm{KPV}}(\psi) = \frac{T_3 p}{g_s} + \frac{T_3 M}{\pi g_s}\left(\sqrt{b_0^4 \sin^4 (\psi) + \left( \frac{\pi p}{M} -\psi +\frac{\sin(2\psi)}{2} \right)^2} -\psi +\frac{\sin(2\psi)}{2} \right)\,,
\end{equation}
where
the NS5-brane tension $\mu_5$ satisfies $T_3=4\pi^2\mu_5$.
The structure of this potential is illustrated in Fig.~\ref{fig:KPVpot_lo}.  
Expanding \eqref{eq:vns5} at small $\psi$ shows that the local minimum sits at
\begin{equation}
    \psi_{\mathrm{min}} \approx \frac{2\pi p}{b_0^{\,4} M}\, .
\end{equation}
As $p/M$ increases, the minimum becomes progressively shallower and eventually merges with the nearby maximum to form an inflection point.  
This marks the loss of metastability, which occurs for 
\begin{equation}\label{eq:KPV_constraint}
    \frac{p}{M} \gtrsim (p/M)_\ast \approx 0.08 \, .
\end{equation}
For larger values of $p/M$, the NS5-brane experiences no barrier and rolls directly toward the supersymmetric configuration. The KPV potential \eqref{eq:vns5} thus provides a simple criterion for assessing whether a warped throat with anti-D3-branes can support a metastable supersymmetry-breaking state, a requirement for many uplift mechanisms in flux compactifications.\footnote{Concerns about the metastability of the anti-D3-brane configuration have been analyzed in \cite{Bena:2009xk} and the subsequent literature: see \cite{VanRiet:2023pnx} for a review with complete references. In our view, the results of \cite{Dymarsky:2011pm,Michel:2014lva,Polchinski:2015bea,Armas:2018rsy} provide an accurate picture.}

We now return to the effective potential sourced by the anti-D3-brane.  
The NS5-brane potential \eqref{eq:vns5} provides a continuous description of the system as a function of the polarization angle~$\psi$ on the tip $S^3$ of the throat.  
At $\psi=0$ the NS5-brane collapses to $p$ anti-D3-branes, recovering their redshifted tension, while at $\psi=\pi$ the NS5-brane 
has swept over the three-sphere and annihilated $p$ units of anti-D3 charge against the background fluxes, yielding the supersymmetric configuration.  
Thus, the NS5-brane 
potential \eqref{eq:vns5} interpolates smoothly between the uplift energy contributed by the anti-D3-branes and the supersymmetric endpoint of brane–flux annihilation.

Evaluating the NS5-brane
potential \eqref{eq:vns5} at $\psi=0$ reproduces the leading contribution of the anti-D3-branes to the four-dimensional effective theory.  
At this point the square root structure in the NS5-brane 
action collapses, giving the familiar expression  
\begin{equation}\label{eq:vd3barsimple}
    V^{\overline{D3}}_{\text{KPV}} \sim \frac{2 p\, \mu_3}{g_s}\, h(0)^{-1}\, ,
\end{equation}
where $h(0)^{-1}$ is the warp factor \eqref{eq:KSwarpItau} at the 
tip.
In compactifications with a single shrinking three-cycle, this warp factor scales as (recall the discussion around Eq.~\eqref{eq:approxwarping})
\begin{equation}
    h(0)^{-1} \sim \exp\left(-\frac{2\pi K}{3 g_s M}\right)\, ,
\end{equation}
with $K$ and $M$ the quantized fluxes on the $B$- and $A$-cycles of the deformed conifold, see Eq.~\eqref{eq:coniKA} and Eq.~\eqref{eq:coniMB}. The overall normalization and its dependence on the moduli in general Calabi-Yau compactifications will be determined more carefully in \S\ref{sec:alignment}.

Before turning to the detailed ingredients of the uplift term, let us first set up the broader framework in which it appears.
In a compactification containing metastable anti-D3-branes, the full scalar potential reads 
\begin{equation}\label{eq:Vfull}
    V = V_F + V_{\overline{D3}}\,,
\end{equation}
where $V_F$ is the $F$-term potential in Eq.~\eqref{eq:vfsum}, and
\begin{equation}
    V_{\overline{D3}} = V^{\overline{D3}}_{\text{KPV}} + \Delta V^{\overline{D3}}_{(\alpha')^2} + \ldots\,,
\end{equation} 
The first contribution, $V^{\overline{\mathrm{D3}}}_{\mathrm{KPV}}$, corresponds to the potential of the metastable configuration computed within the approximations of KPV and extracted from the NS5-brane potential.  
Additional $(\alpha')^2$ corrections to the brane action, summarized in $\Delta V^{\overline{\mathrm{D3}}}_{(\alpha')^2}$, have been computed in \cite{Junghans:2022exo,Junghans:2022kxg,Hebecker:2022zme,Schreyer:2022len,Schreyer:2024pml} and will be discussed in \S\ref{sec:antiD3corr}.  
First, however, we will analyze the normalization and K\"ahler moduli dependence of the leading KPV uplift term. 

For the remainder, we will restrict to the case of a single anti-D3 brane (i.e., $p=1$): as explained in \S\ref{sec:towardsDS_part2}, candidate de Sitter vacua with $p \ge 2$ were beyond the reach of the search methods used in \cite{McAllister:2024lnt}.  The $p=1$ case is not strictly captured by the Myers-effect analysis of \cite{Kachru:2002gs}, but the absence of a polarization effect for $p=1$ potentially \emph{removes} a decay channel, without introducing a new one.\footnote{Related discussions appear in \cite{Michel:2014lva,Polchinski:2015bea}.}  Thus, we expect that --- at least for sufficiently large $g_s M$ --- the true decay rate $\Gamma_{p=1}$ for the $p=1$ case, and the approximation obtained by extrapolating the KPV result, are related as
\begin{equation} \label{eq:kpvp1}
    \Gamma_{p=1}  \le   \Gamma_{\text{KPV}}(p)\Bigr|_{p=1}\,.
\end{equation}
In particular, our use of the right-hand side of \eqref{eq:kpvp1} is a conservative estimate.

\subsection{Anti-D3-brane potential}\label{sec:alignment}

The KPV potential $V^{\overline{D3}}_{\text{KPV}}$ can be written in terms of the overall  Einstein-frame volume $\mathcal{V}_E$ of $X$ as \cite{Kachru:2002gs,Kachru:2003aw,Kachru:2003sx}
\begin{equation}\label{eq:anti-D3-potential00}
    V^{\overline{D3}}_{\text{KPV}} = \frac{c}{\mathcal{V}_E^{4/3}}\,,
\end{equation}
where $c$ is independent of the K\"ahler moduli. We now determine the normalization constant $c$, following Appendix B of \cite{McAllister:2024lnt}.

The K\"ahler potential \eqref{eq:detailedform2} can be written
\begin{equation}
    \mathcal{K} = -\log\bigl(16\mathcal{V}_{E}^2/g_s\bigr)-\log\bigl(||\Omega||^2\bigr)\, 
\end{equation}
where $\mathcal{V}_{E} = \mathcal{V}/g_s^{3/2}$, with $\mathcal{V}$ the volume of $X$ computed in type IIB string frame: see \eqref{eq:detailedform3}. We have introduced the notation $||\Omega||^2\coloneqq -i\int_X \Omega\wedge \overline{\Omega}$. At large complex structure, the volume $\widetilde{\mathcal{V}}$ of the mirror threefold $\widetilde{X}$ computed in type IIA string frame obeys 
\begin{equation}
	||\Omega||^2\approx 8 \widetilde{\mathcal{V}}\,.
\end{equation}

The $F$-term potential evaluated at the supersymmetric vacuum is
\begin{equation}
		|V_F| = \frac{3}{16}\cdot \frac{g_s}{\mathcal{V}_E^2}\cdot \frac{|W_0|^2}{||\Omega||^2}\,.
	\end{equation}
To compute the contribution of an anti-D3-brane, we write the ten-dimensional Einstein frame metric as
(cf.~\cite{Giddings:2005ff,Frey:2008xw})
\begin{equation}
		\mathrm{d}s^2=\frac{\ell_s^2}{4\pi}\frac{\mathrm{e}^{2A}}{t}\mathrm{d}x^2+\ell_s^2 \mathrm{e}^{-2A}\mathrm{d}s^2_{CY}\, ,\quad \mathrm{e}^{-4A}=\mathrm{e}^{-4A_0}+t-t_0\,.
\end{equation}
Here $\mathrm{e}^{-4A_0}$ is a reference solution for the warp factor, $t$ serves as a volume modulus, we have normalized the Calabi-Yau metric $\mathrm{d}s^2_{CY}$ to have unit volume, and we have defined
\begin{equation}
		t_0\coloneqq \int_X \sqrt{g_{CY}}\mathrm{e}^{-4A_0}\, .
\end{equation}
Upon Weyl rescaling and taking $\mathrm{d}x^2$ as the four-dimensional Einstein frame metric,  the four-dimensional reduced Planck mass obeys $M_\text{pl}=1$. For $t\gg 1$, we can relate $t$ and $\mathcal{V}_E$ via
\begin{equation}
		\mathcal{V}_E=\int_X \mathrm{d}^6y\sqrt{g_{CY}}\mathrm{e}^{-6A}\approx t^{\frac{3}{2}}\, ,
	\end{equation}
up to corrections of order $N_{\text{D3}}/\mathcal{V}_E^{\frac{2}{3}}$ --- cf.~\S\ref{sec:10deffects}. Here and below, we will assume that the warp factor obeys
\begin{equation}
		 \mathrm{e}^{3A_{IR}}\ll \frac{1}{\sqrt{\mathcal{V}_E}}\,.
\end{equation}

We now measure the volume of the A-cycle of the conifold in three ways. Using the unit-volume metric $\mathrm{d}s^2_{CY}$ we have
\cite{Koerber:2010bx}  
\begin{equation}\label{eq:cys3vol}
 2^{-3/2}\cdot \text{Vol}(A)_{CY}=\frac{\int_A \Omega}{||\Omega||}=: \frac{|z_{\text{cf}}|}{||\Omega||}=: \hat{z}_{\text{cf}}\, .
	\end{equation}
Using instead the ten-dimensional Einstein-frame metric to measure the physical volume, we find
	\begin{equation}\label{eq:physs3vol}
			\text{Vol}(A)= \mathrm{e}^{-3A_{IR}}	\text{Vol}(A)_{CY}\ell_s^3\, .
	\end{equation}
Finally, the Klebanov-Strassler solution \cite{Klebanov:2000hb} gives
	\begin{equation}\label{eq:ksvol}
		\text{Vol}(A)_{\text{KS}}=\frac{\Sigma^{\frac{3}{4}}}{2\pi\cdot \sqrt{6} } \left(\sqrt{g_s}M\right)^{\frac{3}{2}} \ell_s^3 \, ,
	\end{equation}
	where
	\begin{align}\label{eq:ksint}
    \Sigma&\coloneqq 2^{\frac{2}{3}}\int_0^\infty \mathrm{d}x \frac{x\, \coth(x)-1}{\sinh^2(x)}\bigl(\sinh(2x)-2x\bigr)^{\frac{1}{3}}\approx 1.13983\ldots \, .
	\end{align}
Equating \eqref{eq:cys3vol}, \eqref{eq:physs3vol}, and \eqref{eq:ksvol}, we learn that
	\begin{equation}\label{eq:warpz}
		\mathrm{e}^{3A_{IR}}\approx \gamma \cdot \frac{|\hat{z}_{\text{cf}}|}{(\sqrt{g_s}M)^{\frac{3}{2}}}\, , \quad \gamma\coloneqq \frac{8\pi\cdot \sqrt{3}}{\Sigma^{3/4}}\approx 39.4612\, .
	\end{equation}
We conclude that the warped anti-D3-brane potential is
\begin{equation}\label{eq:vd3barfull}
		V_{\overline{D3}} \approx 2T_{D3}\cdot \Bigl(\frac{\ell_s^2}{4\pi}\Bigr)^2\,\frac{\mathrm{e}^{4A_{IR}}}{t^2}=\frac{1}{4\pi} \frac{\mathrm{e}^{4A_{IR}}}{t^2}\equiv \frac{3\zeta}{32}\cdot\frac{g_s}{\mathcal{V}_E^{4/3}} \left( \frac{\mathcal{V}_E^{4/3}}{t^2} \cdot \frac{|\hat{z}_{\text{cf}}|^{\frac{4}{3}}}{(g_s M)^2}\right)\, ,
	\end{equation} 
where we have introduced
\begin{equation}\label{eq:zetadef}
\zeta \coloneqq  \frac{128}{\Sigma}\cdot\left(\frac{\pi}{3}\right)^{1/3}\approx 114.037\,.
\end{equation}
Thus, the normalization constant $c$ appearing in
\eqref{eq:anti-D3-potential00} is
\begin{equation}
     c = \eta\,\frac{z_{\mathrm{cf}}^{4/3}}{g_sM^2 \widetilde{\mathcal{V}}^{2/3}}
    \; ,\quad \eta \approx 2.6727\, .
\end{equation}

We now define the \emph{alignment} parameter,
\begin{equation}\label{eq:throat_tuning}
    \Xi\coloneqq  \dfrac{V_{\overline{D3}}}{|V_F|} = \frac{|z_{\text{cf}}|^{\frac{4}{3}}}{|W_0|^2}\frac{\mathcal{V}_E^{\frac{2}{3}}\widetilde{\mathcal{V}}^{\frac{1}{3}}}{(g_s M)^2}\cdot \zeta
\,.  
\end{equation}
Compactifications with $V_{\overline{D3}}/|V_F| < 1$ will have negative vacuum energy, while those 
with $V_{\overline{D3}}/|V_F| \gg 1$ will have such a large positive energy that they suffer rapid decompactification instabilities.
The cases of interest are in between, with 
\begin{equation}\label{eq:aligncond}
    1 \lesssim V_{\overline{D3}}/|V_F| \lesssim 3\,,
\end{equation} and we call configurations obeying \eqref{eq:aligncond} \emph{well-aligned}.
 
Equipped with the relation \eqref{eq:warpz}, we can compute two other quantities of interest.
The hierarchy of scales in the warped throat is given,
for $\Xi=1$, by 
\begin{equation}\label{eq:throathier}
        \left(\frac{r_{\text{IR}}}{r_\text{UV}}\right)^3\coloneqq \frac{\mathrm{e}^{3A_{\text{IR}}}}{\mathrm{e}^{3A_{\text{UV}}}}\approx 
        \frac{(3\pi)^{3/4}}{2^{15/4}}
         \times W_0^{3/2}\biggl(\frac{g_s}{\widetilde{\mathcal{V}}}\biggl)^{\frac{3}{4}}  \, .
    \end{equation}
The warped Kaluza-Klein scale is given,
in units where  
$M_{\text{pl}}=1$, by
\begin{equation}\label{eq:mwkk}
    m_{wKK}^2\approx \Sigma^{-\frac{1}{2}} \left(\frac{3}{8\pi}\right)^\frac{1}{3}\frac{|z_{\mathrm{cf}}|^\frac{2}{3}}{(\sqrt{g_s}M)^3\mathcal{V}_E^{\frac{2}{3}}\tilde{\mathcal{V}}^{\frac{1}{3}}}\, .
\end{equation}

\subsection{Corrections to the anti-D3-brane potential}\label{sec:antiD3corr}

In Chapter \ref{chap:quantumEFT} we systematically examined corrections to the data of the $\mathcal{N}=1$ supersymmetric effective theory.  The inclusion of an anti-D3-brane breaks the supersymmetry completely, so one might ask whether the elaborate structures laid out in Chapter \ref{chap:quantumEFT} survive this breaking: for example, why does it make sense to discuss non-perturbative corrections to the superpotential in a theory with anti-D3-brane supersymmetry breaking?

The answer rests on the fact that warping leads to exponential suppression of supersymmetry breaking effects.  The anti-D3-brane potential \eqref{eq:vd3barfull}
is proportional to the string scale,  
and an anti-D3-brane in an unwarped region would correspondingly have a string-scale supersymmetry-breaking energy.\footnote{To simplify this discussion we will set aside the dependence of \eqref{eq:vd3barfull} on $g_s$ and on the volume $\mathcal{V}_E$.}  
However, strings in a warped region have reduced tension, and the supersymmetry-breaking energy of an anti-D3-brane at the tip of a Klebanov-Strassler throat is set by the \emph{warped} string scale.  Correspondingly, loops in the low-energy theory are cut off at the warped string scale.

To understand how supersymmetry breaking propagates through the compactification, one can work in ten-dimensional supergravity: one begins with the supersymmetric ten-dimensional configuration obtained in the absence of an anti-D3-brane 
\cite{Moritz:2017xto,Gautason:2019jwq,Hamada:2019ack,Kachru:2019dvo,Bena:2019mte,Grana:2022nyp}, includes the anti-D3-brane as a local source, and uses the six-dimensional propagator to compute the effects induced elsewhere in the internal space.

A subtlety in this procedure is that because removing an anti-D3-brane changes the D3-brane tadpole, if two configurations $\mathcal{C}_1$ and $\mathcal{C}_2$ are related by adding an anti-D3-brane, with $\mathcal{C}_2:= \mathcal{C}_1+\overline{D3}$, the two cannot both obey Gauss's
law.  In constructing de Sitter vacua we will find cases for which $\mathcal{C}_2$ obeys Gauss's 
law, and is physical, whereas $\mathcal{C}_1$ is an unphysical but  mathematically convenient configuration termed a \emph{precursor} in \cite{McAllister:2024lnt}. We solve the supergravity equations of motion but not Gauss's
law to find the supersymmetric precursor configuration $\mathcal{C}_1$, and then solve the full system of equations, including Gauss's
law, to compute the propagation of supersymmetry breaking.

Because the finite Klebanov-Strassler throat region of a compactification is well-approximated by a finite portion of the
non-compact Klebanov-Strassler solution of supergravity, and Green's functions are computable in the
non-compact model, one can calculate the propagation of effects up and down the throat with excellent accuracy
\cite{Kachru:2003sx,Baumann:2006th,Baumann:2008kq,Flauger:2009ab,Baumann:2010sx,Gandhi:2011id}.  One finds that the perturbations due to the anti-D3-brane are localized in the infrared region \cite{Dymarsky:2011pm}.  

At the same time, the field configuration at the bottom of a Klebanov-Strassler throat can also be affected by perturbations of the throat solution that are sourced in the ultraviolet. Effects in the bulk of the Calabi-Yau, such as gaugino condensation on D7-branes, can source $3$-form flux \cite{Koerber:2007xk,Baumann:2010sx}, which induces perturbations of the throat.  A systematic method for computing such effects was developed in \cite{Baumann:2010sx}: just as one can propagate the effects of the anti-D3-brane toward the ultraviolet, one can propagate bulk effects toward the infrared.

Equipped with these results, one can show that the vacuum energy in a de Sitter vacuum is well-approximated by computing the anti-D3-brane energy and the moduli potential separately, and then adding the results \cite{Kachru:2003sx,Baumann:2010sx,Kachru:2019dvo}.

On the other hand, the supersymmetry-breaking effects display only a very limited degree of sequestering, in the spirit of \cite{Randall:1998uk}: although most of the effects of the anti-D3-brane are encoded in normalizable, infrared-localized perturbations, there are also couplings mediated by the overall volume mode \cite{Kachru:2007xp,Berg:2010ha} (see also \cite{Conlon:2011jq,Aparicio:2014wxa}).  This finding does not limit the validity of the de Sitter vacua discussed here, but it does impact the possibility of building models of a sequestered visible sector.

We return to the $\alpha'$ expansion of the anti-D3-brane potential.
The KPV uplift mechanism, as originally formulated in \cite{Kachru:2002gs}, relies on the leading-order DBI and CS actions of the anti-D3-brane, given schematically in \eqref{equ:DBIXX} and \eqref{eq:NCS}. However, these actions receive an infinite number of higher-derivative $\alpha'$ corrections. Such corrections arise not only from higher-order terms in the worldvolume expansion of the DBI and CS actions themselves, but also from additional bulk couplings involving curvature invariants --- most notably the $R^2$-type operators that we encountered previously in Eq.~\eqref{eq:DBIR2}. When evaluated on an anti-D3-brane sitting at the tip of the KS throat, these higher-derivative contributions modify the effective tension and hence the uplift energy, leading to a potential of the form  
\begin{equation}
    V_{\overline{D3}} = V^{\overline{D3}}_{\text{KPV}} + \Delta V^{\overline{D3}}_{(\alpha')^2} + \ldots\, .
\end{equation}
For example, the modification of the effective uplift energy in a warped throat by the $(\alpha')^2$ curvature terms in \eqref{eq:DBIR2} was computed in \cite{Hebecker:2022zme}, see also \cite{Junghans:2022exo,Junghans:2022kxg}, leading to (up to an overall moduli-dependent factor)  
\begin{equation}\label{eq:antibranecurvecorr}
   V_{\overline{D3}} \supset 2\, \dfrac{\mu_3}{g_s}\, h(0)^{-1} \left[ 1 - \frac{(4 \pi^2 \alpha')^2}{384 \pi^2} R_{a \alpha b}^{\quad \alpha} R_{\;\beta}^{a \;\; b \beta}\right] = 2\,\dfrac{\mu_3}{g_s}\, h(0)^{-1} \left[ 1 - \frac{c_{R^2}}{(g_s M)^2}\right]\, ,
\end{equation}
where $c_{R^2} = 3 \times 1.9747$,
$a,b$ denote directions transverse to the brane, and $\alpha,\beta$ label tangential directions. Additional $(\alpha')^2$ corrections to the KPV setup were analyzed in \cite{Schreyer:2022len,Schreyer:2024pml}, but the picture remains incomplete. In particular, several higher-derivative couplings in the D-brane effective action are only partially known (see e.g. \cite{Robbins:2014ara,Garousi:2014oya,Jalali:2015xca,Garousi:2015mdg,Jalali:2016xtv,BabaeiVelni:2016srs,Mashhadi:2020mzf,Garousi:2022rcv}), especially those involving R–R field strengths such as $F_3$. To address  
this gap, one can employ the M-theory framework of \cite{Bachas:1999um}, which provides a systematic method for deriving higher-derivative corrections to wrapped M5-brane actions. By lifting the problem to M-theory, computing the relevant terms, and then taking the limit to type IIB, one can compute the missing couplings to all orders in the string coupling expansion \cite{Compagnin:2026V1}.
Equipped with these corrections, one can assess their impact on
the structure and stability of the anti-D3-brane configuration.
These questions are the subject of ongoing work \cite{Compagnin:2026V1,Compagnin:2026V2}.

\subsection{Summary of anti-D3-brane uplift}

Subject to the approximations made in \cite{Kachru:2002gs}, the scalar potential for a compactification that contains a metastable anti-D3-brane in a Klebanov-Strassler throat is
\begin{equation}\label{eq:Vfullbis}
    V = V_F + V_{\overline{D3}}\,,
\end{equation}
where the $F$-term potential $V_F$ is given in Eq.~\eqref{eq:vfsum}, with
\begin{equation}\label{eq:anti-D3-potential0}
    V_{\overline{D3}}  = \frac{c}{\mathcal{V}_E^{4/3}}\; ,\quad c = \eta\,\frac{z_{\mathrm{cf}}^{4/3}}{g_sM^2 \widetilde{\mathcal{V}}^{2/3}}
    \; ,\quad \eta \approx 2.6727\, .
\end{equation}
There are corrections to \eqref{eq:anti-D3-potential0} in the $\alpha'$ expansion, and at small $g_s M$ the known corrections are found to be significant.
There are also corrections to the additive structure in \eqref{eq:Vfullbis}, but these are parametrically suppressed.

\chapter{Computational Strategies}\label{ch:CompComp}

The study of de Sitter vacua in flux compactifications of string theory demands computational tools capable of navigating the complexity of high-dimensional moduli spaces, intricate scalar potentials, and the subtle interplay of perturbative and non-perturbative effects. Analytical methods often provide guiding principles, yet they are rarely sufficient on their own. Computational strategies, therefore, become indispensable for testing stability, exploring high-dimensional field spaces, and evaluating complicated functions of the field values.

In this chapter, we present a set of computational approaches tailored to the challenges of string cosmology and de Sitter model building. We begin by outlining general algorithms for scanning large parameter spaces and identifying critical points of the potential. Special attention is given to methods that balance efficiency with robustness, including numerical solvers, stability analyses, and machine-assisted searches. We also discuss the practical limitations of current approaches and highlight strategies for overcoming them, particularly when dealing with the exponential complexity of the string landscape.

Ultimately, computational tools are essential to bringing theory in contact with testable predictions. They allow us to chart otherwise inaccessible regions of the landscape, offering insights into the viability and abundance of de Sitter vacua within string theory.

\section{Computational compactifications}\label{sec:CompComp}

We begin with a brief overview of the computational compactification tools required for constructing explicit examples of string vacua with all moduli stabilized. 

In \S\ref{sec:scheme} we outlined how the moduli fields $\Phi^A$, the superpotential $W(\Phi^A)$, and the K\"ahler potential $\mathcal{K}(\Phi^A,\bar{\Phi}^A)$ can in principle be computed from the geometric data $\mathscr{G}$ of a compactification. As discussed in \S\ref{sec:scheme}, the computation of $\mathscr{L}$ is necessarily carried out to some order in one or more expansion schemes. In Chapter~\ref{chap:quantumEFT} and Chapter~\ref{chap:modulistabilization} we described in detail how this computation proceeds in the parameter regime defined in Eq.~\eqref{eq:control_regime}, cf.~\S\ref{sec:quantum_pert}. Finally, in \S\ref{sec:leadingEFT} we introduced the leading-order EFT and explained how $\mathscr{L}_{\text{l.o.}}$, given in Eq.~\eqref{eq:loldef}, can be computed in terms of discrete data $\mathscr{D}_{\mathbb{Z}}$ summarized in Eq.~\eqref{eq:discrete_data}. But in practice, there is no hope of carrying out such computations by hand.

First of all, the vacuum structures we seek are more prevalent when the number of moduli, $N$, is large.  In particular, a strongly warped but weakly-curved Klebanov-Strassler throat, which can be accurately treated using supergravity, requires a large amount of D3-brane charge dissolved as flux.  At least in the Sen limit, orientifold configurations that can counterbalance such charges occur only when $h^{1,1}+h^{2,1}$ is large.

Second, our list of requirements, such as \eqref{eq:control_regime}, is long, and we aim to fulfill them all by trying many candidate compactifications, i.e., choices of discrete data $\mathscr{D}_{\mathbb{Z}}$, and rejecting those that fail. We will see that --- with the methods available in \cite{McAllister:2024lnt} --- it was necessary to study more than $10^8$ candidates.  Perhaps one can be much more clever in the future, but it is still extremely implausible that a fully-fledged de Sitter vacuum of string theory will arise from a handful of random guesses for $\mathscr{D}_{\mathbb{Z}}$.  Thus, one is faced with carrying out the computation of $\mathscr{L}_{\text{l.o.}}(\Phi_A;\mathscr{D}_{\mathbb{Z}})$ for exponentially many choices of $\mathscr{D}_{\mathbb{Z}}$, and furthermore with searching for de Sitter minima therein, in the resulting $N$-dimensional potentials.

Computerization is clearly essential.  Let us now explain what is involved.  

The first stage involves evaluating $\mathscr{L}_{\text{l.o.}}(\Phi_A;\mathscr{D}_{\mathbb{Z}})$ for a choice of 
$\mathscr{D}_{\mathbb{Z}}$.
Computing the classical data of $\mathscr{L}_{\text{l.o.}}$ at leading order in $\alpha'$, in the setting of Calabi-Yau hypersurfaces in toric varieties, involves constructing suitable triangulations of lattice polytopes, identifying orientifold involutions, computing triple intersection numbers, computing (approximations to) the K\"ahler cone, and evaluating the periods of $\Omega$.  Computing genus-zero worldsheet instanton corrections to the effective theory involves computing the periods of $\widetilde{\Omega}$ on the mirror threefold and extracting Gopakumar-Vafa invariants.  Computing the Euclidean D3-brane superpotential involves computing the Hodge numbers of divisors.
All of the above steps are mathematically well-defined and could in principle be undertaken using various pieces of off-the-shelf computational algebraic geometry software, without much guidance from physical reasoning.  (In contrast, computing string loop corrections in this setting will require major advances in the worldsheet formalism, and will not be discussed here.)  
 
In practice we have found that preexisting software was able to carry out all the above steps only for cases with $N \lesssim 10$ moduli.  In particular, many general-purpose algorithms have cost that is exponential in $N$.  
The software package \texttt{CYTools} was designed to overcome these limitations, and contains many new algorithms that exploit toric structures.  With \texttt{CYTools} one can execute the above steps for any polytope in the Kreuzer-Skarke list, i.e.,~for $h^{1,1}$ or $h^{2,1}$ as large as 491.  Of course, some problems are large for any enumerative algorithm: for example, there are $\gtrsim 10^{276}$ inequivalent triangulations of four-dimensional reflexive polytopes \cite{Big:ICY}, and one cannot study all of them, but one can now efficiently construct and analyze any desired member of this set.

Given a theory  $\mathscr{L}_{\text{l.o.}}(\Phi_A;\mathscr{D}_{\mathbb{Z}})$, the next stage involves surveying choices of  $\mathscr{D}_{\mathbb{Z}}$ in order to realize a desired structure.   Most importantly, one needs to choose quantized fluxes that solve the Diophantine equation \eqref{eqref:pfvcond} defining a PFV, and select cases in which the resulting vacuum is well-controlled.
For this process to succeed one needs to invest heavily in optimizing for speed.

The final stage, for a theory  $\mathscr{L}_{\text{l.o.}}(\Phi_A;\mathscr{D}_{\mathbb{Z}})$ with all the desired structures,
consists of finding a local minimum of the scalar potential.  
This is once again straightforward in principle, and can be attempted through well-known numerical root-finding methods.  However, the basins of attraction for the de Sitter vacua we have constructed are quite small, and we were able to find them only with extensive analytical guidance.  Specifically, we used a series of analytic approximations to compute starting points for root-finding, and by thus starting close enough to an actual solution we were able to find it numerically.

In the remainder of this section we outline computational methods for constructing mirror pairs of Calabi-Yau threefolds via toric geometry, and for numerically stabilizing moduli in such compactifications.

\section{Toric geometry and geometric moduli spaces}\label{sec:toric}
\index{Toric geometry}

Explicit constructions of Calabi–Yau manifolds provide a concrete foundation for realizing type IIB flux compactifications in settings where the full effective theory can be computed. While differential geometry provides the natural starting point, it quickly becomes unwieldy for explicit computations. To make progress toward tractability, we turn to algebraic geometry, where the relevant spaces, known as algebraic varieties, are constructed as the vanishing loci of polynomial equations. A key advantage of this approach is that algebraic varieties are determined by the ring of polynomial functions defined on them.  This allows us to translate geometric and topological questions into problems in commutative algebra, where powerful computational tools and algorithms are available: for instance, those implemented in software like \texttt{Macaulay2} \cite{M2}.

In practice, we often go a step further by focusing on varieties that contain an algebraic torus as a dense open subset. For such spaces, known as toric varieties, the defining polynomials have a particularly simple structure: they can be put in one-to-one correspondence with lattice points inside certain polyhedral cones. Constructing a toric variety then amounts to gluing together these cones along shared faces, a process encoded combinatorially by a fan --- a collection of cones closed under taking faces and intersections. The fan’s combinatorial structure fully determines much of the topological and geometric data of the associated toric variety, including how various components intersect and fit together \cite{cox2011toric,hori2003mirror}.

Constructions of Calabi-Yau threefolds are most tractable as hypersurfaces in compact toric varieties associated to triangulations of reflexive polytopes. A foundational result due to Batyrev~\cite{Batyrev:1993oya} establishes that for any suitable triangulation of a four-dimensional reflexive polytope, one can construct a smooth Calabi-Yau threefold as a hypersurface in the associated toric variety. Building on this insight, Kreuzer and Skarke classified all four-dimensional reflexive polytopes~\cite{Kreuzer:2000xy}, producing what is now known as the KS database: a complete enumeration of 473{,}800{,}776 reflexive polytopes.

The combinatorial richness of this ensemble is immense. It has been estimated that the number of topologically distinct Calabi-Yau threefold hypersurfaces arising from triangulations of the Kreuzer-Skarke polytopes is bounded above by $10^{428}$~\cite{Demirtas:2020dbm}. This vast landscape is dominated by geometries with large Hodge number $h^{1,1}$, and in particular by those with $h^{1,1} = 491$. While it remains an open question whether Calabi-Yau hypersurfaces constructed in this way form a representative sample of all Calabi-Yau threefolds, they constitute by far the largest explicitly known set, and are amenable to detailed analysis via combinatorial and algorithmic methods. The KS database is publicly available, and practical computations, such as triangulating polytopes and extracting topological data of hypersurfaces, can be efficiently performed using the open-source Python package \texttt{CYTools}~\cite{Demirtas:2022hqf}.

\subsection{Toric varieties from polytope triangulations}
\index{Toric varieties}

Below, we collect some elementary definitions and formulae necessary for the construction of Calabi-Yau hypersurfaces from $4$-dimensional reflexive polytopes in the KS database \cite{Kreuzer:2000xy}; see \cite{Altman:2014bfa,Braun:2017nhi,Demirtas:2018akl} for more details.\index{KS database}

\subsubsection*{Toric varieties and fans}

Toric geometry provides a powerful bridge between algebraic geometry and combinatorics, allowing complicated geometric spaces to be described in terms of simple data from convex geometry. A \emph{toric variety} is, by definition, an algebraic variety that contains an algebraic torus $(\mathbb{C}^*)^n$ as a dense open subset, together with an action of the torus that extends to the whole variety. For comprehensive discussions of the construction of toric varieties from combinatorial data, see~\cite{danilov1978geometry,oda1983convex,ewald1996combinatorial,hori2003mirror,cox2011toric,Fulton2016}.

The essential idea is that toric varieties are built from combinatorial data associated with cones in a lattice. One begins with a lattice $N \cong \mathbb{Z}^n$, together with its dual lattice $M = \mathrm{Hom}(N,\mathbb{Z})\cong \mathbb{Z}^n$. A strongly convex rational polyhedral cone\footnote{It is polyhedral since it is generated by finitely many lattice vectors $\{\nu_1,\dots,\nu_r\}$. It is rational because the generators $\nu_i$ lie in the lattice $N$. It is strongly convex in the sense that $\sigma \cap (-\sigma) = \{0\}$, i.e., it contains no non-trivial linear subspaces.} $\sigma \subset N_\mathbb{R}\coloneqq N \otimes_\mathbb{Z} \mathbb{R}$  is a subset of the form
\begin{equation}
    \sigma = \left\{ \sum_{i=1}^r \lambda_i \nu_i \,\middle|\, \lambda_i \in \mathbb{R}_{\geq 0} \right\},
\end{equation}
where each $\nu_i \in N$ is a lattice vector, i.e., has integer coordinates.
Given a cone $\sigma \subset N_\mathbb{R}$, we define its \emph{dual cone} $\sigma^\vee$ in the dual space $M_\mathbb{R} \coloneqq M \otimes_\mathbb{Z} \mathbb{R}$
\begin{equation}
    \sigma^\vee = \{ u \in M_\mathbb{R} \mid \langle u, v \rangle \geq 0 \;\;\forall v \in \sigma \}\, .
\end{equation}
So $\sigma^\vee$ is the set of linear functionals that are nonnegative on $\sigma$.
Intersecting with the dual lattice $M$ gives a semigroup $S_\sigma \coloneqq  \sigma^\vee \cap M$ that is closed under addition. Concretely, this is the set of all lattice points inside the dual cone.

Every such cone determines a semigroup algebra $\mathbb{C}[\sigma^\vee \cap M]$: this is the $\mathbb{C}$-vector space with basis elements $\{\chi^u : u \in S_\sigma\}$, and with multiplication defined by
\begin{equation}
    \chi^u \cdot \chi^{u'} = \chi^{u+u'}\, .
\end{equation}
Thus, the exponents $u \in M$ label monomials, and the cone condition ensures that only nonnegative linear combinations appear, so the ring is well-defined and finitely generated. 
Finally, the affine toric variety associated with $\sigma$ is
\begin{equation}
    U_\sigma = \mathrm{Spec}\,\mathbb{C}[\sigma^\vee \cap M]\, .
\end{equation}
So, in practical terms: the geometry is encoded by the combinatorics of lattice points in the dual cone, and the coordinate ring is generated by the corresponding monomials.

More general toric varieties are obtained by gluing together such affine pieces, corresponding to a collection of cones, denoted by $\Sigma$, that is closed under taking faces and intersections. Such a collection is called a \emph{fan}, and the associated $n$-dimensional toric variety is denoted $\mathbb{P}_\Sigma$.

For each 1-dimensional cone or \emph{ray} $\rho_{i}\in\Sigma$ related to one of the lattice vectors $\nu_{i}$, we introduce a homogeneous coordinate $z_{i}$.
These coordinates satisfy certain equivalence relations
\begin{equation}
    (\lambda^{q_{1}}z_{1},\lambda^{q_{2}}z_{2},\ldots ,\lambda^{q_{n}}z_{n})\sim (z_{1},z_{2},\ldots,z_{n})\kom \lambda\in \mathbb{C}^{*}=\mathbb{C}\setminus\lbrace 0\rbrace 
\end{equation}
which are obtained from conditions satisfied by the lattice vectors
\begin{equation}
    \sum_{i=1}^{r}\, q_{i} \nu_{i}=0\, .
\end{equation}
For $r$ vectors $\nu_{i}$ in an $n$-dimensional ambient space, there are $r-n$ independent linear relations of the above form associated with vectors $Q_i^{\alpha}$, $\alpha=1,\ldots ,r-n$. The matrix obtained from these vectors is called a \emph{weight matrix} (see e.g.~\cite{Kreuzer:2008nu}) or \emph{GLSM charge matrix} (see e.g.~\cite{hori2003mirror}), with the individual entries referred to as \emph{weights} or \emph{charges}.

Thus, to define a toric variety one specifies a fan $\Sigma$ in $N_\mathbb{R}$, and the geometry and topology of $\mathbb{P}_\Sigma$ can then be read off from the combinatorial properties of the cones in $\Sigma$. This correspondence makes toric geometry especially amenable to explicit computation, and it provides a natural framework for constructing Calabi–Yau hypersurfaces \cite{Batyrev:1993oya,Kreuzer:2000xy}, as we now explain.

\subsubsection*{Calabi-Yau threefolds from polytope triangulations}

Since our interest is in Calabi–Yau threefolds, we will restrict attention to this case, while noting that the same construction extends straightforwardly to other dimensions.
One begins with two reflexive polytopes $\Delta$ and $\Delta^{\circ}$ based on two $4$D lattices $M\cong \bZ^{4}$ and $N\cong \bZ^{4}$ with a pairing $\langle\cdot\, ,\cdot\rangle$ so that $\Delta\in M_{\mathbb{R}}=M\otimes \mathbb{R}$ and $\Delta^{\circ}\in N_{\mathbb{R}}=N\otimes \mathbb{R}$ satisfy
\begin{equation}
    \langle\Delta ,\Delta^{\circ}\rangle\geq -1\, .
\end{equation}
Reflexivity of the polytopes means that both $\Delta$ and $\Delta^{\circ}$ are lattice polytopes containing only the origin in their interior. 
To construct a toric variety from the polytope data,
we associate to the polytope $\Delta^{\circ}$ a fan $\Sigma$ in the following way. Reflexivity of $\Delta^{\circ}$ implies that the origin of $N$ is the unique interior lattice point of $\Delta^{\circ}$. We denote all other lattice points of $\Delta^{\circ}$ by $\nu_{i}$. The latter correspond to primitive generators of the rays of the fan $\Sigma$. The cones of $\Sigma$ are given by a triangulation of $\Delta^{\circ}$, i.e., special subsets of the $\nu_{i}$ with each containing the generators of a cone. 
We will focus on so-called \emph{fine, regular, star triangulations}\footnote{A triangulation is \emph{fine} if all points not interior to facets appear as vertices of a simplex.
Further, it is \emph{star} if the origin is a vertex of each full-dimensional simplex.
\emph{Regularity} implies that $\Sigma$ is the normal fan of a polytope and essentially ensures that $\mathbb{P}_{\Sigma}$ and $X$ are projective, see \cite{de2010loera}.} (FRSTs), whose associated fan describes a simplicial toric fourfold denoted $V=\mathbb{P}_{\Sigma}$. One can then introduce weighted, homogeneous coordinates $z_{i}$ on $V$ as described above.

Within the four-dimensional toric variety $V$, a Calabi-Yau threefold $X$ is found as the zero locus of a polynomial
\begin{equation}\label{eq:CYMonomials}
    f=\sum_m\, c_m\, p_m\; ,\quad p_{m}=\prod_{i}\, z_{i}^{\langle m,\nu_{i}\rangle+1}\kom m\in \Delta\cap M\, ,
\end{equation}
where the $p_m$ are monomials in the $z_i$, and the $c_m$ are coefficients that depend on the complex structure moduli.  
The individual monomials $p_{m}$ appearing in $f$ are encoded by $\Delta$, also called the \emph{Newton polytope} of the hypersurface.
The codimension-one subvariety defined by $f=0$ inside $V$ has vanishing first Chern class, and therefore defines a Calabi–Yau hypersurface~\cite{Batyrev:1993oya,Kreuzer:2000qv,Kreuzer:2000xy}. Moreover, although $V$ itself need not be smooth, the \emph{generic} Calabi-Yau hypersurface associated to any FRST is smooth~\cite{Batyrev:1993oya}.

A pair of dual polytopes $(\Delta^\circ, \Delta)$ provides the framework for constructing mirror pairs of Calabi–Yau threefolds. Starting from a triangulation of the polar dual polytope $\Delta$, one defines a four-dimensional toric variety $\widetilde{V}$. The generic anti-canonical hypersurface in $\widetilde{V}$ then yields a Calabi–Yau threefold $\widetilde{X}$, which is the \emph{mirror} of $X$. Mirror symmetry is reflected in the exchange of Hodge numbers: $h^{2,1}(X) = h^{1,1}(\widetilde{X})$ and $h^{1,1}(X) = h^{2,1}(\widetilde{X})$.

\subsubsection*{Example of a 2D reflexive polytope} 

\begin{figure}[t!]
\centering
\includegraphics[scale=0.37]{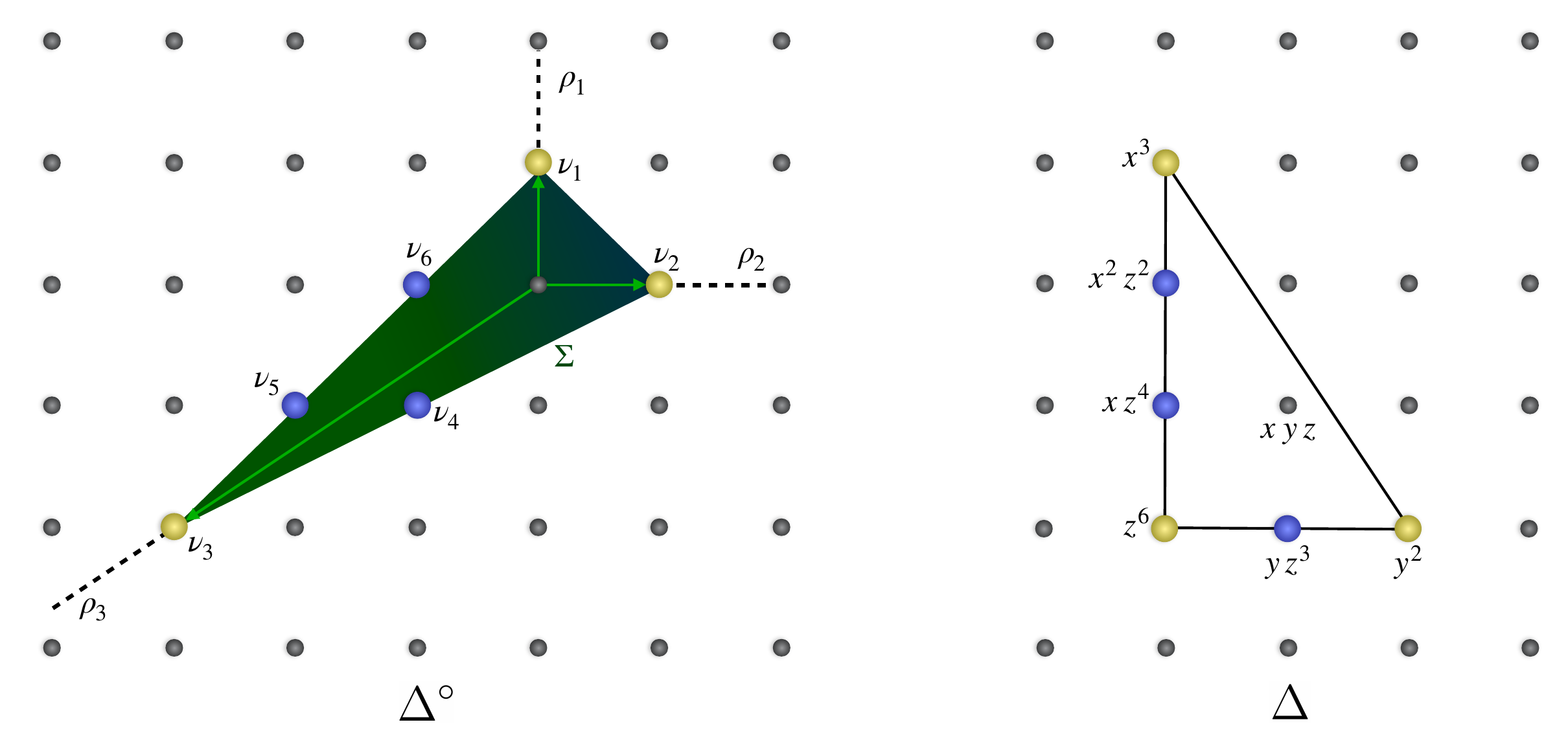}
\caption{Example of a 2D reflexive polytope whose triangulation encodes the weighted projective space $\mathbb{P}[2,3,1]$.}\label{fig:FRST2DExF10} 
\end{figure}
 
Before proceeding, it is useful to illustrate these ideas with a simple example of a triangulated polytope. We consider the two-dimensional reflexive polytope $\Delta^{\circ}$ shown on the left panel of Fig.~\ref{fig:FRST2DExF10}.\footnote{This polytope appears repeatedly as a sub-polytope in the Kreuzer-Skarke database \cite{Kreuzer:2000qv} and in related constructions, see for instance \cite{Huang:2018gpl,Huang:2018esr,Huang:2018vup,Huang:2019pne,Crino:2022zjk}. } 
The polytope $\Delta^{\circ}$ is specified by the lattice points
\begin{equation*}
\lbrace \nu_{1},\ldots, \nu_{6}\rbrace=\biggl \{ \biggl (\begin{array}{c}
        0 \\ [0.1em]
        1
        \end{array} \biggr ),\biggl (\begin{array}{c}
        1 \\ [0.1em]
        0
        \end{array} \biggr ),\biggl (\begin{array}{c}
        -3 \\ [0.1em]
        -2
        \end{array} \biggr ),\biggl (\begin{array}{c}
        -1 \\ [0.1em]
        -1
        \end{array} \biggr ),\biggl (\begin{array}{c}
        -2 \\ [0.1em]
        -1
        \end{array} \biggr ),\biggl (\begin{array}{c}
        -1 \\ [0.1em]
        \hphantom{-}0
\end{array} \biggr )\biggl \}
\end{equation*}
together with its dual $\Delta$ defined by the points
\begin{equation*}
\lbrace\tilde{\nu}_{1},\ldots,\tilde{\nu}_{6}\rbrace=\biggl \{ \biggl (\begin{array}{c}
    \hphantom{-}1 \\ [0.1em]
    -1
    \end{array} \biggr ),\biggl (\begin{array}{c}
    -1 \\ [0.1em]
    -1
    \end{array} \biggr ),\biggl (\begin{array}{c}
    -1\\ [0.1em]
    \hphantom{-}2
    \end{array} \biggr ),\biggl (\begin{array}{c}
    \hphantom{-}0 \\ [0.1em]
    -1
    \end{array} \biggr ),\biggl (\begin{array}{c}
    -1 \\ [0.1em]
    \hphantom{-}0
    \end{array} \biggr ),\biggl (\begin{array}{c}
    -1 \\ [0.1em]
    \hphantom{-}1
\end{array} \biggr )\biggl \}\, .
\end{equation*}
 
A fan $\Sigma$, or equivalently an FRST, generated from the vertices $\{\nu_{1}, \nu_{2}, \nu_{3}\}$ of the polytope is indicated by the shaded region in the left panel of Fig.~\ref{fig:FRST2DExF10}.\footnote{This choice of triangulation leaves residual $\mathbb{Z}_{2}$ and $\mathbb{Z}_{3}$ orbifold singularities in the ambient toric variety $\mathbb{P}_{\Sigma}$; see, for example, \cite{Mayrhofer:2010}.  
These singularities can be resolved by refining the triangulation to include the lattice points interior to the facets, namely $\{\nu_{4}, \nu_{5}, \nu_{6}\}$, which introduces three exceptional $\mathbb{P}^{1}$’s and corresponds to a finer subdivision of the polytope.}
The green arrows in the figure then encode the resulting equivalence relations
\begin{equation}
    q_{1}+2q_{3}=0\kom q_{2}+3q_{3}=0
\end{equation}
which in terms of homogeneous coordinates $z_{i}$ leads us to
\begin{equation}
    \left (\lambda^{2} z_{1},\lambda^{3}z_{2},\lambda z_{3}\right )\sim (z_{1},z_{2},z_{3})\kom \lambda\in \mathbb{C}^{*}\, .
\end{equation}
This is the weighted projective space $\mathbb{P}_{\Sigma}=\mathbb{P}[2,3,1]$.

Using this toric data, we can construct a Calabi--Yau hypersurface in the weighted projective space $\mathbb{P}[2,3,1]$, with the allowed monomials determined by the lattice points of the dual polytope $\Delta$, shown in the right panel of Fig.~\ref{fig:FRST2DExF10}. The most general hypersurface equation compatible with this data is
\begin{equation}\label{eq:HypersurfaceP231} 
    f=c_{1}y^{2}+c_{2}z^{6}+c_{3}x^{3}+c_{4}yz^{3}+c_{5}xz^{4}+c_{6}x^{2}z^{2}\equiv 0\, ,
\end{equation}
where the coefficients $c_i$ parameterize the complex structure moduli of the hypersurface. The compact Calabi-Yau hypersurface defined by \eqref{eq:HypersurfaceP231} is a complex curve of genus one with vanishing first Chern class, and hence describes a two-torus. This becomes manifest after a suitable change of coordinates, which brings the equation into the standard Weierstrass form (see, for example, \cite{Klevers:2014bqa,Abbasi:2025lvn} for details and \cite{Huang:2018gpl,Weigand:2018rez} for general discussions),
\begin{equation}\label{eq:WeierstrassFormGeneral} 
    y^{2}=x^{3}+f\, xz^{4}+g\, z^{6}\, .
\end{equation}
In this representation, the functions $f$ and $g$ encode the complex structure data of the elliptic curve, making explicit its interpretation as an elliptic fibration.

\subsection{Geometric phases and flop transitions}\label{sec:ekc}
\index{Geometric phases}
\index{Flop transitions}

\begin{figure}[t!]
    \centering
    \includegraphics[width=0.3\linewidth]{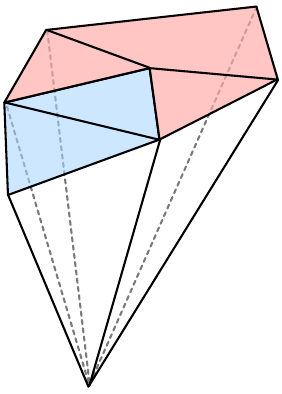}
    \caption{A schematic illustration of an extended K\"ahler cone is shown. Each cone corresponds to the K\"ahler cone of a different Calabi-Yau. Cones shaded in blue denote Calabi–Yau threefolds realized as hypersurfaces in toric varieties, while cones shaded in red correspond to geometries not manifestly obtained through birational transformations of a toric ambient space, but nevertheless accessible via flops of curves within the Calabi-Yau. Figure taken from \cite{Gendler:2022ztv}.}
    \label{fig:extended_kahler}
\end{figure}

To understand the stabilization of K\"ahler moduli, we will first need to map out the K\"ahler moduli space.  In general, the moduli space decomposes into distinct regions, called \emph{phases}, and a central task is to chart the transitions connecting these components.
We will only consider geometric phases, i.e.~those for which a sigma model with Calabi-Yau target space is an appropriate starting point.

We begin by recalling a few standard facts about Calabi–Yau threefolds and their K\"ahler moduli spaces. As explained above, each fine, regular and star triangulation (FRST) of a fixed reflexive polytope $\Delta^\circ$ defines a toric fourfold in which a smooth Calabi-Yau hypersurface $X$ can be obtained via the methods described above. Each geometry obtained in this manner will be referred to as a \emph{FRST phase}. Requiring the Hermitian metric on $X$ to be positive restricts the K\"ahler form $J$, introduced in~\eqref{apA:eqnA.68}, to lie within the \emph{K\"ahler cone} $\mathcal{K}_X \subset H^{1,1}(X) \cap H^2(X,\mathbb{R})$, recall the discussion in \S\ref{sec:kahlercoord}. The dual cone of $\mathcal{K}_X$ is the \textit{Mori cone} $\mathcal{M}_X$ of $X$ containing effective curve classes $\mathcal{C}$ of $X$. The volume of such effective curves $\mathcal{C}$ can be expressed as,
\begin{equation}
	\text{Vol}(\mathcal{C})=\int_\mathcal{C}\, J\, .
\end{equation}
Therefore, on the facets of $\mathcal{K}_X$, one or more effective curves collapse to zero size. For facets of $\mathcal{K}_X$ where only effective curves shrink, but no divisors shrink, this signals the onset of a so-called \emph{birational transition}.\footnote{The shrinking curves are always isolated rational curves \cite{atiyah1958analytic,Candelas:1989js} with normal bundles isomorphic to $\mathcal{O}(-1)\oplus \mathcal{O}(-1)$, $\mathcal{O}\oplus \mathcal{O}(-2)$, or $\mathcal{O}(1)\oplus \mathcal{O}(-3)$ \cite{laufer1972rational,katz1992gorenstein}.}
That is, as we continue past such a facet of $\mathcal{K}_X$, we obtain a birational morphism called a \emph{flop transition}
\begin{equation}\label{eq:flop}
	X\backslash \cup_{\mathcal{C}\in [\mathcal{C}]}\mathcal{C}\rightarrow X'\backslash \cup_{\mathcal{C}'\in [\mathcal{C}']}\mathcal{C}'\, .
\end{equation}
Here, the resulting Calabi-Yau threefold $X'$ is generally topologically distinct, but birationally equivalent to $X$. Since $[\mathcal{C}']\simeq -[\mathcal{C}]$ is now an effective curve class in $X'$, the Mori cone $\mathcal{M}_{X}$ differs from $\mathcal{M}_{X'}$. By systematically performing all possible flops \eqref{eq:flop} and adjoining the resulting K\"ahler cones $\mathcal{K}_X$ and $\mathcal{K}_{X'}$ along their shared facet, one arrives at the \emph{extended K\"ahler cone}\index{extended K\"ahler cone}
\begin{align}\label{eq:Kext}
	\mathcal{K}_{\star} \coloneqq \bigcup_{X\in [X]_\text{b}} \mathcal{K}_X \, ,
\end{align}
where $[X]_\text{b}$ is the birational equivalence class of $X$. The extended K\"ahler cone $\mathcal{K}_{\star}$ describes the K\"ahler moduli space of birationally equivalent Calabi-Yau threefolds, and takes the form of a $h^{1,1}$-dimensional cone. This cone is assembled from a collection of smaller cones, each representing the K\"ahler cone of a distinct geometric phase and connected to its neighbors by birational transformations, as illustrated in Fig.~\ref{fig:extended_kahler}. As $h^{1,1}$ increases, the number of phases grows rapidly, and constructing $\mathcal{K}_{\star}$ becomes increasingly challenging once $h^{1,1} > 2$. 

\begin{figure}[t!]
    \centering
    \includegraphics[width=0.99\linewidth]{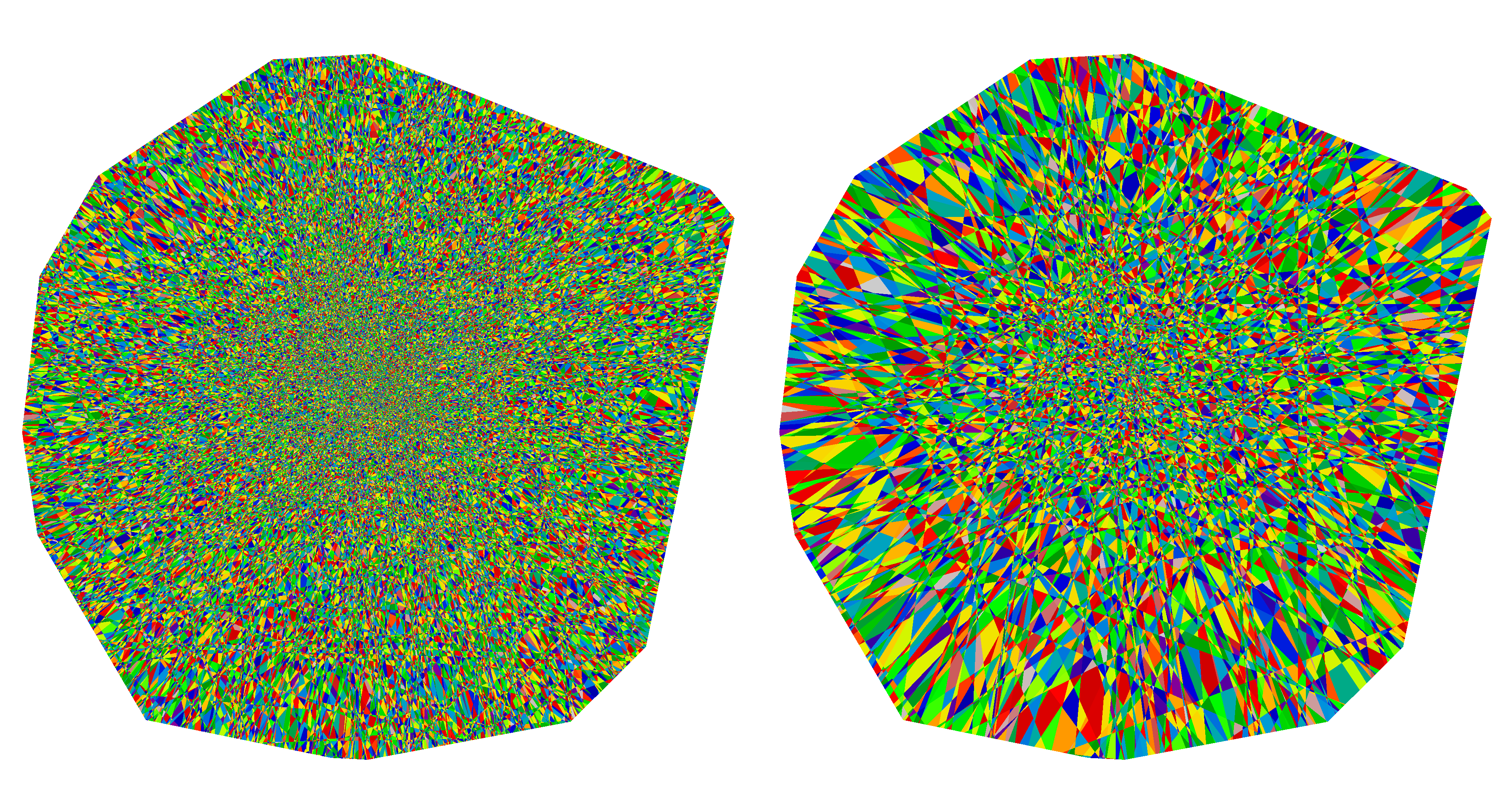}
    \caption{Shown on the left is a two-dimensional cross section of the FRST subregion of the extended K\"ahler cone $\mathcal{K}_{\star}$ of the largest four-dimensional reflexive polytope, and on the right the result after combining regions that yield the same Calabi-Yau. Each colored region on the left (respectively right) corresponds to a distinct FRST (respectively Calabi-Yau). The outermost boundaries mark the region beyond which the triangulation ceases to be an FRST. Figure taken from \cite{Demirtas:2020dbm}.}
    \label{fig:491fan}
\end{figure}

We display a two-dimensional cross section of the extended K\"ahler cone $\mathcal{K}_{\star}$ of the largest polytope $\Delta^\circ$ in the KS database, with $h^{1,1}=491$, in Fig.~\ref{fig:491fan}. 
On the left, each colored region corresponds to a distinct FRST of $\Delta^\circ$. However, not every FRST gives rise to a distinct Calabi–Yau geometry: substantial redundancy arises when passing from polytope triangulations to Calabi–Yau hypersurfaces, see e.g.~\cite{MacFadden:2023cyf,Gendler:2023ujl,MacFadden:2024him}. Wall’s theorem\index{Wall’s theorem} \cite{Wall1966} states that the homotopy type of a compact, simply connected Calabi–Yau threefold with torsion-free homology is completely determined by its Hodge numbers, triple intersection numbers, and second Chern class. For hypersurfaces in toric varieties, the Hodge numbers follow directly from polytope data, while the triple intersection numbers and second Chern class depend only on the induced triangulations of the two-faces. Consequently, FRSTs of $\Delta^\circ$ that agree on their two-face triangulations yield topologically equivalent Calabi–Yau threefolds. By identifying chambers of the extended K\"ahler cone $\mathcal{K}_{\star}$ that correspond to equivalent Calabi–Yau geometries, we arrive at the simplified plot shown on the right of Fig.~\ref{fig:491fan}. For practical purposes, such as optimizing within moduli space or searching for minima of the scalar potential \eqref{eq:Vfull}, it is advantageous to work in this reduced space, where trivial redundancies have been removed. 

Computing all FRSTs for a given $\Delta^{\circ}$, and hence constructing all FRST phases, 
becomes 
increasingly expensive as
$h^{1,1}(X)$ --- and correspondingly, the number of lattice points in $\Delta^{\circ}$ --- increases.
All FRSTs for polytopes with $h^{1,1}(X) \le 6$ were obtained in \cite{Altman:2014bfa}.
Using the software \texttt{CYTools} \cite{Demirtas:2022hqf}, and removing redundancies by the methods of \cite{MacFadden:2023cyf}, one can efficiently compute all inequivalent FRSTs of larger polytopes, including some cases with $h^{1,1}$ as large as 30.   
In any event, for constructing flux vacua, computing all triangulations is often unnecessary, and a more relevant benchmark is the ability to compute and explore many triangulations of a polytope with $h^{1,1} \sim \mathcal{O}(100)$, which is now extremely fast with
\texttt{CYTools}.

In general, exploring toric flops connecting FRST phases does not capture the entirety of $\mathcal{K}_{\star}$.
First of all,
given a reflexive polytope $\Delta^{\circ}$, FRST phases can adjoin more general toric phases that
arise as hypersurfaces in toric varieties that are defined not by FRSTs of the points of $\Delta^{\circ}$, but by suitable triangulations of the associated vector configuration.
Following \cite{Berglund:2016yqo,MacFadden:2025ssx}, we call these
\emph{vex phases}.
Second, there exist can flops intrinsic to the threefold that do not admit an embedding into a pair of birationally equivalent toric fourfolds; we refer to these as \emph{non-toric flops}. Such flops can lead to new Calabi–Yau threefolds without any known hypersurface embedding in a toric variety,\footnote{Often, however, one can find a toric description at higher codimension.} which we correspondingly call \emph{non-toric phases}.
Constructing the extended K\"ahler cone by assembling and adjoining both toric and non-toric phases is a central result of \cite{Gendler:2022ztv} (see Fig.~\ref{fig:extended_kahler}).
For the purpose of these lectures, FRST phases will suffice.

In what follows we restrict attention to the subcone of the extended K\"ahler cone $\mathcal{K}_{\star}$ generated by toric flops connecting FRST phases. While the extension of our analysis to the full cone is conceptually straightforward, it requires the development of additional tools. For example, under a flop transition the intersection numbers and second Chern classes of $X'$ can be computed from those of $X$ via
\begin{align}\label{eq:flop_formulas_kappa_c2}
	&\kappa_{ijk}'=\kappa_{ijk}-\sum_{\mathcal{C}\in [\mathcal{C}]}\,\mathcal{C}_i \mathcal{C}_j \mathcal{C}_k\; ,\quad c'_i=c_i+2\sum_{\mathcal{C}\in [\mathcal{C}]}\,\mathcal{C}_i\, ,
\end{align}
where $\mathcal{C}_i \coloneqq \int_X [\mathcal{C}]\wedge [\omega_i]$ with $\{\omega_i\}_{i=1}^{h^{1,1}(X)}$ of $H^2(X,\mathbb{Z})$, recall \eqref{eq:jdef}.\footnote{We note that  $H^2(X,\mathbb{Z})\simeq H^2(X',\mathbb{Z})$ and $H_2(X,\mathbb{Z})\simeq H_2(X',\mathbb{Z})$ because divisors in $X$ are identified with divisors in $X'$.}
By contrast, computing GV invariants in non-FRST phases would require a 
generalization 
of~\cite{Hosono:1993qy}, implemented in the computational framework of~\cite{Demirtas:2023als}.

\subsection{Prime toric divisors and their topologies}\label{sec:DivTops}

A key object for the remainder of these notes are generalizations of codimension-one subvarieties, namely the \emph{divisors}\index{Divisors}.\footnote{For a brief overview of their properties, see Appendix A of~\cite{Altman:2014bfa}.} For our purposes it suffices to focus on a special class of divisors, namely \emph{prime toric divisors}\index{Prime toric divisors}\index{Toric divisors} $\tilde{D}_{i}$, each of which is defined by the vanishing of a homogeneous coordinate, $\tilde{D}_{i} = \{z_{i} = 0\}$ (see e.g. \cite{Demirtas:2018akl} for details). The subset of prime toric divisors that intersect $X$ transversely corresponds to lattice points lying in faces of $\Delta^{\circ}$ of dimension $\leq 2$. Intersecting these loci with the Calabi–Yau hypersurface $X$ defines divisors $D_i \in H_4(X,\mathbb{Z})$, each representing a four-cycle in $X$ dual to a  
$2$-form $\omega^i\in H^{1,1}(X,\mathbb{Z})$. As we restrict to favorable polytopes and geometries, all such prime toric divisors remain irreducible on $X$. Consequently, $H_{4}(X,\mathbb{Z})$ is generated by any basis constructed from the set $\{D_i\}$, with $i = 1, \ldots, h^{1,1}(X) + 4$.

The Hodge numbers of a divisor $D$ of a threefold are $h^{1,1}(D)$ and
\begin{equation}
h^{\bullet}(D)\coloneqq\Bigl(h^{0,0}(D), h^{0,1}(D), h^{0,2}(D)\Bigr)\,.
\end{equation}
A rigid divisor $D_{\text{rig}}$ is one obeying
\begin{equation}
h^{\bullet}(D_{\text{rig}})=(1,0,0)\,,
\end{equation} 
and the rigidity condition alone does not determine 
$h^{1,1}(D_{\text{rig}})$.

To compute the Hodge numbers, we adopt the method developed in~\cite{Braun:2015pza,Braun:2017nhi}, which we briefly summarize here. As discussed above, each prime toric divisor $D_i$ of the Calabi-Yau hypersurface corresponds to a lattice point $\nu_i \in \Delta^\circ$. The Hodge numbers $h^{0,p}(D_i)$ are determined by the combinatorial properties of $\nu_i$, in particular by its location within the polytope $\Delta^\circ$. Depending on whether $\nu_i$ lies at a vertex, in the interior of an edge, or in the interior of a higher-dimensional face of $\Delta^\circ$, one can read off some of the corresponding Hodge data. The relevant formulas, originally derived in~\cite{Batyrev:1993oya,danilov1987newton}, are
\begin{enumerate}
\item \emph{Rigid divisors:} A toric divisor $D_i$ is said to be rigid if it admits no holomorphic deformations which is equivalent to
\begin{equation}
    \ell^{*}(\Theta)=0\:,
\end{equation}
where $\ell^{*}(\Theta)$ denotes the number of lattice points in the interior of the face $\Theta$ of the dual polytope $\Delta$.  
Here, $\Theta$ is the face dual to the face of $\Delta^{\circ}$ that contains the lattice point $\nu_i$ associated with the divisor $D_i$.
\item \emph{Deformation divisors:} Divisors that possess complex structure deformations but no Wilson-line moduli are characterized by $h^{0,2}(D_{i})>0$ and $h^{0,1}(D_{i})=0$. Such divisors correspond to lattice points $\nu_i$ that are vertices of the polytope $\Delta^{\circ}$. Their Hodge numbers are determined by the combinatorics of the dual polytope and are given by
\begin{equation}
    h^{0,1}(D_{i})=0\kom h^{0,2}(D_{i})=\ell^{*}(\Theta^{[3]})\:,
\end{equation}
where $\Theta^{[3]}$ denotes the three-dimensional face of $\Delta$ dual to the vertex $\nu_i = \Theta^{\circ[0]}$.
\item \emph{Wilson divisors:} Finally, divisors $D_i$ associated with lattice points $\nu_i$ lying in the interior of a one-dimensional face $\Theta^{\circ[1]}$ of the polytope $\Delta^{\circ}$ carry Wilson-line moduli but no complex structure deformations. Their Hodge numbers are determined by the dual two-dimensional face $\Theta^{[2]}$ of $\Delta$ and satisfy
\begin{equation}
    h^{0,1}(D_{i})=\ell^{*}(\Theta^{[2]})\kom h^{0,2}(D_{i})=0\, .
\end{equation}
\end{enumerate}
The conditions discussed above can be checked straightforwardly using \texttt{Sage} \cite{sagemath}. Once $h^{0,1}(D_i)$ and $h^{0,2}(D_i)$ have been determined, the remaining Hodge numbers follow from the Euler characteristic $\chi(D_i)$ and the arithmetic genus $\chi_0(D_i)$, which are given by
\begin{align}
    \label{eq:EulerDiv} \chi(D_i)&=\int_{D_i} c_2(D_i)=2h^{0,0}-4h^{0,1}+2h^{0,2}+h^{1,1}\, ,\\
    \label{eq:AGDiv} \chi_{0}(D)&=\dfrac{1}{12}\int_{D_i}\left (c_1(D_i)^2+c_2(D_i) \right)=h^{0,0}-h^{0,1}+h^{0,2}\, .
\end{align}
The right-hand sides of these expressions can be computed directly from the Calabi--Yau data using the adjunction formula.  
In particular, one has $c_{2}(X)=c_{2}(D_i)-c_{1}(D_i)^{2}$ and $c_{1}(D_i)=-\iota^*D_i$, which implies
\begin{equation}
    \int_{D_i} c_2(D)_i= \int_{D_i}\, \bigl(D_i^{2}+c_{2}(X)\bigr) \kom \int_{D_i}\bigl(c_1(D_i)^2+c_2(D_i)\bigr)=\int_{D_i}\, \bigl(2D_i^{2}+c_{2}(X)\bigr)\:.
\end{equation}
By introducing
\begin{equation}
    \kappa_{iii} =  \int_{X}\, D_i^3\; ,\quad  c_i = \int_{D_i}c_{2}(X)\, ,
\end{equation}
triple self-intersection number and second Chern class of $D_i$,
the Euler characteristic and arithmetic genus of $D_i$ can be written as
\begin{equation}\label{eq:chiDexp}
    \chi(D_{i}) = \kappa_{iii}+c_i\kom \chi_0(D_{i}) = \dfrac{1}{12}\bigl (2\kappa_{iii}+c_i\bigl )
\end{equation}
These relations can then be solved for the remaining Hodge numbers, $h^{0,0}$ and $h^{1,1}$ as
\begin{align}
    h^{0,0}=\chi_{0}(D_{i})+h^{0,1}-h^{0,2}\kom h^{1,1}=\chi(D_{i})-2\chi_{0}(D_{i})+2h^{0,1}\, .
\end{align}
Finally, we note that the Euler characteristic $\chi(D_i)$ enters directly into the definition of the leading-order K\"ahler coordinates $T_i^{\text{l.o.}}$ through the tree-level $(\alpha')^2$ correction appearing in \eqref{eq:detailedform4}.

\section{Stabilizing complex structure moduli}\label{sec:csms}

Given a Calabi–Yau orientifold $X/\mathcal{I}$ with a chosen conifold curve $\mathcal{C}_{\mathrm{cf}}$, the next step is to specify flux vectors that stabilize the complex structure moduli exponentially close to the conifold point, while yielding a small flux superpotential $W_0$. Following~\cite{Demirtas:2020ffz,McAllister:2024lnt}, this is achieved by constructing conifold PFVs, as defined in \S\ref{sec:conipfv}.

To reiterate, a conifold PFV is specified by vectors $\vec{M}\in\mathbb{Z}^{h^{2,1}(X)}$, $\vec{K}\in\mathbb{Z}^{h^{2,1}(X)}$, $\vec{p}\in\mathbb{Q}^{h^{2,1}(X)}$ that satisfy \eqref{eq:coniPFV}.  The conditions for solutions can be expressed in terms of $N_{\alpha\beta}\coloneqq M^a \kappa_{a\alpha\beta}$, $p^\alpha\coloneqq  N^{\alpha\beta} K_\alpha$, and $p^a\coloneqq (0,p^\alpha)$, as
\begin{subequations}
\begin{align}
    \label{eq:detN_conition}
    &\det N\neq 0\, ,\\
    \label{eq:p_in_Kcf}
    &\vec{p}\in \mathcal{K}_{\mathrm{cf}}\, ,\\
    \label{eq:Diophantine_eq}
    &K_{\alpha}p^\alpha=0\, ,\\
    \label{eq:integrality_condition_I}
    & \mathbb{A}_{\alpha b}M^b\in 2\mathbb{Z}\, ,\\
    \label{eq:integrality_condition_II}
    & \tilde{c}'_a M^a\in 24\mathbb{Z}\, .
\end{align}
\end{subequations} 
Concretely, \eqref{eq:p_in_Kcf} requires that $\vec{p}$ lies in a facet $\mathcal{K}_{\mathrm{cf}}$ of the K\"ahler cone of the mirror threefold $\widetilde{X}$ along which the conifold curve $\mathcal{C}_{\text{cf}}$ collapses; the entries of $\vec{M}$ must obey the integrality conditions~\eqref{eq:integrality_condition_I} and~\eqref{eq:integrality_condition_II}; and the vectors $\vec{K}$ and $\vec{M}$ must together satisfy the Diophantine relation~\eqref{eq:Diophantine_eq}.

We now present an efficient algorithm for enumerating solutions to these constraints, following~\cite{McAllister:2024lnt}.\footnote{We use a variant of an algorithm developed in related work by M. Demirtas and implemented by A. Rios-Tascon, and we thank them for permission to include it here.} First, we notice that, due to \eqref{eq:integrality_condition_I} and \eqref{eq:integrality_condition_II}, $\vec{M}$ has to lie in some full-dimensional sub-lattice. A convenient basis for this lattice can be obtained  via the Euclidean algorithm. Writing $\vec{M}$ as an integer linear combination of these basis vectors automatically ensures that \eqref{eq:integrality_condition_I} and \eqref{eq:integrality_condition_II} are satisfied.

Next, to solve \eqref{eq:Diophantine_eq}, we collect a set of 
of lattice points $p^a$ lying in the interior of the facet $\mathcal{K}_{\mathrm{cf}}$, a convex polyhedral cone of dimension $h^{2,1}(X)-1$. Once either the generators of this cone or the defining dual hyperplanes are known, the enumeration of lattice points reduces to a problem in integer linear programming.  

For any such lattice point $p^a$, the Diophantine condition~\eqref{eq:Diophantine_eq} can be expressed as
\begin{equation}
     \kappa_{a\alpha\beta}\, M^a \,p^\alpha\, p^\beta  = 0 \, .
\end{equation}
This gives a linear integer constraint on $\vec{M}\in \mathbb{Z}^{h^{2,1}(X)}$, allowing us to determine the sub-lattice on which it is identically satisfied. We introduce a lattice basis ${e_\mu}^a$, with $\mu = 1,\ldots, h^{2,1}(X)-1$, for the sub-lattice on which \eqref{eq:Diophantine_eq}, \eqref{eq:integrality_condition_I}, and \eqref{eq:integrality_condition_II} all hold, and then expand
\begin{equation}
    M^a = \sum_{\mu} m^\mu {e_\mu}^a\, ,\quad \vec{m}\in \mathbb{Z}^{h^{2,1}(X)-1}\, .  
\end{equation}  
For each $\vec{m}$, we obtain a one-parameter family of solutions
\begin{equation}\label{eq:Kprime_def}
    K_a = \kappa_{ab\gamma}M^bp^\gamma-K' \delta_{a1}\, ,
\end{equation}
where $K'\in \mathbb{Q}$ is constrained by Gauss's 
law \eqref{eq:D3Tadpole}, since the D3-brane charge in fluxes can be expressed as
\begin{equation}\label{eq:D3_flux_splitting}
    Q_{\mathrm{flux}}=-M^a M^b \kappa_{ab\gamma}p^\gamma+M K' \, .
\end{equation}
The quantity $K'$ was previously introduced in \eqref{eq:Kprime}, and determines
the effective D3-brane charge in the throat $Q_{\text{flux}}^{\text{throat}}\approx M K'>0$, and correspondingly affects the VEV of the conifold modulus \eqref{eq:conifold_vev}.

More generally, for any integral $p^\alpha$ one may instead take a rational multiple $p^\alpha \to p^\alpha/k$ with $k > 1$, $k \in \mathbb{Z}$. One can then post-select those solutions in which $K_a$ is divisible by $k$, while $K'$ need not be integral for $k \neq 1$. As no analytic bound is known for the largest admissible $k$, in practice we fix a reasonably large cutoff $k_{\mathrm{max}}$ and restrict to solutions with $k \leq k_{\mathrm{max}}$.

Once the fluxes $\vec{M},\vec{K}$ have been selected, we numerically search for the vacuum expectation values for the complex structure moduli $z^a$ and the axio-dilaton $\tau$. Our procedure works as follows:
\begin{enumerate}[1)]
    \item Identify a solution $\langle \tau \rangle_{\mathrm{PFV}}$ to the effective theory defined by a perturbatively flat vacuum, in which the conifold modulus has been integrated out via the relationship \eqref{eq:conifold_vev} and the flux superpotential vanishes along a complex one-dimensional locus.
    \item Use the point $\tau = \langle \tau \rangle_{\mathrm{PFV}},\ z^\alpha = p^\alpha \langle \tau \rangle_{\mathrm{PFV}}$ as an initial guess for a numerical search of the $F$-term conditions. At this stage, we no longer rely on the approximations underlying the PFV effective theory, but continue to integrate out the conifold modulus analytically using the relation \eqref{eq:conifold_vev}.
    \item Finally, solve the complete $F$-term conditions --- including the dynamics of the conifold modulus --- without expanding the flux superpotential to linear order in $z_{\text{cf}}$ as in \eqref{eq:WExpConi}. The modulus $z_{\text{cf}}$ is now fixed numerically.
\end{enumerate}
Step 2) is typically efficient and yields reliable solutions, provided the PFV approximation is valid.\footnote{In practice one often encounters flux configurations for which the PFV approximation is not very accurate, and in such cases the solution obtained numerically, while trustworthy, can be far in field space from the location predicted by the PFV initial guess.}
Step 3) is often straightforward when the analytic expression \eqref{eq:conifold_vev} remains a good approximation, but becomes essential in intermediate regimes where the accuracy of \eqref{eq:conifold_vev} diminishes.
 
Following this procedure, we determine the vacuum expectation values of the axio-dilaton, the bulk complex structure moduli, and the conifold modulus within the full theory, treating this sector in isolation and neglecting both the K\"ahler moduli and any sources of supersymmetry breaking,
\begin{align}\label{eq:promotedvevs}
    \tau &\rightarrow  \langle \tau \rangle_{\text{F}}\kom z^{\alpha} \rightarrow \langle z^{\alpha} \rangle_{\text{F}}\kom z_{\text{cf}}  \rightarrow  \langle z_{\text{cf}} \rangle_{\text{F}}\,,
\end{align}
and we define the flux superpotential evaluated at this minimum as
\begin{equation}
    W_0 \coloneqq \left\langle |W_{\text{flux}}| \right\rangle_{\text{F}} \equiv \left| W_{\text{flux}} \left( \langle \tau \rangle_{\text{F}}, \langle z^{\alpha} \rangle_{\text{F}}, \langle z_{\text{cf}} \rangle_{\text{F}} \right) \right|,.
\end{equation}
This value $W_0$ enters directly into the scalar potential for the K\"ahler moduli, and plays a central role in determining whether controlled AdS$_4$ or uplifted dS$_4$ vacua can be realized.

\medskip

Before proceeding, it is useful to pause and comment on an important conceptual point.  
Recall from \eqref{eq:coniFH} that for a conifold PFV the flux vectors entering the GVW superpotential \eqref{eq:GVW} take the form
\begin{equation}
    \vec{f} = \left(\frac{1}{24}M^a \tilde{c}'_a,0,M^a \tilde{a}_{a\beta},0,M^a\right)^\top \, ,\; \vec{h} = \left(0,K_a,0,0^a\right)^\top\,.
\end{equation}
With the normalization conventions adopted here, the flux quanta are required to be integers. A natural question is whether these integers can be chosen freely, or whether additional consistency conditions further restrict their allowed values.  At present, the only universally established constraint is Gauss’ law, namely the D3-brane tadpole cancellation condition \eqref{eq:D3tadpole_compact}.

In certain toroidal orientifolds, however, Frey and Polchinski showed that consistency imposes a stronger requirement: the fluxes must take even integer values \cite{Frey:2002hf}.  
Whether an analogous constraint applies more generally to Calabi-Yau orientifolds, namely in the form
\begin{equation}\label{eq:evenflux}
    \vec{f},\; \vec{h} \in 2\mathbb{Z}\,,
\end{equation}
remains an open question. Resolving it likely requires new conceptual input, as existing arguments do not straightforwardly extend beyond the toroidal setting.

For the sake of completeness, one could impose the restriction \eqref{eq:evenflux} by hand. While this substantially reduces the available flux space and makes systematic searches considerably more expensive, it does not alter the qualitative structure of the problem. Indeed, supersymmetric AdS vacua satisfying \eqref{eq:evenflux} were exhibited in \cite{McAllister:2024lnt}. However, constructing candidate de Sitter vacua under the same restriction appears to require a significantly larger search effort, and progress in this direction will be reported in \cite{EvenSmaller}.

Throughout these lectures, we adopt the minimal and commonly used assumption that the flux integers satisfy the tadpole constraint \eqref{eq:D3tadpole_compact} but are otherwise unconstrained.

\section{Stabilizing K\"ahler moduli}\label{sec:stabKah}
 
In \S\ref{sec:KKLTintro} we sketched the K\"ahler moduli stabilization procedure in a toy model with $h^{1,1}=1$, but extending this process to geometries with $h^{1,1}\gg1$ requires much additional work. First, we need to find enough rigid divisors supporting non-perturbative contributions to the superpotential,
\begin{equation}\label{eq:Wnum}
    W = W_0+ \sum_{D}\, \mathcal{A}_D\,\exp\left (-\tfrac{2\pi}{c_D}\, T_D\right )\, .
\end{equation}
At least $h^{1,1}$ such divisors are required to generate sufficiently many terms in the non-perturbative superpotential \eqref{eq:Wnpexact}. Otherwise, one or more directions remain flat in the scalar potential. Intuitively, each exponential lifts one combination of moduli directions in field space.

Second, as explained in \S\ref{sec:ekc}, when $h^{1,1} \gg 1$, the K\"ahler cone $\mathcal{K}_{\star}$ typically decomposes into exponentially many chambers $\mathcal{K}_X$ connected to each other via flop transitions, recall Fig.~\ref{fig:extended_kahler} and Fig.~\ref{fig:491fan}. In a compactification with a KKLT superpotential, a randomly chosen triangulation of $\Delta^{\circ}$ will typically correspond to a chamber $\mathcal{K}_X$ that does \emph{not} admit any vacuum. Consequently, one must search through the extended K\"ahler cone $\mathcal{K}_{\star}$ to identify a triangulation in which such a vacuum exists. Because the number of chambers grows exponentially with $h^{1,1}$, a brute-force search is generally impractical.

To make matters worse, the scalar potential depends on $2h^{1,1}(X) + 2h^{2,1}(X) + 2$ real variables, which makes an unguided numerical search equally impractical. To address this challenge, we adopt a more structured strategy. The key steps of this procedure, described in detail below, can be summarized as follows:\footnote{For alternative approaches to numerical K\"ahler moduli stabilization, see \cite{AbdusSalam:2020ywo,AbdusSalam:2025twp}.}  
\begin{enumerate}
    \item In all examples, we initially find a supersymmetric AdS$_4$ vacuum at $T_{\mathrm{AdS}}$ in K\"ahler moduli space following \cite{Demirtas:2021nlu}.
    \item In examples containing a conifold, we then reinstate the anti-D3-brane potential and use $T_{\mathrm{AdS}}$ from the previous step as the starting point for a numerical search for a de Sitter minimum $T_{\mathrm{dS}}$ of the full potential~\eqref{eq:Vfull}, see \cite{McAllister:2024lnt}. 
    In practice, the uplift induces shifts in the VEVs of the complex structure moduli and the axio-dilaton, as explained below.
\end{enumerate}

In explicit compactifications that contain conifold singularities and a resulting warped KS throat (recall \S\ref{sec:KS}), Euclidean D3-branes wrapping divisors that intersect the conifold locus experience a large warp factor. As shown in \cite{Baumann:2006th}, the corresponding non-perturbative contributions are then exponentially suppressed relative to those from Euclidean D3-branes in the bulk. Such contributions cannot materially participate in K\"ahler moduli stabilization. 
In examples where conifolds are present, we therefore identify the prime toric divisors that intersect the conifold curve in the singular limit. In the relevant examples of Chapter~\ref{chap:deSitter}, there is always exactly one such divisor,
Its contribution needs to be omitted from the sum in \eqref{eq:Wnum}, and we require that, in examples containing a conifold and after excluding all conifold-intersecting divisors, there remain at least $h^{1,1}$ pure rigid divisors.  

In compactifications without conifold singularities, or with conifolds that do not lie on any rigid divisor (as e.g. in \cite{Crino:2020qwk}), no such deletion is required.

\subsection*{SUSY stabilization}

We now seek to solve the $F$-term conditions for the K\"ahler moduli,
\begin{equation}\label{eq:FtermsT}
    D_{T_i}W=0\; ,\quad \forall\, i=1,\ldots,h^{1,1}\,,
\end{equation}
where $T_i$, $i=1,\ldots,h^{1,1}$, are a basis of complexified K\"ahler coordinates, and the superpotential is given by \eqref{eq:Wnum}. In the single-modulus toy example of \S\ref{sec:KKLTintro}, we found the solution \eqref{eq:twapprox},
\begin{equation} 
    2\pi T=\log(|W_0|^{-1})+\ldots \, .
\end{equation}
We will now generalize this result to the $h^{1,1}>1$.

Beyond leading order in the $\alpha'$ and string loop expansions, the system \eqref{eq:FtermsT} becomes a coupled set of nonlinear equations, which is most effectively approached numerically. Nevertheless, it is useful to first make analytic progress by expanding in a small parameter and constructing controlled iterative solutions, as in \cite{Demirtas:2021nlu}.

\paragraph{Explicit form of the $F$-terms.} The $F$-term conditions can be written as
\begin{equation}\label{eq:Kahler_Fterms}
    D_{T_i}W(T)=-\frac{2\pi}{c_i}\mathcal{A}_i \mathrm{e}^{-\frac{2\pi}{c_i}T_i}-g_s \,\frac{t^i}{2\mathcal{V}}\left(W_0+\sum_j\mathcal{A}_j \mathrm{e}^{-\frac{2\pi}{c_j}T_j}\right)\, ,
\end{equation}
The first term encodes the non-perturbative contribution associated with the divisor $D_i$, while the second term arises from the K\"ahler derivative acting on the tree-level K\"ahler potential. The balance between these two competing effects stabilizes the K\"ahler moduli.

We define the small parameter
\begin{equation}
    \epsilon^i \coloneqq - g_s\mathcal{A}_i^{-1} \frac{t^i}{2\mathcal{V}} \frac{c_i}{2\pi}\, ,
\end{equation}
which satisfies
\begin{equation}
    |\epsilon^i| \lesssim g_s \sim \log(W_0)^{-1} \ll 1\, .
\end{equation}
The smallness of $\epsilon^i$ ensures that the system can be solved iteratively, order by order in this parameter. Indeed, the solution to the $F$-term equations~\eqref{eq:Kahler_Fterms} can be written as
\begin{equation}\label{eq:$F$-term-solution}
    T_i=\frac{c_i}{2\pi}\log(W_0^{-1})-\frac{c_i}{2\pi}\log\left[\epsilon^i\left(1+\sum_{j}\mathcal{A}_j \epsilon^j+\sum_{k,j}\mathcal{A}_{j}\mathcal{A}_k\epsilon^j\epsilon^k+\ldots\right)\right]\, ,
\end{equation}
and hence
\begin{equation}
T_i = T_i^{(0)} + \delta T_i\; ,\quad T_i^{(0)} \coloneqq \frac{c_i}{2\pi}\log(W_0^{-1})\, ,
\end{equation}
with a relative correction 
\begin{equation}\label{eq:relcorr}
    \frac{\delta T_i}{T_i^{(0)}} = \mathcal{O}\left(\frac{\log[\log(W_0^{-1})]}{\log(W_0^{-1})}\right) \ll 1 \, .
\end{equation}
The iterative structure of \eqref{eq:$F$-term-solution} reflects the algorithmic approach of \cite{Demirtas:2021nlu}: one begins with the zeroth-order guess $T_i^{(0)}$, plugs this into the right-hand side of \eqref{eq:FtermsT}, and refines the solution step by step. Each iteration incorporates additional non-perturbative effects, suppressed by higher powers of $\epsilon^i$, and converges rapidly because $|\epsilon^i| \ll 1$.

We now turn to constructing solutions of the form \eqref{eq:$F$-term-solution} and verifying that they lie in a region of theoretical control, where the assumptions underlying \eqref{eq:$F$-term-solution}, in particular \eqref{eq:control_regime}, are justified \textit{a posteriori}. In particular, we describe an effective algorithm for identifying points in the extended K\"ahler cone $\mathcal{K}_{\star}$ where a chosen basis of divisors $\{D_i\}$, $i=1,\ldots,h^{1,1}$, attains the desired values as given by \eqref{eq:$F$-term-solution}.

\paragraph{Initial guess.} 
As explained above, the $F$-flatness conditions for the K\"ahler moduli are at leading order solved by
\begin{equation}\label{eq:fflatsol0}
    \mathrm{Re}(T_i^{\text{tree}})=\frac{1}{g_s} \frac{1}{2} \kappa_{ijk} t^j t^k \approx \frac{c_i}{2\pi} \log\bigl(W_0^{-1}\bigr) \quad \forall i \,,
\end{equation}
using the tree-level expression \eqref{eq:Ttree} for the K\"ahler coordinates.
Since both $\log(W_0)$ and $g_s$ appear only as overall factors in the solution \eqref{eq:fflatsol0}, we can equivalently solve
\begin{equation}\label{eq:fflatsol00}
    \tau_i = \frac{1}{2} \kappa_{ijk} t^j t^k = c_i \quad \forall i \,,
\end{equation}
which is independent of the specific choice of fluxes. The true solution \eqref{eq:fflatsol0} is then easily obtained by an appropriate rescaling, and is trivially contained in the corresponding chamber $\mathcal{K}_X$ of the extended K\"ahler cone $\mathcal{K}_{\star}$.

Our aim is to identify a point in $\mathcal{K}_{\star}$ where a chosen basis of $h^{1,1}$ linearly independent prime toric divisors $\{D_i\}$, $i=1,\dots,h^{1,1}$, satisfies \eqref{eq:fflatsol00}, while ensuring that the volumes of all remaining divisors are greater than or equal to one. 
The relevant choices of basis consist of suitable subsets of $h^{1,1}$ prime toric divisors that contribute to the superpotential.  The number of such choices is typically small enough that one can test all of them.  

Once a suitable basis is identified, the next step is to determine the K\"ahler parameters $t_\star$ that yield the desired values $\tau_\star = (c_1,\ldots, c_{h^{1,1}})$ for the divisor volumes $\tau_i = \tfrac{1}{2} \kappa_{ijk} t^j t^k$,  
i.e., that solve \eqref{eq:fflatsol00}. We begin by selecting a random point $t_{\text{init}}$ in the K\"ahler cone $\mathcal{K}_X$ of a  
Calabi-Yau hypersurface $X$ determined by a random triangulation of $\Delta^\circ$.
We consider any point along the straight line connecting the associated basis divisor volumes $\tau_{\text{init}}$ and the target values $\tau_\star$
\begin{equation}
    \tau_\alpha = (1-\alpha)\, \tau_{\text{init}} + \alpha\, \tau_\star, \quad 0 \leq \alpha \leq 1\, .
\end{equation}
The corresponding path between $t_{\text{init}}$ and $t_\star$ is not a straight line, because the divisor volumes $\tau(t)$ are quadratic functions of the K\"ahler parameters $t^i$, with coefficients $\kappa_{ijk}$ that can jump across different phases. Nevertheless, the functions $\tau(t)$ are continuous and once differentiable, producing a smooth path without cusps. This property allows us to efficiently track the path from $t_{\text{init}}$ to $t_\star$.

Our final task is to implement a numerical algorithm that follows this continuous path. To do so, we subdivide the path into $N \gg 1$ small segments by defining points at $\alpha = m/N$, for $m = 0, \dots, N$. We denote the corresponding points in $\mathcal{K}_X$ by $t_m$, respectively. Following the path then reduces to iteratively moving from $t_m$ to $t_{m+1}$.
Let us write $t_{m+1}^i = t_m^i + \varepsilon^i$. Then, the divisor volumes at step $m+1$ are given by
\begin{equation}
    \tau_{m+1}^i = \tau_m^i + \kappa_{ijk} t_m^j \varepsilon^k + \mathcal{O}(\varepsilon^2)\; , \quad \tau_m^i = \frac{1}{2} \kappa_{ijk} t_m^j t_m^k\, .
\end{equation}
Determining $\epsilon$ therefore reduces to solving the linear system
\begin{equation}
    \kappa_{ijk} t_m^j \varepsilon^k = \tau_{m+1}^i - \tau_m^i\, .
\end{equation}
Along this path, whenever $t_{m+1}^i$ is not contained in the current chamber $\mathcal{K}_X$, we have to explicitly perform the flop transition across some facet of $\mathcal{K}_X$ to a new phase $X'$ with K\"ahler cone $\mathcal{K}_{X'}$.
This provides a practical step-by-step procedure to move along the path in the extended K\"ahler cone $\mathcal{K}_{\star}$ towards $t_\star$. Once a solution to \eqref{eq:fflatsol00} has been found, the corresponding $t^i$ must be rescaled by a factor
\begin{equation}\label{eq:tsrescaling}
    c_\tau^{-1/2} = \biggl (g_s\frac{\log (W_0^{-1})}{2\pi} \biggl )^{1/2} \, .
\end{equation}

\paragraph{Including corrections.}
Next, the perturbative and non-perturbative corrections in \eqref{eq:detailedform4} must be incorporated systematically. This procedure is only meaningful within the radius of convergence of the type IIB worldsheet instanton expansion, where at most finitely many curves provide significant contributions. 

We begin with the zeroth-order solution $t_{(0)} = t_{\star}$ obtained in the previous step, which satisfies, after the rescaling with \eqref{eq:tsrescaling},
\begin{equation}
    \frac{1}{2} \kappa_{ijk} t^j_{(0)} t^k_{(0)} = \frac{c_i}{c_\tau}\, .
\end{equation}
For $n > 0$, we define $t^i_{(n)}$ recursively as the solution to the quadratic equation
\begin{equation}\label{eq:recur}
    \frac{1}{2} \kappa_{ijk} t^j_{(n)} t^k_{(n)} = \frac{c_i}{c_\tau} + \frac{\chi(D_i)}{24} - \frac{1}{(2\pi)^2} \sum_{\mathbf{q} \in \mathcal{M}_X} q_i\, \mathscr{N}_{\mathbf{q}}\, \mathrm{Li}_2\Bigl((-1)^{\mathbf{\gamma} \cdot \mathbf{q}} \mathrm{e}^{-2\pi \mathbf{q} \cdot \mathbf{t}_{(n-1)}}\Bigr)\, ,
\end{equation}
where the right-hand side depends on $t^i_{(n-1)}$. At each step $n$, the instanton sum can be truncated by retaining only those terms whose contribution exceeds a small threshold. If a solution exists, this iterative procedure allows one to converge to it with arbitrary precision by increasing $n$.

We note that in our analysis we incorporate many sub-leading corrections to the K\"ahler coordinates $T_i$, as captured in \eqref{eq:detailedform4},
but at the same time we can safely omit most of the corrections \eqref{eq:detailedform3}
to the K\"ahler potential itself --- specifically, we can omit the worldsheet instanton series in \eqref{eq:detailedform3} --- 
as these terms produce subdominant effects in the parameter regimes of interest.  Since \eqref{eq:detailedform3} can be evaluated to any desired accuracy, one can directly quantify the approximation resulting from the above omission.

\paragraph{Full numerical solution.}
Having constructed a controlled approximate solution $T_i \approx T_i^{(0)}$ including all relevant perturbative and non-perturbative corrections, we are now in a position to solve the full system \eqref{eq:FtermsT} numerically. Concretely, we promote the complexified moduli $T_i$ to independent real variables and treat the real and imaginary parts of \eqref{eq:FtermsT} as a set of $2h^{1,1}(X)$ coupled nonlinear equations. Using the corrected solution from the previous step as the initial guess typically ensures rapid convergence of standard multidimensional root-finding algorithms, such as Newton's method. This procedure yields the fully stabilized K\"ahler moduli $\langle T_i \rangle_{\mathrm{F}}$ in a regime of theoretical control, completing the construction of a supersymmetric AdS$_4$ vacuum.

\subsection*{de Sitter uplift}

So far, our analysis has been carried out in the absence of anti-D3-branes, and the resulting solution is a supersymmetric AdS vacuum characterized by a negative $F$-term potential, $V_F < 0$. The final step in the construction is to add the contribution of the anti-D3-brane sector, denoted by $V_{\overline{\mathrm{D3}}}$ in Eq.~\eqref{eq:anti-D3-potential00}, and to analyze the full scalar potential \eqref{eq:Vfull}. This requires searching for critical points of the complete potential, which now includes both the supersymmetric F-term contributions and the explicit supersymmetry-breaking uplift, by solving
\begin{equation}
    \partial_I V = 0
\end{equation}
for the K\"ahler moduli, complex structure moduli, and axio-dilaton.

An important subtlety is that the uplift potential $V_{\overline{D3}}$ depends 
on the complex structure moduli $z^a$ and the axio-dilaton $\tau$ through the warping at the location of the anti-D3-brane, but also depends on the K\"ahler moduli $T_i$. This induces a backreaction between the complex structure and K\"ahler moduli sectors, which are otherwise approximately decoupled at tree level. Concretely, recall from Eq.~\eqref{eq:anti-D3-potential0} that the anti-D3-brane tension contributes as
\begin{equation}
    V_{\overline{D3}} \sim    \frac{z_{\mathrm{cf}}^{4/3}}{g_sM^2\, {\widetilde{\mathcal{V}}}^{2/3}\mathcal{V}_E^{4/3}}\, .
\end{equation}
When the K\"ahler moduli shift in response to the uplift, the volume changes, and the warp factor may also shift indirectly through changes in the complex structure sector that minimize the total potential. Thus, adding $V_{\overline{D3}}$ generally displaces not only the K\"ahler moduli but also the complex structure moduli $z^a$ and $\tau$.

This backreaction can spoil the delicate balance of the original supersymmetric solution if it is not small. In practice, to ensure metastability, one must verify that the eigenvalues of the mass matrix of the complex structure and axio-dilaton remain positive when the uplift is turned on. This is typically justified when the hierarchy $m_{z^a}, m_\tau \gg m_{T_i}$ holds, so that the heavy complex structure sector adjusts adiabatically to the changes in the lighter K\"ahler sector. This hierarchy is indeed realized in the standard KKLT scenario, because the masses of the complex structure moduli and the axio-dilaton are set by the flux superpotential $|W_0|$ at tree level, while the K\"ahler moduli acquire their masses only through non-perturbative effects and are therefore exponentially lighter. In this regime, the complex structure moduli approximately track their original minima during the uplift, leading to small corrections that can be computed perturbatively.

However, in the context of PFVs introduced in \S\ref{sec:pfv} and \S\ref{sec:conipfv}, one complex structure direction remains almost massless at leading order and is stabilized only by exponentially suppressed mirror worldsheet instantons. Because this direction is so light, it is highly susceptible to backreaction effects. Consequently, the usual two-step procedure of first stabilizing complex structure moduli and then uplifting is not valid here: the uplifted minimum must be found by solving for all moduli simultaneously, and the stability of the resulting vacuum is considerably more delicate.
 
In our numerical procedure, this interplay can be incorporated by allowing all moduli to vary when solving $\partial_I V = 0$. As an initial point, we take the VEVs $\langle T_i \rangle_{\mathrm{F}}$, $\langle \tau \rangle_{\mathrm{F}}$, $\langle z^\alpha \rangle_{\mathrm{F}}$, and $\langle z_{\mathrm{cf}} \rangle_{\mathrm{F}}$ obtained by solving the $F$-term conditions without the anti-D3-brane potential. 
This configuration defines what we call an \emph{AdS precursor}\index{AdS precursor}: a supersymmetric $\mathrm{AdS}_4$ vacuum in which the K\"ahler moduli are stabilized by non-perturbative effects, while the complex structure moduli and the axio-dilaton are fixed by the classical flux superpotential.  
Such AdS vacua serve as natural starting points for constructing metastable de Sitter solutions, since their detailed properties determine whether an uplift by an anti-D3-brane in a warped throat can lead to a controlled positive-energy minimum.
To implement the uplift, we add the anti--D3-brane contribution gradually by introducing a continuous parameter $\epsilon$, 
\begin{equation}\label{eq:Vfulleps}
    V^{\epsilon} = V_F +\epsilon\, V_{\overline{D3}}\,.
\end{equation}
which is increased adiabatically from $\epsilon=0$ to $\epsilon=1$.  
For sufficiently small increments, the moduli shift only slightly at each step, allowing the minimum to be efficiently tracked using Newton’s method.  
If all eigenvalues of the Hessian remain positive throughout this process and the vacuum energy becomes positive at $\epsilon=1$, the endpoint defines a candidate de Sitter vacuum, with vacuum expectation values denoted $\langle T_i \rangle_{\mathrm{dS}}$, $\langle \tau \rangle_{\mathrm{dS}}$, etc.

\section{Convergence tests}\label{sec:conv}

Given a candidate solution at a particular point $T_i$ in K\"ahler moduli space, one needs to ensure that worldsheet instanton corrections  have been consistently incorporated. In \eqref{eq:recur} --- see also \eqref{eq:detailedform4}) --- the sum over $\mathbf{q} \in \mathcal{M}_X$ formally includes all curves in the Mori cone, but as we noted, when carrying out an actual numerical solution the sum is necessarily truncated.

To test for consistency, one should verify that the candidate solution at $T_i$ lies within the radius of convergence of the worldsheet instanton expansion.  At $h^{1,1} \gg 1$, this radius is not usually analytically computable, but can be determined by computing genus-zero Gopakumar–Vafa invariants to high degree using mirror symmetry. Specifically, one expands the period vector of the mirror $\widetilde{X}$ at large complex structure, and employs the method of \cite{Hosono:1993qy}, as formulated and implemented in \cite{Demirtas:2023als}.

Equipped with the Gopakumar-Vafa invariants of an effective curve $\mathbf{q}$ and its multiples $n\mathbf{q}$ for $n=2,\ldots, n_{\text{max}}$, with the maximum degree $n_{\text{max}}$ determined by available computational resources, one can test for convergence of the worldsheet instanton series for this curve by computing 
\begin{equation}
    \xi_n(\mathbf{q}) \coloneqq  \mathscr{N}_{n\mathbf{q}}\mathrm{e}^{-2\pi n\,\mathbf{q}\cdot\mathbf{t}}\,.
\end{equation}
If $\Xi_{n}(\mathbf{q})$ decays exponentially with $n$, the convergence test is passed, and higher multiples $n$ of the curve $\mathbf{q}$ apparently\footnote{This test is not a proof because in principle one could find $\Xi_{n}(\mathbf{q})$ to decrease with $n$ for $1 \le n \le n_{\text{mid}}$, and to increase thereafter.  However, we have not encountered this behavior in the examples reported here.} make ever-decreasing contributions to the effective theory.

To evaluate convergence, one should thus compute $\Xi_n(\mathbf{q})$ for the most dangerous curves $\mathbf{q}$. For this purpose we classify curves into two categories. If $\mathscr{N}_{n\mathbf{q}} \neq 0$ for infinitely many $n \in \mathbb{N}$, i.e.~if the series of Gopakumar-Vafa invariants of multiples of $\mathbf{q}$ is infinite, we call $\mathbf{q}$ a \emph{potent curve}, and we call the ray in $\mathcal{M}_X$ generated by $\mathbf{q}$ a potent ray.  All other curves --- those for which $\mathscr{N}_{n\mathbf{q}} \neq 0$ for only finitely many $n$ --- are called \emph{nilpotent}.

We refer to the closure of the cone generated by all potent rays as $\mathcal{M}_\infty(X)$. Nilpotent curves outside $\mathcal{M}_\infty(X)$ are collapsible, and their finitely-many contributions to \eqref{eq:detailedform4} can be explicitly incorporated in \eqref{eq:recur}.  Such flop curves can in general make important contributions to the K\"ahler coordinates $T_i$, but for such a curve to be small does not indicate  breakdown of the worldsheet instanton expansion.

On the other hand, a sufficiently small potent curve does herald such a breakdown. We therefore need to verify that, at a candidate solution $T_i$ all potent curves $\mathbf{q}$ are large enough so that $\Xi_n(\mathbf{q})$ decreases with $n$. Computing a Hilbert basis for curve classes in $\mathcal{M}_\infty(X)$ is presently out of reach, so we sample facets of $\mathcal{M}_\infty(X)$ and attempt to collect as many linearly-independent potent curves as possible. 

An analogous test can be performed for type IIA worldsheet instantons on the mirror $\widetilde{X}$, which contribute to the flux superpotential. In particular, for the flux choices under consideration, the flux superpotential takes a racetrack form (see \eqref{eq:Widef}), making it essential to verify the convergence of the corresponding series with care. Because in our examples $h^{2,1} \ll h^{1,1}$, the computation of  Gopakumar-Vafa invariants of curves in $\widetilde{X}$ is far less expensive than for curves in $X$, and is easily taken to extremely high degrees.

\begin{figure}[!t]
\centering
\includegraphics[width=\linewidth]{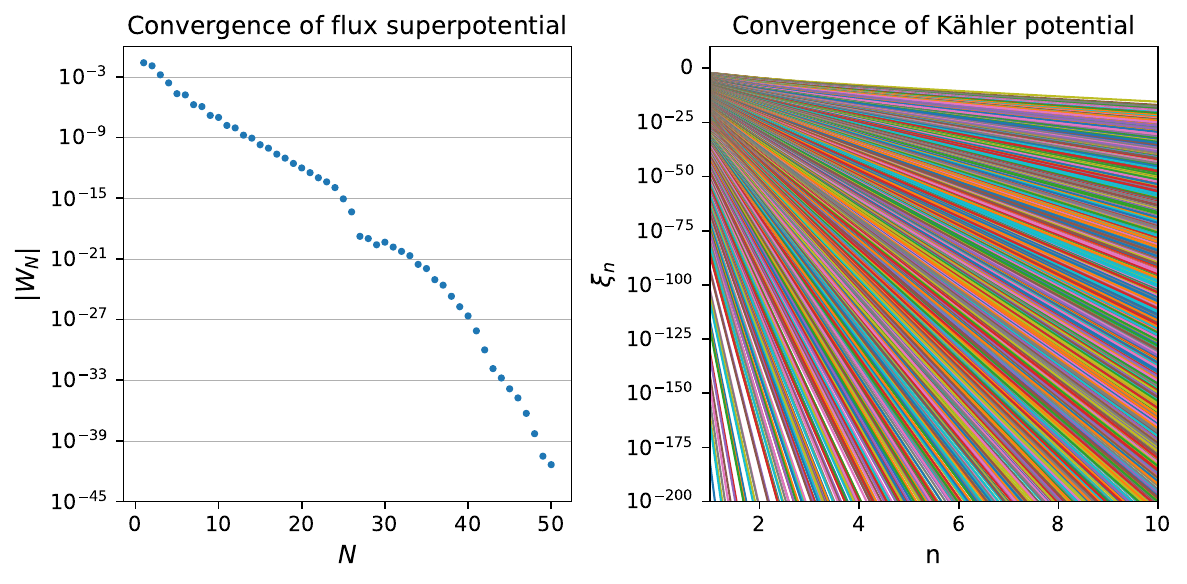}
\caption{Convergence results of the vacuum in \S\ref{sec:manwe}. \emph{Left:} Convergence of the flux superpotential \eqref{eq:Widef}. 
\emph{Right:} Convergence of the worldsheet instanton corrections from a sample of $2{,}643$ potent rays.
}\label{fig:manwe_W_rainbow}
\end{figure}

In each candidate vacuum, we display the results of the convergence tests as shown in Figure \ref{fig:manwe_W_rainbow}. These tests provide a direct probe of whether the instanton sums behave as expected in the regime of interest, and they allow us to verify that higher-order contributions decrease with the appropriate rate.  
Consistent convergence across a broad range of degrees offers evidence that the instanton expansion is under control and that the corresponding vacuum lies within a trustworthy region of the effective theory. 

Having discussed the validity of truncating the infinite series of worldsheet instanton corrections to $\mathcal{K}$ and $T_i$, and of IIA worldsheet instanton contributions to $W_{\text{flux}}$, we turn to the non-perturbative superpotential.
Here too the sum is formally infinite, running over all effective divisors $D$, i.e.~classes $[D] \in H_4(X,\mathbb{Z}) \cap \mathcal{E}_2(X)$, with $\mathcal{E}_2(X)$ the effective cone.
However, some effective $D$ may support more than two fermion zero modes and so do not yield a contribution to $W$, and, more importantly, at a point in moduli space where $\langle T_i \rangle \gtrsim 1$, nearly all $[D] \in H_4(X,\mathbb{Z})$ are so large that Euclidean D3-branes wrapping them can be neglected.
The most important contributions are those from the smallest\footnote{One should also account for the dual Coxeter number, not just the volume, when considering contributions from condensing gauge groups on seven-branes.} effective divisors.  

Thus, given K\"ahler parameters $t_i$, we seek to find the $[D] \in H_4(X,\mathbb{Z}) \cap \mathcal{E}_2(X)$ with the smallest volumes.  
At first sight this would appear to involve computing the generators of the semigroup of effective divisors, i.e.~the Hilbert basis of the effective cone.
However, one can show \cite{holes}
that the only Hilbert basis elements of the inherited effective cone --- i.e., of the cone of effective divisors $\widehat{D}$ on the parent toric variety $V$, restricted to $X$ --- that are actually effective, and so can in principle support superpotential terms, are the prime toric divisors themselves.  
We conclude that in addition to evaluating the volumes of prime toric divisors, one should consider autochthonous divisors, i.e.~effective divisors $D$ on $X$ that are not of the form $\widehat{D} \cap X$.  No efficient algorithm for enumerating all autochthonous divisors is presently available, but classes of autochthonous divisors have been found in various settings, and among these the smallest-known volumes occur for the so-called `min-face factorized' divisors \cite{Demirtas:2021nlu,Demirtas:2021gsq,Gendler:2022qof}.  Thus, in the end, we evaluate the truncation of the non-perturbative superpotential by ensuring that all min-face factorized autochthonous divisors are very large compared to the prime toric divisors that we have retained in the sum.

\section{Towards explicit examples}\label{sec:twee}

In the preceding sections we have developed the computational environment needed to construct and analyze candidate vacua in large ensembles of Calabi-Yau orientifolds at the level of the leading order EFT introduced in \S\ref{sec:leadingEFT}. We now summarize the assumptions and simplifications that make the search for explicit solutions feasible in practice, and record the numerical choices that enter our computations.

Our starting point is a class of \emph{trilayer orientifolds} $X/\mathcal{O}$, 
in which $h^{1,1}_{-}=h^{2,1}_{+}=0$, and the D7-brane tadpole is
canceled locally by placing D7-branes on top of O7-planes, leading to 
$\mathfrak{so}(8)$ gauge sectors (recall \S\ref{sec:indD3ch}).  The topological data of these geometries imply a simple and universal formula for the induced D3-brane charge from localized objects,
\begin{equation}
    Q_{\mathrm{O}} = 2 + h^{1,1} + h^{2,1}\,,
\end{equation}
which is typically of order ${\cal O}(10^2)$ in the range of Hodge numbers we study. Such values allow for substantial flux scanning while still remaining comfortably inside the tadpole bound.

We adopt the normalization \eqref{eq:K0} for the Pfaffian prefactors, and set
\begin{equation}\label{eq:pfaffvalue}
    n_{D}=1 \qquad \Longrightarrow \qquad 
    \mathcal{A}_{D}=\sqrt{\frac{2}{\pi}}\frac{1}{4\pi^{2}}\,,
\end{equation}
for every rigid prime toric divisor $D$ contributing to the superpotential. This leads to the working expression
\begin{equation}\label{eq:wlo4}
    W = \sqrt{\tfrac{2}{\pi}}\,\vec{\Pi}^{\top}\cdot\Sigma\cdot(\vec{f}-\tau\vec{h})
    + \sqrt{\tfrac{2}{\pi}}\,\frac{1}{4\pi^{2}} \sum_{D} \exp\left(-\frac{2\pi}{c_{D}}\,T_{D}\right),
\end{equation}
where the sum runs over the set of rigid prime toric divisors.
As discussed above, these divisors control the dominant non-perturbative contributions to the superpotential.

To quantify the sensitivity of our solutions to the actual values of the Pfaffians, we scale all $\mathcal{A}_{D}$ by a common factor $n_D$ and test the stability of vacua as $n_{D}$ is varied. As detailed in Chapter~\ref{chap:deSitter}, all of the de~Sitter candidates presented there persist throughout the wide range
\begin{equation}\label{eq:ndrange}
    10^{-3} \;\le\; n_{D} \;\le\; 10^{4}\,.
\end{equation}
Persistence beyond this interval is commonly observed as well, though the detailed behavior is strongly model-dependent and we do not attempt to systematize it.  For supersymmetric AdS solutions the situation is even more robust, as small shifts in the Pfaffians perturb the vacuum energy only at sub-leading order.

With this computational machinery in place, we now turn to explicit constructions. Chapter~\ref{chap:deSitter} presents three representative classes of solutions:
\begin{enumerate}
    \item supersymmetric AdS vacua with exponentially small cosmological constant (from \cite{Demirtas:2021nlu}),
    \item a compactification exhibiting both a warped throat and a metastable, potentially
    inflationary direction (from \cite{McAllister:2024lnt}), and
    \item several examples admitting de~Sitter minima of the leading-order effective theory (from \cite{McAllister:2024lnt}).
\end{enumerate}
Each of these illustrates different aspects of the mechanisms developed in the preceding chapters, and together they demonstrate that the computational strategy laid out in this chapter can indeed produce fully explicit, controlled vacua in large ensembles of Calabi-Yau orientifolds.

\chapter{Candidate Vacua} \label{chap:deSitter} 

In this chapter, we present a class of candidate de Sitter vacua constructed in flux compactifications of type IIB string theory on Calabi-Yau orientifolds. These solutions, recently developed in \cite{McAllister:2024lnt}, implement the KKLT scenario in explicit models: the complex structure moduli and the axio-dilaton are stabilized at tree level by fluxes, the K\"ahler moduli are fixed supersymmetrically via non-perturbative effects, and the resulting AdS vacua are uplifted to metastable de Sitter vacua using anti-D3-branes located at the bottom of Klebanov-Strassler throats. Using the key ingredients reviewed in the previous chapter, we now examine representative examples that demonstrate the viability of the mechanism in concrete compactifications. We also comment on the control conditions for these constructions.

\begin{table}[h]
\begin{centering}
\resizebox{1.\textwidth}{!}{
\begin{tabular}{|c|c|c|c|c|c|c|c|c|c|}\hline
& & & & & & & & &\\[-1.em]
\S & $h^{2,1}$ & $h^{1,1}$ & Type & Ref & $g_s$ & $W_0$ & $g_s M$ & $|z_{\text{cf}}|$ & $V_0$ \\[0.1em]
\hline
\hline
& & & & & & & & &\\[-1.em]
\ref{sec:SUSYAdS} & 5 & 113 & SUSY AdS & \cite{Demirtas:2021nlu} & 0.011 & $6.46\times 10^{-62}$ & N/A &  N/A &  $-1.68 \times 10^{-144}$ \\[0.1em]\hline 
& & & & & & & & &\\[-1.em]
\ref{sec:KKLMMTexPartA}& 8 & 150 & SUSY AdS & \cite{McAllister:2024lnt} &  $0.075$ & $0.032$ & $1.05$ & $9.06\times 10^{-7}$ & $-1.49\times 10^{-17}$\\[0.1em]\hline
& & & & & & & & &\\[-1.em]
\ref{sec:manwe} & 8 & 150 & dS & \cite{McAllister:2024lnt} & 0.066 & 0.012 & 1.05 &  2.82$\times10^{-8}$ &  +1.937$\times10^{-19}$ \\[0.1em]\hline
& & & & & & & & &\\[-1.em]
\ref{sec:aule} & 5 & 93 & dS & \cite{McAllister:2024lnt} & 0.040 & 0.054 & 0.81 & 1.97$\times10^{-6}$ &  +2.341$\times10^{-15}$ \\[0.1em]\hline 
\end{tabular}}
\caption{Overview of control parameters for the explicit candidate vacua discussed throughout this chapter.}
\label{tab:summary}
\end{centering}
\end{table}

\vfill

\newpage

\section{Small cosmological constants in AdS$_4$}\label{sec:SUSYAdS}

In this section, we aim to construct supersymmetric AdS$_4$ vacua in type IIB flux compactifications. Our strategy involves identifying solutions to the conditions for unbroken supersymmetry, which correspond to critical points of the four-dimensional scalar potential where the $F$-terms $D_{\Phi^A} W=0$ vanish for all moduli $\Phi^A$. As mentioned before, a key requirement in this mechanism is that $|W_0| \ll 1$, ensuring that the non-perturbative effects dominate over the tree-level term and allow for stabilization at parametrically large volumes and weak coupling. In practice, such small values of $|W_0|$ can be achieved through fine-tuned flux choices\footnote{See \cite{Chauhan:2025rdj} for recent advances.} or via the mechanism of PFVs, which we introduced in detail in \S\ref{sec:pfv}. To construct explicit examples of these vacua, we rely on purpose-built software tools summarized in \S\ref{sec:CompComp} that leverage techniques from computational algebraic geometry, enabling precise control over the geometry, flux choices, and moduli dynamics in Calabi-Yau compactifications. These supersymmetric AdS$_4$ vacua form the essential starting point for the construction of metastable de Sitter vacua in Chapter~\ref{chap:deSitter}.

\medskip

Following \cite{Demirtas:2021nlu}, we focus on a concrete setup involving a mirror pair of smooth Calabi-Yau threefolds $X$ and $\widetilde{X}$ realized as hypersurfaces in four-dimensional toric varieties $V$ and $\widetilde{V}$, respectively, and satisfying
\begin{equation}
    h^{1,1}(X)=h^{2,1}(\widetilde{X})=113\, ,\quad h^{2,1}(X)=h^{1,1}(\widetilde{X})=5\, .
\end{equation}
All necessary data to construct those examples can be found in the supplementary material of \cite{Demirtas:2021nlu}.

A type IIB O3/O7 orientifold of $X$ is obtained by the involution 
\begin{equation}
    \mathcal{I}:\,x_1\mapsto -x_1\,,
\end{equation}
where $x_1$ is a particular toric coordinate.
For this orientifold, $h^{1,1}_-(X)=h^{2,1}_+(X)=0$. There are 25 O7-planes wrapping rigid divisors and 48 O3-planes at the triple intersections of divisors. There is a single O7-plane on a non-rigid divisor $D_1$ with $h^\bullet(D_1,\mathcal{O}_{D_1})=(1,0,2)$. Each O7-plane on a rigid divisor hosts a confining $\mathcal{N}=1$ pure Yang-Mills theory with gauge algebra $\mathfrak{so}(8)$. The D3-brane tadpole \eqref{eq:QOf} for this orientifold is equal to 
\begin{equation}\label{eq:5113tadpole}
    Q_O = h^{1,1}(X)+h^{2,1}(X)+2=120\, .
\end{equation}
The sizable D3-brane charge makes it possible to turn on sufficiently rich quantized $3$-form fluxes to find a fully stabilized vacuum.  

\medskip

We follow a two-step strategy. First, we engineer a perturbatively flat vacuum by choosing quantized 3-form fluxes such that the classical flux superpotential $W_0$ vanishes along a complex line in moduli space, as described in section \S\ref{sec:pfv}. In the second step, we incorporate non-perturbative effects to stabilize the K\"ahler moduli by solving the corresponding $F$-term conditions $D_{T_i}W=0$. The resulting vacuum is fully stabilized and supersymmetric, with a negative cosmological constant characteristic of AdS$_4$ space.

\subsubsection*{Complex structure moduli stabilization}

First, we need to find suitable flux choices of the form described in \S\ref{sec:pfv}. 
In a suitable basis $H_4(\widetilde{X})$, one can now search for flux vacua in the LCS regime \eqref{eq:LCS} as introduced in \S\ref{sec:CMS}. The second Chern classes are
\begin{equation}\label{eq:c2_5113}
    c_2(\widetilde{X})=\begin{pmatrix} 146 & -4  & 24 & 24 & 14 \end{pmatrix}^\top\, ,
\end{equation}
and the non-trivial triple intersection numbers are
\begin{align}\label{eq:kappa_5113}
&\widetilde{\kappa}_{1ab}=\begin{pmatrix}
    89 &0 &16 &12 &\hphantom{-}7  \\
    & 0& 0& 0& \hphantom{-}0\\
    & & 0&3 &\hphantom{-}0  \\
    & & & 0& \hphantom{-}3 \\
    & & & &  -3\\
\end{pmatrix}\, ,\quad \widetilde{\kappa}_{2ab}=\begin{pmatrix}
    8& -2& -2& -2  \\
    & \hphantom{-}0&  \hphantom{-}1& \hphantom{-}0  \\
    & &    \hphantom{-}0& \hphantom{-}1 \\
    & & &    \hphantom{-}0\\
\end{pmatrix}\, ,\quad \widetilde{\kappa}_{555}=-1\, ,
\end{align}
where for fixed $a'=1,\ldots,5$ we display only the non-trivial $\widetilde{\kappa}_{a' ab}$ with $a'\leq a \leq b$.

The following flux choice
\begin{equation}\label{5113flux}
    \vec{M}=\begin{pmatrix} 0 & 2 & 4 & 11 & -8 \end{pmatrix}^\top
    \, ,\quad 
    \vec{K}=\begin{pmatrix} 8 & -15 & 11 & -2 & 13 \end{pmatrix}^\top\, ,
\end{equation}
satisfies the conditions \eqref{eq:PFV} for a perturbatively flat vacuum. From Eq.~\eqref{eq:PvecDef}, we find that the complex structure moduli and the axio-dilaton are related as
\begin{equation}\label{5113flat}
    \vec{z}=\vec{p}\,\tau
    \, ,\quad 
    \vec{p}= \tfrac{1}{116}\times\begin{pmatrix} 14 & 30 & 101 & 302 & -13 \end{pmatrix}\, .
\end{equation}
The D3-brane charge induced by the fluxes \eqref{5113flux} amounts to
\begin{equation}
    -\frac{1}{2}M_aK^a=56\, .
\end{equation}
Comparing with the total D3-brane charge in Eq.~\eqref{eq:5113tadpole}, four mobile D3-branes have to be present the cancel the D3-brane tadpole. 

To stabilize the remaining modulus $\tau$, we follow the discussion in \S\ref{sec:pfv} and compute the GV invariants of $\widetilde{X}$ systematically, and find that the leading instantons give rise to the effective flux superpotential
\begin{equation}\label{5113wf}
    W_{\text{eff}}(\tau) = \frac{8}{2^{3/2}\pi^{5/2}} \, \left(-2\,\mathrm{e}^{2\pi \I  \tau \cdot \frac{7}{29}}+252\,\mathrm{e}^{2\pi \I  \tau \cdot \frac{7}{28}}\right)+\mathcal{O}\left(\mathrm{e}^{2\pi \I  \tau \cdot \frac{43}{116}}\right) \, .
\end{equation} 
One then finds for the string coupling and the VEV of the flux superpotential
\begin{equation}\label{5113w02}
    g_s  \approx \frac{2\pi }{116\log(261/2)} \approx 0.011 \, , \quad W_0\approx  0.526 \times \left(\frac{2}{252}\right)^{29} \approx 6.46\times 10^{-62}\,.
\end{equation}
The exponential smallness of $W_0$ stems from the hierarchy in the two leading GV invariants $(-2,252)$, raised to a power $29$ determined by the racetrack appearing in \eqref{5113wf}. While the GV invariants are intrinsically determined by the geometry, the exponents $(7/28,7/29)$ in the racetrack superpotential \eqref{5113wf} come from the choice of fluxes in \eqref{5113flux}.

\subsubsection*{K\"ahler moduli stabilization}

Our next objective is to stabilize the K\"ahler moduli. As a prerequisite, we must determine the full non-perturbative superpotential \eqref{eq:Wnpexact}. In this geometry, non-perturbative contributions arise from Euclidean D3-branes wrapping rigid divisors as well as from gaugino condensation on seven-branes. We begin with the 25 rigid O7-planes, each supporting an $\mathfrak{so}(8)$ gauge sector. Every such stack yields a gaugino condensate superpotential term, characterized by the dual Coxeter number $c_i=6$. In addition to the O7-plane divisors, the Calabi-Yau $X$ contains 91 additional rigid prime toric divisors. Among these, 
83 are pure rigid regardless of triangulation, while 8 have the property that $h^{2,1}(\widehat{D})$ depends on the triangulation, cf.~\S\ref{sec:pure_rigid}, and 
can jump across phases. A careful analysis, carried out in \cite{Demirtas:2021nlu}, shows that in the specific triangulation realizing the relevant AdS vacuum, 6 of these 8 divisors are in fact pure rigid. 

Collecting all contributions, we thus obtain $25+83+6=114$ distinct non-perturbative superpotential terms with constant Pfaffians. Since all of these exponentials appear at comparable order, they jointly contribute to the scalar potential for the K\"ahler moduli. Moreover, because  
$114>113=h^{1,1}$, the system is in principle sufficiently overconstrained to allow stabilization of all K\"ahler moduli using only the pure rigid divisors. As a result, omitting the remaining (non-)rigid divisors is both justified and self-consistent within the effective theory.

We now specify the non-perturbative superpotential as\footnote{As noted above, the precise values of the Pfaffians are currently unknown, although their overall normalization can be constrained using duality arguments \cite{McAllister:2024lnt}, see \eqref{eq:K0}. In the present vacuum, we adopt the conventions of \cite{Demirtas:2021nlu} and set all Pfaffians to $\mathcal{A}_D = 1$. It was shown in \cite{Demirtas:2021nlu} that the corresponding minimum of the scalar potential persists for a wide range of values, specifically $\mathcal{A}_D \in [10^{-4}, 10^{4}]$.}
\begin{equation}
    W = W_0 + \sum_{i}\, \mathrm{e}^{-2\pi T_{i}/c_{i}}
\end{equation}
where the overall prefactor of the sum encodes the (constant) Pfaffians associated with each pure rigid divisor. Here, $W_0$ is given in \eqref{5113w02} as specified by the specific flux choice \eqref{5113flux}. With the full superpotential in hand, we apply the stabilization algorithm described in \S\ref{sec:stabKah}. This iterative procedure solves the $F$-term conditions $D_{T_i}W=0$ while simultaneously ensuring that the putative minimum lies inside the extended K\"ahler cone $\mathcal{K}_{\star}$.  Because the K\"ahler cone of a hypersurface in a toric variety is typically a union of many adjacent chambers (cf.~Fig.~\ref{fig:extended_kahler}), separated by walls corresponding to flops of curves, the search for a vacuum may require moving between distinct toric phases. Starting from the default (Delaunay) triangulation (see e.g. \cite{Demirtas:2020dbm}), we indeed find that the minimization path traverses several such chambers of the extended K\"ahler cone. The algorithm ultimately identifies a consistent supersymmetric solution $\mathbf{t}_{\star}^{\mathrm{AdS}}$ in the interior of the cone, ensuring that all curve volumes remain positive and that the leading-order EFT is trustworthy in the neighborhood of the minimum.

The outcome is a fully specified KKLT-type compactification with a controllable non-perturbative superpotential and a supersymmetric AdS$_4$ vacuum.
At the corresponding point  in K\"ahler moduli space,
the string frame volume and Einstein-frame volume of $X$ in string units are
\begin{equation}
    \mathcal{V}\approx 944.5\; ,\quad \mathcal{V}_E =g_s^{-3/2}\, \mathcal{V}  \approx 8.1 \times 10^5\, .
\end{equation}
The divisors wrapped by Euclidean D3-branes, which generate the leading non-perturbative contributions, have Einstein-frame volumes of order $22$, while those supporting gaugino condensation are larger by a factor of six, reflecting the dual Coxeter number of $\mathfrak{so}(8)$ that appears in the exponent.
Evaluating the scalar potential at the minimum yields the vacuum energy
\begin{equation}\label{eq:V05113}
    V_0=-3\mathrm{e}^{\mathcal{K}}|W|^2\approx -1.68 \times 10^{-144}M_{\mathrm{pl}}^4\, ,
\end{equation}
demonstrating that the vacuum is indeed AdS with an exceptionally small cosmological constant. This example illustrates how a large, explicitly identifiable set of rigid divisors can be leveraged to achieve full K\"ahler moduli stabilization in a controlled and calculable supersymmetric compactification with a small cosmological constant.

Let us briefly summarize how the exponentially small cosmological constant \eqref{eq:V05113} arises. At the heart of this mechanism lies a highly non-trivial cancellation among a set of 
polynomials 
with integer coefficients determined by quantized fluxes threading the internal cycles of the compactification manifold. The leading-order $F$-term potential \eqref{eq:vfsum} depends on these flux quanta in a polynomial fashion, yet for suitable choices of fluxes \eqref{eq:PFV} described in \S\ref{sec:pfv}, there are regions of moduli space where these contributions conspire to cancel almost exactly. As a consequence, the leading terms in the effective potential are no longer polynomial but instead exponentially suppressed, enabling the emergence of large hierarchies in a controlled and technically natural way.

The structure of the mechanism is intrinsically tied to quantum gravity, and to supersymmetry. Flux quantization originates from the compactness of the internal manifold $X$ and the corresponding Dirac quantization conditions on higher-form gauge fields integrated over topological cycles of $X$.
The form of the superpotential is determined by holomorphy, and by axion shift symmetries descending from ten-dimensional gauge symmetries.  Moreover, computing the flux superpotential is made possible by mirror symmetry.  
This interplay between discrete flux data and non-perturbative physics exemplifies how deep ultraviolet features of the theory dictate the structure of the infrared effective 
action. 

\subsubsection*{Control analysis}

We now assess the degree of control over the leading-order effective theory in the neighborhood of the supersymmetric vacuum $\mathbf{t}_{\star}^{\mathrm{AdS}}$. Our first task is to analyze the spectrum of effective curves and their contributions to the worldsheet instanton expansion following the convergence tests outlined in \S\ref{sec:conv}. At the solution point, we identify 238 curves that arise as complete intersections of toric divisors and have string frame volumes $\leq 1$. These curves have been fully incorporated into the algorithmic framework described in \S\ref{sec:stabKah}. By computing GV invariants systematically, we further verified that this set includes \emph{all} effective curves with volume below $0.05$, i.e.,~precisely those that can make non-negligible contributions to the K\"ahler potential and to the holomorphic coordinates appearing in \eqref{eq:detailedform3} and \eqref{eq:detailedform4}. Given the observed distribution of curve volumes, it is highly plausible that no additional effective curves with volume $\lesssim 0.5$ have been missed.

To gauge the numerical relevance of these contributions, note that
\begin{equation}
    \frac{\mathrm{Li}_2(\mathrm{e}^{-\pi})}{(2\pi)^2} \approx 0.0011\, ,
\end{equation}
so curves with volume $\gtrsim 1$ contribute at or below the sub-percent level. Thus, provided the vacuum lies within the radius of convergence of the worldsheet instanton expansion, the set of incorporated curves suffices to capture all relevant corrections.

\begin{figure}[t]
	\centering
	\includegraphics[keepaspectratio,width=17cm]{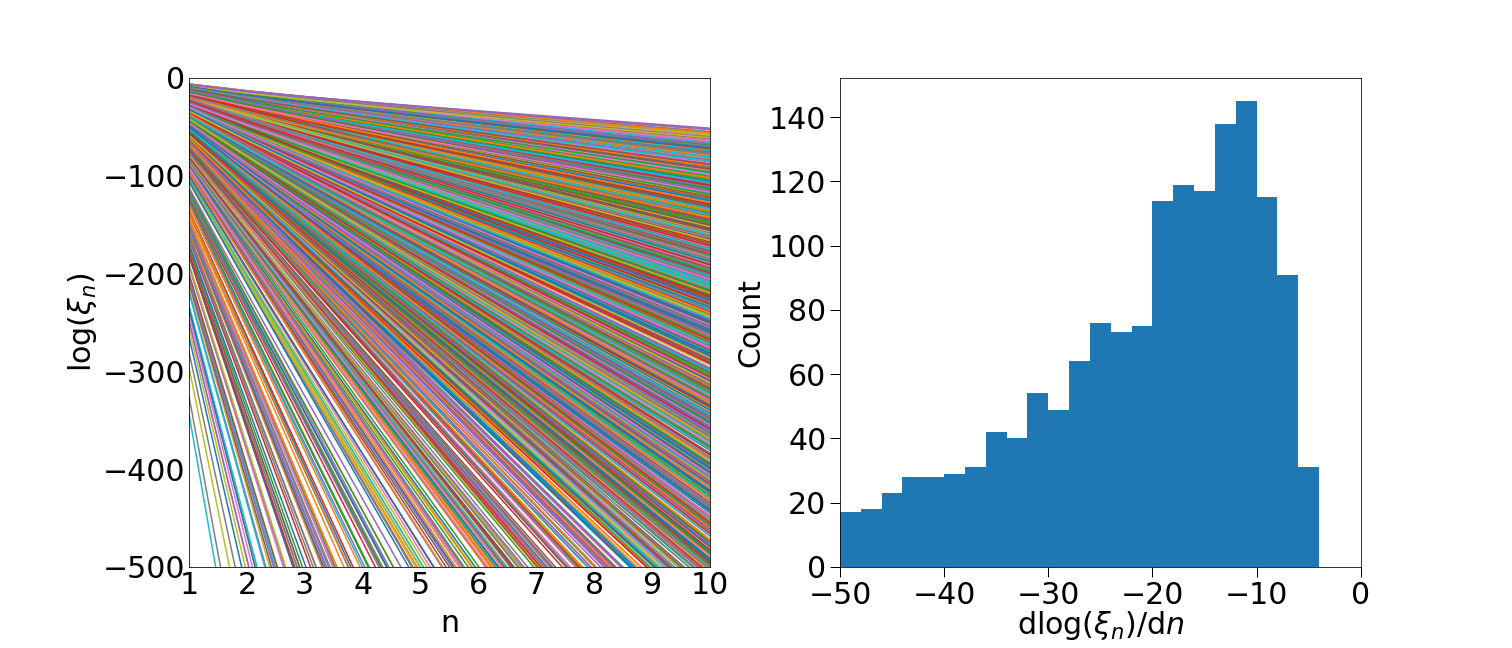}
	\caption{Convergence of the worldsheet instanton sum for $(h^{2,1},h^{1,1}) = (5,113)$. \emph{Left:} Log-magnitude $\log(\xi_n)$, cf.~\eqref{eq:xi}, for the $n$-th instanton term along 1{,}728 potent rays spanning a $101$-dimensional cone. \emph{Right:} Histogram of slopes of $\log(\xi_n)$ vs.~$n$ for the same rays, demonstrating clear exponential convergence.}\label{fig:convergence_5-113-4627}
\end{figure}

While it is not feasible to compute GV invariants along every ray in the Mori cone $\mathcal{M}_X$, we conducted a systematic analysis using a large sample of 1{,}728 randomly generated potent rays and spanning a $101$-dimensional subcone. Along each ray, GV invariants were computed to very high degree. In every case, the corresponding instanton series exhibited clear and rapid convergence, with contributions that diminish exponentially. This behavior is illustrated in Fig.~\ref{fig:convergence_5-113-4627}, where we plot the log-magnitude of the $n$-th term,
\begin{equation}\label{eq:xi}
    \xi_n \coloneqq \mathscr{N}_{n\mathbf{q}}\, \mathrm{e}^{-2\pi n\, \mathbf{q}\cdot \mathbf{t}}\, ,
\end{equation}
for the sample of potent rays. The slopes of $\log(\xi_n)$ confirm the exponential decay of the series and the numerical stability of the growth rate up to very high degree ($n \leq 100$). In particular, the most significant curve has
\begin{equation}
    t_{\mathrm{min}} \approx 1.19\; , \quad \mathscr{N} = 3\, ,
\end{equation}
and contributes
\begin{equation}
    3 \times \frac{\mathrm{Li}_2(\mathrm{e}^{-2\pi\cdot 1.19})}{(2\pi)^2}
    \approx 4.3 \times 10^{-5}\, ,
\end{equation}
which is entirely negligible for the purposes of moduli stabilization.

Next, let us consider the impact of perturbative $\alpha'$ effects and worldsheet instantons on the axio-dilaton $F$-term. We find that their effect on $D_\tau W$ is minimal because
\begin{equation}
    \partial_\tau K = \mathrm{i} g_s \left( 2 - \frac{\mathcal{T}_i t^i}{2\mathcal{V}} \right)\approx 0.0056\,\mathrm{i}\, ,
\end{equation}
which is suppressed by the small string coupling $g_s \approx 0.01$. As a result,
\begin{equation}
    D_\tau W \approx \partial_\tau W\, ,
\end{equation}
i.e.,~the corrections to the K\"ahler potential have negligible influence on the dilaton $F$-term.

Similarly, worldsheet instantons contribute only marginally to the total volume,
\begin{equation}
    \mathcal{V} 
    = \mathcal{V}^{(0)} + \delta\mathcal{V}_{(\alpha')^3}+ \delta\mathcal{V}_{\mathrm{WSI}}\; ,
    \quad
    \mathcal{V}^{(0)} \approx 945.18\; ,
    \qquad
     \delta\mathcal{V}_{(\alpha')^3}+ \delta\mathcal{V}_{\mathrm{WSI}} \approx -0.23\, ,
\end{equation}
showing that their cumulative effect is at the $10^{-4}$ level.

Furthermore, the coefficients
\begin{equation}
    g_{\mathcal{N}=1}^X \approx 0.0069\; ,
    \quad
    \max_i g_{\mathcal{N}=1}^{\omega_i} \approx 0.014\, ,
\end{equation}
which measure the strength of unknown warping corrections to the K\"ahler potential (see \eqref{eq:Demirtas:2021nlu417} and \eqref{eq:Demirtas:2021nlu418}), are also small. This confirms these uncomputed corrections are strongly suppressed at the vacuum.

Finally, because the flux-induced superpotential is extremely small, the largest sub-leading correction to the K\"ahler-moduli expectation values, estimated via \eqref{eq:relcorr}, remains modest:
\begin{equation}
    \frac{\log\left[ \log(W_0^{-1}) \right]}{\log(W_0^{-1})}
    \approx 0.04\, .
\end{equation}
This again falls well within the regime of perturbative control.

Taken together, these results demonstrate that worldsheet instantons, perturbative $\alpha'$ corrections, and residual $\mathcal{N}=1$ effects are all parametrically suppressed in the vacuum. The effective theory used to determine $\mathbf{t}_{\star}^{\mathrm{AdS}}$ is therefore 
well-controlled.

\subsubsection*{The supersymmetric cosmological constant problem and perturbative stability}

A fundamental question for any construction exhibiting an exponentially small vacuum energy is whether such a hierarchy is stable under perturbative corrections. In the present compactification, the string coupling takes the moderately weak value $g_s \simeq 0.01$, so the primary concern is whether loop corrections of order $g_s$ could destabilize the delicate structure responsible for the small cosmological constant. In a generic non-supersymmetric effective theory, corrections to vacuum energies scale as $g_s \Lambda^4$ for some ultraviolet scale $\Lambda$, and even tiny loop corrections would easily overwhelm an exponentially suppressed vacuum energy. However, the situation is far more favorable in an $\mathcal{N}=1$ supersymmetric theory.

The key observation is that the $\mathcal{N}=1$ vacuum energy is exactly
\begin{equation}
    \rho_{\mathrm{vac}}\Bigl |_{\mathcal{N}=1} = -3\mathrm{e}^{\mathcal{K}}\, |W|^2\, ,
\end{equation}
where $W$ is the holomorphic superpotential and $\mathcal{K}$ is the (real) K\"ahler potential. Supersymmetry ensures that the superpotential is protected against perturbative renormalization (cf.~\S\ref{sec:non-renormalization}); loop effects can only enter through corrections to the K\"ahler potential $\mathcal{K}$. This drastically restricts the ways in which perturbative corrections can affect the vacuum energy: they can shift $\mathcal{K}$ and the location of the moduli vacuum, but they cannot alter the holomorphic structure of $W$ nor generate new perturbative contributions that compete with its exponentially small value, as we now explain.

To proceed, we parameterize the loop-corrected K\"ahler potential as
\begin{equation}
    \mathcal{K}(\Phi,\overline{\Phi}) = \mathcal{K}_{\mathrm{l.o.}}(\Phi,\overline{\Phi})+g_s\delta\mathcal{K}(\Phi,\overline{\Phi})
\end{equation}
where $\mathcal{K}_{\mathrm{l.o.}}$ denotes the leading order K\"ahler potential \eqref{eq:detailedform2} including all the tree-level $\alpha'$ corrections and $\delta\mathcal{K}$ encodes the full set of perturbative string loop corrections. Because the moduli now minimize the loop-corrected potential, they shift from their leading-order value $\Phi_0$ to\footnote{For simplicity, we suppress the indices for the fields $\Phi^A$ in what follows, noting that the expressions below can be suitably covariantized.}
\begin{equation}\label{eq:PhiVacSusy}
    \Phi_{\mathrm{vac}} = \Phi_{0} + \delta\Phi\, .
\end{equation}
The vacuum must satisfy the corrected supersymmetry condition
\begin{equation}\label{eq:DWSPhi}
    D_{\Phi}W=\partial_{\Phi}W+\mathcal{K}_{\Phi} W=0\, .
\end{equation}
At leading order, the stabilization algorithm ensured that $\Phi_0$ obeyed
\begin{equation}
    |W(\Phi_{0})|\ll 1\; ,\quad |\partial_{\Phi}W(\Phi_{0})|\ll 1\; ,\quad \biggl (\partial_{\Phi}W+(\mathcal{K}_{\mathrm{l.o.}})_{\Phi} W\biggl )\bigl |_{\Phi=\Phi_{0}}=0\, .
\end{equation}
which already required delicate cancellations among the quantized flux polynomials. Including loop corrections modifies the K\"ahler-covariant derivative in \eqref{eq:DWSPhi}, but by continuity, 
for sufficiently small\footnote{In principle one should check that the resulting solution lies inside the moduli space.} $\delta \mathcal{K}$, one can find a nearby solution $\Phi_{\mathrm{vac}}$ satisfying
\begin{equation}
    |W(\Phi_{\mathrm{vac}})|\ll 1\; ,\quad\Bigl(\partial_{\Phi}W+(\mathcal{K}_{\mathrm{l.o.}}+g_s\delta \mathcal{K})_{\Phi} W
    \Bigl)\bigl |_{\Phi=\Phi_{\mathrm{vac}}}=0\, .
\end{equation}
Plugging in \eqref{eq:PhiVacSusy} and expanding the corrected $F$-term condition around $\Phi_0$ to linear order in $\delta \Phi$, we obtain 
\begin{equation}
    \delta \Phi= -g_s\dfrac{W(\Phi)(\delta\mathcal{K})_{\Phi}}{\partial_{\Phi}D_{\Phi}W}\biggl |_{\Phi=\Phi_{0}}=\mathcal{O}(g_s)\, .
\end{equation} 
Thus, $\delta\Phi$ is parametrically suppressed.

Therefore the value of the superpotential in the corrected vacuum becomes
\begin{equation}
    W(\Phi_{\mathrm{vac}}) = W(\Phi_{0})\bigl [1+\mathcal{O}(g_s)\bigl ]
\end{equation}
and inserting this into the vacuum energy gives
\begin{equation}\label{eq:SUSYrhovaccor}
    \rho_{\mathrm{vac}}\bigl |_{\Phi=\Phi_{\mathrm{vac}}} = -3\mathrm{e}^{\mathcal{K}_{\mathrm{l.o.}}+g_s\delta \mathcal{K}}\, |W(\Phi_{0})|^2\times \bigl [1+\mathcal{O}(g_s)\bigl ] =\rho_{\mathrm{vac}}\bigl |_{\Phi=\Phi_{0}} \times \bigl [1+\mathcal{O}(g_s)\bigl ]\, .
\end{equation}
Here, the vacuum energy in the leading-order EFT
\begin{equation}
    \rho_{\mathrm{vac}}\bigl |_{\Phi=\Phi_{0}} = -3\mathrm{e}^{\mathcal{K}_{\mathrm{l.o.}}}\, |W(\Phi_{0})|^2
\end{equation}
is known to be exponentially small. Thus loop corrections enter only as a mild multiplicative factor in \eqref{eq:SUSYrhovaccor} and do not jeopardize the exponentially suppressed vacuum energy. This robustness is a direct consequence of $\mathcal{N}=1$ supersymmetry, flux quantization, and the non-renormalization of the superpotential.

Before concluding, it is important to emphasize the scope and limitations of the construction. The vacua we have identified remain exactly supersymmetric, and although the resulting vacuum energy is exponentially small, it is nevertheless negative. From the perspective of cosmology, this is insufficient: addressing the cosmological constant problem would require achieving a small \emph{positive} vacuum energy after supersymmetry breaking, a feature that is not realized within the present setup.  

Thus, while the mechanism outlined above provides a remarkably controlled realization of supersymmetric AdS$_4$ vacua with parametrically small vacuum energy and fully stabilized moduli, it does not by itself resolve the cosmological constant problem. Any phenomenologically viable construction would require additional ingredients that break supersymmetry and uplift the vacuum energy to a metastable de Sitter value.  In this sense, the present analysis should be viewed as establishing a precise and stable starting point upon which such further ingredients, whether explicit uplifts, or dynamical supersymmetry-breaking mechanisms, would need to be incorporated. We will come back to these issues in \S\ref{sec:towardsDS_part2}.

\section{A vacuum with a Klebanov-Strassler throat}\label{sec:KKLMMTexPartA}

In the previous section, we constructed a supersymmetric AdS$_4$ vacuum featuring exponentially small vacuum energy, achieved by engineering perturbatively flat vacua and stabilizing the K\"ahler moduli through non-perturbative effects. We now turn to a different but complementary goal: constructing a compactification that incorporates a warped deformed conifold region. The exponential warping in such a region correspondingly suppresses physical scales, and this structure will be the key to controlled supersymmetry breaking in \S\ref{sec:KKLMMTex}-\S\ref{sec:aule}.

For context, we remark that an example of a Klebanov-Strassler region in a warped flux compactification was constructed long ago in \cite{Giddings:2001yu}.  However, in \cite{Giddings:2001yu} the complex structure moduli and dilaton of the bulk compactification were not explicitly stabilized, though a plausible argument was made that these fields would be stabilized somewhere.  More importantly, the K\"ahler moduli were not stabilized in \cite{Giddings:2001yu}: indeed, no concrete mechanism for their stabilization had been invented.  The first example of a Calabi-Yau orientifold flux compactification containing a Klebanov-Strassler region \emph{in a vacuum with stabilized moduli} appeared in \cite{McAllister:2024lnt}.

\medskip

Following \cite{McAllister:2024lnt}, we consider a mirror pair of smooth Calabi-Yau threefolds $X$ and $\widetilde{X}$ realized as hypersurfaces in four-dimensional toric varieties $V$ and $\widetilde{V}$, respectively, and satisfying
\begin{equation}
    h^{1,1}(X)=h^{2,1}(\widetilde{X})=150\, ,\quad h^{2,1}(X)=h^{1,1}(\widetilde{X})=8\, .
\end{equation}
All data necessary to construct the examples can be found in a dedicated GitHub repository \github{https://github.com/AndreasSchachner/kklt_de_sitter_vacua}.
There exists a sign-flip orientifold of $X$ with $h^{1,1}_-(X/\mathcal{I})=h^{2,1}_+(X/\mathcal{I})=0$ \cite{Moritz:2023jdb}, and we   
will compactify on this orientifold.
There are 35 O7-planes wrapping rigid divisors, a single O7-plane on a non-rigid divisor $D_1$, and 60 O3-planes at the triple intersections of divisors.  
Each O7-plane on a rigid divisor hosts a confining $\mathcal{N}=1$ pure Yang-Mills theory with gauge algebra $\mathfrak{so}(8)$.
The induced D3-brane charge \eqref{eq:QOf} from orientifolds 
is
\begin{equation}
    Q_O = \frac{\chi_f}{2}=h^{1,1}(X)+h^{2,1}(X)+2=160\, .
\end{equation}

\medskip

Our strategy in this section parallels that of the previous example, but now with the added feature of a conifold modulus associated with a warped throat region. As before, we begin by choosing quantized 3-form fluxes to engineer a perturbatively flat vacuum, ensuring that the classical flux superpotential vanishes along a complex line in moduli space. In this case, the PFV includes a conifold point, allowing us to realize a warped deformed conifold geometry in the internal space. Once the complex structure moduli and axio-dilaton are stabilized by the fluxes, we proceed to stabilize the K\"ahler moduli through non-perturbative effects, solving the corresponding $F$-term equations.

\subsubsection*{Complex structure moduli stabilization}

We start by selecting fluxes in $H^{3}(X,\mathbb{Z})$ that lead to a conifold PFV, as introduced in \S\ref{sec:conipfv}. One of the generators of the Mori cone $\mathcal{M}_{\widetilde{X}}$ of $\widetilde{X}$ is a conifold curve with  $n_{\text{cf}}=2$. We work in a basis for $H_2(\widetilde{X},\mathbb{Z})$ in which the conifold curve is $\mathcal{C}_{\text{cf}}=(1,0,\ldots,0)$, and the second Chern class of $\widetilde{X}$ is
\begin{equation}
     c_2(\widetilde{X}) = \begin{pmatrix}  0& 184& 112&  10&  10&  26&   0&  -4 \end{pmatrix} ^\top\, .
\end{equation}
The following choice of flux vectors
\begin{align}
    \vec{M} &= \begin{pmatrix} 14 & -23&  39&  -8&  2&  -6&  -5&  -13  \end{pmatrix}^\top\, , \\[0.4em]
    \vec{K}&=\begin{pmatrix} -4 & -1 &  -3 &  -1 &  -2 &  1 &  -1 &  1\end{pmatrix}^\top\,,
\end{align}
give rise to a conifold PFV satisfying \eqref{eq:coniPFV}, with 
\begin{equation}
    \vec{p} = \frac{1}{48}\times \begin{pmatrix} 0 & 8 &  -3 &   2 &  -4 &  -6 &  -3 & -4\end{pmatrix}^\top\,.   
\end{equation}
The conifold flux quanta \eqref{eq:coniMdef} and \eqref{eq:Kprime} are then given by
\begin{equation}
    M = 14\,  \quad  \text{and }\quad  K' = 4\,.
\end{equation}
Moreover, we obtain
\begin{equation}
    -\vec{M}\cdot\vec{K} = 160 = 2+h^{1,1}(X)+h^{2,1}(X)\,,
\end{equation}
and so Gauss's 
law \eqref{eq:D3Tadpole} is satisfied.

Along the PFV locus, the bulk flux superpotential admits a controlled expansion whose leading contributions take the form
\begin{equation}\label{eq:WfKKLMMT}
    W_{\text{eff}} = \frac{1}{\sqrt{8\pi^5}}\biggl(28\,\mathrm{e}^{2\pi \I \tau \cdot \frac{1}{48}} - 22\,\mathrm{e}^{2\pi \I\tau \cdot \frac{1}{24}} +
    76\,\mathrm{e}^{2\pi \I \tau \cdot \frac{1}{16}} +\ldots
    \biggr)\,.
\end{equation}
We use the approximate PFV minimum obtained by retaining only the first two terms in \eqref{eq:WfKKLMMT} as an initial seed for a numerical solution of the full set of F-term conditions, $D_{a}W_{\text{flux}}=D_{\tau}W_{\text{flux}}=0$. The numerical minimization converges to an exact $F$-flat vacuum of the complete flux superpotential \eqref{eq:WfluxPeriods} with
\begin{equation}
    g_s = 0.0746\, , \ z_{\text{cf}} = 9.06\times 10^{-7}\, , \ W_0 = 0.032\, , \ \text{and}\ g_sM = 1.05\,.
\end{equation}

\subsubsection*{K\"ahler moduli stabilization}

Once the complex structure moduli have been fixed, the remaining task is to stabilize the K\"ahler sector using non-perturbative effects. As a first step, it is necessary to assemble the full non-perturbative superpotential \eqref{eq:Wnpexact}. In total, there are 152 rigid prime toric divisors of $X$, and their contributions to the superpotential fall into two qualitatively different classes:
\begin{itemize}
    \item 35 divisors are wrapped by O7-planes carrying $\mathfrak{so}(8)$ SYM theories. Strong coupling dynamics in these sectors leads to gaugino condensation, producing non-perturbative superpotential terms of the form \eqref{eq:wlo3} with coefficient $c_D = 6$.
    \item The remaining 117 rigid divisors are wrapped by Euclidean D3-branes and generate non-perturbative superpotential contributions with $c_D = 1$.
\end{itemize}
Among the Euclidean D3-brane divisors, one intersects the conifold locus.  
As discussed in \S\ref{sec:stabKah}, instantons supported on such divisors do not give rise to significant contributions and can therefore be neglected. After excluding this single contribution, the non-perturbative superpotential contains $116+35=h^{1,1}(X) + 1$ independent terms.

A key structural property of this divisor set is that all 151 contributing divisors are pure rigid in every FRST phase.\footnote{Because the dual polytope $\Delta^{\circ}$ contains two genus-one two-faces, the computation of $h^{2,1}(\widehat{D})$ in \eqref{eq:dpurerigid} requires some care. Using the techniques developed in \cite{Kim:2021lwo}, we have verified explicitly that all divisors entering the superpotential are pure rigid in all phases.} We fix the Pfaffian prefactors of all 151 contributions according to \eqref{eq:pfaffvalue}.

The outcome is a fully specified KKLT-type compactification with a controllable non-perturbative superpotential and a supersymmetric AdS$_4$ vacuum. At the minimum, the string-frame and Einstein-frame volumes are
\begin{equation}
    \mathcal{V} \approx 650.31\,, \qquad
    \mathcal{V}_E = g_s^{-3/2}\, \mathcal{V} \approx 3.192\times 10^{4}\,.
\end{equation}
Evaluating the scalar potential at the minimum yields
\begin{equation}
    V_0=-3\mathrm{e}^{\mathcal{K}}|W|^2\approx -1.49\times 10^{-17}M_{\mathrm{pl}}^4\, ,
\end{equation}
confirming that the vacuum is AdS.

\section{Towards inflation in string theory}\label{sec:KKLMMTex}
 
In \S\ref{sec:KKLMMTexPartA} we presented a supersymmetric AdS$_4$ vacuum with a strongly warped KS throat, in which all moduli are stabilized.  In this vacuum,
Gauss’ law \eqref{eq:D3tadpole_compact} is satisfied by $3$-form fluxes alone, so it is not possible to introduce an anti-D3-brane.  However, a D3-brane---anti-D3-brane pair carries no net charge, and can be included consistent with Gauss's 
law.

The compactification of \S\ref{sec:KKLMMTexPartA} is particularly special because it is \emph{well-aligned}, with $\Xi \approx 3.38$ --- recall Eq.~\eqref{eq:aligncond} and the discussion in \S\ref{sec:alignment}.  As a result, adding a single
brane/anti-brane pair does not trigger immediate decompactification. Instead, it uplifts the supersymmetric vacuum to a configuration 
with positive energy, in which the closed string moduli are still in the basin of attraction of their supersymmetric minimum.
This positive-energy configuration is not necessarily a vacuum, i.e.~a critical point of the scalar potential: instead, dynamical evolution will generically occur.  

The nature of the evolution depends on the initial conditions. If the D3-brane and anti-D3-brane are introduced within a string length of each other, then an open string tachyon instability will lead to rapid 
annihilation. However, suppose that instead the anti-D3-brane is initially near the throat region, and the D3-brane is far away in the compactification.  Then the anti-D3-brane will quickly fall to the tip, under the influence of the strong classical force of the fluxes.\footnote{Recall that the Klebanov-Strassler throat carries positive D3-brane charge, so the anti-D3-brane is pulled just as it would be to a stack of D3-branes.  For a discussion of the potential the anti-D3-brane experiences in the angular directions of the $S^3$ at the tip, see \cite{DeWolfe:2007hd}.}  The D3-brane, on the other hand, preserves the same supersymmetries as the no-scale ISD flux background, and is subject to much weaker forces: it experiences a Coulomb attraction to the anti-D3-brane, and is also affected by the non-perturbative effects that stabilize the K\"ahler moduli \cite{Kachru:2003sx}.

Just such a setting --- a D3-brane/anti-D3-brane pair in a Klebanov-Strassler region of a KKLT compactification --- was proposed in \cite{Kachru:2003sx} as a model of \emph{inflation in string theory}.  The idea is that the D3-brane's position in the compact space can correspond to an inflaton field,\footnote{Strictly speaking, the D3-brane position corresponds to six real inflaton fields.  The radial separation from the anti-D3-brane along the Klebanov-Strassler throat was singled out in \cite{Kachru:2003sx} because of the role of the Coulomb interaction, but as shown in \cite{Baumann:2007ah,Agarwal:2011wm,McAllister:2012am}, the angular coordinates on $T^{1,1}$ are also relevant in general.}
with motion toward the anti-D3-brane occurring as inflation proceeds, and with brane-antibrane annihilation through tachyon condensation providing a hybrid exit.\footnote{The proposal of \cite{Kachru:2003sx} thus provided a concrete scheme for  brane-antibrane inflation \index{brane-antibrane inflation}\index{inflation} \cite{Burgess:2001fx,Dvali:2001fw} in a compactification 
in which the closed string moduli can be stabilized in a KKLT vacuum.}
However, inflation in this setting requires an underlying stable AdS$_4$ or metastable de Sitter vacuum, and this necessary foundation was unavailable before  \cite{McAllister:2024lnt}.  For this reason, to date no totally explicit compactification realizing warped D-brane inflation has been constructed. 

The compactification of  \S\ref{sec:KKLMMTex} is thus a perfect setting for an incarnation of the warped D-brane inflation scenario of \cite{Kachru:2003sx}.  Let us now outline the steps that will be required to achieve this goal, referring to \cite{Baumann:2014nda} for details.

At first glance, the D3-brane position appears to receive its dominant potential from the Coulomb potential, which pulls the D3-brane toward the anti-D3-brane at the bottom of the throat.  Because the warp factor redshifts the interaction energy, the resulting potential is highly suppressed and can become exceptionally flat \cite{Kachru:2003sx}. In the absence of additional corrections, this warped Coulomb potential therefore provides a natural setting for slow-roll inflation driven by the motion of the D3-brane toward the anti-D3-brane.

However, as already emphasized in \cite{Kachru:2003sx}, the inclusion of K\"ahler moduli stabilization introduces a significant complication. The non-perturbative effects that stabilize the K\"ahler moduli also induce corrections to the inflaton potential, through the dependence of the superpotential and K\"ahler potential on the D3-brane position.  These corrections generically cause the inflationary $\eta$-parameter to be of order unity, and thus spoil the conditions required for slow-roll inflation. To achieve viable inflationary dynamics, a fine-tuning at roughly the percent level is required to cancel these contributions and restore sufficient flatness of the inflaton potential. 

For such a fine-tuning to be possible --- for example, by rejection sampling of models --- one needs to know the potential to sufficient accuracy.  Specifically, one needs to be able to compute inflaton potential terms up to dimension-six Planck-suppressed contributions, and to percent-level accuracy. Significant effort has gone into characterizing contributions to the inflaton potential. The impact of the inflaton-dependence of Pfaffian prefactors in the non-perturbative superpotential was computed in \cite{Baumann:2006th} (building on \cite{Berg:2004ek}) and \cite{Baumann:2007ah}, while contributions from the bulk of the compactification were characterized in \cite{Baumann:2008kq} and \cite{Baumann:2010sx}.\footnote{The resulting phenomenology was explored in \cite{Agarwal:2011wm,McAllister:2012am}; see \cite{Baumann:2014nda} for a detailed review.}
 
Computing the potential for a D3-brane inflaton in the compactification of \S\ref{sec:KKLMMTexPartA} will require substantial numerical work, but may be within reach.  Key unknowns include the effects of gluing the Klebanov-Strassler throat into the bulk, as well as the backreaction on the flux background of the four-cycles supporting non-perturbative effects \cite{Koerber:2007xk,Baumann:2010sx,Dymarsky:2010mf}. Determining whether the D3-brane trajectory leads to slow-roll inflation, rather than a rapid fall followed by tachyon condensation, would at the very least require applying the analysis of \cite{Baumann:2010sx}, and more realistically, a full numerical supergravity solution. This remains a natural direction for future work.

The inflationary model of \cite{Kachru:2003sx} thus illustrates both the promise of embedding inflation in string theory, and the challenges posed by the interplay between moduli stabilization and inflaton dynamics.

\section{The search for de Sitter vacua}\label{sec:towardsDS_part2}

Up to this point, our analysis has focused primarily on constructing supersymmetric $\mathrm{AdS}_4$ vacua. We now turn to the question of uplifting such solutions to de Sitter space by introducing anti-D3-branes at the bottom of a warped throat. This requires working with a slightly different set of flux configurations. In particular, tadpole cancellation \eqref{eq:D3tadpole_compact} demands that the quantized fluxes $\vec{f}$ and $\vec{h}$ satisfy
\begin{equation}\label{eq:qqplus2}
     Q_{\text{flux}} = Q_{\text{O}} + 2p
\end{equation}
where $p$ denotes the number of anti-D3-branes placed at the tip of the throat (recall \S\ref{sec:KPV}). Thus, compared to the supersymmetric case, constructing upliftable vacua requires adjusting the fluxes such that the total flux-induced D3-brane charge exceeds the orientifold contribution $Q_{\text{O}}$ by an even integer.

A controlled uplift further requires identifying \emph{well-aligned} supersymmetric AdS vacua that satisfy the tuning condition~\eqref{eq:aligncond}. This criterion is sensitive to several parameters of the compactification, so it is natural to ask which regions of parameter space can be realized in explicit models. The condition~\eqref{eq:aligncond} demands that the conifold modulus develops a small expectation value, $|z_{\text{cf}}| \ll 1$. From \eqref{eq:conifold_vev}, this vacuum expectation value is determined by the amount of D3-brane charge stored in the throat, given in \eqref{eq:D3_charge_in_throats}, which obeys the bound
\begin{equation}\label{eq:throatisless}
    Q_{\text{flux}}^{\text{throat}} \le  Q_{\text{flux}} = Q_{\text{O}} + 2p\,,
\end{equation}
in terms of the D3-brane tadpole from orientifolds, $Q_{\text{O}}$, and the total D3-brane charge from fluxes, $Q_{\text{flux}}$, which we defined in~\eqref{eq:D3tadpole_compact} and \eqref{eq:QOtrilayer}.
Using this inequality together with~\eqref{eq:conifold_vev} yields the lower bound
\begin{equation}\label{eq:zbound}
    \langle |z_{\mathrm{cf}}| \rangle \gtrsim \frac{1}{2\pi} \exp \left(-\frac{2\pi Q_{\text{O}}}{g_s M^2\, n_{\mathrm{cf}}}\right)\, .
\end{equation}
In our examples, the orientifold tadpole $Q_{\text{O}}$ is fixed entirely by the Hodge numbers of the Calabi-Yau threefold $X$ (recall \eqref{eq:QOtrilayer}), so the right-hand side of \eqref{eq:zbound} depends only on $(g_s, M, n_{\mathrm{cf}})$.

As emphasized in Refs.~\cite{Bena:2018fqc,Bena:2020xrh,Gao:2022fdi}, the condition~\eqref{eq:aligncond} may equivalently be interpreted as a \emph{lower bound on the flux superpotential} $W_0$. For a (conifold-)PFV one typically has $\widetilde{\mathcal{V}} \sim g_s^{-3}$,\footnote{For a (conifold-)PFV, $z^a = \tau p^a$, which implies $\widetilde{\mathcal{V}} = \widetilde{\kappa}_{abc}\, \mathrm{Im}(z^a)\mathrm{Im}(z^b)\mathrm{Im}(z^c)/6 \sim g_s^{-3}$.} and empirically we find $\mathcal{V}_E^{2/3} \sim Q_{\text{O}}/g_s$.  
Using these relations and setting $n_{\mathrm{cf}} = 2$, we obtain the parametric bound
\begin{equation}\label{eq:cookedness}
    W_0 \gtrsim \frac{1}{g_s^2M} \sqrt{\frac{\zeta Q_{\text{O}}}{(2\pi)^{4/3}}}\, \exp\left(-\frac{2\pi Q_{\text{O}}}{3g_sM^2}\right)\, .
\end{equation}
At large ’t~Hooft coupling $g_s M$, the exponential suppression in~\eqref{eq:cookedness} becomes severe, greatly restricting the range of achievable control parameters. For example, taking $g_s = 1/(g_s M) = 0.2$ and using the maximal value $Q_{\text{O}} \le 504$ in our dataset, one finds that upliftable vacua require $W_0 > 0.015$.

For completeness, we remark that if a metastable supersymmetry-breaking state existed directly within the Klebanov-Strassler gauge theory \cite{Klebanov:2000hb}, i.e., at small ’t~Hooft coupling $g_s M$, it could in principle accommodate de Sitter vacua with exponentially small $W_0$. Determining whether such a metastable state can be realized in the gauge theory regime remains an important open problem and a compelling avenue for future work.

\begin{table}
    \centering
    \begin{tabular}{c|c|c}
   \textbf{Condition}      & \textbf{Number of configurations} & \textbf{Explanation} \\ \hline 
    & & \\[-0.8em]
   consistent models  &  416 conifolds  & \S\ref{sec:towardsDS_part2} \\[0.4em]
   fluxes giving conifold PFV\index{conifold PFV}   &  240{,}480{,}253 conifold PFVs  & \S\ref{sec:conipfv} \\[0.4em]
    two-term racetrack    &  141{,}594{,}222  racetrack PFVs  & \S\ref{sec:conipfv}, \eqref{weffis} \\[0.4em]
   $M>12$; one anti-D3-brane     & 33{,}371  anti-D3-brane PFVs  & \S\ref{sec:towardsDS_part2}, \eqref{eq:qqplus2} \\[0.4em]
   de Sitter vacuum & 30 de Sitter vacua & \S\ref{sec:manwe}, \S\ref{sec:aule}
    \end{tabular}
    \caption{Count of configurations retained after each step. Each line incorporates all requirements listed in the preceding rows, so that the constraints accumulate as one moves downward. In the final step, involving flux choices that generate conifold PFVs (recall \S\ref{sec:conipfv}), the exploration was thorough but not guaranteed to capture every admissible configuration.}
    \label{tab:scan_summary}
\end{table}

\medskip

Next, let us describe our systematic search for metastable de Sitter vacua in type~IIB Calabi-Yau orientifold compactifications.  Our objective is to identify explicit geometries and flux choices in which all complex structure moduli, the axio-dilaton, and the K\"ahler moduli are stabilized, and in which the resulting scalar potential admits a controlled de Sitter minimum sourced by a single anti-D3-brane. In what follows, we outline the sequence of constraints used to narrow the landscape of possibilities to a tractable subset of promising candidates.
Our selection procedure consists of the following steps:
\begin{enumerate}
    \item We begin with Calabi-Yau threefold hypersurfaces $X$ realized in toric varieties obtained from fine, star triangulations of reflexive polytopes.  
    We restrict to threefolds whose mirrors $\widetilde{X}$ contain a toric flop curve that can shrink to produce a conifold singularity.

    \item We choose an orientifold projection inherited from a toric involution such that $h^{1,1}_-(X)=h^{2,1}_+(X)=0$.  
    For these orientifolds the total D3-brane charge of the O-planes is fixed by the topology of $X$ through $Q_{\mathrm{O}} = 2 + h^{1,1}(X) + h^{2,1}(X)$.

    \item We retain only those cases in which the orientifold fixed locus does not intersect the conifold locus of $X$.  
    This ensures that the KS throat can form consistently within the compactification.

    \item We require that at least $h^{1,1}(X)$ rigid divisors that do not meet the conifold locus are present.  
    These divisors must support pure non-perturbative contributions to the superpotential of the form \eqref{eq:Wnpexact}.

    \item We select quantized three-form fluxes satisfying $Q_{\mathrm{flux}} = Q_{\mathrm{O}} + 2$, so that Gauss’ law \eqref{eq:D3tadpole_compact} demands the presence of exactly one anti-D3-brane.

    \item Among all such fluxes, we keep only those that stabilize the complex structure moduli and the axio-dilaton near a conifold point, at weak string coupling, and with a small flux superpotential vev.  Concretely, we require  
    \begin{equation}
        \langle |z_{\mathrm{cf}}|\rangle_{\!F} \ll 1,\qquad g_s = (\mathrm{Im}(\langle\tau\rangle_{\!F}))^{-1} \ll 1,\qquad |W_0| \ll 1\, ,
    \end{equation}
    consistent with the conditions listed in \eqref{eq:promotedvevs}.

    \item We impose the metastability criterion for the anti-D3-brane in the KS throat by demanding that the flux $M$ on the conifold $A$-cycle (cf.\ \eqref{eq:coniMdef}) obeys $M>12$.  
    This ensures compatibility with the KPV condition \eqref{eq:KPV_constraint}.

    \item We then determine whether the full scalar potential
    \begin{equation}
        V(\Phi)= V_F(\Phi)\,+\,V_{\overline{\mathrm{D3}}}(\Phi)\, ,
    \end{equation}
    viewed as a function of the complete set of moduli $\Phi$, possesses a local de Sitter minimum. A configuration is accepted if there exists a point $\Phi_{\star}$ with 
    \begin{equation}
        V(\Phi_{\star})>0,\qquad \partial_{\Phi}V(\Phi_{\star})=0,\qquad 
        \text{Hess}\,V(\Phi_{\star})>0\, .
    \end{equation}
    \item Finally, we verify that all truncations are self-consistent at the candidate vacuum.  
    In particular, the omitted divisors must contribute negligibly to the non-perturbative superpotential \eqref{eq:Wnpexact}, and all curves excluded from the instanton expansions \eqref{eq:detailedform3} and \eqref{eq:detailedform4} must yield subdominant contributions compared with those kept in the analysis.
\end{enumerate}
A configuration satisfying all of these conditions constitutes a bona fide KKLT-type de Sitter vacuum of the leading-order effective  theory.

\medskip

The successive constraints and the number of configurations that survive each filter are summarized in Table~\ref{tab:scan_summary}. At the level of reflexive polytopes, our scan includes all entries in the Kreuzer-Skarke database compatible with the imposed geometric and orientifold criteria. For each such threefold, conifolds associated with toric flop curves were fully analyzed; potential conifolds arising from non-toric curves were not included.

\begin{figure}[!t]
\centering
\includegraphics[width=\linewidth]{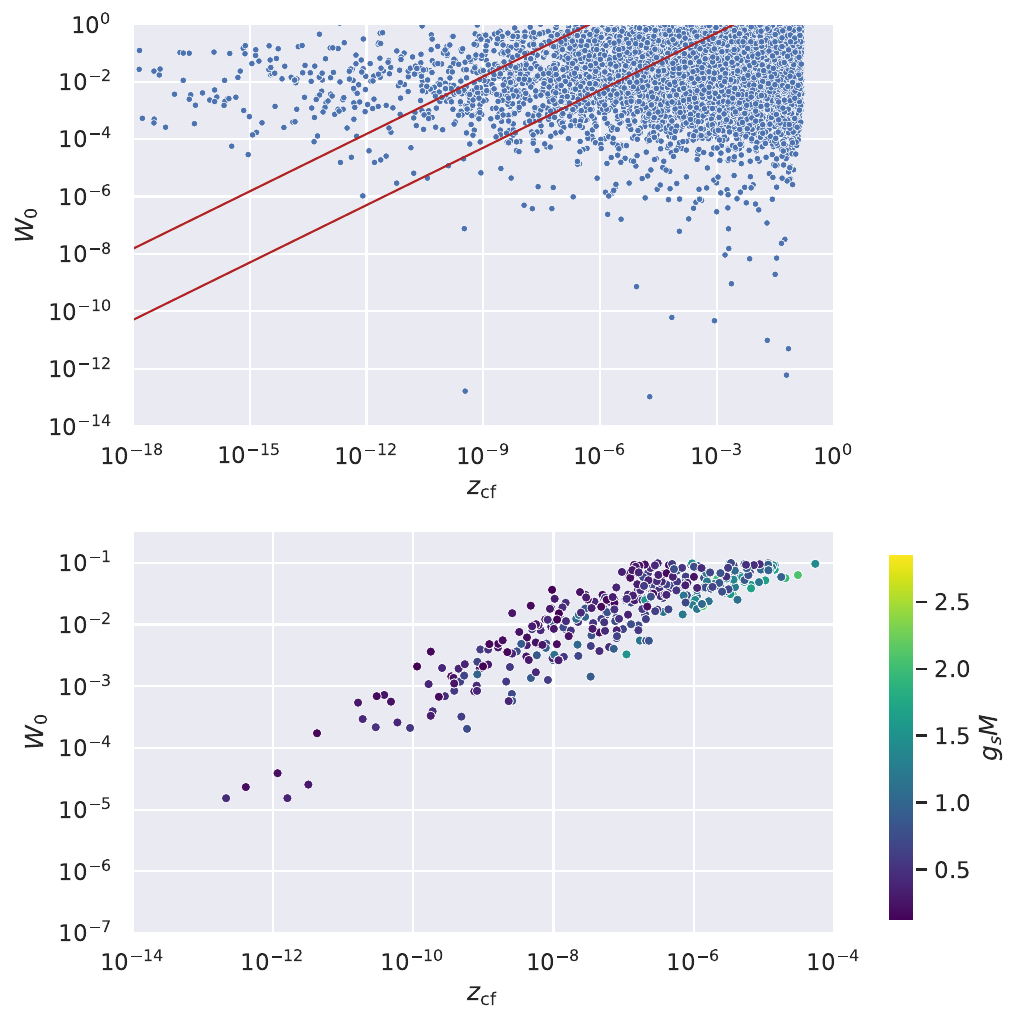}
\caption{Scatter plot of $W_0$ versus the conifold modulus $z_{\mathrm{cf}}$ for the $33{,}371$ anti-D3-brane PFVs identified in our scan. The lines in the upper panel indicate the alignment bounds $0.1 \leq \Xi \leq 10$, where $\Xi$ is defined in \eqref{eq:throat_tuning}. The lower panel focuses on the 396 configurations that satisfy the alignment condition together with $g_s < 0.4$ and $W_0 < 0.1$.}
\label{fig:W0_zcf_comb}
\end{figure}

Implementing the above algorithm across $416$ orientifold geometries, we enumerated $240{,}480{,}253$ flux vectors satisfying 
\begin{equation}\label{eq:flux_ellipsoid2}
    Q_{\text{flux}}^{\text{bulk}}\approx - m^\mu m^\nu \kappa_{\mu\nu\gamma}p^\gamma\leq Q_O+2(N_{\overline{\text{D3}}}-N_{\text{D3}})\, ,
\end{equation}
with $N_{\overline{\mathrm{D3}}}=1$, $N_{\mathrm{D3}}=0$, and with $k_{\mathrm{max}}=500$. Among these, $141{,}594{,}222$ admit two-term racetrack structures, and after imposing
\begin{equation}\label{eq:flux_ellipsoid}
    Q_{\text{flux}}^{\text{bulk}}\approx - m^\mu m^\nu \kappa_{\mu\nu\gamma}p^\gamma\overset{!}{=}Q_O+2(N_{\overline{\text{D3}}}-N_{\text{D3}})- M K'\, .
\end{equation}
together with $M>12$, we obtain a set of \emph{anti-D3-brane PFVs}\index{anti-D3-brane PFV}, in total $33{,}371$ such vacua.  
These results are displayed in Fig.~\ref{fig:W0_zcf_comb}.

\section{A candidate de Sitter vacuum}\label{sec:manwe} 

As our first illustration of a candidate de Sitter vacuum, we examine a pair of smooth Calabi-Yau threefolds $X$ and $\widetilde{X}$ that arise as hypersurfaces in four-dimensional toric varieties $V$ and $\widetilde{V}$, respectively,\footnote{We note that the Calabi-Yau geometries analyzed in \S\ref{sec:KKLMMTex} and \S\ref{sec:manwe} originate from FRSTs of the same reflexive polytope~$\Delta^\circ$, as documented in the associated GitHub repository \github{https://github.com/AndreasSchachner/kklt_de_sitter_vacua}.}
with Hodge numbers
\begin{equation}
    h^{1,1}(X)=h^{2,1}(\widetilde{X})=150\, ,\quad h^{2,1}(X)=h^{1,1}(\widetilde{X})=8\, .
\end{equation}
All toric and geometric input required to build these compactifications is available in a dedicated public repository at \github{https://github.com/AndreasSchachner/kklt_de_sitter_vacua}. For this geometry, one can choose an orientifold involution satisfying \cite{Moritz:2023jdb} 
\begin{equation}
    h^{1,1}_-(X/\mathcal{I}) = 0\,, \qquad 
    h^{2,1}_+(X/\mathcal{I}) = 0\, .
\end{equation}
There are $35$ O7-planes wrapping divisors that are 
pure 
rigid in all toric phases, as well as $60$ O3-planes located at triple intersections of toric divisors. In addition, one O7-plane wraps a non-rigid divisor. Every O7-plane supported on a rigid divisor gives rise to an $\mathcal{N}=1$ pure Yang-Mills sector with gauge algebra $\mathfrak{so}(8)$. The total D3-brane charge sourced by the orientifold planes and D-branes, entering the tadpole condition \eqref{eq:D3Tadpole}, is given by
\begin{equation}
    Q_{\text{O}} = \frac{\chi_f}{2}
    = h^{1,1}(X) + h^{2,1}(X) + 2
    = 160\,.
\end{equation}
This value sets the flux budget available for complex structure moduli stabilization and constrains the range of admissible $3$-form fluxes. The mirror Calabi-Yau hypersurface $\widetilde{X}$ has a Mori cone $\mathcal{M}_{\widetilde{X}}$ that includes as one of its generators a conifold curve $\mathcal{C}_{\text{cf}}$ with $n_{\text{cf}} = 2$. In this example, and in those that follow, we choose a basis of $H_{2}(\widetilde{X},\mathbb{Z})$ in which the conifold class is represented in the simple form  
\begin{equation}\label{eq:coni_basis}
    \mathcal{C}_{\text{cf}} = (1,0,\ldots,0)\, .
\end{equation}

\subsubsection*{Complex structure moduli stabilization}

We have now constructed a Calabi–Yau orientifold $X/\mathcal{I}$ whose mirror admits a conifold curve $\mathcal{C}_{\text{cf}} \subset \widetilde{X}$. Next, we choose flux vectors $\vec{f}, \vec{h} \in H^{3}(X,\mathbb{Z})$ so that Gauss’ law \eqref{eq:D3tadpole_compact} requires the presence of a single anti-D3-brane. The stabilization of the complex structure moduli and the axio-dilaton by these fluxes follows the procedure described in \S\ref{sec:csms}: starting from the effective theory for the perturbatively flat vacuum as defined in \S\ref{sec:conipfv}, we search for a supersymmetric minimum with $\tau = \langle \tau \rangle_{\mathrm{PFV}}$ and $z^\alpha = p^\alpha \langle \tau \rangle_{\mathrm{PFV}}$, and then refine this solution via numerical root-finding. When successful, this yields vacuum expectation values \eqref{eq:promotedvevs} for the complex structure moduli and axio-dilaton at which their $F$-terms vanish. If $W_0$, $g_s$, and $\langle z_{\mathrm{cf}}\rangle_{\mathrm{F}}$ are small, the resulting solution lies at weak coupling and is exponentially close to the conifold point.

Working in the basis introduced in \eqref{eq:coni_basis}, the second Chern class of $\widetilde{X}$ takes the form
\begin{equation}
     c_2(\widetilde{X}) = 
     \begin{pmatrix}  
        -2& -184& -112&  -10&  -10&  -26&   -2&    6
     \end{pmatrix} ^\top\, .
\end{equation}
We choose $3$-form fluxes in $H^{3}(X,\mathbb{Z})$ that realize a conifold PFV in the sense described in \S\ref{sec:conipfv}. The following choice
\begin{align}
    \vec{M} &= \begin{pmatrix} 16& 10& -26& 8& 32& 30& 18& 28  \end{pmatrix}^\top\, , \\[0.5em]
    \vec{K}&=\begin{pmatrix}-6& -1& 0& 1& -3& 2& 0& -1\end{pmatrix}^\top\,,
\end{align} 
indeed furnishes a conifold PFV, with 
\begin{equation}
    \vec{p} = \frac{1}{40}\begin{pmatrix}0& -8&  0& -2&  4&  5&  5&  4\end{pmatrix}^\top\,.   
\end{equation}
The conifold flux integers introduced in \eqref{eq:coniMdef} and \eqref{eq:Kprime} are
\begin{equation}
    M = 16\,  \quad  \text{and }\quad  K' = \frac{26}{5}\,.
\end{equation}
Moreover, one verifies that the tadpole contribution from fluxes is
\begin{equation}
    -\vec{M}\cdot\vec{K} = 162 = 4+h^{1,1}(X)+h^{2,1}(X)\,,
\end{equation}
and therefore one must introduce a single anti-D3-brane in order to satisfy Gauss's 
law \eqref{eq:D3tadpole_compact}.

The dominant contributions to the bulk flux superpotential along the PFV locus take the form
\begin{equation}\label{eq:manwew}
    W_{\text{PFV}} = \frac{1}{\sqrt{8\pi^5}}\biggl(14\,\mathrm{e}^{2\pi \I  \tau \cdot \frac{1}{40}} - 80\,\mathrm{e}^{2\pi \I \tau \cdot \frac{2}{40}} +
    118\,\mathrm{e}^{2\pi \I \tau \cdot \frac{3}{40}} +\ldots
    \biggr)\,.
\end{equation} 
Starting from the approximate PFV minimum obtained by keeping only the first two terms in \eqref{eq:manwew}, we use this point as the initial guess for a full numerical minimization of $D_{z^a}W_{\text{flux}}$ and $D_{\tau}W_{\text{flux}}$. This procedure converges to a genuine $F$-flat solution of the complete flux superpotential \eqref{eq:GVW}, i.e.,~a point where $DW_{\text{flux}} = 0$, with
\begin{equation}
    g_s = 0.0732\, , \ z_{\text{cf}} = 1.390\times10^{-7}\, , \ W_0 = 0.0103\, , \ \text{and}\ g_sM = 1.171\,.
\end{equation}
The numerical accuracy quoted above is adequate for our purposes, though one may achieve substantially higher precision by including GV invariants to higher degree.

\subsubsection*{K\"ahler moduli stabilization}

With the complex structure sector stabilized, we next turn to the stabilization of the K\"ahler moduli. Non-perturbative contributions arise from the 152 rigid prime toric divisors of $X$, which fall into two distinct categories:
\begin{itemize}
    \item 35 of these divisors are wrapped by O7-planes supporting $\mathfrak{so}(8)$ gauge theories. Their gaugino condensates generate superpotential terms of the form \eqref{eq:wlo3}, with coefficient $c_D = 6$.
    \item The remaining 117 rigid prime toric divisors are wrapped by Euclidean D3-branes, producing non-perturbative contributions with $c_D = 1$.
\end{itemize}
One of the Euclidean D3-brane divisors intersects the conifold locus. As explained in \S\ref{sec:stabKah}, such a divisor yields a negligible non-perturbative effect, and so we exclude it from the superpotential. After this removal, we are left with $h^{1,1}(X) + 1 = 151$ non-perturbative terms. Every one of these 151 contributing divisors is pure rigid in all toric phases: none of them acquires complex structure deformations in any phase.\footnote{The dual polytope $\Delta^{\circ}$ contains two two-dimensional faces of genus one, making the computation of $h^{2,1}(\widehat{D})$ in \eqref{eq:dpurerigid} non-trivial. Using the methods of \cite{Kim:2021lwo}, we have verified explicitly that all relevant rigid divisors are indeed pure rigid across all phases.} For consistency across the entire divisor set, we adopt the Pfaffian normalization \eqref{eq:pfaffvalue} for each of the 151 surviving non-perturbative contributions.

We begin by locating a candidate AdS minimum $T_{\text{AdS}}$ for the K\"ahler moduli. For this purpose, we reiterate that we work with the full K\"ahler potential \eqref{eq:detailedform2}, which incorporates both the $\alpha'^3$ correction \eqref{eq:Valphap3} and the worldsheet instanton contributions \eqref{eq:VWSI} to the Calabi-Yau volume \eqref{eq:detailedform3}. The corresponding corrections to the holomorphic K\"ahler coordinates, given in \eqref{eq:detailedform4}, are included as well.
 
Within the torically extended K\"ahler cone as defined in Eq.~\eqref{eq:Kext}, we identify a single KKLT-type critical point. Using this point as the initial seed for a numerical minimization of the full scalar potential, we obtain one consistent AdS precursor geometry. The fully corrected string-frame volume at $T_{\text{AdS}}$ is given by\footnote{We use units with $\ell_s=1$ throughout this chapter.}
\begin{equation}
     \mathcal{V} = \mathcal{V}^{(0)} + \delta\mathcal{V}_{\alpha'^3} + \delta\mathcal{V}_{\text{WSI}} = 665.447 - 0.344  -0.578 = 664.525\,.
\end{equation}
The corresponding volume in Einstein frame is  
\begin{equation}
    \mathcal{V}_{\text{E}} = g_s^{-3/2}\mathcal{V} = 3.304 \times 10^4\,.  
\end{equation}

\begin{figure}[!t]
\centering 
\includegraphics[width=.85\linewidth]{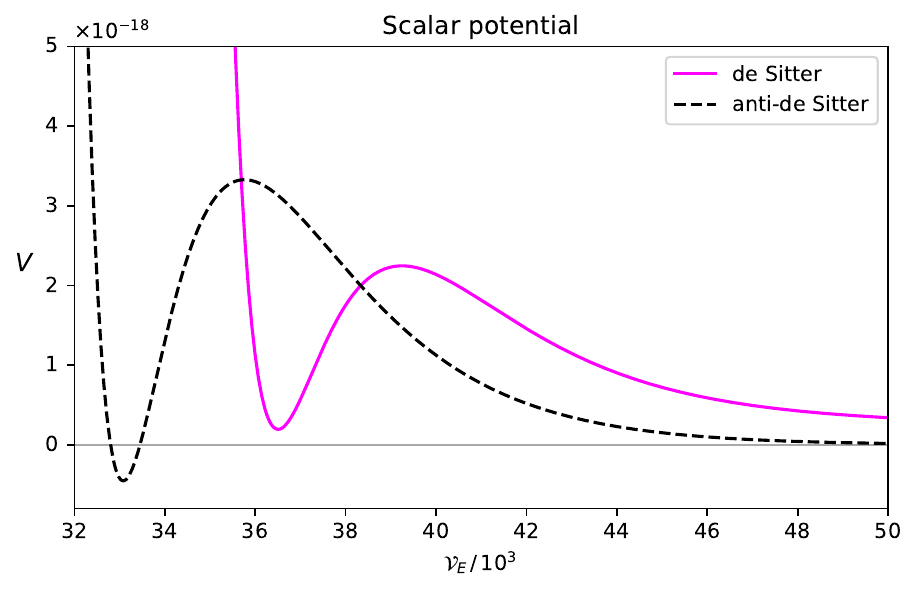}
\caption{Scalar potential for the K\"ahler moduli in the example of \S\ref{sec:manwe}, shown both prior to and after the inclusion of the uplift term.}
\label{fig:manwe_uplift}
\end{figure} 

At this point all moduli have been stabilized in a supersymmetric AdS configuration, and we may introduce supersymmetry breaking. Following the procedure outlined in \S\ref{sec:stabKah}, we uplift the complex structure sector and the K\"ahler sector in turn, iterating the two steps until the resulting minima are mutually compatible and the full system converges. Applying this method to the present example yields a non-supersymmetric local minimum with positive vacuum energy, i.e.,~a candidate de Sitter solution. Supersymmetry is broken in both the complex structure and K\"ahler sectors, and the resulting vacuum energy is
\begin{equation}
    V_{\text{dS}} = +1.937\times 10^{-19} M_{\text{pl}}^4\,.
\end{equation}
At the de Sitter point, the stabilized complex structure parameters take the values
\begin{equation}
    g_s = 0.0657\, , \ z_{\text{cf}} = 2.822\times10^{-8}\, , \ W_0 = 0.0115\, , \ \text{and}\ g_sM = 1.051\,.
\end{equation}

At the de Sitter point $T_{\text{dS}}$ in K\"ahler moduli space, the corrected string-frame volume receives contributions from the classical volume and from the $\alpha'^3$ and worldsheet instanton corrections. Evaluating these terms at the uplifted minimum gives
\begin{equation}
     \mathcal{V}  = \mathcal{V}^{(0)} + \delta\mathcal{V}_{\alpha'^3} + \delta\mathcal{V}_{\text{WSI}} = 614.834 - 0.344  -0.579 = 613.911\, .
\end{equation}
Converting to Einstein frame using the updated value of the string coupling $g_s$ yields
\begin{equation}
    \mathcal{V}_{\text{E}} = g_s^{-3/2}\mathcal{V} \approx 3.646 \times 10^4\,. 
\end{equation}
The impact of the uplift term on the K\"ahler moduli potential is visualized in Figure~\ref{fig:manwe_uplift}. From the perspective of the original single-modulus KKLT scenario \cite{Kachru:2003aw}, the appearance of a de Sitter critical point directly within the AdS potential might seem unexpected. However, in models with a large number of K\"ahler moduli, such behavior appears to be quite common. Even in simplified one-modulus toy models, a de Sitter stationary point can arise when the non-perturbative superpotential contains two competing contributions with opposite signs. The present example illustrates how this mechanism can naturally arise in the multi-modulus setting.

\begin{figure}[!t]
\centering
\includegraphics[width=0.85\linewidth]{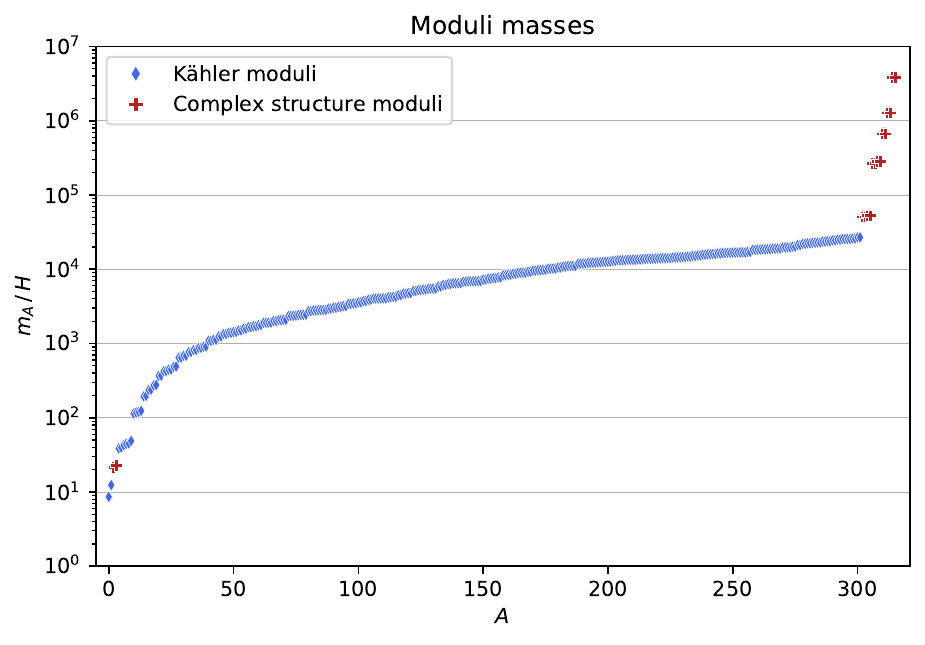}
\caption{Mass spectrum of the K\"ahler moduli, the bulk complex structure moduli, and the axio-dilaton in the de Sitter vacuum described in \S\ref{sec:manwe}. For comparison, the warped Kaluza-Klein scale \eqref{eq:mwkk} is $m_{wKK} \approx 10^{4} H_{\text{dS}}$.} 
\label{fig:manwe_masses}
\end{figure}

In this vacuum, all scalars have positive masses. 
The lightest state is a K\"ahler modulus with
\begin{equation}
    m_{\text{min}} = 8.616 H_{\text{dS}}\, ,
\end{equation} 
where the Hubble scale of the de Sitter solution is
\begin{equation}
    H_{\text{dS}} = \sqrt{\tfrac{1}{3}V_{\text{dS}}} = 2.5 \times 10^{-10} M_\text{pl}\,.
\end{equation}
The resulting mass spectra for the K\"ahler moduli, the bulk\footnote{The conifold modulus $z_{\text{cf}}$ is not displayed, as in the warped throat it is replaced by a redshifted tower of localized states with masses of order the warped Kaluza-Klein scale $m_{wKK}$ given in \eqref{eq:mwkk}.} complex structure moduli, and the axio-dilaton are shown in Figure~\ref{fig:manwe_masses}.

\subsubsection*{Control analysis}

We now assess the validity of the approximations used in constructing the solution. The first check concerns the flux superpotential, which receives an infinite series of corrections from type~IIA worldsheet instantons on the mirror $\widetilde{X}$, as encoded in \eqref{eq:Widef}. The left panel of Fig.~\ref{fig:manwe_W_rainbow} shows that these contributions decrease rapidly, confirming that the series converges well at the vacuum.

A second issue arises from the fact that, although the overall Calabi-Yau volume at $T_{\text{dS}}$ is relatively large, many individual two-cycle volumes in string frame are not.  As a result, the convergence of the worldsheet instanton expansions entering the K\"ahler potential and the holomorphic K\"ahler coordinates, cf.~\eqref{eq:detailedform3} and \eqref{eq:detailedform4}, is not guaranteed \emph{a priori} and must be checked explicitly.

To verify that the $\alpha'$ expansion remains under quantitative control, we performed a broad sampling of potent rays, that is, rays that support infinitely many non-vanishing GV invariants, as defined in \cite{Demirtas:2021nlu}. In total, we generated $2{,}643$ such rays contained in low-dimensional faces of the Mori cone $\mathcal{M}_X$. These rays collectively span a $144$-dimensional subcone, providing a representative probe of the instanton sector. For each potent ray $\{ n \mathcal{C} \,|\, n \in \mathbb{Z}_{+} \}$ generated by an effective curve $\mathcal{C}$ with charge vector $\mathbf{q} \in \mathcal{M}_X \cap H_{2}(X,\mathbb{Z})$, we evaluate
\begin{equation}\label{eq:xin}
    \xi_n(\mathbf{t},\mathcal{C}) \coloneqq  \bigl |\mathscr{N}_{n\mathbf{q}}\, \mathrm{e}^{-2\pi n\,  \mathbf{q}\cdot\mathbf{t}}\bigl |\, 
\end{equation}
for a wide range of degrees $n$. The results, displayed on a logarithmic scale in the right panel of Fig.~\ref{fig:manwe_W_rainbow}, show that the quantities $\xi_n$ fall off exponentially with $n$. This behavior provides strong evidence that the worldsheet instanton sector is well behaved and that the $\alpha'$ expansion remains under control in this vacuum.

The smallest volume potent curve identified in our scan, denoted $\mathcal{C}_{\text{min}}$, has
\begin{equation}
    \mathrm{Vol}_s(\mathcal{C}_{\text{min}}) \approx 0.971\,,\; \mathscr{N}_{\mathcal{C}_{\text{min}}} = 3\,,\; \mathscr{N}_{\mathcal{C}_{\text{min}}}\dfrac{\mathrm{Li}_2\bigl (\mathrm{e}^{-2\pi\mathrm{Vol}_s(\mathcal{C}_{\text{min}})}\bigl )}{(2\pi)^2} \approx 1.705\times 10^{-4} \, ,
\end{equation} 
where the final quantity characterizes the size of the correction contributed by $\mathcal{C}_{\text{min}}$ to the holomorphic K\"ahler coordinates \eqref{eq:detailedform4} (see \S\ref{sec:conv} for details).
The magnitude of this correction is very small, indicating that worldsheet instanton effects from potent rays are negligible in the vacuum under consideration. By contrast, instantons associated with curves that collapse at finite distance in moduli space can produce corrections to \eqref{eq:detailedform4} of non-negligible size; however, these contributions are already included at string tree level in our expressions, as discussed in \S\ref{sec:leadingEFT}.

Next, we examine the size of additional corrections inherited from the $\mathcal{N}=2$ parent theory. These include string loop effects as well as contributions from Euclidean D($-1$)-branes and Euclidean D1-branes, which enter the correction $\delta\mathcal{K}_{\mathcal{N}=2}^{(g_s)}$ defined in \eqref{eq:kformalcorrforKahler3P1}, see \cite{Robles-Llana:2006hby,Robles-Llana:2007bbv}. Among these, the dominant term arises from Euclidean~D1-branes wrapping the smallest two-cycles. In our vacuum, this correction is suppressed relative to the corresponding worldsheet-instanton effect already included in $\delta\mathcal{K}_{\mathcal{N}=2}^{\text{tree}}$. To see this, we define for loop corrections \eqref{eq:LoopCorN2} the control parameter
\begin{equation}\label{eq:epsN2def}
    \epsilon_{g_s}^{\mathcal{N}=2} \coloneqq \max_{\textrm{curves}} \left(\frac{I_{m\neq 0}}{I_{m=0}} \right)\, ,
\end{equation}
which measures the size of the leading $\delta T^{\mathcal{N}=2,(g_s)}_i$ contribution relative to the corresponding worldsheet instanton term already included in our analysis. For this example, we find
\begin{equation}
    \epsilon_{g_s}^{\mathcal{N}=2} \approx 0.005\, ,
\end{equation}
which confirms that the omitted $\mathcal{N}=2$ corrections are safely sub-leading.

Finally, let us estimate the effects of $\mathcal{N}=1$ string loop corrections in this model. Since no general first-principles expression for $\delta T^{\mathcal{N}=1,(g_s)}_i$ is currently available for Calabi-Yau orientifolds, our aim is to bound their size using the parametrizations introduced in \S\ref{sec:N1loop} and conservative choices for the undetermined $\mathcal{O}(1)$ coefficients. These estimates are sufficient to assess whether such corrections can be treated perturbatively in the region of moduli space relevant for the vacuum. Using the model \eqref{eq:ModelForCorrectionsToTdiv} and keeping only the leading term in the expansion, we estimate the largest relative divisor self-correction by setting $k^i_{\text{self},1}=1$,
\begin{equation}\label{eq:ModelForCorrectionsToTdiv1}
   \dfrac{\delta \mathcal{T}^{\mathcal{N}=1,(g_s)}_{i,\text{divisor}}}{\mathcal{T}_{i,s}^{\text{l.o.}}}\biggl |_{n=1} =  k^{i}_{\text{self},1} \,\Biggl(\frac{c_{D_i}}{4\pi}\Biggr)^2 \cdot \frac{g_s}{(\mathcal{T}_{i,s}^{\text{l.o.}})^2}\lesssim  0.00609  \,.
\end{equation}
Potentially larger contributions arise from corrections associated with small curves. Using the estimate \eqref{eq:potloopest} and setting $k_{\mathcal{C}}=1$ yields
\begin{equation}\label{eq:potloopest1}
    \dfrac{\delta \mathcal{T}^{\mathcal{N}=1,(g_s)}_{i,\text{curve}}}{\mathcal{T}_{i,s}^{\text{l.o.}}} = \frac{ k_{\mathcal{C}}\,g_s}{\mathcal{T}_{i,s}^{\text{l.o.}}\, \vol_s{\mathcal{C}}} \approx 0.0511\, .
\end{equation}
We therefore conclude that $\mathcal{N}=1$ loop corrections do not seem to jeopardize the validity of the effective field theory description in this compactification.

To summarize, within the effective theory specified by the approximations of \S\ref{sec:leadingEFT} and \S\ref{sec:KS}, we have constructed an explicit and fully controlled de Sitter vacuum in a Calabi-Yau orientifold compactification with background fluxes.

\section{A geometry with many candidate de Sitter vacua}\label{sec:aule}

In the example discussed in \S\ref{sec:manwe}, each geometry admitted only a single critical point.\footnote{More precisely, the computational procedure described in Chapter~\ref{ch:CompComp} located only a single critical point of the scalar potential. We do not, however, have a formal proof that no further minima exist in this geometry for the chosen set of parameters.} 
We now present an example that exhibits a much richer structure, with a single specification of Pfaffians and fluxes leading to several distinct AdS precursors. As we will see, the large number of AdS precursors found in this geometry leads to a substantial landscape of candidate de Sitter vacua.

\medskip

We consider the mirror pair $(X,\widetilde{X})$ characterized by
\begin{equation}
    h^{1,1}(X)=h^{2,1}(\widetilde{X})=93\, ,\quad h^{2,1}(X)=h^{1,1}(\widetilde{X})=5\, ,
\end{equation}
for which there exists a sign-flip orientifold with $h^{1,1}_-(X/\mathcal{I})=h^{2,1}_+(X/\mathcal{I})=0$ \cite{Moritz:2023jdb}. Of the $h^{1,1}(X)+4=97$ prime toric divisors, $h^{1,1}(X)+3=96$ are rigid. As in the previous examples, all relevant data needed to construct this geometry are available on GitHub \github{https://github.com/AndreasSchachner/kklt_de_sitter_vacua}.

\subsubsection*{Complex structure moduli stabilization}

One can identify a Calabi-Yau threefold $\widetilde{X}$ whose Mori cone includes, as one of its generators, a conifold curve characterized by $n_{\text{cf}} = 2$.  
In this case, a conifold PFV is generated by the following set of vectors
\begin{align}\label{eq:flux_aule}
    \vec{M} &= \begin{pmatrix}  20&   4&   8& -18& -20  \end{pmatrix}^\top\; , \quad \vec{K} =\begin{pmatrix}-5& -1&  0&  1& -1\end{pmatrix}^\top\,,
\end{align}
which lead to
\begin{equation}
    \vec{p} = \begin{pmatrix}0& 1 & 2 & 1 &0\end{pmatrix}^\top\,\times \frac{1}{48}\, ,
\end{equation}
with 
\begin{equation}
    M = 20\, \quad \ \text{and }\ \quad K' = \frac{17}{5}\,.
\end{equation}
Moreover, the induced D3-brane charge from fluxes,
\begin{equation}
    -\vec{M}\cdot\vec{K} = 102 = 4+h^{1,1}(X)+h^{2,1}(X)\,,
\end{equation}
allows for a single anti-D3-brane, which becomes necessary to fulfill Gauss's
law \eqref{eq:D3tadpole_compact}. 

Given this choice of flux quanta, the bulk flux superpotential restricted to the PFV locus takes the form
\begin{equation}\label{eq:aule_w} 
    W_{\text{PFV}} = \frac{1}{\sqrt{8\pi^5}}\biggl(108\,\mathrm{e}^{2\pi \I \tau \cdot \frac{1}{48}} - 1120\,\mathrm{e}^{2\pi \I\tau \cdot \frac{2}{48}} +
    60\,\mathrm{e}^{2\pi \I\tau \cdot \frac{3}{48}} +\ldots
    \biggr)\,.
\end{equation} 
We use the truncated expression \eqref{eq:aule_w} as an initial approximation for solving the $F$-term equations in the space of bulk complex structure moduli.
Starting from this controlled expansion, we obtain a supersymmetric solution characterized by
\begin{equation}
    g_s = 0.0410\, , \ z_{\text{cf}} = 2.369\times10^{-6}\, , \ W_0 = 0.0525\, , \ \text{and}\ g_sM = 0.821\,.
\end{equation}
The small value of the conifold modulus $z_{\text{cf}}$ confirms that the vacuum lies deep in the highly warped region of moduli space, while the moderate value of $g_s$ ensures perturbative control.

\subsubsection*{K\"ahler moduli stabilization}

In this model, we identify a total of $95=h^{1,1}+2$ rigid divisors that remain disjoint from the conifold locus. These divisors are in fact pure rigid across all FRSTs --- recall \S\ref{sec:pure_rigid}. In comparison to the geometry studied in \S\ref{sec:manwe}, the present model exhibits a far greater multiplicity of supersymmetric solutions. In the earlier case, we only obtained a single KKLT point, and therefore only one AdS precursor suitable for subsequent uplifting. In the present example, there is a much richer vacuum landscape: in \cite{McAllister:2024lnt} we found \emph{36 distinct KKLT critical points} of the K\"ahler moduli potential.  The root cause of this proliferation of critical points is that there are $h^{1,1}+2$ contributing rigid divisors, allowing a more intricate structure in the non-perturbative superpotential, whereas in \S\ref{sec:manwe} there were only $151=h^{1,1}+1$ such divisors.
After imposing our control criteria, \emph{29} of the critical points persist as viable AdS precursors. 
This example illustrates how the detailed topology of the Calabi-Yau can dramatically enhance the vacuum multiplicity even within a fixed choice of flux quanta.

\begin{figure}[!t]
\centering
\includegraphics[width=0.85\linewidth]{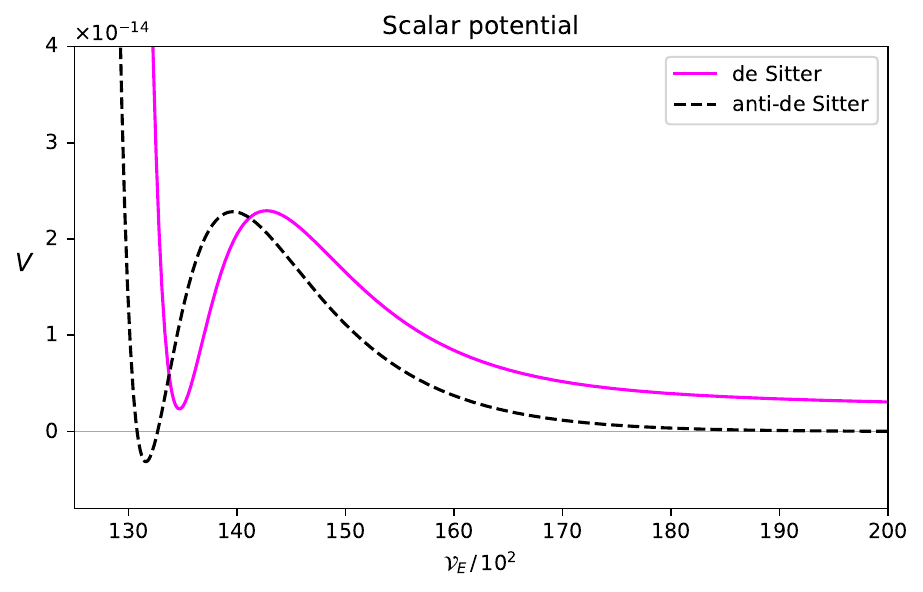}
\caption{Scalar potential for the K\"ahler moduli, shown both before and after the uplift, for the example discussed in \S\ref{sec:aule}.}\label{fig:aule_uplift}
\end{figure}

\begin{figure}[!t]
\centering
\includegraphics[width=0.85\linewidth]{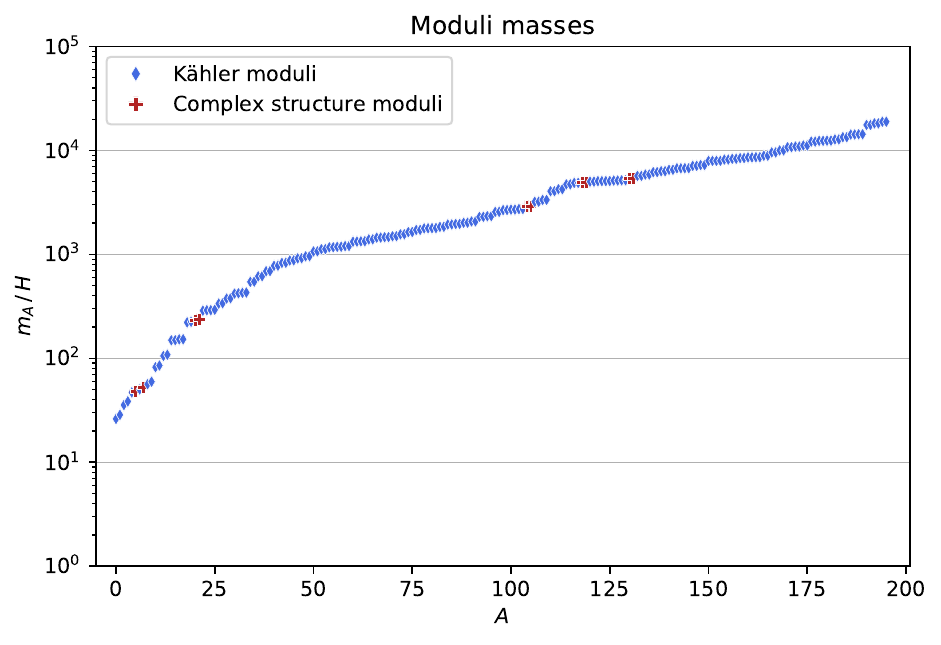}
\caption{The mass spectrum of the K\"ahler moduli, the bulk complex structure moduli, and the axio-dilaton in the de Sitter vacuum constructed in \S\ref{sec:aule}.}\label{fig:aule_masses}
\end{figure}

For concreteness, we concentrate on a particular AdS precursor drawn from the broader ensemble of solutions discussed above. This vacuum sits at a fully corrected Einstein-frame volume of
\begin{equation}
    \mathcal{V}_E = 1.310 \times 10^4\,.   
\end{equation}
Once supersymmetry breaking from the anti-D3-brane is included, the complex structure sector adjusts slightly and settles into a non-supersymmetric minimum with
\begin{equation}
    g_s = 0.0404\, , \ z_{\text{cf}} = 1.965\times10^{-6}\, , \ W_0 = 0.0539\, , \ \text{and } \ g_sM = 0.808\,.
\end{equation}
A corresponding non-supersymmetric 
minimum at
$T_{\text{dS}}$ is also obtained for the K\"ahler moduli. The vacuum energy for this metastable de Sitter vacuum is given by
\begin{equation}
    V_{\text{dS}} = +2.341\times10^{-15}\, M_{\text{pl}}^4\,.
\end{equation}
At this de Sitter point, the fully corrected Einstein-frame volume shifts only mildly, taking the value
\begin{equation}
    \mathcal{V}_E = 1.340 \times 10^4\,. 
\end{equation}
The modification of the K\"ahler moduli potential induced by the uplift is illustrated in Figure~\ref{fig:aule_uplift}. The resulting de Sitter vacuum is perturbatively stable: no tachyons are present, and the lightest scalar mode satisfies
\begin{equation}
    m_{\text{min}} = 26.157\, H_{\text{dS}}\,,
\end{equation}
comfortably above the Hubble scale. The complete mass spectra for both the K\"ahler and complex structure sectors are displayed in Figure~\ref{fig:aule_masses}.

\begin{figure}[!t]
\centering
\includegraphics[width=0.85\linewidth]{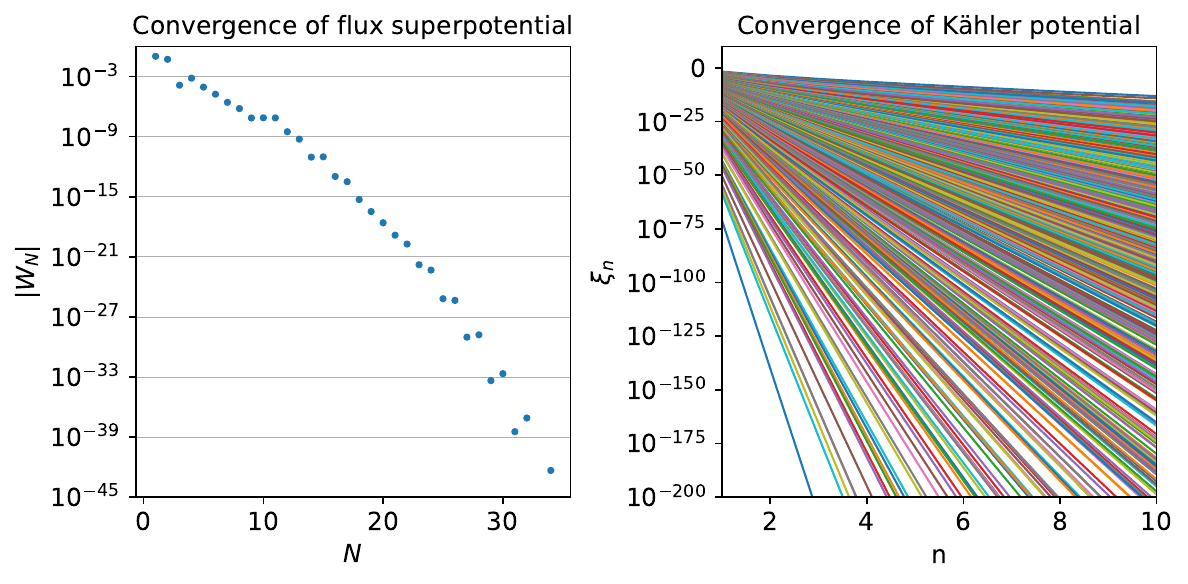}
\caption{
Convergence tests for the de Sitter example presented in \S\ref{sec:aule}.  
\emph{Left:} Convergence behavior of the bulk complex structure superpotential \eqref{eq:Widef} evaluated at the de Sitter vacuum.  
\emph{Right:} Convergence of the worldsheet instanton expansion associated with 1{,}201 potent rays spanning a $93$-dimensional subcone of $\mathcal{M}_X$. 
}\label{fig:aule_rainbow}
\end{figure}

\subsubsection*{Control analysis}

Finally, we turn to the question of control over the non-perturbative corrections by studying the convergence behavior of the worldsheet instanton expansions on both $X$ and its mirror $\widetilde{X}$. On the mirror side, the individual instanton contributions entering the flux superpotential \eqref{eq:Widef} are shown in the left panel of Fig.~\ref{fig:aule_rainbow}; these terms decrease rapidly, providing a clear indication that the PFV expansion remains well within its domain of validity in the vacuum of interest. For the K\"ahler moduli of $X$, we perform a detailed analysis of the worldsheet instanton contributions arising from effective curves with small volumes. Our scan identifies a set of 1{,}201 potent rays whose charge vectors span a rank-$93$ lattice, matching the dimension of the K\"ahler cone.  The associated instanton sums display robust convergence, as illustrated in Figure~\ref{fig:aule_rainbow}. This behavior is essential: it ensures that the non-perturbative corrections incorporated into the scalar potential are under quantitative control and that higher-order contributions are genuinely sub-leading. The smallest potent curve, denoted $\mathcal{C}_{\mathrm{min}}$, provides a measure of the accuracy of the instanton expansion. Its properties and contribution to the K\"ahler potential and coordinates are
\begin{equation}
    \mathrm{Vol}_s(\mathcal{C}_{\text{min}}) \approx 0.795\,,\;  \mathscr{N}_{\mathcal{C}_{\text{min}}} = 3\,,\; \mathscr{N}_{\mathcal{C}_{\text{min}}}\dfrac{\mathrm{Li}_2\bigl (\mathrm{e}^{-2\pi\mathrm{Vol}_s(\mathcal{C}_{\text{min}})}\bigl )}{(2\pi)^2} \approx 5.168\times 10^{-4} \, .
\end{equation}

We conclude by estimating the size of the $\mathcal{N}=1$ string loop effects in this compactification, using the parametrization introduced in \S\ref{sec:N1loop}. Adopting the model \eqref{eq:ModelForCorrectionsToTdiv} and truncating to the leading contribution, we obtain an estimate of the largest relative self-correction to a divisor by setting $k^i_{\text{self},1}=1$,
\begin{equation}\label{eq:ModelForCorrectionsToTdiv2}
   \dfrac{\delta \mathcal{T}^{\mathcal{N}=1,(g_s)}_{i,\text{divisor}}}{\mathcal{T}_{i,s}^{\text{l.o.}}}\biggl |_{n=1} =  k^{i}_{\text{self},1} \,\Biggl(\frac{c_{D_i}}{4\pi}\Biggr)^2 \cdot \frac{g_s}{(\mathcal{T}_{i,s}^{\text{l.o.}})^2}\lesssim  0.0147  \,.
\end{equation}
Using the estimate \eqref{eq:potloopest} and again choosing $k_{\mathcal{C}}=1$, we find for corrections associated with small curves
\begin{equation}\label{eq:potloopest2}
    \dfrac{\delta \mathcal{T}^{\mathcal{N}=1,(g_s)}_{i,\text{curve}}}{\mathcal{T}_{i,s}^{\text{l.o.}}} =  \frac{k_{\mathcal{C}}\,g_s}{\mathcal{T}_{i,s}^{\text{l.o.}}\, \vol_s{\mathcal{C}}} \approx 0.0543\, .
\end{equation}
Both estimates are comfortably below unity, indicating that the $\mathcal{N}=1$ loop corrections do not seem to compromise the validity of the four-dimensional effective field theory.

\begin{figure}[!t]
\centering
\includegraphics[width=0.8\linewidth]{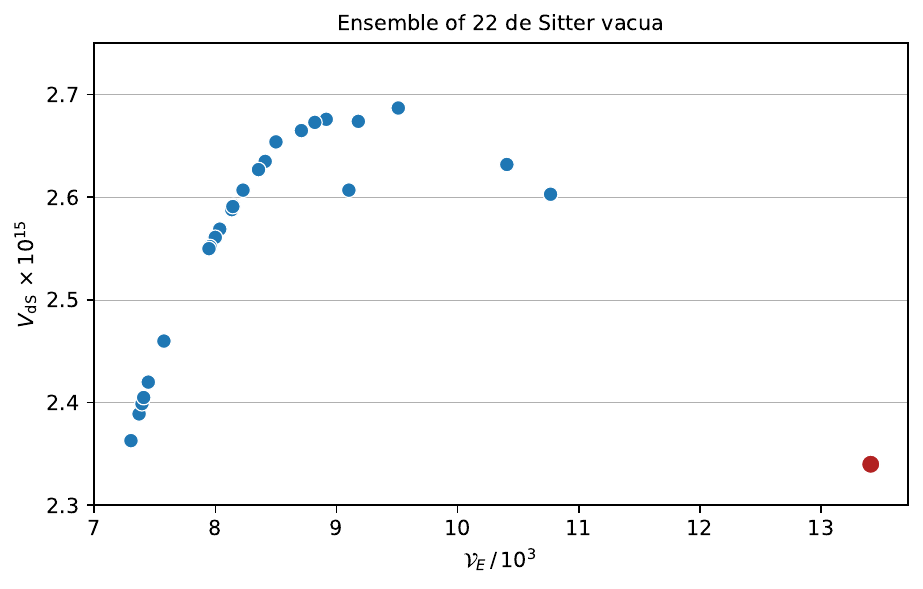}
\caption{
Twenty-one additional de Sitter vacua (blue points) identified in the same moduli space and arising from the same flux configuration \eqref{eq:flux_aule} as the benchmark solution presented in \S\ref{sec:aule} (red point). 
} \label{fig:aule_others}
\end{figure}

\subsubsection*{A landscape of candidate de Sitter solutions from multiple AdS precursors}

Up to this point, our discussion has centered on the de Sitter vacuum obtained by uplifting a particular AdS precursor.  However, this example represents only one corner of a much larger landscape accessible within the same flux choice.  As noted earlier, there are 28 additional AdS precursors that satisfy our control criteria and therefore merit analysis. When each of these precursors is uplifted in the same manner, the resulting scalar potentials generically develop non-supersymmetric de Sitter critical points. In total, we identify \emph{21 further de Sitter vacua} emerging from these alternative starting points.  The distribution of these additional solutions, including their positions in moduli space and their vacuum energies, is displayed in Figure~\ref{fig:aule_others}. The appearance of many controlled de Sitter vacua highlights a simple but important point: even when the orientifold background and the fluxes are held fixed, the interplay between the K\"ahler moduli and the instantons can produce a surprisingly large number of metastable solutions.

\chapter{Future Compactifications} \label{chap:future}

The constructions presented in these lectures are a step toward cosmology in Calabi-Yau compactifications, but many further advances will be required to fulfill the aims of the subject.  Let us first comment on some of the most immediate steps that could be taken, and then speculate about longer-term prospects.  

\section{Corrections and extensions}\label{sec:correctionsfuture}

One conceptually straightforward but computationally expensive task is to search for more de Sitter vacua along the exact lines of \cite{McAllister:2024lnt}, but to find solutions with better control parameters.  For example, larger values of $g_s M$ than those reported here are no doubt possible in compactifications with large $Q\coloneqq h^{1,1}+h^{2,1}+2$.  Performing a search at the necessary scale requires significant improvements in speed, but no fundamental change in strategy. Moreover, supersymmetric vacua with even smaller values of $W_0$ can be found by exploiting structures in the lattice of fluxes \cite{EvenSmaller}.

A more promising but still incremental approach is to develop techniques that reveal vacua with the same structures, but in qualitatively new regimes. For instance, the vacua we have described occur for K\"ahler parameters corresponding to triangulations of polytopes, as laid out by Batyrev.  More general Calabi-Yau hypersurfaces in toric varieties are possible \cite{Berglund:2016yqo,MacFadden:2025ssx}, and support new vacua \cite{dSWIP} --- potentially including cases that allow much larger $g_s M$ for a given $Q$.

More ambitiously, one could try to replace the supersymmetry-breaking mechanism. The Klebanov-Strassler gauge theory at weak 't Hooft coupling might have a metastable supersymmetry-breaking state.  Working with such a state would bypass the difficulties we have faced from the
fact that a supergravity description of the supersymmetry-breaking state of \cite{Kachru:2002gs} requires achieving weak curvatures in the warped region.
One could also seek to employ more general supersymmetry-breaking mechanisms in gauge theories arising in Calabi-Yau cones. 

Going further, one could try to achieve metastable supersymmetry breaking at an exponentially low scale neither from a warped throat in supergravity, nor from a dual gauge theory with a dynamically-generated scale, but instead from a competition among instanton effects in the bulk string theory.  An intricate racetrack of worldsheet instantons is one possibility \cite{deSitterpaper}. Another possibility is a winding uplift, as in \cite{Hebecker:2020ejb}. In either case, the underlying structure would involve $F$-term breaking in the spirit of \cite{Saltman:2004sn}.

Advances in computing string loop corrections and $\alpha'$ corrections in Calabi-Yau flux compactifications would at a minimum be very informative, and could conceivably transform the problem.  Even limited information about such corrections might suffice to invalidate the particular examples given here, and to point the way to more secure parameter regimes.  More optimistically, if we were equipped with detailed knowledge of corrections we could cast aside our caution and explore a much wider range of vacua, rather than surviving on the margins of moduli space. As we noted in \S\ref{sec:sft}, advances in string field theory \cite{Sen:2014pia,Cho:2018nfn,Cho:2023mhw,Kim:2024dnw,Sen:2024nfd,Mazel:2025fxj,Frenkel:2025wko} may open a path to computing the necessary corrections. In parallel, Schwinger-type techniques in M-theory \cite{Bachas:1999um} offer a complementary route to deriving $\alpha'$-corrected D-brane actions \cite{Compagnin:2026V1}, and one can then analyze the impact of such corrections
on the stability of the KPV uplift \cite{Compagnin:2026V2}, as discussed in \S\ref{sec:antiD3corr}.

\section{More general constructions of de Sitter vacua}\label{sec:alternatives}

Let us next step back from the specific constructions presented in these lectures, and consider the broader context of string compactifications. 
The purpose of this section is not to give a comprehensive survey of all known approaches to de Sitter model-building in string theory. Rather, our goal is to highlight several complementary strategies and to convey a qualitative picture that has emerged from recent work.

First of all, in these lectures we have restricted our attention to type IIB flux compactifications on Calabi-Yau orientifolds because of the limitations of our own expertise.  However, string theory and M-theory admit a far broader and richer landscape of solutions.  Particularly promising approaches to constructing candidate de Sitter vacua in other regimes include flux compactifications of M-theory on hyperbolic manifolds \cite{DeLuca:2021pej},
as well as orientifold compactifications of type IIA string theory 
\cite{Cordova:2018dbb,Cordova:2019cvf}.\footnote{See also the recent works \cite{ValeixoBento:2025yhz,ValeixoBento:2025qih} on flux compactifications of M-theory on non-supersymmetric Riemann-flat manifolds.}  

Returning to type IIB flux compactifications on Calabi-Yau orientifolds,
much of the literature over the past twenty years has focused on two benchmark scenarios: the KKLT construction \cite{Kachru:2003aw} and the Large Volume Scenario (LVS) \cite{Balasubramanian:2005zx}\index{LVS}. Both provide internally consistent mechanisms for stabilizing moduli and, in principle, for obtaining de Sitter vacua. However, the examples studied in these lectures illustrate that full-fledged stabilized compactifications can deviate in important ways from such simplified models. 

In particular, the candidate de Sitter vacua presented in Chapter \ref{chap:deSitter} differ from the single-modulus picture laid out in \cite{Kachru:2003aw}. 
First, the PFV mechanism of \cite{Demirtas:2019sip} yields structures in the complex structure and axio-dilaton sector that were not anticipated in \cite{Kachru:2003aw}, and leads to a correlation between the string coupling $g_s$ and the flux superpotential $W_0$.  
In turn, this correlation causes worldsheet instanton corrections to the K\"ahler coordinates to be significant --- though, as we have explained, 
these contributions are computable \cite{Demirtas:2023als}, and 
we have shown by explicit computation that their effects are under control: see Fig.~\ref{fig:manwe_W_rainbow}. 
Moreover, the presence of many K\"ahler moduli and of numerous non-perturbative contributions to the superpotential leads to scalar potentials whose structure is far more complex than that of the minimal KKLT toy model.  Indeed, as we have emphasized repeatedly, a single choice of flux quanta can give rise to a large number of distinct candidate de Sitter vacua. The explicit solutions discussed in \S\ref{sec:aule} provide concrete evidence for this multiplicity, echoing similar phenomena observed in other geometries in \cite{McAllister:2024lnt}.

Related behavior has also appeared in other works such as \cite{AbdusSalam:2020ywo,AbdusSalam:2025twp}, where it was found that multiple supersymmetric and non-supersymmetric minima coexist within a single effective potential, including so-called hybrid vacua that share features of different stabilization mechanisms. 
Thus, in
sufficiently complicated compactifications, one can expect a spectrum of solutions that interpolate between, or partially realize, the defining features of the original KKLT and LVS proposals.
 
A second set of alternatives concerns the mechanism by which a negative vacuum energy is lifted to positive values. 
These lectures have focused on uplift via anti-D3-branes in warped throats, 
and in \S\ref{sec:correctionsfuture} we indicated a few other possibilities that one could try to realize in compactifications like those of Chapter \ref{chap:deSitter}.
However, there is a much wider spectrum of possibilities in the literature. 
In addition to the $F$-term uplift from fluxes \cite{Saltman:2004sn} mentioned above, 
there are the possibilities of K\"ahler uplift \cite{Balasubramanian:2004uy,Westphal:2005yz,Westphal:2006tn}, or $D$-term uplifts \cite{Burgess:2003ic}, including T-brane configurations \cite{Cicoli:2015ylx}. Each of these approaches offers potential advantages, but they also come with significant challenges. In particular, controlling supersymmetry-breaking effects remains a central open problem. As explained in \S\ref{sec:antiD3corr}, accurately computing such effects 
requires a careful accounting of backreaction, loop effects, and higher-order corrections. In practice, achieving such control is often the most demanding aspect of any uplift construction.

A broader lesson is that the landscape of flux compactifications is considerably richer than suggested by early toy models. As more explicit and computationally controlled examples become available, it is increasingly clear that moduli stabilization and uplift should be viewed as problems in a high-dimensional, highly structured potential, rather than as isolated mechanisms with unique outcomes. 
The constructions presented in these lectures provide concrete evidence for this perspective. 

At the same time, the work of ensuring control and validating vacua requires a 
total commitment to computing whatever is necessary, often including the development of new technology.  The reason is simple: since the Dine-Seiberg paper \cite{Dine:1985he} forty years ago, it has been clear that finding parametrically controlled vacua will be challenging, and as explained in \S\ref{sec:hier}, in any finite region of the landscape, arbitrarily good parametric control is impossible.  This fundamental result does not imply that vacua do not exist, or that they cannot be understood: instead, it implies that  \emph{establishing the existence of a vacuum generally requires computing quantum corrections}.
Thus, to make progress one should develop the capability to compute subleading corrections to high accuracy.  In this way one can give strong evidence for or against a candidate vacuum, in terms of control parameters that are finite but small.  The works reviewed in these lectures \cite{Demirtas:2019sip, Demirtas:2020ffz, Demirtas:2021nlu, Demirtas:2021ote, Demirtas:2022hqf, Demirtas:2023als, McAllister:2024lnt} are part of a programmatic effort directed at developing and deploying such capabilities, in the   setting of type IIB flux compactifications on Calabi-Yau orientifolds.  Corresponding efforts in other corners of the duality web will illuminate a wider swath of the landscape.

\section{Applications}

Looking ahead, one would hope to study cosmological dynamics through direct computation of off-shell quantities, rather than being limited to vacua. 
A first step in this direction is to study dynamics around the vacua we have found.  In \S\ref{sec:KKLMMTex} we presented a flux compactification that plausibly admits a dynamical evolution during which the closed string moduli evolve to some extent, but remain in the basin of attraction of a vacuum, while a D3-brane moves toward an anti-D3-brane, as in \cite{Kachru:2003sx}.  Whether prolonged inflation occurs along this trajectory could be determined by computing the ten-dimensional field configuration numerically, and applying the methods of \cite{Baumann:2010sx}. Dynamical solutions that involve more general evolution of the moduli are a future target.

The methods reviewed in these lectures are rooted in the spacetime effective field theory, but string theory can be approached from other angles that provide powerful new perspectives.  A critical task for the future is to arrive at a description of de Sitter vacua of string theory in terms of the worldsheet CFT. Much of the path forward on this problem is unclear to us, but a practical approach involves on the one hand constructing de Sitter vacua of the effective theory in ever-increasing depth, and at the same time developing improved methods for computing string amplitudes in flux backgrounds. In time these two efforts may reinforce each other.

Another difficult and important task is to use a holographically dual CFT to analyze  vacua like those described here. One could aim to begin by finding a supersymmetric CFT dual to the supersymmetric AdS$_4$ vacua of \cite{Demirtas:2021nlu} reviewed in \S\ref{sec:SUSYAdS}, and then treat the de Sitter configuration as arising from spontaneous supersymmetry breaking in the cascading field theory dual to the KS throat.  Even the first step is non-trivial: as originally observed in \cite{Lust:2022lfc}, one simple proposal for the holographic dual of the vacua of \cite{Demirtas:2021nlu} fails to achieve a sufficiently large central charge.  In our view, the task is to find a more intricate D-brane configuration (along lines first laid out in \cite{Silverstein:2003jp}) from which a dual CFT can be extracted.

One application of a holographically dual description, if one were available, would be to approach the measure problem from a new direction.  Specifically, one 
could aspire to sharpen the problem by constructing landscapes of inflating solutions, and de Sitter vacua, in such an ultraviolet complete setting, and then study the resulting dynamics. 


Turning to the cosmological constant problem, we explained in \S\ref{sec:SUSYAdS} that the vacua of \cite{Demirtas:2021nlu} provide a striking example in which discrete parameters in quantum gravity determine the vacuum energy at exponentially lower energies.  Such vacua illustrate one solution to a sort of `supersymmetric cosmological constant problem'.  However, they do not yet hint at an approach to the actual cosmological constant problem in our Universe, which involves vacuum energy that is far smaller than the scale of supersymmetry breaking.  We expect that major conceptual advances will be required to resolve this question.

Our hope is that in the long run, the pursuit of cosmological solutions of string theory will shed light on the deep mysteries of cosmology.

\section*{Acknowledgments}

We thank  Federico Carta, Byron Chen, Federico Compagnin, Mike Douglas, Sebastian Vander Ploeg Fallon, Naomi Gendler,  Jim Halverson, Arthur Hebecker,  Ben Heidenreich, Shamit Kachru, Severin Lüst, Nate MacFadden, David J.E.~Marsh, Fernando Quevedo,  Tom Rudelius, Simon Schreyer, Elijah Sheridan, Michael Stepniczka, Mike Stillman, Sandip Trivedi, and Xi Yin for discussions of related topics. We are particularly grateful to Mehmet Demirtas, Manki Kim, Jakob Moritz, Richard Nally, and Andres Rios-Tascon for 
their 
collaboration in creating the works reviewed in these lectures. We thank Manki Kim, Richard Nally, and Fernando Quevedo for comments on a draft, and we thank
S.~Schreyer for sharing Figs.~\ref{fig:KSsolution}-\ref{fig:KPVpot_lo}. 
This research was supported by NSF grant PHY-2309456.

\addcontentsline{toc}{chapter}{Bibliography}
\bibliographystyle{utphys}
\bibliography{refs}

\end{document}